\documentclass[12pt]{mydiss}
\usepackage[english]{babel}
\usepackage[dvips]{graphics}
\usepackage{epsfig,psfrag,color,graphicx,here}
\usepackage{xspace,array,delarray,tabularx,calc,pifont,rotating}
\usepackage{wrapfig,units,nicefrac}
\usepackage{amscd,amsmath,amssymb,subeqn,bbm}

\input xy
\xyoption{all}

\newcommand{\clearemptydoublepage}{\newpage{\pagestyle{empty}\cleardoublepage}}

\newcommand{\HRule}{\rule{\linewidth}{0.5mm}}
\usepackage{fancyheadings}
\renewcommand{\chaptermark}[1]{\markboth{#1}{}}

\numberwithin{equation}{chapter}

\newtheorem{Def}{Definition}[chapter]
\numberwithin{Def}{chapter}

\numberwithin{Cla}{chapter}

\newtheorem{Pro}{Proposition}[chapter]
\numberwithin{Pro}{chapter}

\numberwithin{Cor}{chapter}

\numberwithin{Lem}{chapter}

\newtheorem{Thm}{Theorem}[chapter]
\numberwithin{Thm}{chapter}

\newcommand{\PreserveBackslash}[1]{\let\temp=\\#1\let\\=\tem}

{\begin{list}{}{%
\settowidth{\labelwidth}{#1}%
\setlength{\leftmargin}{\labelwidth+\labelsep}}}%
{\end{list}}

\DeclareFontFamily{U}{rsf}{}
\DeclareFontShape{U}{rsf}{m}{n}{
  <5> <6> rsfs5 <7> <8> <9> rsfs7 <10-> rsfs10}{}
\DeclareMathAlphabet\Scr{U}{rsf}{m}{n}

\def\cC{{\Scr C}}
\def\cD{{\Scr D}}
 
\def\cH{{\Scr H}}

\def\cT{{\Scr T}}

\def\cV{{\Scr V}}

\def\cG{{\Scr G}}

\def\cDb{{\overline{\Scr D}}}

\newcommand{\CA}{\mathcal{A}}
\newcommand{\CB}{\mathcal{B}}
\newcommand{\CC}{\mathcal{C}}

\newcommand{\CE}{\mathcal{E}}
\newcommand{\CF}{\mathcal{F}}
\newcommand{\CG}{\mathcal{G}}
\newcommand{\CH}{\mathcal{H}}
\newcommand{\CI}{\mathcal{I}}
\newcommand{\CJ}{\mathcal{J}}
\newcommand{\CK}{\mathcal{K}}
\newcommand{\CL}{\mathcal{L}}
\newcommand{\CM}{\mathcal{M}}
\newcommand{\CN}{\mathcal{N}}
\newcommand{\CO}{\mathcal{O}}
\newcommand{\CP}{\mathcal{P}}

\newcommand{\CS}{\mathcal{S}}
\newcommand{\CT}{\mathcal{T}}
\newcommand{\CU}{\mathcal{U}}
\newcommand{\CV}{\mathcal{V}}
\newcommand{\CW}{\mathcal{W}}
\newcommand{\CX}{\mathcal{X}}
\newcommand{\CY}{\mathcal{Y}}
\newcommand{\CZ}{\mathcal{Z}}

\newcommand{\bCO}{{\overline{\mathcal O}}}

\newcommand{\hCU}{\hat{\mathcal{U}}}
\newcommand{\hCV}{\hat{\mathcal{V}}}
\newcommand{\hCW}{\hat{\mathcal{W}}}

\newcommand{\IR}{\mathbbm{R}}
\newcommand{\IC}{\mathbbm{C}}

\newcommand{\IF}{\mathbbm{F}}
\newcommand{\IT}{\mathbbm{T}}
\newcommand{\IL}{\mathbbm{L}}
\newcommand{\IM}{\mathbbm{M}}
\newcommand{\IN}{\mathbbm{N}}
\newcommand{\IP}{\mathbbm{P}}
\newcommand{\IZ}{\mathbbm{Z}}

\newcommand{\fA}{\mathfrak{A}} 
\newcommand{\fC}{\mathfrak{C}} 
 
\newcommand{\fH}{\mathfrak{H}} 

\newcommand{\fJ}{\mathfrak{J}}
\newcommand{\fP}{\mathfrak{P}} 
\newcommand{\fS}{\mathfrak{S}} 
\newcommand{\fU}{\mathfrak{U}} 
\newcommand{\fV}{\mathfrak{V}}
\newcommand{\fW}{\mathfrak{W}}

\newcommand{\hfU}{\hat{\mathfrak{U}}} 
\newcommand{\hfV}{\hat{\mathfrak{V}}} 
\newcommand{\hfW}{\hat{\mathfrak{W}}} 

\newcommand{\dpar}{\partial}
\newcommand{\pbar}{{\bar{\partial}}}
\newcommand{\e}{{\rm e}}

\renewcommand{\i}{\mathrm{i}}
\newcommand{\bz}{{\bar{z}}}
\newcommand{\bw}{{\bar{w}}}
\newcommand{\bE}{{\bar{E}}}
\newcommand{\bW}{{\bar{W}}}
\newcommand{\bV}{{\bar{V}}}
\newcommand{\hl}{{\hat{\lambda}}}
\newcommand{\bl}{{\bar{\lambda}}}
\newcommand{\T}{\Theta}
\newcommand{\bT}{{\bar{\Theta}}}
\renewcommand{\l}{{\lambda}}
\renewcommand{\a}{{\alpha}}
\newcommand{\ad}{{\dot{\alpha}}}
\renewcommand{\o}{{\omega}}
\renewcommand{\O}{{\Omega}}
\renewcommand{\b}{{\beta}}
\newcommand{\bd}{{\dot\beta}}
\newcommand{\g}{{\gamma}}

\newcommand{\gd}{{\dot\gamma}}
\renewcommand{\d}{{\delta}}
\newcommand{\dd}{{\dot\delta}}
\newcommand{\ed}{{\dot\epsilon}}

\newcommand{\tN}{{\widetilde N}}
\newcommand{\tP}{{\widetilde P}}
\newcommand{\tQ}{{\widetilde Q}}
\newcommand{\tK}{{\widetilde K}}
\newcommand{\tJ}{{\widetilde J}}
\newcommand{\tD}{{\widetilde D}}
\newcommand{\tA}{{\widetilde A}}

\newcommand{\bnab}{{\bar\nabla}}
\newcommand{\te}{\theta}

\newcommand{\Ad}{{\dot A}}
\newcommand{\Bd}{{\dot B}}
\newcommand{\Cd}{{\dot C}}

\newcommand{\Ac}{\overset{\circ}{\CA}}
\newcommand{\Bc}{\overset{\circ}{\CB}}
\newcommand{\Wc}{\overset{\circ}{W}}
\newcommand{\cc}{\overset{\circ}{\chi}}
\newcommand{\pc}{\overset{\circ}{\phi}}
\newcommand{\psc}{\overset{\circ}{\psi}}
\newcommand{\Pc}{\overset{\circ}{\Phi}}
\newcommand{\Gc}{\overset{\circ}{G}}
\newcommand{\fc}{\overset{\circ}{f}}
\newcommand{\cnab}{\overset{\circ}{\nabla}}

\newcommand{\tr}{{\rm tr}}
\newcommand{\dt}{{\rm d}}

\newcommand{\der}[1]{\frac{\partial}{\partial #1}}
\newcommand{\derr}[2]{\frac{\partial#1}{\partial #2}}

\def\contra#1#2{\,{\buildrel \,\hbox{$
    \vrule height 3pt width 1pt depth 0pt
    \vrule height 3pt width #1pt depth -2pt
    \vrule height 3pt width 1pt depth 0pt $}
    \over {#2} }\,}

\newcommand{\eqn}[2]{\begin{equation}\label{#1}#2\end{equation}}
\newcommand{\eqna}[2]{\begin{equation}
                      \label{#1}
                       \begin{aligned}
                          #2
                        \end{aligned}
                        \end{equation}
                     }

\newcommand{\eqarr}[2]{\begin{eqnarray}\label{#1}#2\end{eqnarray}}

\begin{document}


\thispagestyle{empty}

\parbox{15cm}{
   \vspace*{-.8cm}
   \begin{center}
 
    \HRule\\[.2cm]
    \sffamily\Large\bf
       On \\
       Supertwistor Geometry \\ 
       And Integrability In Super Gauge Theory\\[-.2cm]
    \HRule\\[2cm]

\large\sc  
       Von der Fakult\"at f\"ur Mathematik und Physik\\ der
       Gottfried Wilhelm Leibniz
       Universit\"at Hannover\\
       Zur Erlangung des Grades\\
       Doktor der Naturwissenschaften\\ 
       Dr. rer. nat.\\
       genehmigte Dissertation\\ [1.5cm]
    \end{center}
}

\parbox{15cm}{
    \begin{center}
     \large  von\\[0.05in]
       Dipl.-Phys. Martin Wolf\\[0.05in]
       geboren am 19. Februar 1979 in Zwickau (Sachsen)\\[3cm]
       Hannover 2006
    \end{center}
}

\vfill
\begin{center}
\begin{figure}[H]
\includegraphics[width=10cm]{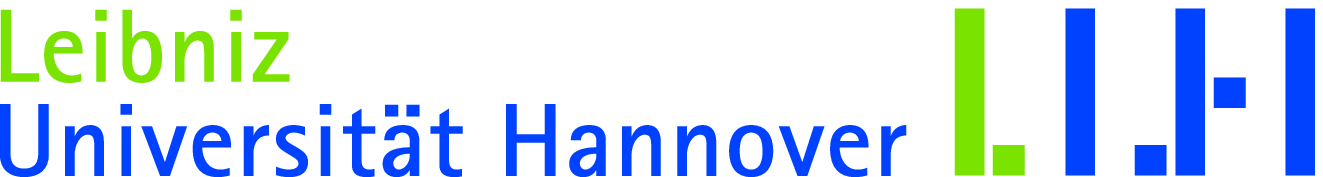}
\end{figure}
\end{center}

\clearpage


\thispagestyle{empty}

{$ $}
\vfill
\noindent
\hspace{1pt} Date: July 21$^{\rm st}$ 2006\\[.3in]
\noindent
\let\PBS=\PreserveBackslash
\setlength{\extrarowheight}{4pt}
 \begin{tabular}{ >{\PBS\raggedright\hspace{0pt}}p{1in} 
  	  >{}p{4in}l }
   Supervisors: & Prof. Dr. Olaf Lechtenfeld and Dr. Alexander D. Popov 
                   \\
   1$^{\rm st}$ referee: & Prof. Dr. Olaf Lechtenfeld 
                    \\
   2$^{\rm nd}$ referee: & Prof. Dr. Holger Frahm 
 \end{tabular}

\clearpage


\voffset 15mm
\hoffset 0mm

\thispagestyle{empty}
{$ $}
\vfill
\begin{center}
Copyright \copyright\ 2006 by Martin Wolf.
All rights reserved.\\
ITP--UH--18/06\\
Institut f\"ur Theoretische Physik\\
Gottfried Wilhelm Leibniz Universit\"at Hannover\\
Appelstra{\ss}e 2, 30167 Hannover, Germany\\
{\ttfamily wolf@itp.uni-hannover.de}
\end{center}

\clearemptydoublepage


\thispagestyle{empty}

\vspace*{3in}
\begin{center}
 {\it\Large To My Family}
\end{center}

\clearemptydoublepage


\linespread{1.2}\selectfont
{\small 
\thispagestyle{empty}
\begin{center}
 {\bf Acknowledgments}
\end{center}
\noindent

This work could not have been carried out without the support of 
many people. It is therefore a pleasure to acknowledge their help.

First of all, I would like to thank Olaf Lechtenfeld for allowing 
me to be a member of his research group and for giving me the 
opportunity to do my Ph.D. studies here in Hannover. I am grateful 
to him for his help with advice and expertise, and for numerous 
helpful discussions. I am also deeply indebted to Alexander Popov 
who was always a brilliant advisor. I would like to express my 
gratitude towards him for patiently teaching me a lot of mathematics 
and physics, for our countless conversations on various topics which 
crossed the way, and for always helping with advice on any sort of 
problem. He has always encouraged my interest in mathematical physics 
since the time I came to Hannover. Furthermore, I would like to thank 
Alexander Popov and Christian S\"amann for very enjoyable 
collaborations.

I am also grateful to Holger Frahm for sparing his invaluable time 
reviewing the manuscript of my thesis, and to Elmar Schrohe for kindly 
agreeing to be the head of the examination board. In addition, I am 
thankful to the following people for their help and for numerous 
discussions: Chong-Sun Chu, Louise Dolan, Norbert Dragon, Klaus Hulek, 
Alexander Kling, Niall J. MacKay, Stefan Petersen, Ronen Plesser,
Elmar Schrohe, Christian S\"amann, Richard Szabo, Sebastian Uhlmann, 
Kirsten Vogeler and Robert Wimmer.

Furthermore, I would like to thank a number of wunderful people for 
having provided such an enjoyable time here in Hannover. Therefore, 
many thanks to all of them: Hendrik Adorf, Alexandra De Castro, Henning 
Fehrmann, Carsten Grabow, Matthias Ihl, Alexander Kling, Michael Klawunn, 
Michael K\"ohn, Marco Krohn, Carsten Luckmann, Andreas Osterloh, Klaus 
Osterloh, Guillaume Palacios, Leonardo Quevedo, Christian S\"amann, 
Sebastian Uhlmann and Carsten von Zobeltitz. In particular, I have to
thank Stefan Petersen, Kirsten Vogeler and Robert Wimmer for a wonderful
friendship. They made it an unforgetable time here in Hannover.
In addition, our countless discussions, their comments, suggestions and 
ideas were always a great source of inspiration to me. Furthermore, I am 
also very grateful to Ingmar Glauche and Cindy Rockstroh for always helping 
and supporting me. Thank you!

I want to express my deepest gratitude towards those this thesis has been
dedicated to, i.e., to my parents Bettina and Hans-Peter Wolf and to my
brother Henrik Wolf. I would not have been in the position to carry out 
this work at all without their support and encouragement.

Finally, I would like to thank the Deutsche Forschungsgemeinschaft (DFG) 
for continued support of my graduate work through the Graduiertenkolleg No. 
282 entitled {\it Quantenfeldtheoretische Methoden in der Teilchenphysik,
Gravitation, Statistischen Physik und Quantenoptik}.
}
\clearemptydoublepage


\thispagestyle{empty}
{\small 
\thispagestyle{empty}
\begin{center}
 {\bf Abstract}
\end{center}
In this thesis, we report on different aspects of integrability in
supersymmetric gauge theories. Our main tool of investigation is 
supertwistor geometry. In the first chapter, we briefly review the 
basics of twistor geometry. Afterwards, we discuss 
self-dual super Yang-Mills (SYM) theory and some of its relatives. 
In particular, a detailed twistor description of self-dual 
SYM theory is presented. Furthermore, we introduce certain
self-dual models which are, in fact, obtainable from self-dual SYM theory
by a suitable reduction. Some of them can be interpreted within the
context of topological field theories.
To provide a twistor description of these models,
we propose weighted projective superspaces as twistor spaces.
These spaces turn out to be Calabi-Yau supermanifolds. Therefore,
it is possible to write down appropriate action principles, as well. In chapter
three, we then deal with the twistor formulation of a certain supersymmetric
Bogomolny model in three space-time dimensions. The nonsupersymmetric
version of this model describes static Yang-Mills-Higgs monopoles in the 
Prasad-Sommerfield limit. In particular, we consider a supersymmetric
extension of mini-twistor space. This space is in turn a part of 
a certain double fibration. It is then possible to formulate
a Chern-Simons type theory on the correspondence space of this
fibration. As we explain, this theory describes partially holomorphic
vector bundles. It should be noticed that the correspondence
space can be equipped with a Cauchy-Riemann structure. Moreover, we formulate 
holomorphic BF theory on mini-supertwistor space. We then prove that the
moduli spaces of all three theories are bijective. In addition,  
complex structure deformations on mini-supertwistor space are investigated
eventually resulting in a twistor correspondence
involving a supersymmetric Bogomolny model with massive fields.
In chapter four, we review the twistor formulation of non-self-dual
SYM theories. The remaining chapter is devoted to a more detailed
investigation of (classical) integrability of self-dual 
SYM theories. In particular, we explain the twistor construction of
infinite-dimensional algebras of hidden symmetries. Our discussion 
is exemplified by deriving affine extensions
of internal and space-time symmetries. Furthermore,
we construct self-dual SYM hierarchies and their truncated versions. 
These hierarchies describe an
infinite number of flows on the respective solution space. The lowest
level flows are space-time translations. The existence of such
hierarchies allows us to embed a given solution to the equations
of motion of self-dual SYM
theory into an infinite-parameter family of new solutions.
The dependence of the self-dual SYM fields 
on the additional moduli can be recovered by 
solving the equations of the hierarchy. We in addition derive
infinitely many nonlocal conservation laws.

\vfill
}
\clearemptydoublepage 


\pagenumbering{roman}
\linespread{1.3}\selectfont

\lhead[\fancyplain{}{\bfseries \thepage}]{\fancyplain{}{\bfseries Motivation and introduction}}
\rhead[\fancyplain{}{\bfseries Motivation and introduction}]{\fancyplain{}{\bfseries\thepage}}
\cfoot{}

\chapter*{Motivation and introduction}
\thispagestyle{empty}

\vspace*{-.5cm}

\centerline{\sc Twistor geometry}\vskip 5pt

In 1967, Penrose \cite{Penrose:1967wn} introduced the 
foundations of twistor geometry. The corner stone
of twistor geometry is the substitution of space-time
as background for physical processes by some new background
manifold -- called twistor space, and furthermore the reinterpretation
of physical theories on this new space.
As originally proposed, twistor space, or rather the projectivization thereof, 
which is associated with complexified four-dimensional Minkowski space, is the
complex projective space $\IC P^3$. Geometrically, this space
parametrizes all isotropic two-planes in complexified
Minkowski space. Upon this correspondence,
it is possible to reinterpret solutions to zero rest mass 
free field equations on Minkowski space 
in terms of certain cohomology groups on twistor space
\cite{Penrose:1967wn}--\cite{Penrose:1973}. For instance,
if $U$ is some open subset in 
complexified Minkowski space and if $\CZ_{\pm h}$ denotes
the sheaf of solutions to the helicity $\pm h$ zero rest mass
field equations, then there is an isomorphism between 
the two cohomology groups $H^1(U',\CO_{U'}(\mp 2h-2))$ and
$H^0(U,\CZ_{\pm h})$, where $U'$ is an appropriate region on
twistor space related to $U$. Here, $\CO_{U'}(\mp 2h-2)$ is the sheaf of sections
of a certain holomorphic line bundle (over $U'$) having 
first Chern class $\mp 2h-2$.
The map between representatives of these cohomology groups
has been termed Penrose transform. Putting it differently,
any solution to zero rest mass free field equations can be
represented by certain holomorphic ``functions" on twistor space,
which are ``free" of differential constraints. 
For a detailed discussion of this correspondence in terms
of \v Cech cohomology, we refer the reader to
\cite{Eastwood:1981jy}.
Let us mention in passing that,
as was shown by Eastwood in \cite{Eastwood:1981}, 
it is also possible to 
generalize this description to the case of massive free
fields. 

Besides giving insights into the geometric nature of solutions to
linear field equations, the ideas of twistor geometry turned out to be 
extremely powerful for studying various nonlinear classical 
field equations. 
Twistor methods have successfully been applied to subclasses of the Einstein's
equations of General Relativity and of the (super) Yang-Mills equations,
to name just the most prominent ones.
Indeed, it is possible to associate with any self-dual 
oriented Riemannian four-dimensional manifold $X$, that is, a manifold
with self-dual Weyl tensor, a 
complex three-dimensional twistor space \cite{Penrose:1976js,Atiyah:wi}.
All the information about the conformal structure of $X$ is
encoded in the complex structure of this ``curved" twistor space. Some
additional data then
allows for the construction of self-dual metrics and 
conformal structures on $X$.
Among such metrics, there are interesting subclasses as, e.g., self-dual
Einstein metrics, K\"ahler metrics with vanishing scalar curvature
and Ricci-flat self-dual metrics. For constructions, see
Refs. \cite{Ward:1978,Tod:1979tt,Ward:1980,Penrose:1980,Hitchin:1982,Hitchin:1984,Jones:1985,
Hitchin:1986ea,Woodhouse:1988,Fletcher:1990,LeBrun:1991,Tod:1994,Hitchin:1995,Mason:2001nu,LeBrun:2005qf}, 
for instance. Furthermore, the question when such a 
twistor space is in addition K\"ahler was answered by Hitchin
\cite{Hitchin:1981}. He showed that given a compact oriented self-dual
Riemannian four-dimensional manifold $X$, the corresponding twistor
space admits a K\"ahler structure if and only if $X$ is 
conformally equivalent to either $S^4$ or $\IC P^2$. Besides
questions related to Einstein's equations, twistor theory has given 
striking advances in our understanding of the properties of gauge theories.
In an important paper \cite{Ward:1977ta}, Ward has proven a one-to-one correspondence
between gauge equivalence classes of self-dual Yang-Mills fields on complexified 
four-dimensional space-time and equivalence classes of holomorphic vector 
bundles over twistor space which obey certain triviality conditions. 
Upon imposing additional conditions on the bundles over twistor space,
one may also reduce 
the structure group of the bundles on space-time to some real form
of the general linear group and in addition, the discussion
to, for instance, an Euclidean setting.
Furthermore, replacing twistor space by ambitwistor space -- the space
of complex light rays (null lines) in complexified four-dimensional Minkowski 
space -- one can 
discuss general Yang-Mills fields, as well. 
In fact, Isenberg et al. 
\cite{Isenberg:1978kk}
and Witten \cite{Witten:1978xx} showed that it is possible to 
represent gauge equivalence classes of solutions to the full
Yang-Mills equations in terms of equivalence classes of certain
holomorphic vector bundles
on formal neighborhoods of ambitwistor space. Khenkin et al.
\cite{Khenkin:1980ff} discussed a generalization thereof by giving
an interpretation in terms of certain cohomology groups. A detailed
discussion of the cohomology interpretation of solutions to
the full Yang-Mills equations can be found in \cite{Buchdahl:1985,Pool:1987}.
Of course, the
discussion can also be reduced to real Yang-Mills fields and 
to, e.g., Minkowski space-time.
A map between gauge equivalence classes of 
solutions to Yang-Mills equations on some space-time and 
equivalence classes of holomorphic vector bundles over a twistor
space associated with the space-time under consideration
is called a Penrose-Ward transform \cite{MasonRF}. 

Furthermore, in this respect 
it is also worth mentioning that,
by considering the space of all complex null geodesics of
some general four-dimensional complex space-time, one can generalize
ambitwistor space to a curved setting. As was shown
by LeBrun \cite{LeBrun:1983}, this ``curved" ambitwistor 
space is a five-dimensional complex manifold. Moreover,
this construction then forms the basis for a twistorial description 
of four-dimensional
conformal gravity \cite{Baston:1987av,LeBrun:1991jh}.

Besides Penrose-Ward transforms involving the already 
mentioned twistor and ambitwistor spaces, one may establish 
such correspondences on quite generic ground. This was 
considered, e.g., by Eastwood \cite{Eastwood:1985}. In fact, suppose
we are given three complex manifolds $X$, $Y$ and $Z$, where
$Y$ is simultaneously fibered over $X$ and $Z$, that is,
we consider a double fibration of the form
$$\begin{picture}(50,40)
  \put(0.0,0.0){\makebox(0,0)[c]{$Z$}}
  \put(64.0,0.0){\makebox(0,0)[c]{$X$}}
  \put(34.0,33.0){\makebox(0,0)[c]{$Y$}}
  \put(7.0,18.0){\makebox(0,0)[c]{$\pi_1$}}
  \put(55.0,18.0){\makebox(0,0)[c]{$\pi_2$}}
  \put(25.0,25.0){\vector(-1,-1){18}}
  \put(37.0,25.0){\vector(1,-1){18}}
 \end{picture}
$$
together with two suitable holomorphic projections $\pi_{1,2}$.
Here and in the
following, we shall refer to $X$ as space-time, to $Y$ as 
correspondence space and to $Z$ as twistor space, respectively.
Clearly, such a double fibration induces a correspondence
between $X$ and $Z$, that is, there is a relation
between points and subsets in either manifold. In 
particular, a point $x\in X$ gives a submanifold
$\pi_1(\pi_2^{-1}(x))\hookrightarrow Z$, and conversely
a point in twistor space, $z\in Z$, yields a
submanifold $\pi_2(\pi_1^{-1}(z))$ embedded into
space-time $X$. In addition, this correspondence can
be used to transfer data given on $Z$ to data on $X$
and vice versa. Indeed, given some analytic
object, Ob$_Z$,  on twistor space (e.g., certain cohomology 
classes or holomorphic vector bundles), one may 
relate it
to some object Ob$_X$ on space-time. The latter will be 
constrained by certain partial differential equations 
since by definition, the pull-back
$\pi^*_1$Ob$_Z$ must be constant along the fibers of 
$\pi_1\,:\,Y\to Z$. Under suitable topological 
conditions, the maps
$$\operatorname{Ob}_Z\ \mapsto\ \operatorname{Ob}_X\qquad{\rm and}\qquad
  \operatorname{Ob}_X\ \mapsto\ \operatorname{Ob}_Z$$ 
define a bijection between equivalence classes $[$Ob$_Z]$ and $
[$Ob$_X]$ (in general, the objects under consideration
will only be defined up to equivalence). The correspondence
thus established gives a way of studying objects
on space-time obeying differential
equations in terms of objects on twistor space which
are ``free" of such differential constraints. 
As before, we shall refer to such a map as Penrose-Ward 
transform. 

Suppose now that the objects in question are holomorphic
vector bundles. 
If we let $\O^1_{\pi_1}(Y):=\O^1(Y)/\pi_1^*\O^1(Z)$ be
the sheaf of relative differential one-forms on the correspondence space, i.e., the dual
of the vertical tangent vectors spanning the tangent
spaces of $\pi_1\,:\,Y\to Z$, we have the following
theorem \cite{Eastwood:1985}:

\noindent
{\bf Theorem.}\ 
{\it Suppose the fibers of $\pi_1\,:\,Y\to Z$ are 
     simply connected and $\pi_1(\pi_2^{-1}(x))\hookrightarrow Z$ 
     is compact and connected for all $x\in X$. If in addition
     $\O^1(X)\cong\pi_{2*}\O^1_{\pi_1}(Y)$, then there is a one-to-one 
     correspondence between equivalence classes of holomorphic vector bundles
     $E\to Z$ being holomorphically trivial on any
     $\pi_1(\pi_2^{-1}(x))\hookrightarrow Z$, equivalence classes of 
     holomorphic vector bundles on $Y$ equipped with 
     flat relative connection and equivalence classes of 
     holomorphic vector bundles $E'\to X$ equipped with a connection
     flat on each $\pi_2(\pi_1^{-1}(z))\hookrightarrow X$ for 
     $z\in Z$.\\[8pt]}
\noindent
In particular, if $Z$ is twistor space then 
$\pi_1(\pi_2^{-1}(x))\cong\IC P^1_x$ and the submanifolds 
$\pi_2(\pi_1^{-1}(z))$ represent
isotropic two-planes in complexified Minkowski space. In case of
ambitwistor space, one has 
$\pi_1(\pi_2^{-1}(x))\cong(\IC P^1\times\IC P^1)_x$
and $\pi_2(\pi_1^{-1}(z))$ are complex light rays. To jump ahead of our
story a bit, such double fibrations and correspondences between
vector bundles
will play the basis of all the discussion presented in this thesis.

The twistor approach to field theories also provides 
a suitable framework for studying certain supersymmetric field
theories. One simply needs to replace the manifolds appearing
in double fibrations like the one discussed above, by 
so-called supermanifolds. Originally, the concept of supermanifolds 
goes back to the work by Berezin, 
Kostant and Leites \cite{Berezin:1975,Kostant,Leites:1980}.
The idea is basically to define a supermanifold as
a particular ringed space $(X,\CO_X)$, that is, a topological
space $X$ together with a sheaf of supercommutative rings 
 $\CO_X$ (the structure sheaf)
obeying certain local properties. 
A thorough description of supermanifolds, 
as proposed by the above authors, has been developed
by Manin \cite{Manin}. In our subsequent discussion,
we shall adopt this approach to supermanifolds.
Let us also mention that there
are alternative ways to introduce supermanifolds. For
instance, a different one has been 
given by DeWitt \cite{DeWittCY}.\footnote{There is another, third
approach due to Rogers \cite{Rogers:1980}.} It is based on the concept
of supernumbers which are elements of a certain
Gra{\ss}mann algebra. We refer also to the book by 
Bartocci et al. \cite{Bartocci} (and references therein), where 
all existing approaches
are described and compared among each other.

As indicated above, the twistor approach facilitates the
studies of supersymmetric field theories.
For instance, by combining the
ideas of formal neighborhoods with those of supermanifolds, 
Witten has shown \cite{Witten:1978xx} that it is possible
to recover the full $\CN=3$ super Yang-Mills theory in
four dimensions
by discussing holomorphic vector bundles over a certain
superambitwistor space, which we shall denote by $\IL^{5|6}$. In fact, 
$\IL^{5|6}$ is the space
of super light rays $\IC^{1|6}$ in complexified $\CN=3$ 
Minkowski superspace $\IC^{4|12}$. Furthermore, it can be
viewed as a certain degree two hypersurface in the direct product
$\IC P^{3|3}\times\IC P^{3|3}$.
In this respect, 
it is also important to stress that the 
$\CN=3$ and $\CN=4$ theories are physically equivalent. 
The only thing appearing
to be different is that the $\CN=3$ formulation makes only
the $U(1)\times SU(3)$ subgroup of the R-symmetry group
$SU(4)$ manifest. 
In this sense, Witten's approach can be understood as a twistorial
formulation of $\CN=4$ super Yang-Mills theory.
Furthermore, Manin \cite{Manin} 
generalized Witten's discussion to
theories with less supersymmetry. In particular, he established
a Penrose-Ward transform between solutions to the $\CN$-extended
super Yang-Mills equations and holomorphic vector bundles
over $\CN$-extended superambitwistor space -- the space
parametrizing super light rays $\IC^{1|2\CN}$ in $\IC^{4|4\CN}$ --
which admit an extension to a $(3-\CN)$-th formal neighborhood in
superambitwistor space.

\vskip 15pt
\centerline{\sc The twistor string}\vskip 5pt

Within the last three years, the twistorial studies of supersymmetric 
field theories received a somewhat unexpected renaissance due to
string theory. In 2003, Witten \cite{Witten:2003nn} 
proposed a string theory whose target manifold is the supertwistor space
$\IC P^{3|4}$. His idea is based on three facts:
i) holomorphic vector bundles over twistor space and similarly
over supertwistor space are related to gauge and, respectively, super gauge 
theories on four-dimensional space-time, ii) supertwistor space $\IC P^{3|4}$
is a Calabi-Yau supermanifold, that is, its first Chern class
vanishes and in addition it admits a Ricci-flat metric and
iii) the existence of a string theory -- the open topological
B model  -- whose space-time effective action 
is holomorphic Chern-Simons theory \cite{Witten:1992fb}. 
More specifically, Witten showed that the open topological B model
on supertwistor space $\IC P^{3|4}$ with a stack of $r$ 
D5-branes\footnote{These
D5-branes are not quite space-filling and defined by the condition
that all open string vertex operators do not depend on
antiholomorphic Gra{\ss}mann coordinates on $\IC P^{3|4}$.} is
equivalent to holomorphic Chern-Simons theory on the same
space. This theory describes holomorphic structures on a rank $r$
complex vector bundle $\CE$ over $\IC P^{3|4}$ which are given by
the (0,1) part $\CA^{0,1}$ of a connection one-form $\CA$ on
$\CE$. The components of $\CA^{0,1}$ appear as the excitations of
open strings ending on the D5-branes. Furthermore, the spectrum 
of physical states
contained in $\CA^{0,1}$ is the same as that of $\CN=4$ super
Yang-Mills theory, but the interactions of both theories
differ. In fact, by analyzing the linearized \cite{Witten:2003nn}
and the full \cite{Popov:2004rb} field equations, it was shown
that holomorphic Chern-Simons theory on $\IC P^{3|4}$ is equivalent 
to $\CN=4$ self-dual super Yang-Mills theory in four dimensions as
introduced by Siegel \cite{SiegelZA}. Note that this theory can be considered 
as a truncation of the full $\CN=4$ super Yang-Mills theory.

It was conjectured by Witten that perturbative amplitudes of
full $\CN=4$ super Yang-Mills theory are recovered by including
into the B model so-called
D-instantons which wrap certain holomorphic curves in 
supertwistor space.  The presence of these
D1-branes leads to additional fermionic states from the
strings stretching between the D5- and D1-branes. 
Scattering amplitudes are then computed in terms of currents
constructed from these additional fields, which localize on the
D1-branes, by integrating certain correlation functions over the
moduli space of these D1-branes in $\IC P^{3|4}$. This proposal
generalizes an earlier construction of maximally
helicity-violating amplitudes by Nair \cite{Nair:bq}. Thus, by
incorporating D1-branes into the topological B model, one can
complement $\CN=4$ self-dual super Yang-Mills theory to the full 
theory. However, soon after this proposal
it was realized that this construction works only at tree-level,
as already at one-loop level amplitudes of super Yang-Mills
theory mix with those of conformal supergravity 
\cite{Berkovits:2004jj}. Recently, Boels et al. \cite{Boels:2006ir} proposed 
a twistor approach to $\CN=4$ super Yang-Mills theory which
reproduces perturbative Yang-Mills theory without conformal
supergravity.\footnote{In \cite{Abou-Zeid:2006wu}, a family of twistor 
string theories has been constructed yielding Einstein supergravity
coupled to super Yang-Mills theory on space-time.} 
For related papers regarding twistor actions,
see also Refs. \cite{Mason:2005zm,Mason:2005kn}. After all,
twistor string theory resulted in striking advances in 
computing gauge theory scattering amplitudes -- even beyond tree-level,
as powerful methods inspired by the twistor string have been
developed; see, e.g., 
Ref. \cite{Cachazo:2005ga,Xiao:2006vr} for reviews and also \cite{webpage} for
more references.

In addition to twistor string theory on supertwistor space,
Witten \cite{Witten:2003nn} also mentioned the possibility of formulating
a twistor string theory, i.e., the open topological B model
on the superambitwistor space which is associated with 
$\CN=3$ complexified Minkowski superspace. 
Within such a formulation, no D-instantons would be needed as already
at classical level, holomorphic Chern-Simons theory on superambitwistor
space gives the full $\CN=3$ (respectively, $\CN=4$) 
super Yang-Mills theory in four dimensions 
\cite{Witten:1978xx}. Therefore,
the mechanism for reproducing perturbative 
super Yang-Mills theory would be completely
different compared to the supertwistor space approach. However, even
though this particular superambitwistor space is a Calabi-Yau supermanifold, it 
is not entirely clear how to formulate the B model on it. This
is basically because of the difficulty in making sense of an appropriate
measure on this space.
Only recently, some progress in this direction has been made by Mason et al.
\cite{Mason:2005kn}. Therein,
the authors formulated a Chern-Simons type theory (whose 
interpretation on four-dimensional space-time is super Yang-Mills theory) 
on a certain codimension
$2|0$ Cauchy-Riemann submanifold in superambitwistor space. 
Their description is based on Euclidean signature (instead of Minkowski
signature) but as far as perturbation
theory is concerned, this would not be of importance.

Since Witten's proposal, a variety of
questions associated with supertwistor theory and twistor
string theory has been investigated.  
For instance, an interesting property of Calabi-Yau supermanifolds 
is that Yau's theorem does not apply, that is, a K\"ahler 
supermanifold with vanishing first Chern class does not automatically
admit a Ricci-flat metric. This fact was already observed by Sethi
\cite{Sethi:1994ch} and after Witten's paper  
explored in more detail by the authors of
\cite{Rocek:2004bi,Zhou:2004su,Rocek:2004ha,Saemann:2004tt,
Lindstrom:2005uh}.
Another direction triggered by twistor string theory is the
discussion of mirror symmetry between Calabi-Yau supermanifolds.
For instance, it has been argued that for a certain limit of
the K\"ahler moduli, supertwistor space $\IC P^{3|4}$ and
superambitwistor space $\IL^{5|6}\hookrightarrow\IC P^{3|3}\times\IC P^{3|3}$
are mirror manifolds \cite{Neitzke:2004pf,Aganagic:2004yh}. 
For further surveys on this material, we refer to 
\cite{Kumar:2004dj,Ahn:2004xs,Belhaj:2004ts,Policastro:2005dm,
Ricci:2005cp,Laamara:2006rk}. Furthermore, various other target
spaces for twistor string theory have been discussed by the authors of
\cite{Popov:2004nk,Saemann:2004tt,Park:2004bw,Giombi:2004xv,
Chiou:2005jn,Saemann:2005ji,Chiou:2005pu,Popov:2005uv,Lechtenfeld:2005xi} 
leading to, e.g., certain dimensional reductions of
self-dual super Yang-Mills and full super Yang-Mills theories.
Moreover, alternative formulations of twistor string theories 
have been proposed in Refs.
\cite{Berkovits:2004hg,Berkovits:2004tx,Siegel:2004dj,Lechtenfeld:2004cc,
Bandos:2006af}.
In addition, issues related to gravity were investigated, as well. See, e.g., 
\cite{Giombi:2004ix,Ahn:2004yu,Ahn:2004ua,
Bedford:2005yy,Bjerrum-Bohr:2005xx,Abe:2005se,Abou:2005,Abou-Zeid:2006wu}. For other
aspects, see also Refs. 
\cite{Abe:2004ep,Sinkovics:2004fm,Kulaxizi:2004pa,
Seki:2005hx,Saemann:2006tt,Seki:2006cj}.

\vskip 15pt
\centerline{\sc Integrability}\vskip 5pt

Besides twistor string theory, $\CN=4$ super Yang-Mills theory
appears to be connected with another string theory. Approximately
ten years ago, a correspondence -- nowadays known as AdS/CFT
correspondence -- was discovered. Maldacena conjectured an
equivalence between $\CN=4$ super Yang-Mills
theory in four dimensions and type IIB superstring theory on the curved
background AdS$_5\times S^5$
\cite{Maldacena:1997re,Gubser:1998bc,Witten:1998qj}. 
Besides matching of global symmetries, this conjecture claims
full dynamical agreement of both theories. For instance, one
of the statements is that the spectrum of the scaling dimensions
of the gauge theory should coincide with the energy spectrum
of the string states. However, due to the weak-strong nature
of this correspondence, that is, the weak coupling regime of
either theory gets mapped to the strong coupling regime of the other,
it is extremely difficult to test or rather prove this conjecture.
Therefore, one is searching for appropriate tools which may 
help to clarify this relation on general grounds.

One such tool which emerged in recent years 
is integrability. As already indicated, $\CN=4$ super
Yang-Mills theory appears to be integrable at classical level in
the sense of admitting a certain twistor formulation (thus yielding 
a Lax pair formulation). 
Quantum integrable structures in 
$\CN=4$ super Yang-Mills theory have first been 
discovered by Minahan et al. \cite{Minahan:2002ve} 
being inspired by the work of 
Berenstein et al. \cite{Berenstein:2002jq}.\footnote{For an ealier account 
of integrable structures in QCD, see, e.g., Refs. 
\cite{Lipatov:1993yb,Faddeev:1994zg,Belitsky:1999qh,Braun:1999te,Belitsky:1999ru,Belitsky:1999bf}.} 
They found that the planar dilatation operator,
which measures the planar scaling dimension of local
operators, in a
certain sector of the theory can be interpreted
at one-loop level as Hamiltonian of an integrable quantum spin chain.
Based on this observation, 
it has then been shown that it is indeed possible 
to interpret the complete one-loop dilatation 
operator as Hamiltonian of an integrable quantum spin chain; 
see, e.g., 
\cite{Beisert:2004ry,Belitsky:2004cz} and 
references therein. 
For discussions beyond leading order, see
also \cite{Beisert:2003tq,Belitsky:2004sf,Staudacher:2004tk,Beisert:2005fw,
McLoughlin:2005gj,Plefka:2005bk,Berenstein:2005jq,Beisert:2005cw,
Zwiebel:2005er,Frolov:2006cc,Eden:2006rx,Belitsky:2006av}.
Another development pointing
towards integrable structures was initiated by Bena et al. 
\cite{Bena:2003wd}. Their investigation is based on the observation that the 
Green-Schwarz formulation of the superstring on AdS$_5\times S^5$
can be interpreted as a coset theory, where the fields take values in the
supercoset space $$PSU(2,2|4)/(SO(4,1)\times SO(5))$$ 
\cite{Metsaev:1998it,Kallosh:1998zx,Roiban:2000yy}.
Although this is not a symmetric space (the denominator group is too small)
\cite{Mandal:2002fs}, this coset theory
admits an infinite set of conserved
nonlocal charges, quite similar to those that exist in two-dimensional
field theories.\footnote{These charges were also independently found by Polyakov
\cite{Polyakov:2004br}.} For the construction of nonlocal conserved charges in 
two-dimensional sigma models, we refer to \cite{LuscherRQ,Luscher2}.
Such charges are in turn related to Kac-Moody algebras 
\cite{DolanFQ,Dolan2,Schwarz:1995td} and generate Yangian
algebras \cite{DrinfeldRX,DrinfeldRX2} as has been discussed, e.g., in 
\cite{BernardJW}. For reviews 
of Yangian algebras see Refs. \cite{Bernard:1992ya,MacKay:2004tc}.
Some time later,  
the construction of an analogous set of nonlocal conserved charges 
using the pure spinor formulation of the superstring 
\cite{Berkovits:2000fe,Berkovits:2000yr,Vallilo:2002mh} on 
AdS$_5\times S^5$ was given in \cite{Vallilo:2003nx}.
For further developments, we refer the reader to Refs.
\cite{Alday:2003zb,Arutyunov:2004xy,Hou:2004ru,Arutyunov:2004vx,
Hatsuda:2004it,Berkovits:2004jw,Swanson:2004qa,Beisert:2005bm,
Alday:2005gi,Young:2005jv,Chen:2005uj,
Frolov:2005ty,Frolov:2005dj,Alday:2005jm,
Mann:2005ab,Kagan:2005wt,Hatsuda:2006ts,Roiban:2006yc,Bianchi:2006im}, for example.\footnote{See also
Refs. \cite{Georgiou:2004by,Brandhuber:2004yw,Khoze:2004ba,Su:2004ym}.}
Clearly, the question that arises of what could be the
charges in super Yang-Mills theory.
Within the spin chain approach,
Dolan et al. \cite{Dolan:2003uh,Dolan:2004ps,Dolan:2004ys}
related these nonlocal charges
for the superstring to a corresponding set of nonlocal charges in the 
superconformal gauge 
theory in the extreme weak coupling limit
(see also \cite{Arutyunov:2003rg,Agarwal:2004sz,
Mikhailov:2004ca,Agarwal:2005jj}).

\vskip 15pt
\centerline{\sc Outline and main results}\vskip 5pt

This thesis is devoted to studying different questions related to supersymmetric
gauge theories. The main tool we shall be using is twistor theory.

In the first chapter, we begin by reviewing some of the basic aspects and
properties of twistor spaces which correspond to flat space-times.
No prior knowledge of twistor geometry is assumed.
The unifying framework for our discussion 
is complex flag manifolds, which are natural
generalizations of complex projective spaces.
Furthermore, after having presented the concepts of supermanifolds, supervector
bundles, etc., we explain the extension of twistor geometry 
to supertwistor geometry. As otherwise it would carry us too far afield
from the main thread of development, we present 
the material only to that extent which is of need in later applications.
For detailed expositions on the subject, we refer to the books 
\cite{Huggett,Manin,MasonRF,PenroseJW,Ward}. 

In the second chapter,
we give a first application by discussing self-dual super Yang-Mills
theory and some related self-dual models. In particular, we start by describing the
equivalence of the \v Cech and Dolbeault approaches to holomorphic
vector bundles over supermanifolds. This lies somehow in the heart of 
the twistor approach to gauge theories. After that, we give a detailed
explanation of the Penrose-Ward transform for $\CN$-extended self-dual
super Yang-Mills theory including appropriate superfield expansions, etc. By replacing
supertwistor space by certain weighted projective superspaces, we develop
Penrose-Ward transforms for truncations of self-dual super
Yang-Mills theories. As the spaces under consideration will always
be Calabi-Yau supermanifolds, we are also able to write down appropriate
action principles. The discussion of the second part of this 
chapter is based on the work done together
with Alexander Popov \cite{Popov:2004nk}. 

The third chapter is 
devoted to the twistor construction of certain supersymmetric 
Bogomolny models in three space-time dimensions. In \cite{Chiou:2005jn},
it was shown that
scattering amplitudes of $\CN=4$ super Yang-Mills theory which are 
localized on holomorphic curves in supertwistor space can be
reduced to amplitudes of $\CN=8$ super Yang-Mills theory in three dimensions
which are localized on holomorphic curves in a supersymmetric
extension of mini-twistor space. Note that the simplest of 
such curves in mini-twistor space is the Riemann sphere 
which coincides with the spectral curve of the BPS $SU(2)$
monopole. Note also that every static $SU(2)$ monopole of charge $k$
may be constructed from an algebraic curve in mini-twistor space
\cite{Hitchin:1982gh}, and an $SU(r)$ monopole is defined by
$r-1$ such holomorphic curves \cite{Murray:1985ji}. The
corresponding string theory after this reduction is the
topological B model on mini-supertwistor space 
with $r$ not quite space-filling D3-branes (defined analogously to
the D5-branes in the six-dimensional case) and additional
D1-branes wrapping holomorphic cycles in mini-supertwistor space. It is
reasonable to assume that the latter correspond to monopoles and
substitute the D-instantons in the case of
supertwistor space. In \cite{Chiou:2005jn}, also a twistor string theory
corresponding to a certain massive super Yang-Mills theory in three dimensions
was described. The target space of the underlying 
B model is a Calabi-Yau supermanifold obtained from 
mini-supertwistor space by a deformation of its complex
structure. 
The goal of this chapter is to complement the discussion given
in \cite{Chiou:2005jn}.
In particular, we show that the B model with only the
D3-branes included corresponds to a field
theory on mini-supertwistor space obtained by a
reduction of holomorphic Chern-Simons theory on supertwistor
space. We show that this field theory 
is a holomorphic BF-type theory\footnote{These
theories were introduced in \cite{Popov:1999cq} and considered,
e.g., in \cite{Ivanova:2000xr,Baulieu:2004pv}.} which in turn is
equivalent to a supersymmetric Bogomolny model. This model can be
understood as the BPS equations of $\CN=8$ super Yang-Mills theory in three
dimensions. 
The action functional of holomorphic
BF theory on mini-supertwistor space is not of
Chern-Simons type, but one can introduce a Chern-Simons type action on
the correspondence space of mini-supertwistor space.
This
space admits a so-called Cauchy-Riemann
structure. After enlarging the integrable distribution
defining this Cauchy-Riemann structure by one real direction to a 
distribution
$\cT$, one is led to the notion of $\cT$-flat vector bundles over
the correspondence space. These bundles take over the role of holomorphic
vector bundles, and they can be defined by a $\cT$-flat connection
one-form $\CA_\cT$ \cite{Rawnsley}.
The condition of
$\cT$-flatness of $\CA_\cT$ can be derived as the equations of
motion of a theory we shall call partially holomorphic
Chern-Simons theory. This theory can be obtained by
a dimensional reduction of holomorphic Chern-Simons theory on 
supertwistor space. We prove that there are one-to-one 
correspondences
between equivalence classes of holomorphic vector
bundles subject to certain triviality
conditions over mini-supertwistor space,
equivalence classes of $\cT$-flat vector bundles
over the correspondence space and gauge equivalence classes of solutions to
supersymmetric Bogomolny equations in three dimensions. In other words, the
moduli spaces of all three theories are bijective. 
By deforming the complex structure on mini-supertwistor space, 
which in
turn induces a deformation of the Cauchy-Riemann structure on the
correspondence space, we
obtain a similar correspondence but with additional
mass terms for fermions and scalars in the supersymmetric
Bogomolny equations.
The twistorial description of the supersymmetric Bogomolny
equations has the nice feature of yielding novel
methods for constructing explicit solutions. For simplicity, we
restrict our discussion to solutions where only fields with
helicity $\pm 1$ and a Higgs field are nontrivial. The
corresponding Abelian configurations give rise to the Dirac
monopole-antimonopole systems. For the non-Abelian case, we
present two ways of constructing solutions: first, by using a
dressed version of the Penrose transform and second, by
considering a nilpotent deformation of the holomorphic vector
bundle corresponding to an arbitrary seed solution of the ordinary
Bogomolny equations. The third chapter is based on the work
done together with Alexander Popov and Christian S\"amann
\cite{Popov:2005uv}.

The fourth chapter deals with a discussion of full super
Yang-Mills theories on four-dimensional space-time. In particular,
we review Witten's \cite{Witten:1978xx} and Manin's \cite{Manin}
constructions of Penrose-Ward
transforms relating gauge equivalence classes of solutions to 
the $\CN$-extended super Yang-Mills equations to certain equivalence
classes of holomorphic vector bundles over superambitwistor space
which admit an extension to a $(3-\CN)$-th order formal
neighborhood. 

As indicated above, there exist infinitely many nonlocal conserved charges
in $\CN=4$ super Yang-Mills theory. So far, they have been formulated
within the spin chain approach \cite{Dolan:2003uh,Dolan:2004ps,Dolan:2004ys}
which is perturbative in nature. Note that field theoretic expressions 
for these charges at zero 't Hooft coupling are known and were also given 
in \cite{Dolan:2003uh,Dolan:2004ps}. 
It is therefore reasonable to consider the problem 
of constructing such charges from first principles.
As a modest step, one may first study a simplification of the theory -- namely 
its self-dual truncation. It will be the
purpose of the fifth chapter to use twistor theory 
for studying infinite-dimensional algebras of
hidden symmetries
in $\CN$-extended self-dual super Yang-Mills theories. Here, we  
generalize the results known for the self-dual 
Yang-Mills equations \cite{PohlmeyerYA,ChauGI1,ChauGI2,ChauGI3,UENO,Dolan,
Crane,Popov:1995qb,Popov:1998,Popov:1998fb,Ivanova:1997cu,MasonRF}.
In particular, we will first consider deformation theory of 
holomorphic vector bundles over supertwistor space. Using the 
Penrose-Ward transform, we relate these infinitesimal 
deformations to symmetries of the self-dual super Yang-Mills
equations. After some general words on Riemann-Hilbert problems,
hidden symmetries and related
algebras and using also a cohomological classification, we exemplify
our discussion by constructing Kac-Moody symmetries which come from 
affine extensions of the gauge algebra. In addition, we 
consider affine extensions of the superconformal algebra to obtain super
Kac-Moody-Virasoro-type symmetries. The existence of such algebras
originates from the fact that the full group of continuous transformations
acting on the space of holomorphic vector bundles over supertwistor space
is a semi-direct product of the group of local holomorphic automorphisms
of the supertwistor space and of the group of one-cochains with respect
to a certain covering with values in the sheaf of holomorphic
maps of supertwistor space into the gauge group. See \cite{Popov:1998}
for a discussion in the purely bosonic setting. 
By focusing on a certain
Abelian subalgebra  
of the affinely extended superconformal algebra, 
we introduce a family of generalized supertwistor spaces. These manifolds,
being parametrized by certain integers, 
then allow us to introduce truncated self-dual super Yang-Mills hierarchies. 
Such a hierarchy consists
of a finite system of partial differential equations, where the
self-dual super Yang-Mills equations are embedded in. The lowest level
flows of such a hierarchy represent space-time translations.
Furthermore, one may also consider 
the asymptotic limit to obtain the full hierarchy. Indeed, 
the existence of such a hierarchy allows us
to embed a given solution into an
infinite-parameter family of new solutions.
This generalizes the results known for the self-dual Yang-Mills 
equations \cite{Mason,Takasaki,Mason:1992vd,Ablowitz,IvanovaZT,MasonRF}. 
We remark that such symmetries of the latter 
equations are intimately connected with one-loop maximally helicity violating 
amplitudes \cite{BardeenGK,Cangemi1,Cangemi2,Rosly:1996vr,Gorsky:2005sf}. 
As for certain values of the
parameters the generalized supertwistor spaces become Calabi-Yau,
we are also able to give action principles for the truncated hierarchies. 
In addition, we also construct infinitely many nonlocal conservation laws
in self-dual super Yang-Mills theory which are associated with the symmetries in question 
(see \cite{Ablowitz} in the context of the self-dual Yang-Mills hierarchy).
The discussion of the fifth chapter is based on \cite{Wolf:2004hp,Wolf:2005sd}.   

Finally, we give a short summary of the results derived in this thesis and 
close with an outlook.

\vskip 15pt
\centerline{\sc Conventions}\vskip 5pt

In the sequel, we shall be using the standard abbreviations
YM theory and SYM theory for Yang-Mills theory and super
Yang-Mills theory, respectively. We also abbreviate
holomorphic Chern-Simons theory by hCS theory. 
In addition, the reader may find a list with
symbols which are most frequently used throughout this work
at the end of this thesis.  

\vskip 15pt
\centerline{\sc Disclaimer}\vskip 5pt

I apologize in advance to those people whose names were either
mentioned incorrectly or not mentioned at all. 
I have tried to track down all the literature related to the 
topics discussed below but nevertheless because of the broad 
fields of twistor theory, integrable systems, etc., it might 
have happened that some of the works slipped through my hands. 
If so then simply because of the fact that I was not aware of them.

\begin{flushright}
Hannover, Summer 2006\\[.1in] Martin Wolf
\end{flushright}

\clearemptydoublepage
\clearemptydoublepage


\lhead[\fancyplain{}{\bfseries\thepage}]{\fancyplain{}{\bfseries\rightmark}}
\rhead[\fancyplain{}{\bfseries\leftmark}]{\fancyplain{}{\bfseries\thepage}}
\cfoot{}

\tableofcontents
\clearemptydoublepage
\newpage
\pagenumbering{arabic}


\chapter{Supertwistor geometry}\label{STG-Chapter}
\HRule\\

{\Large T}his first chapter is intended to give some of the
main ideas of supertwistor geo-\linebreak[1] metry. The material 
is presented only to that extent which is needed in later applications. 
For pedagogical reasons, we begin with ordinary twistor 
geometry. As we will be discussing the twistorial approach to 
supersymmetric field theories, we need to generalize this 
framework to also include so-called supermanifolds -- manifolds 
parametrized by $\IZ_2$-graded coordinates. We will shortly 
realize, however, that for developing these generalizations, 
it is necessary to work in a bit more abstract mathematical setting.
In due course, we therefore also present some 
definitions and notions to simplify understanding. 
The reader unfamiliar with the underlying mathematics may 
consult, e.g., Refs. \cite{Bartocci,Griffith,Manin,Wells}. 
A thorough introduction into twistor geometry can be found in 
the books \cite{Huggett,Manin,MasonRF,PenroseJW,Ward}.

\section{Twistor spaces}\label{TS}

\paragraph{Flag manifolds.}
Let us begin by recalling the definition of a flag manifold. 
Consider the complex space $\IC^n$. Then we define its flag 
manifolds by
\eqn{flag}{F_{d_1\cdots d_m}(\IC^n)\ :=\ 
     \{(S_1,\ldots,S_m)\,|\,S_i\subset 
     \IC^n,\, {\rm dim}_\IC S_i=d_i,\, S_1\subset S_2
     \subset\cdots\subset S_m\},
}
where the $S_i$s are subspaces of $\IC^n$. Typical examples of 
such flag manifolds are the projective space $F_1=\IC P^{n-1}$ 
and the Gra{\ss}mannian $F_k=G_{k,n}(\IC)$. Note that all of 
these manifolds are compact complex manifolds. Moreover, they 
have an equivalent representation as homogeneous spaces. That 
is, consider the decomposition
$$ \IC^{d_1}\oplus\IC^{d_2-d_1}\oplus\IC^{d_3-d_2}\oplus
   \cdots\oplus\IC^{d_m-d_{m-1}}\oplus\IC^{n-d_m}
$$
of $\IC^n$.
The subgroup of $U(n)$ which preserves this decomposition is
$$ U(d_1)\times U(d_2-d_1)\times\cdots\times U(d_m-d_{m-1})
   \times U(n-d_m)\ \subset\ U(n)$$
and hence we may give an equivalent definition by the quotient
\eqn{flag2}{F_{d_1\cdots d_m}(\IC^n)\ :=\ 
     \frac{U(n)}{U(d_1)\times U(d_2-d_1)
     \times\cdots\times U(d_m-d_{m-1})\times U(n-d_m)}.
}
From this it rather straightforwardly follows that their 
dimensionality is given by the formula
$$ {\rm dim}_\IC F_{d_1\cdots d_m}(\IC^n)\ =\ d_1(n-d_1)+
   (d_2-d_1)(n-d_2)+\cdots+(d_m-d_{m-1})(n-d_m). 
$$

\paragraph{Twistor space.}
We now want to use the notion of flag manifolds to introduce  
fundamental complex manifolds considered in twistor geometry. 
In order to describe field theories in four dimensions,
we restrict ourselves to four dimensions. So, let $\IT$ be a 
fixed complex four-dimensional vector space which we call 
twistor space. In due course, we shall also endow $\IT$ with 
various real structures. Furthermore, we have a natural fibration 
in terms of flag manifolds 
\eqna{DF1}{
 \begin{picture}(50,40)
  \put(0.0,0.0){\makebox(0,0)[c]{$F_1(\IT)$}}
  \put(64.0,0.0){\makebox(0,0)[c]{$F_2(\IT)$}}
  \put(34.0,33.0){\makebox(0,0)[c]{$F_{12}(\IT)$}}
  \put(7.0,18.0){\makebox(0,0)[c]{$\pi_1$}}
  \put(55.0,18.0){\makebox(0,0)[c]{$\pi_2$}}
  \put(25.0,25.0){\vector(-1,-1){18}}
  \put(37.0,25.0){\vector(1,-1){18}}
 \end{picture}
}
together with the canonical projections $\pi_i(S_1,S_2)=S_i$ 
for $i=1,2$. In fact, the double fibration \eqref{DF1} is the 
first in a row of important double fibrations in twistor geometry 
we will encounter throughout this thesis. We shall now develop 
some of its basic properties. We follow the literature and define 
\eqna{TM1}{\IP^3\ &:=\ F_1(\IT)\ =\ \IC P^3,\\ 
       \IM^4\ &:=\ F_2(\IT)\ =\ G_{2,4}(\IC),\\
       \IF^5\ &:=\ F_{12}(\IT),
} 
and call them projective twistor space, compactified complexified 
four-dimensional space-time and correspondence space, respectively. 
At this point we stress that Minkowski and Euclidean spaces can 
naturally be realized as four-dimensional subsets of the complex 
four-dimensional manifold $\IM^4$ (see Ref. \cite{Ward} for a 
detailed discussion). By virtue of the double fibration \eqref{DF1},
geometric data is transferred from $\IM^4$ to $\IP^3$ and vice versa
according to the subsequent proposition (see, e.g., \cite{Ward}):
{\Pro\label{GTC} There is the following correspondence between 
points and subsets:
\vspace*{-2mm}
\begin{center} 
 \begin{tabular}{cccc}
   {\rm (i)} & point in $\IP^3$ & $\quad\longleftrightarrow\quad$ & 
                                    $\IC P^2\subset\IM^4$\\
   {\rm (ii)} & $\IC P^1\subset\IP^3$&$\quad\longleftrightarrow
                   \quad$ & point in $\IM^4$
 \end{tabular}
\end{center}}\vskip 4mm

\paragraph{Local coordinates.}\label{someLC}
Next we introduce local coordinates. For this, we let
\eqn{LCR}{x\ =\ (x^{\a\ad})\ \in\ {\rm Mat}(2,\IC)\ 
     \overset{\varphi}{\mapsto}\
     \left[\begin{matrix} x^{\a\ad}\\ \mathbbm{1}_2\end{matrix}\right]
}
be a coordinate mapping for $\IM^4$. The brackets denote, as usual, 
the span. Then we define a coordinate chart on $\IM^4$ by
\eqn{CM1}{\CM^4\ :=\ \varphi({\rm Mat}(2,\IC))\ \cong\ \IC^4.}
We call $\CM^4$ affine complexified four-dimensional space-time. 
Note that it is simply one of six possible choices of standard 
coordinate charts for $\IM^4$. Using the projections $\pi_{1,2}$, 
we may naturally define the affine parts of $\IF^5$ and $\IP^3$ 
according to
\eqn{}{\CF^5\ :=\ \pi_2^{-1}(\CM^4)\qquad{\rm and}\qquad
       \CP^3\ :=\ \pi_1(\pi_2^{-1}(\CM^4)).
}
Then we have 
\eqn{le:LL-ch1}{\CF^5\ \cong\ \CM^4\times\IC P^1.}
To see this, let $[\l_\ad]=[\l_{\dot1},\l_{\dot2}]$ be homogeneous 
coordinates on $\IC P^1$ and $x$ as above. Then consider the mapping
$$ \begin{aligned}
      (x^{\a\ad},[\l_\ad])\ &\mapsto\ 
      \left(\left[\begin{matrix} x^{\a\ad}\\ 
      \mathbbm{1}_2\end{matrix}\right]\l_\ad,
      \left[\begin{matrix} x^{\a\ad}\\ \mathbbm{1}_2\end{matrix}
      \right]\right)\ =\ \left(\left[\begin{matrix} x^{\a\ad}\l_\ad\\ 
      \l_\ad\end{matrix}\right],\left[\begin{matrix} x^{\a\ad}\\ 
      \mathbbm{1}_2\end{matrix}\right]\right)\\
      &=\ \left(S^{x,\l}_1,S^{x,\l}_2\right)\ \in\ \IF^5,
  \end{aligned}
$$
which in fact proves \eqref{le:LL-ch1}.

It then follows by our above discussion that the projection 
$\pi_1\,:\,\CF^5\to\CP^3$ is given by
\eqn{}{(x^{\a\ad},[\l_\ad])\ \overset{\pi_1}{\mapsto}\ 
\left[\begin{matrix} x^{\a\ad}\l_\ad\\ \l_\ad\end{matrix}\right]
       \in\ \IP^3.}
Therefore, in terms of these coordinates our double fibration 
\eqref{DF1} takes the following form:
\eqna{doublefibrationI}{
 \begin{picture}(50,40)
  \put(0.0,0.0){\makebox(0,-35)[c]{$\left[\begin{matrix} 
          x^{\a\ad}\l_\ad\\ \l_\ad\end{matrix}\right]\in\IP^3$}}
  \put(74.0,0.0){\makebox(0,-35)[c]{$\left[\begin{matrix} 
           x^{\a\ad}\\ 
           \mathbbm{1}_2\end{matrix}\right]\in\IM^4$}}
  \put(40.0,33.0){\makebox(0,0)[c]{$(x^{\a\ad},[\l_\ad])\in\IF^5$}}
  \put(7.0,18.0){\makebox(0,0)[c]{$\pi_1$}}
  \put(55.0,18.0){\makebox(0,0)[c]{$\pi_2$}}
  \put(25.0,25.0){\vector(-1,-1){18}}
  \put(37.0,25.0){\vector(1,-1){18}}
 \end{picture}
\hspace*{4cm}
 \begin{picture}(50,40)
  \put(0.0,0.0){\makebox(0,0)[c]{$\CP^3$}}
  \put(64.0,0.0){\makebox(0,0)[c]{$\CM^4$}}
  \put(34.0,33.0){\makebox(0,0)[c]{$\CF^5$}}
  \put(7.0,18.0){\makebox(0,0)[c]{$\pi_1$}}
  \put(55.0,18.0){\makebox(0,0)[c]{$\pi_2$}}
  \put(25.0,25.0){\vector(-1,-1){18}}
  \put(37.0,25.0){\vector(1,-1){18}}
 \end{picture}
}

\vspace*{35pt}\noindent
Altogether, these considerations allow us to introduce 
affine coordinates according to:
\begin{itemize}
\setlength{\itemsep}{-2mm}

\item $\CM^4$: $x^{\a\ad}$,
\item $\CF^5$: $x^{\a\ad}$ and $\l_\pm$ with 
      $\l_+=\l_-^{-1}$ on $U_+\cap U_-$, where $\{U_+,U_-\}$ denotes 
      the canonical covering of $\IC P^1$,
\item $\CP^3$: $z^\a_\pm$ and $z^3_\pm$; 
      $\pi_1:(x^{\a\ad},\l_\pm)\mapsto (z^\a_\pm=x^{\a\ad}\l_\ad^\pm,
      z^3_\pm=\l_\pm)$,
      with $\l_\ad^+=\l_+\l_\ad^-$ and 
      $(\l_\ad^+):=\ \!^t(1,\l_+)$.
\end{itemize}
The last point deserves a little more attention. Let us denote the
covering of $\CP^3$ by $\fU=\{\CU_+,\CU_-\}$.
On the intersection $\CU_+\cap\CU_-$
the coordinates $(z_\pm^\a,z^3_\pm)$ satisfy
\eqn{TF1}{z_+^\a\ =\ \frac{1}{z^3_-}z^\a_-\qquad{\rm and}\qquad
       z^3_+\ =\ \frac{1}{z^3_-}.}
Hence, the twistor space\footnote{By a slight abuse of terminology, 
we shall also refer to $\CP^3$ as the twistor space.} $\CP^3$ 
is a rank two holomorphic vector bundle over the Riemann sphere. 
In more details, it is the total space of
\eqn{holofib}{\CO_{\IC P^1}(1)\oplus\CO_{\IC P^1}(1)\ \to\ \IC P^1,}
where $\CO_{\IC P^1}(1)$ denotes the hyperplane line bundle
(the dual of the tautological line bundle) over $\IC P^1$. Moreover, 
the equations 
\eqn{IR1}{z_\pm^\a\ =\ x^{\a\ad}\l^\pm_\ad,}
are also called incidence relations. In a given trivialization, 
holomorphic sections of the bundle $\CP^3\to\IC P^1$ are of the form
\eqref{IR1} and are parametrized by the moduli $x^{\a\ad}$.
In other words, Eqs. \eqref{IR1} make Prop.~\ref{GTC} transparent:
a fixed point $p=(z^\a_\pm,z^3_\pm)\in\CP^3$ corresponds to a null 
two-plane $\IC_p^2\subset\CM^4$ and furthermore, a fixed point 
$x=(x^{\a\ad})\in\CM^4$ 
corresponds to a holomorphic embedding of a rational degree one curve 
$\IC P^1_x\hookrightarrow\CP^3$. To see that these two-planes are 
indeed null,
we solve \eqref{IR1} for a generic $p=(z^\a_\pm,z^3_\pm=\l_\pm)$,
\eqn{}{x^{\a\ad}\ =\ \hat{x}^{\a\ad}+\mu^\a_\pm\l^\ad_\pm,}
where $\mu^\a_\pm$ is arbitrary. Here, $\hat{x}^{\a\ad}$ denotes a
particular solution to \eqref{IR1}.
By recalling that any null four-vector
$k^{\a\ad}$ can be decomposed as $k^{\a\ad}=v^\a w^\ad$ for two
generic commuting spinors $v^\a$ and $w^\ad$, we have thus shown 
that $\IC^2_p$ is totally null. Hence, $\CP^3$ is the space of all
null two-planes in $\CM^4\cong\IC^4$.
  
\paragraph{Remark.}\label{RI} 
Two remarks are in order.
By definition, the projective twistor space $\IP^3$ is the same as 
$\IC P^3$. On this space we may introduce homogeneous coordinates
$[z^\a,\pi_\ad].$ Then consider
$\IC P^3\setminus\IC P^1$, where $\IC P^1\subset\IC P^3$ is defined by 
setting $\pi_\ad=0$ and $z^\a\neq 0$. We can cover 
$\IC P^3\setminus\IC P^1$ 
by two coordinate patches, say $\CU_+$ and $\CU_-$, for which 
$\pi_{\dot1}\neq0$ and $\pi_{\dot2}\neq0$, respectively, and introduce 
the coordinates 
\eqna{LC-1}{z^\a_+\ &:=\ \frac{z^\a}{\pi_{\dot1}}\qquad{\rm and}\qquad
    z^3_+\ :=\ \frac{\pi_{\dot2}}{\pi_{\dot1}}\qquad 
      {\rm on}\qquad\CU_+, \\
       z^\a_-\ &:=\ \frac{z^\a}{\pi_{\dot2}}\qquad{\rm and}\qquad
    z^3_-\ :=\ \frac{\pi_{\dot1}}{\pi_{\dot2}}\qquad 
       {\rm on}\qquad\CU_-,
}
which are related on $\CU_+\cap\CU_-$ by \eqref{TF1}. This shows that 
$\IC P^3\setminus\IC P^1$ is biholomorphic to 
$\CP^3=\CO_{\IC P^1}(1)\oplus\CO_{\IC P^1}(1)$.

There is yet another interpretation of $\CP^3\to\IC P^1$. The Riemann
sphere $\IC P^1$ can be emdedded into $\IC P^3$. The normal bundle of $\IC P^1$ inside
$\IC P^3$ is $\CO_{\IC P^1}(1)\oplus\CO_{\IC P^1}(1)$.
Kodaira's theorem on relative deformation theory \cite{Kodaira:1962,Kodaira} states that
if $Y$ is a compact complex submanifold of a complex manifold -- not 
necessarily compact, and if $H^1(Y,N_{Y|X})\cong0$, where $N_{Y|X}$ is the normal
sheaf of $Y$ in $X$, then there
exists a $d$-parameter family of deformations
of $Y$ inside $X$, where $d={\rm dim}_\IC H^0(Y,N_{Y|X})$. In our example,
$$H^1(\IC P^1,\CO_{\IC P^1}(1)\oplus\CO_{\IC P^1}(1))\ \cong\ 0$$ and 
$d=4$.\footnote{We 
shall often denote a vector bundle
and the corresponding sheaf of sections by the same letter; cf. also
footnote \ref{fo}. In addition, we shall sometimes write $N_Y$ instead
of $N_{Y|X}$.}

\paragraph{Two other twistor spaces.}\label{ATTS}
Above, we have introduced (projective) twistor space in terms of a 
particular
double fibration. However, this is not the only possibility as there is 
a variety of other twistor spaces which turn out to be extremely 
important in later applications. In the remainder of this section,
we pick two examples which again are based on flag manifolds.
Let us consider the following two correspondences:
\eqna{DF2}{
 \begin{picture}(50,40)
  \put(0.0,0.0){\makebox(0,0)[c]{$F_3(\IT)$}}
  \put(64.0,0.0){\makebox(0,0)[c]{$F_2(\IT)$}}
  \put(34.0,33.0){\makebox(0,0)[c]{$F_{23}(\IT)$}}
  \put(7.0,18.0){\makebox(0,0)[c]{$\pi_1$}}
  \put(55.0,18.0){\makebox(0,0)[c]{$\pi_2$}}
  \put(25.0,25.0){\vector(-1,-1){18}}
  \put(37.0,25.0){\vector(1,-1){18}}
 \end{picture}
\hspace*{4cm}
\begin{picture}(50,40)
  \put(0.0,0.0){\makebox(0,0)[c]{$F_{13}(\IT)$}}
  \put(64.0,0.0){\makebox(0,0)[c]{$F_2(\IT)$}}
  \put(34.0,33.0){\makebox(0,0)[c]{$F_{123}(\IT)$}}
  \put(7.0,18.0){\makebox(0,0)[c]{$\pi_1$}}
  \put(55.0,18.0){\makebox(0,0)[c]{$\pi_2$}}
  \put(25.0,25.0){\vector(-1,-1){18}}
  \put(37.0,25.0){\vector(1,-1){18}}
 \end{picture}
}  
The space $F_2(\IT)=\IM^4$ is, of course, compactified complexified
four-dimensional space-time. Note that the space $F_3(\IT)$, that 
is, the space of all three-dimensional subspaces in $\IT$, is
naturally dual to $F_1(\IT^*)$ as we have a natural duality between
lines and hyperplanes in a vector space. For that reason we call
$F_3(\IT)$ dual projective twistor space and denote it by $\IP^3_*$, 
in the following.
The twistor manifold $F_{13}(\IT)$ is called projective ambitwistor
space as it inherits aspects of the projective and the dual projective
twistor spaces. We shall denote it by $\IL^5$. 
Then one can prove the following proposition:
{\Pro\label{GTC2} There are the following correspondences: 
\vspace*{-2mm}
\begin{center} 
 \begin{tabular}{lcccc}
   {\rm (a)} &\kern.5cm {\rm (i)} & point in $\IP^3_*$ & 
              $\quad\longleftrightarrow\quad$ & 
                                    $\IC P^2\subset\IM^4$\\
   & \kern.5cm{\rm (ii)} & $\IC P^1_*\subset\IP^3_*$&$\quad
                      \longleftrightarrow\quad$ & 
                                    point in $\IM^4$\\
 {\rm (b)}&\kern.5cm {\rm (i)} & point in $\IL^5$ & $\quad
        \longleftrightarrow\quad$ & 
                                    $\IC P^1\subset\IM^4$\\
   &\kern.5cm {\rm (ii)} & $\IC P^1\times\IC P^1_*\subset\IL^5$&$\quad
               \longleftrightarrow\quad$ & 
                                    point in $\IM^4$
 \end{tabular}
\end{center}
}\vskip 4mm

Note that for all these manifolds we have coordinate representations 
similar to \eqref{LCR}, and we shall develop a suitable notation for 
these coordinates as they are needed. We want to conclude this section 
by stressing that ambitwistor 
space $\IL^5$ can be regarded as a submanifold in $\IP^3\times\IP^3_*$ 
as there is an embedding
\eqna{}{\IL^5\ &\hookrightarrow\ \IP^3\times\IP^3_*,\\
           (S_1\subset S_3)\ &\mapsto\ (S_1,S_3).         
} 
In fact, $\IL^5$ is a degree two hypersurface in $\IP^3\times\IP^3_*$
since this embedding is given by the zero locus
\eqn{QC}{z^\a\rho_\a-w^\ad\pi_\ad\ =\ 0,}
where $[z^\a,\pi_\ad]$ are homogeneous coordinates on $\IP^3$ and
$[\rho_\a,w^\ad]$ on $\IP^3_*$, respectively. Eq. \eqref{QC} is just
the orthogonality relation between the vectors characterizing $S_1$ 
and the ones characterizing $S_3$, the latters being normal vectors
to the hyperplanes. The minus sign in \eqref{QC} has been chosen
for convenience.

\section{Supermanifolds}

So far, we have discussed twistor spaces in the purely even (bosonic)
setting. In the sequel, we extend the discussion to supertwistor
spaces. To do this, let us first present some preliminaries. 

\paragraph{Graded rings and modules.}\label{Rings}
The first notions we need are
$\IZ_2$-rings and modules and some relations among them. 
So, let $R\cong R_0\oplus R_1$ be a $\IZ_2$-graded
ring, that is, $R_0R_0\subset R_0$, $R_1R_0\subset R_1$, 
$R_0R_1\subset R_1$
and $R_1R_1\subset R_0$. We call elements of $R_0$ 
even and elements of $R_1$ odd. 
An element of $R$ is said to be homogeneous if it is either even or odd. 
The 
degree (or parity) of a homogeneous element is defined to be zero if it 
is even and one if it is odd,
respectively. We denote the degree of a homogeneous element $r\in R$ by
$p_r$ ($p$ for parity). We define the supercommutator, 
$[\cdot,\cdot\}\,:\,R\times R\to R$, by
\eqn{}{[r_1,r_2\}\ :=\ r_1r_2-(-)^{p_{r_1}p_{r_2}}r_2r_1,}
for all homogeneous elements $r_{1,2}\in R$.
The ring $R$ is called supercommutative if the supercommutator
vanishes for all of the ring's elements.
An important example of such $\IZ_2$-graded rings 
is the Gra{\ss}mann algebra over $\IC^n$,
\eqn{GA}{R\ =\ \Lambda^\bullet \IC^n\ :=\ \bigoplus_p\Lambda^p\IC^n,}
with the $\IZ_2$-grading being
\eqn{}{R\ =\ \bigoplus_p\Lambda^{2p}\IC^n\oplus \bigoplus_p
       \Lambda^{2p+1}\IC^n.}

An $R$-module $M$ is a $\IZ_2$-graded bimodule which satisfies
\eqn{}{rm\ =\ (-)^{p_rp_m}mr,}
for all homogeneous 
$r\in R$, $m\in M$, with $M\cong M_0\oplus M_1$. An additive
mapping of $R$-modules, $\varphi\,:\,M\to N$, is called an 
even morphism if it preserves the grading and is $R$-linear. We
denote the group of such morphisms by ${\rm Hom}_0(M,N)$. On the
other hand, we call an additive mapping of $R$-modules odd if
it reverses the grading, $p_{\varphi(m)}=p_m+1$, and is
$R$-linear, that is, $\varphi(rm)=(-)^{p_r}r\varphi(m)$ and
$\varphi(mr)=\varphi(m)r$. The group of such morphisms is denoted by 
${\rm Hom}_1(M,N)$. Then we set
$${\rm Hom}(M,N)\ :=\ {\rm Hom}_0(M,N)\oplus{\rm Hom}_1(M,N)$$ 
and it can be given an $R$-module structure.

Furthermore, there is a natural mapping $\Pi$ -- called the parity 
map -- defined by  
\eqn{}{(\Pi M)_0\ :=\ M_1\qquad{\rm and}\qquad(\Pi M)_1\ :=\ M_0}
and by requiring that i) addition in $\Pi M$ is the same as in $M$,
ii) right multiplication by $R$ is the same as in $M$ and iii) left
multiplication differs by a sign, i.e., $r\Pi(m)=(-)^{p_r}\Pi(rm)$
for $r\in R$, $m\in M$ and $\Pi(m)\in\Pi M$. Corresponding to the
morphism $\varphi\,:\,M\to N$, we let $\varphi^\Pi\,:\,\Pi M\to\Pi N$
be the morphism which agrees with $\varphi$ as a mapping of 
sets.
Moreover, corresponding to the morphism $\varphi\,:\,M\to N$, we
can find morphisms
\eqna{}{\Pi\varphi\,:\,M\ \to\ \Pi N,\qquad&{\rm with}&\qquad
        (\Pi\varphi)(m)\ &:=\ \Pi(\varphi(m)),\\
        \varphi\Pi\,:\,\Pi M\ \to\ N,\qquad&{\rm with}&\qquad
        (\varphi\Pi)(\Pi m)\ &:=\ \varphi(m),}
and hence $\varphi^\Pi=\Pi\varphi\Pi$. 
We stress that $R$ is an $R$-module itself, and as such $\Pi R$ 
is, as well. However, $\Pi R$ is no longer a $\IZ_2$-graded ring 
since
$(\Pi R)_1(\Pi R)_1\subset(\Pi R)_1$, for instance.

Next we need the notion of free $R$-modules. 
A free $R$-module of rank $m|n$ is defined to be an $R$-module
isomorphic to
\eqn{}{R^{m|n}\ :=\ R^m\oplus(\Pi R)^n.} 
This has a free system of generators, $m$ of which are even and $n$ of
which are odd, respectively. We stress that the decomposition of
$R^{m|n}$ into $R^{m|0}$ and 
$R^{0|n}$ has, in general, no invariant meaning and
does not coincide with the decomposition into even and odd parts,
$$ [R_0^m\oplus(\Pi R_1)^n]\oplus[R_1^m\oplus(\Pi R_0)^n].$$
Only when $R_1=0$, these decompositions are the same. 

\paragraph{Supermatrices, supertranspose, supertrace and superdeterminant.}
Let $R$ be a supercommutative ring and $R^{m|n}$ be a freely generated
$R$-module. Just as in the commutative case, morphisms between free
$R$-modules can be given by matrices. The standard matrix format
is 
\eqn{SF}{A\ =\ \begin{pmatrix} A_1&A_2\\ A_3&A_4\end{pmatrix},}
where $A$ is said to be even (respectively, odd) if $A_1$ and $A_4$ are
filled with even (respectively, odd) elements of the ring while
$A_2$ and $A_3$ are filled with odd (respectively, even) elements.
Furthermore, $A_1$ is a $p\times m$-, $A_2$ a $q\times m$-, $A_3$ 
a $p\times n$- and $A_4$ a $q\times n$-matrix. The set of matrices
in standard format with elements in $R$ is denoted by 
${\rm Mat}(m|n,p|q,R)$. It forms a $\IZ_2$-graded module which,
with the usual matrix multiplication, is naturally isomorphic
to ${\rm Hom}(R^{m|n},R^{p|q})$.  
We denote by $GL(m|n,R)$ the even invertible automorphisms of $R^{m|n}$.

The supertranspose of $A\in{\rm Mat}(m|n,p|q,R)$ is defined according to
\eqn{}{^{{\rm st}}\!\!A\ :=\ 
                            \begin{pmatrix}
                              ^{\rm t}\!A_1& (-)^{p_A}\ \!^{\rm t}\!A_3\\
                              -(-)^{p_A}\ \!^{\rm t}\!A_2 &^{\rm t}\!A_4
                            \end{pmatrix},
}
where the superscript ``t" denotes the usual transpose. The 
supertransposition satisfies 
$^{\rm st}\!(A+B)=\ \!\!^{\rm st}\!\!A+\ \!\!^{\rm st}\!B$ and
$^{\rm st}\!(AB)=(-)^{p_Ap_B}\ \!^{\rm st}\!B\ \!^{\rm st}\!\!A$.

We shall use the following definition of the supertrace of 
$A\in{\rm Mat}(m|n,p|q,R)$:
\eqn{}{{\rm str}A\ :=\  \tr A_1-(-)^{p_A}\tr A_4.
}
Clearly, the supertrace of the supercommutator vanishes and the
supertrace of a supertransposed matrix is the same as the supertrace
of the matrix one has started with. 

Finally, let $A\in GL(m|n,R)$. The superdeterminant is given by
\eqn{}{{\rm sdet}A\ :=\ \det(A_1-A_2A_4^{-1}A_3)\det A_4^{-1},}
where the right-hand side is well-defined for $A_1\in GL(m|0,R_0)$
and $A_4\in GL(n|0,R_0)$. Furthermore, it belongs to $GL(1|0,R_0)$,
that is, to $R_0$. The superdeterminant satisfies also the
usual rules,
$${\rm sdet}(AB)\ =\ {\rm sdet}A\ {\rm sdetB}\qquad{\rm and}\qquad
{\rm sdet}\,^{\rm st}\!A\ =\ {\rm sdet}A
$$
for $A,B\in GL(m|n,R)$.

\paragraph{Supermanifolds.}\label{SM} 
We have now all necessary ingredients to give the definition of 
a supermanifold. As it will be most convenient for us,
we shall follow Manin \cite{Manin} and 
consider the graded space approach to supermanifolds.\footnote{We note that
the systematic theory of smooth supermanifolds goes back to the 
work by Kostant \cite{Kostant} and Leites \cite{Leites:1980}.}
The idea is roughly to extend the structure sheaf of a manifold to 
a sheaf of supercommutative rings. However, note that 
there are other approaches as, e.g., 
the one by DeWitt \cite{DeWittCY,Bartocci} (see also 
\cite{Cartier:2002zp} for a review). Furthermore, we stress in
advance that the subsequent definition will not be the most
generic one. For a more 
general discussion, we refer the interested reader to Ref. \cite{Manin}.

\pagebreak[4]
{\Def A complex supermanifold of complex dimension $m|n$ is a ringed space
$X^{m|n}:=(X,\CO_X)$, where $X$ is a topological space of real dimension
$2m$, and $\CO_X$ is a sheaf of supercommutative rings on $X$ 
such that, if we let $\CN$ be the ideal subsheaf in $\CO_X$ of all
nilpotent elements in $\CO_X$, the following is fulfilled:
\vspace*{-2mm}\begin{itemize}
\setlength{\itemsep}{-1mm}
\item[{\rm (i)}] $X_{\rm red}=(X,\CO_{\rm red}:=\CO_X/\CN)$ is a 
complex manifold.\footnote{Recall that a ringed space $(X,\CO_X)$ 
with the property that for each $x\in X$ 
there is a neighborhood $U \ni x$ such that there is a ringed space 
isomorphism $(U,\CO_X|_U)\cong(V,\CO_V)$, where $V\subset\IC^m$ and 
$\CO_V$ is the sheaf of holomorphic functions on $V$ can be given 
the structure of a complex manifold. Moreover, any 
complex manifold arises in this manner.} 
\item[{\rm (ii)}] For each point $x\in X$ there is a neighborhood 
       $U\ni x$ such that
      $$ \CO_X|_U\ \cong\ \CO_{\rm red}(\Lambda^\bullet\CE)|_U, $$
       where $\CE$ is a locally free sheaf of $\CO_{\rm red}$-modules 
       of purely even rank $n|0$ on $X$ and $\Lambda^\bullet$ denotes 
       the exterior algebra, i.e.,
       the tensor algebra modulo the ideal generated by the 
       superanticommutator.
\end{itemize}
}\vskip 4mm

We shall call $X_{\rm red}$ the body of $(X,\CO_X)$. 
If no confusion arises, we will not mention the sheaf $\CO_X$ 
explicitly. 
The latter is also called the sheaf of holomorphic 
superfunctions on $X$ or simply structure sheaf of the supermanifold.
Note that a similar definition can also be given for  
differentiable supermanifolds. Furthermore, as for purely even complex 
manifolds, it will turn out to be useful to consider the sheaf 
$\CS_X$ of smooth complex-valued superfunctions on $X$, which
is defined in a similar fashion. Clearly, $\CO_X$ is a subsheaf
of $\CS_X$. In the sequel, we shall loosely speak of functions but
always mean superfunctions. Moreover, we will often
suppress the explicit appearance of the dimensionality and
simply write $X$ instead of $X^{m|n}$. 

Let  $(z^1,\ldots,z^m)$ be local coordinates on $U\subset X$ 
and $(\eta_1,\ldots,\eta_n)$ be basis sections of
$\CE$. Then $(z^1,\ldots,z^m,\eta_1,\ldots,\eta_n)$ is an even-odd
system of coordinates for the space 
$(U,\CO_{\rm red}(\Lambda^\bullet\CE)|_U)$. Any
$f\in\Gamma(U,\CO_{\rm red}(\Lambda^\bullet\CE))$ can thus be 
expressed as
\eqn{}{f(z,\eta)\ =\ \sum_I\eta^I f_I(z),}
where $I$ is a multiindex and the $f_I$s are local functions on 
$X_{\rm red}$. 

Let us give a first example of a supermanifold:
let $E\to X$ be a
rank $r$ holomorphic vector bundle over an ordinary complex 
manifold $X$ and $\CE$ the sheaf of sections of 
$E$.$^{\rm \ref{fo}}$ Then $(X,\CO_{\rm red}(\Lambda^\bullet\CE))$ 
is a supermanifold by the very definition, since $\CE$ is locally 
free. Such a supermanifold is called globally split.
A particular example is $\IC^{m|n}$ defined by 
$\IC^{m|n}=(\IC^m,\CO_{\rm red}(\Lambda^\bullet\IC^n))$.
Note that due to a theorem by Batchelor \cite{Batchelor}, 
any differentiable supermanifold is globally split.
This is basically because of the existence of a partition of unity. 
The reader should be warned that, in general, 
complex supermanifolds are not globally split.

\paragraph{Vector bundles.}\label{VectBund-1}
As for ordinary manifolds, we need the notions of tangent and cotangent
bundles. Since these are just special vector bundles and in view of our
later discussion it is certainly worth stating some more general words 
on vector bundles. The idea
to define a holomorphic supervector bundle over a complex
supermanifold $(X,\CO_X)$ is to 
use the notion of a locally free sheaf
of rank $r|s$, which is defined to be a sheaf of $\CO_X$-modules
locally isomorphic to 
$\CO_X^{r|s}:=\CO_X^{\oplus r}\oplus(\Pi\CO_X)^{\oplus s}$.\footnote{\label{fo}Remember
the following fact: Let $X$ be a purely even complex manifold. 
Then there is a one-to-one
correspondence between rank $r$ locally free sheaves $\CE$ 
of $\CO_X$-modules
on $X$ and rank $r$ holomorphic vector bundles $E$ over $X$. 
In fact, if $\{U_i\}$ is an open
covering of $X$ and if $\{U_i\}$ trivializes $E\to X$, then the 
group of
holomorphic sections of $E$ over $U_i$ is $\CO_X^{\oplus r}|_{U_i}$. 
Conversely,
if $\CE$ is a locally free sheaf of $\CO_X$-modules and if 
$\phi_i\,:\,\CE|_{U_i}\to\CO_X^{\oplus r}|_{U_i}$ is the explicit 
isomorphism, then on any $U_i\cap U_j$
we may define $r\times r$ matrices of holomorphic functions according to
$\phi_j\circ\phi_i^{-1}\,:\,\CO_X^{\oplus r}|_{U_i\cap U_j}\to
\CO_X^{\oplus r}|_{U_i\cap U_j}$ which give the transition functions of
a holomorphic vector bundle $E\to X$.}    
In the following, we shall suppress the prefix super: thus, vector bundle
means supervector bundle.
Next let $\varphi\,:\,(Y,\CO_Y)\to(X,\CO_X)$ be a holomorphic mapping of 
complex supermanifolds, that is, $\varphi$ is defined to be the
pair $(\phi,\tilde{\phi})$, where $\phi\,:\,Y\to X$ is a continuous 
mapping of
topological spaces and $\tilde{\phi}\,:\,\CO_X\to\varphi_*\CO_Y$ is a 
morphism
of sheaves of rings: for any open subset $U$ in $X$ there is a
morphism $\tilde{\phi}_U\,:\,\CO_X|_U\to\CO_Y|_{\phi^{-1}(U)}$. Note that
$\varphi_*\CO_Y$ is also called the zeroth direct image sheaf of the
sheaf $\CO_Y$. By the pull-back
via $\varphi$ of a vector bundle $\CE$ over $X$, we mean the
locally free sheaf of $\CO_Y$-modules
\eqn{}{\varphi^*\CE\ :=\ \CO_Y\otimes_{\varphi^{-1}\CO_X}\varphi^{-1}\CE,}
where $\varphi^{-1}\CO_X$ (respectively, $\varphi^{-1}\CE$) denotes the
topological inverse sheaf of $\CO_X$ (respectively, of $\CE$). 
For instance, $\varphi^{-1}\CO_X$ is the
sheaf on $Y$ defined by the presheaf
$$Y\supset V\ {\rm open}\ \mapsto\ \Gamma(\phi(V),\CO_X)$$
with the obvious restriction mappings. It is characterized
by the property that the stalk $(\varphi^{-1}\CO_X)_y=\CO_X|_{\phi(y)}$
for all $y\in Y$. For a complex vector bundle, replace 
$\CO_X$ by $\CS_X$. 

The tangent sheaf $TX$ of a complex supermanifold $X$ is then the sheaf 
of local vector fields, that is, the superderivations of the ring
of functions. It is locally free -- in fact, it is a locally free sheaf
of $\CO_X$-modules -- and of rank equal to the dimension of
$X$. Let $(z^i,\eta_j)$ be local coordinates on $X$. Then $TX$ is
freely generated by its sections 
$(\partial/\partial z^i,\partial/\partial\eta_j)$.
The cotangent sheaf $\O^1(X)$ is defined to be 
\eqn{}{\O^1(X)\ :=\ T^*X\ =\ \cH om_{\CO_X}(TX,\CO_X),}
where $\cH om_{\CO_X}(TX,\CO_X)$ is the sheaf of local morphisms from 
$TX\to\CO_X$. 
There is also an obvious differential ${\rm d}\,:\,\CO_X\to\O^1(X)$ with 
the property $X\lrcorner\,{\rm d}f=Xf$ for $X\in TX$.
\paragraph{Remark.}
In the sequel, we shall only be dealing with
locally free sheaves. Therefore, we will often not make any notational
distinction between such sheaves and the vector bundles corresponding
to them, and we also use the two objects interchangeably. 

\section{Supertwistor spaces}\label{sec:STS-ch1}

Now we want to use the above machinery for generalizing
the twistor spaces of Sec. \ref{TS} to supertwistor spaces.

\paragraph{Flag supermanifolds.} For the sake of
concreteness, let us consider $\IC^{m|n}$. Then we 
define the flag superspace\footnote{See also Ref. \cite{Howe:1995md}.} 
$F_{d_1\cdots d_k}(\IC^{m|n})$ to be the
set of all $k$-tuples $(S_1,\ldots,S_k)$ of free submodules of 
$\IC^{m|n}$
satisfying $S_1\subset\cdots\subset S_k\subset\IC^{m|n}$ and 
$d_i:={\rm rank}\,S_i=p_i|q_i$. This naturally generalizes our 
definition \eqref{flag}. In fact, one can also introduce suitable 
coordinate systems and thus a suitable structure sheaf which makes 
this set into a complex supermanifold. 
As in the purely even setting, we can give
an equivalent definition by considering the decomposition
$$\IC^{d_1}\oplus\IC^{d_2-d_2}\oplus\IC^{d_3-d_2}\oplus
  \cdots\oplus\IC^{d_k-d_{k-1}}\oplus\IC^{m|n-d_k}$$
of $\IC^{m|n}$, where $d_i-d_j=(p_i-p_j)|(q_i-q_j)$.
Let now $U(m|n)\subset GL(m|n,\IC)$ be the unitary automorphisms
of $\IC^{m|n}$.\footnote{Hermitian conjugation is defined by 
composing supertransposition and complex conjugation.}
The subgroup of $U(m|n)$ which preserves this decomposition is
$$U(d_1)\times U(d_2-d_1)\times\cdots\times U(d_k-d_{k-1})\times
U(m|n-d_k)\ \subset\ U(m|n)$$
and hence we may write the quotient
\eqn{flag3}{F_{d_1\cdots d_k}(\IC^{m|n})\ :=\ 
  \frac{U(m|n)}{U(d_1)\times U(d_2-d_1)
  \times\cdots\times U(d_k-d_{k-1})\times U(m|n-d_k)}.}
One can actually show that $U(m|n)$ has real dimension
$m^2+n^2|2mn$. Then it is a rather straightforward exercise
to determine the dimension of a flag supermanifold. So,
we leave it to the interested reader. Note that flag 
supermanifolds are, in general, not globally split \cite{Manin}.

\paragraph{Supertwistor space.}\label{STS}
Similarly to our discussion in Sec. \ref{TS}, 
we let $\IT$ be of the form $\IC^{4|\CN}$. 
We call this space 
$\CN$-extended supertwistor space \cite{Ferber:1977qx}. Then we introduce
\eqna{}{\IP^{3|\CN}\ &:=\ F_{1|0}(\IT),\\
        \IM^{4|2\CN}_R\ &:=\ F_{2|0}(\IT),\\
        \IF^{5|2\CN}_R\ &:=\ F_{1|0,2|0}(\IT).} 
This time we call them projective supertwistor space, 
compactified complexified four-dimensional antichiral superspace-time 
and correspondence space, respectively. Therefore, the double fibration
\eqref{DF1} becomes:
\eqna{DF3}{
 \begin{picture}(50,40)
  \put(0.0,0.0){\makebox(0,0)[c]{$\IP^{3|\CN}$}}
  \put(64.0,0.0){\makebox(0,0)[c]{$\IM^{4|2\CN}_R$}}
  \put(34.0,33.0){\makebox(0,0)[c]{$\IF^{5|2\CN}_R$}}
  \put(7.0,18.0){\makebox(0,0)[c]{$\pi_1$}}
  \put(55.0,18.0){\makebox(0,0)[c]{$\pi_2$}}
  \put(25.0,25.0){\vector(-1,-1){18}}
  \put(37.0,25.0){\vector(1,-1){18}}
 \end{picture}
}
Note that $\IP^{3|\CN}=\IC P^{3|\CN}=(\IP^3,\CO_{\rm red}
(\Lambda^\bullet(\IC^\CN\otimes\CO_{\IP^3}(-1)))),$
where $\CO_{\IP^3}(-1)$ is the tautological sheaf on $\IP^3$.
Furthermore, Prop. \ref{GTC} generalizes accordingly.
{\Pro There is the following geometric correspondence:
\vspace*{-2mm}
\begin{center} 
 \begin{tabular}{cccc}
   {\rm (i)} & point in $\IP^{3|\CN}$ & $\quad
       \longleftrightarrow\quad$ & 
             $\IC P^{2|\CN}\subset\IM^{4|2\CN}_R$\\
   {\rm (ii)} & $\IC P^{1|0}\subset\IP^{3|\CN}$
             &$\quad\longleftrightarrow\quad$ & 
                        point in $\IM^{4|2\CN}_R$
 \end{tabular}
\end{center}
}\vskip 4mm

In the following, we shall abbreviate $X^{m|0}\equiv X^m$ for 
any ordinary manifold $X$.
As before, we may introduce local coordinates. Eventually, one
finds
\eqna{DF4}{
 \begin{picture}(50,40)
  \put(0.0,0.0){\makebox(0,0)[c]{$\CP^{3|\CN}$}}
  \put(64.0,0.0){\makebox(0,0)[c]{$\CM^{4|2\CN}_R$}}
  \put(34.0,33.0){\makebox(0,0)[c]{$\CF^{5|2\CN}_R
            \cong\CM^{4|2\CN}_R\times
                \IC P^1$}}
  \put(7.0,18.0){\makebox(0,0)[c]{$\pi_1$}}
  \put(55.0,18.0){\makebox(0,0)[c]{$\pi_2$}}
  \put(25.0,25.0){\vector(-1,-1){18}}
  \put(37.0,25.0){\vector(1,-1){18}}
 \end{picture}
}
together with
\begin{itemize}
\setlength{\itemsep}{-2mm}

\item$\CM^{4|2\CN}_R\cong\IC^{4|2\CN}$: $x_R^{\a\ad}$ and $\eta^\ad_i$,
                           where $i=1,\ldots,\CN$,
\item $\CF^{5|2\CN}_R$: 
            $x_R^{\a\ad}$, $\eta^\ad_i$ and $\l_\pm$ with 
            $\l_+=\l_-^{-1}$ on
           $U_+\cap U_-$, where $\{U_+,U_-\}$ denotes again the 
           canonical 
          covering of $\IC P^1$,
\item $\CP^{3|\CN}$: $z^\a_\pm$, $z^3_\pm$ and $\eta^\pm_i$; 
          $\pi_1:(x_R^{\a\ad},\l_\pm,\eta^\ad_i)
                \mapsto(z^\a_\pm=x_R^{\a\ad}\l_\ad^\pm,
                  z^3_\pm=\l_\pm,\eta^\pm_i=\eta^\ad_i\l^\pm_\ad)$,
          with $\l_\ad^+=\l_+\l_\ad^-$ and $(\l_\ad^+):=\ \!^t(1,\l_+)$.
\end{itemize}

Again, the last point shows that $\CP^{3|\CN}$ 
is a rank $2|\CN$ holomorphic vector bundle
\eqn{holofibI}{\CP^{3|\CN}\ =\ \CO_{\IC P^1}(1)\otimes\IC^2
              \oplus\Pi\CO_{\IC P^1}(1)
              \otimes\IC^\CN.}
Moreover, the relations
\eqn{IRII}{z^\a_\pm\ =\ x_R^{\a\ad}\l_\ad^\pm\qquad{\rm and}\qquad
       \eta_i^\pm\ =\ \eta^\ad_i\l^\pm_\ad}
explicitly say that a fixed point 
$p=(z^\a_\pm,z^3_\pm,\eta^\pm_i)\in\CP^{3|\CN}$ corresponds to a null
$2|\CN$-dimensional subspace of $\CM^{4|2\CN}_R$. By analogy with
the purely even setting, we shall refer to those as  
super null planes of dimension $2|\CN$ in $\CM^{4|2\CN}_R$. 
Hence, $\CP^{3|\CN}$ is the space of super null planes of
$\CM^{4|2\CN}_R\cong\IC^{4|2\CN}$.
On the other hand, a fixed point
$(x_R,\eta)=(x_R^{\a\ad},\eta^\ad_i)\in\CM^{4|2\CN}_R$ 
corresponds to a holomorphic embedding of a rational curve 
$\IC P^1_{x_R,\eta}\hookrightarrow\CP^{3|\CN}$. Before closing this
paragraph, we would like to stress that the argumentation of \ref{RI}~~ 
also applies here. 

\paragraph{Two other supertwistor spaces.}\label{ATTSII}
Above, we have introduced the (projective) supertwistor 
space. Let us now also generalize the other two twistor spaces
given in \ref{ATTS}\kern6pt:
the dual projective supertwistor space and the projective
superambitwistor space, respectively. So, 
let us consider the following two double fibrations:
\eqna{DF5}{
 \begin{picture}(50,40)
  \put(0.0,0.0){\makebox(0,0)[c]{$F_{3|\CN}(\IT)$}}
  \put(64.0,0.0){\makebox(0,0)[c]{$F_{2|\CN}(\IT)$}}
  \put(34.0,33.0){\makebox(0,0)[c]{$F_{2|\CN,3|\CN}(\IT)$}}
  \put(7.0,18.0){\makebox(0,0)[c]{$\pi_1$}}
  \put(55.0,18.0){\makebox(0,0)[c]{$\pi_2$}}
  \put(25.0,25.0){\vector(-1,-1){18}}
  \put(37.0,25.0){\vector(1,-1){18}}
 \end{picture}
\hspace*{4cm}
\begin{picture}(50,40)
  \put(0.0,0.0){\makebox(0,0)[c]{$F_{1|0,3|\CN}(\IT)$}}
  \put(64.0,0.0){\makebox(0,0)[c]{$F_{2|0,2|\CN}(\IT)$}}
  \put(34.0,33.0){\makebox(0,0)[c]{$F_{1|0,2|0,2|\CN,3|\CN}(\IT)$}}
  \put(7.0,18.0){\makebox(0,0)[c]{$\pi_1$}}
  \put(55.0,18.0){\makebox(0,0)[c]{$\pi_2$}}
  \put(25.0,25.0){\vector(-1,-1){18}}
  \put(37.0,25.0){\vector(1,-1){18}}
 \end{picture}
}  
The space $\IM^{4|2\CN}_L:=F_{2|\CN}(\IT)$ is compactified 
complexified
four-dimensional chiral superspace-time, which is naturally 
dual to $F_{2|0}(\IT^*)$. Note also that the space 
$F_{3|\CN}(\IT)$ is naturally dual to $F_{1|0}(\IT^*)$. 
For that reason we call $F_{3|\CN}(\IT)$ dual projective 
supertwistor space and denote it by $\IP^{3|\CN}_*$. 
Furthermore, $\IM^{4|4\CN}:=F_{2|0,2|\CN}(\IT)$ is compactified
complexified superspace-time and obviously,
we have the natural fibration:
\eqna{DF6}{
 \begin{picture}(50,40)
  \put(0.0,0.0){\makebox(0,0)[c]{$\IM_R^{4|2\CN}$}}
  \put(64.0,0.0){\makebox(0,0)[c]{$\IM_L^{4|2\CN}$}}
  \put(34.0,33.0){\makebox(0,0)[c]{$\IM^{4|4\CN}$}}
  \put(7.0,18.0){\makebox(0,0)[c]{$\pi_1$}}
  \put(55.0,18.0){\makebox(0,0)[c]{$\pi_2$}}
  \put(25.0,25.0){\vector(-1,-1){18}}
  \put(37.0,25.0){\vector(1,-1){18}}
 \end{picture}
}
The manifold $F_{1|0,3|\CN}(\IT)$ is called projective 
superambitwistor space and we shall denote it by $\IL^{5|2\CN}$. 
Then one can show: 
{\Pro\label{GTC3} There are the following correspondences:
\vspace*{-2mm}
\begin{center} 
 \begin{tabular}{lcccc}
   {\rm (a)} &\kern.5cm {\rm (i)} & 
     point in $\IP^{3|\CN}_*$ & $\quad\longleftrightarrow\quad$ & 
                     $\IC P^{2|\CN}\subset\IM^{4|2\CN}_L$\\
   & \kern.5cm{\rm (ii)} & 
    $\IC P^1_*\subset\IP^{3|\CN}_*$&$\quad\longleftrightarrow\quad$ & 
                                    point in $\IM^{4|2\CN}_L$\\
 {\rm (b)}&\kern.5cm {\rm (i)} & 
      point in $\IL^{5|2\CN}$ & $\quad\longleftrightarrow\quad$ & 
                         $\IC P^{1|2\CN}\subset\IM^{4|4\CN}$\\
   &\kern.5cm {\rm (ii)} & 
        $\IC P^1\times\IC P^1_*\subset\IL^{5|2\CN}$&
    $\quad\longleftrightarrow\quad$ & 
                        point in $\IM^{4|4\CN}$
 \end{tabular}
\end{center}
}\vskip 4mm

Note that for all these manifolds we can construct coordinate 
representations similar to those given in \ref{STS}\kern6pt. 
As before, we shall develop suitable notation for these 
coordinates as we need them. 

Finally, we want to stress that also the projective 
superambitwistor space
$\IL^{5|2\CN}$ can be viewed as a degree two hypersurface in 
$\IP^{3|\CN}\times\IP^{3|\CN}_*$. In fact, if we let
$[z^\a,\pi_\ad,\eta_i]$ be homogeneous coordinates on $\IP^{3|\CN}$ 
and $[\rho_\a,w^\ad,\theta^i]$ on $\IP^{3|\CN}_*$, respectively, 
then $\IL^{5|2\CN}$
is given by $\IL^{5|2\CN}=(\IL^5,\CO_{\IL^{5|2\CN}}=
\CO_{\IP^{3|\CN}\times\IP^{3|\CN}_*}/\CI)$, with 
\eqn{}{\CO_{\IP^{3|\CN}\times\IP^{3|\CN}_*}\ 
         =\ \CO_{\rm red}(\Lambda^\bullet(
       \IC^\CN\otimes\CO_{\IP^{3}\times\IP^{3}_*}(-1,0)\oplus 
       \IC^{\CN*}\otimes\CO_{\IP^{3}\times\IP^{3}_*}(0,-1)))}
and $\CI$ is the ideal subsheaf in 
$\CO_{\IP^{3|\CN}\times\IP^{3|\CN}_*}$ given by
\eqn{eq:is-ch1}{\CI\ =\ 
       \langle z^\a\rho_\a-w^\ad\pi_\ad+2\theta^i\eta_i\rangle.}
Furthermore, we have abbreviated
\eqn{}{\CO_{\IP^3\times\IP^3_*}(m,n)\ :=\ 
     {\rm pr}_1^*\CO_{\IP^3}(m)\otimes{\rm pr}_2^*\CO_{\IP^3_*}(n),}
where ${\rm pr_1}\,:\,\IP^3\times\IP^3_*\to\IP^3$ and
${\rm pr_2}\,:\,\IP^3\times\IP^3_*\to\IP^3_*$, respectively.
 
\section{Connections and curvature}\label{sec:CAC-Chap1}

So far, we have been dealing with just supermanifolds and 
vector bundles. Next we
want to introduce some additional geometric structure, such as  
connections. This will then allow us to talk about
curvature and characteristic classes. 

\paragraph{Connections and curvature.}\label{CCAAT}
As we have already mentioned in \ref{VectBund-1}\kern6pt, the notion
of a holomorphic vector bundle over a complex supermanifold $(X,\CO_X)$
is equivalent to the notion of a locally free sheaf $\CE$ of $\CO_X$-modules. 
Then a connection $\nabla$ is an even morphism of sheaves  
\eqn{}{\nabla\,:\,\CE\ \to\ \O^1(X)\otimes\CE}
satisfying the Leibniz formula 
\eqn{Leibniz}{\nabla(f\sigma)\ =\ {\rm d}f\otimes\sigma+
     f\nabla\sigma,}
where $f$ is a local holomorphic function on $X$ and $\sigma$ 
is a local section of $\CE$.
Let $(Z^I)=(z^i,\eta_j)$ be local coordinates on $X$ and
$TX$ be generated by 
$(\partial/\partial Z^I)=
(\partial/\partial z^i,\partial/\partial\eta_j)$.
Therefore,
\eqn{}{{\rm d}\ =\ {\rm d}Z^I\partial_I
              \ =\ {\rm d}z^i\frac{\partial}{\partial z^i}+
                   {\rm d}\eta_j\frac{\partial}{\partial\eta_j}}
and \eqref{Leibniz} reads as
\eqn{}{\nabla_I(f\sigma)\ =\ (\partial_If)\sigma+(-)^{p_Ip_f}
      f\nabla_I\sigma.}
Locally, $\nabla$ has the form
\eqn{}{\nabla\ =\ {\rm d}+\CA,}
where $\CA\in\Gamma(X,\O^1(X)\otimes {\rm End}\,\CE)$. 

As usual, $\nabla^2$ induces the curvature 
\eqn{}{\CF\ \in\ \Gamma(X,\Lambda^2\O^1(X)\otimes {\rm End}\,\CE),}
where $\Lambda^\bullet$ denotes the exterior algebra, i.e., it 
is the tensor algebra modulo the ideal generated by
the superanticommutator. Here, we have just introduced one
version of a holomorphic de Rham complex on supermanifolds. 
There are
other possible ways; see, e.g., Ref. \cite{Manin} for details. 
Note that the above definitions carry naturally over to 
complex vector bundles.

\paragraph{Integral forms and Berezin integral.}\label{par:IF-BI-ch1}
In the purely even setting, differential forms are objects which 
can be integrated over. However, in the case of supermanifolds, the 
situation is more subtle. First, let us introduce the holomorphic 
Berezinian. Let $X$ be a complex
supermanifold with tangent sheaf $TX$. The holomorphic Berezinian 
line bundle, 
${\rm Ber}\,X$, or holomorphic Berezinian for short, 
is defined to be the line bundle over $X$ having holomorphic super 
Jacobians as transition functions. 
Thus, it can be considered as the natural extension of the 
canonical sheaf (sheaf of sections of the canonicle bundle) on a 
purely even complex manifold.  
Then integral forms are defined to be sections of the sheaf
\eqn{}{\Sigma^\bullet(X)\ :=\ {\rm Ber}\,X\otimes\Lambda^\bullet(TX)
       \ =\ \cH om_{\CO_X}(\Lambda^\bullet\O^1(X),{\rm Ber}\,X).}
Thus, ${\rm Ber}\,X=\Sigma^0(X)$ and if $X$ is $m|n$ dimensional 
with local coordinates $(z^i,\eta_j)$, then $\o\in{\rm Ber}\,X$ is
locally of the form
\eqn{}{\o\ =\ {\rm d}z^1\wedge\cdots\wedge{\rm d}z^m{\rm d}\eta_1
               \cdots{\rm d}\eta_n\otimes f,}
where $f$ is some local section of $\CO_X$. Sections of ${\rm Ber}\,X$ 
are also called holomorphic volume forms.
A complexification $T_\IC X$ of $TX$ splits into a direct sum 
$T_\IC X\cong T^{1,0}_\IC X\oplus T^{0,1}_\IC X$. Clearly, this 
then implies that
\eqna{splitt}{\Sigma^k_\IC(X)\ &=\ {\rm Ber}_\IC\,X\otimes
                                 \Lambda^k(T_\IC X)\\
        &=\ \bigoplus_{p+q=k}{\rm Ber}_\IC^{1,0}\,X
        \otimes\Lambda^p(T_\IC^{1,0}X)
        \otimes {\rm Ber}_\IC^{0,1}\,X\otimes
        \Lambda^q(T_\IC^{0,1}X)\\
        &=:\ \bigoplus_{p+q=k}\Sigma^{p,q}_\IC(X).
}

Let now $X$ be a differentiable supermanifold with local coordinates
$(x^i,\eta_j)$. Furthermore, suppose that $X_{\rm red}$ is connected and
given an orientation. Let ${\rm Ber}_0\,X=\Gamma_0(X,{\rm Ber}\,X)$, that
is, the volume forms with compact support. Then $\o\in{\rm Ber}_0\,X$
is locally
\eqna{}{\o\ &=\ {\rm d}x^1\wedge\cdots\wedge{\rm d}x^m{\rm d}\eta_1\cdots
              {\rm d}\eta_n\otimes\sum_I\eta^If_I(x)\\
            &=\ {\rm d}x^1\wedge\cdots\wedge{\rm d}x^m{\rm d}\eta_1\cdots
              {\rm d}\eta_n\otimes
            \sum_{\{|I_i|\leq1\}}\eta_1^{I_1}\cdots\eta_n^{I_n}
            f_{I_1\cdots I_n}(x), }
where the $f_I$s are local functions on $X_{\rm red}$. We define
the Berezin integral as
\eqn{}{\int_X\o\ :=\ (-)^{\frac{n}{2}(n-1)}\int_{X_{\rm red}}
            {\rm d}x^1\cdots{\rm d}x^mf_{1\cdots 1}(x).}
For a more general volume form with compact support, we define the
integral by using a partition of unity and the additivity property of 
the integral. Furthermore,
if we let $\Sigma^k_0(X)$ be the integral $k$-forms with compact support,
then such a form can be hooked into a differential $k$-form with compact 
support, i.e., an element of 
$\Lambda^k\O^1_0(X)$, to give an element of ${\rm Ber}_0\,X$ which,
by virtue of the above formula, can be integrated over $X$. 

\paragraph{Remark.}
Two things are worth mentioning. First of all, integral forms can be 
given the structure of a complex, since one may introduce a mapping
$\d\,:\,\Sigma^\bullet(X)\to\Sigma^\bullet(X)$ with $\d^2=0$. Therefore, 
one can discuss cohomology of integral forms \cite{Manin}. 
Another issue concerns the
integration theory on supermanifolds. In general, one is interested
in objects which can be integrated over immersed sub(super)manifolds $Y$
of some supermanifold $X$. One way of doing this is to introduce
so-called $k$-densities. They are sections of $({\rm Ber}\,\Pi\CT^*)^*$,
where $\CT$ is the tautological sheaf on the relative Gra{\ss}mannian
$G_X(k,TX)\to X$. Then if $\varphi\,:\,Y\to X$ is an immersion of
supermanifolds and $\o$ a density on $X$, one can canonically define 
a volume form $\varphi^*(\o)$ on $Y$. We shall not go into further
details at this point
and refer the reader to the book by Manin \cite{Manin} for a nice 
discussion
of these aspects. We also refer to the work by Bernstein and Leites
\cite{Bernstein:1977A,Bernstein:1977B} who generalized the integration
theory of volume forms for the first time.

\paragraph{Formal Calabi-Yau supermanifolds.}\label{FCYS}
Given a rank $r|s$ complex vector bundle $(\CE,\nabla)$ over a complex
supermanifold $X$, we define the $k$-th Chern class of $\CE$ to be
\eqn{}{c_k(\CE)\ :=\ \frac{1}{k!}\left.\frac{\dt^k}{\dt t^k}\right|_{t=0}
            {\rm sdet}\left(\mathbbm{1}+t\,\o\right)
            \qquad{\rm for}\qquad k\ \leq\ r+s,}
where $\o:=\frac{1}{2\pi\i}\CF$ is (up to an overall factor) the curvature 
of $\nabla$. The first few Chern classes
are given by:
\eqna{}{c_0(\CE)\ &=\ 1,\\
        c_1(\CE)\ &=\ {\rm str}\, \o,\\
        c_2(\CE)\ &=\ \tfrac{1}{2}(({\rm str}\,\o)^2-{\rm str}\,\o^2),\\
                 &\kern6pt\vdots
}
The total Chern class is then $c(\CE)=\sum_{k=0}^{r+s}c_k(\CE)$.
Note that in a similar fashion, one may also introduce the $k$-th 
Chern character according to
\eqn{}{ch_k(\CE)\ :=\ \left.\frac{\dt^k}{\dt t^k}\right|_{t=0}
            {\rm str}\,\exp(t\,\o)
            \qquad{\rm for}\qquad k\ \leq\ r+s.}
As in the purely even setting, one may prove the following useful
result: for a short exact sequence of complex vector bundles over $X$,
\eqn{}{0\ \to\ \CF\ \to\ \CE\ \to\ \CG\ \to\ 0,}
we have 
\eqn{}{c(\CE)\ =\ c(\CF)c(\CG).}
In particular, this formula yields
\eqn{}{c_1(\CE)\ =\ c_1(\CF)+c_1(\CG).}
More details about Chern classes, Chern characters, etc. can be
found, e.g., in the book by Bartocci et al. \cite{Bartocci}.

Furthermore, the total Chern class  of a
complex supermanifold $X$ is defined to be the total Chern class
of $TX$. In this case, we shall simply write $c(X)$. 
In the purely even case, we have a relation between the first
Chern class of a vector bundle and its determinant line bundle.
In fact, both agree (up to a sign). The question is whether we 
have an extension of this relation to our present setting. 
The analog of the determinant line bundle is, as we have already
seen above, the superdeterminant
line bundle -- the Berezinian line bundle. Using splitting principle
arguments, one may indeed deduce that the first Chern class
of ${\rm sdet}\,\CE$ for some vector bundle $\CE$
coincides, again up to a sign, with the first Chern class of $\CE$, 
\eqn{}{c_1({\rm sdet}\,\CE)\ =\ \mp c_1(\CE),}
where the sign depends on whether ${\rm sdet}\,\CE$ is of
rank $1|0$ or $0|1$, respectively.
When talking about $TX$, we shall use our old notation
${\rm Ber}\,X$. 
{\Def 
Let $X$ be a complex supermanifold. Then $X$ is called a
formal Calabi-Yau supermanifold if it fulfills the following
equivalent conditions:
 \vspace*{-2mm}\begin{itemize}
\setlength{\itemsep}{-1mm}
 \item[{\rm (i)}] The first Chern class of $X$ vanishes.
 \item[{\rm (ii)}] The holomorphic Berezinian of $X$ is trivial.
 \item[{\rm (iii)}] There exists a globally defined and
             nowhere vanishing holomorphic volume form.
\end{itemize}
}\vskip 4mm

Before we will discuss some examples, let us point out
an important issue: in contrast to ordinary Calabi-Yau manifolds,
formal Calabi-Yau supermanifolds do not necessarily admit 
Ricci-flat metrics -- even if one assumes compactness. 
For some expositions on this
issue, see Refs. \cite{Sethi:1994ch} and 
\cite{Rocek:2004bi,Zhou:2004su,Rocek:2004ha,Saemann:2004tt,
Lindstrom:2005uh}.

Let us now discuss some examples.
First, consider the projective superspace
$\IC P^{m|n}=(\IC P^m,\CO_{\IC P^{m|n}})$ with 
$\CO_{\IC P^{m|n}}=\CO_{\rm red}
(\Lambda^\bullet(\IC^n\otimes\CO_{\IC P^m}(-1)))$.
To compute the total Chern class $c(\IC P^{m|n})$, we use the short 
exact sequence
\eqn{Euler}{0\ \to\ \CO_{\IC P^{m|n}}\ \to\ 
     \CO_{\IC P^{m|n}}(1)\otimes\IC^{m+1}
         \oplus\Pi\CO_{\IC P^{m|n}}(1)\otimes\IC^n\ 
\to\ T\IC P^{m|n}\ \to\ 0,}
as we have to take equivalence classes with respect to overall 
rescalings. Here, $\CO_{\IC P^{m|n}}(1)$ denotes the dual 
tautological sheaf on $\IC P^{m|n}$. Note that the above sequence 
is nothing but a $\IZ_2$-graded extension of the Euler sequence.\footnote{The
proof of \eqref{Euler} is a straightforward extension of the one given
in the purely even setting. The proof of the Euler sequence for $\IC P^m$ can
be found in, e.g., \cite{Griffith}.} 
Using our relations given in \ref{FCYS}\kern6pt, we find
$$ c(\IC P^{m|n})\ =\ c(\CO_{\IC P^{m|n}}(1)\otimes\IC^{m+1}
         \oplus\Pi\CO_{\IC P^{m|n}}(1)\otimes\IC^n)
$$
which immediately implies
$$ c_1(\IC P^{m|n})\ =\ c_1(\CO_{\IC P^{m|n}}(1)\otimes\IC^{m+1})
         -c_1(\CO_{\IC P^{m|n}}(1)\otimes\IC^n)
$$
and hence
\eqn{}{c_1(\IC P^{m|n})\ =\ (m+1-n)x,}
where $x:=c_1(\CO_{\IC P^{m|n}}(1))$.
Thus, we may conclude that our supertwistor space 
$\IP^{3|\CN}=\IC P^{3|\CN}$ becomes
a formal Calabi-Yau supermanifold if and only if $\CN=4$. 

A similar argumentation can be given for superambitwistor 
space $\IL^{5|2\CN}$ as introduced in \ref{ATTSII}\kern6pt. 
There, it was shown that $\IL^{5|2\CN}$ can be
realized as a hypersurface in $\IP^{3|\CN}\times\IP^{3|\CN}_*$. 
Thus, we have a short exact sequence of sheaves
\eqn{eq:ses-ch1}{0\ \to\ T\IL^{5|2\CN}\ \to\ \CO_{\IL^{5|2\CN}}
          \otimes T(\IP^{3|\CN}\times\IP^{3|\CN}_*)\ 
          \to\ N_{\IL^{5|2\CN}}\ \to\ 0,}
where $N_{\IL^{5|2\CN}}$ is the normal sheaf of $\IL^{5|2\CN}$ in 
$\IP^{3|\CN}\times\IP^{3|\CN}_*$.
In fact, $N_{\IL^{5|2\CN}}\cong\CO_{\IL^{5|2\CN}}(1,1)=\CO_{\IP^{3|\CN}
\times\IP^{3|\CN}_*}(1,1)/\CI$, where
$\CI$ is the ideal subsheaf \eqref{eq:is-ch1}.\footnote{See also 
our discussion given in \ref{pa:FN-ch4}\kern6pt.}
Then a short calculation reveals that \eqref{eq:ses-ch1} implies
\eqn{}{c_1(\IL^{5|2\CN})\ =\ (3-\CN)x+(3-\CN)y,} 
where $x:=c_1(\CO_{\IL^{5|2\CN}}(1,0))$ and 
$y:=c_1(\CO_{\IL^{5|2\CN}}(0,1))$, respectively. Hence,
$\IL^{5|6}$ is a formal Calabi-Yau supermanifold. Furthermore, it 
has been shown \cite{Neitzke:2004pf,Aganagic:2004yh} 
that $\IL^{5|6}$ and $\IP^{3|4}$ are related in some sense by mirror
symmetry. For related aspects of mirror symmetry see also Refs.
\cite{Kumar:2004dj,Ahn:2004xs,Belhaj:2004ts,Policastro:2005dm,
Ricci:2005cp,Laamara:2006rk}.

Before coming to our next topic, let us illustrate a final example 
given by LeBrun \cite{LeBrun04}. Let $X$ be some ordinary complex 
manifold. As we have seen in \ref{SM}\kern6pt,
$(X,\CO_{\rm red}(\Lambda^\bullet\CE^*))$ is a complex supermanifold, 
where $\CE$ is a rank $r|0$ locally free sheaf of 
$\CO_{\rm red}$-modules. The holomorphic Berezinian of 
$(X,\CO_{\rm red}(\Lambda^\bullet\CE^*))$ 
is then given by 
${\rm Ber}\,(X,\CO_{\rm red}(\Lambda^\bullet\CE^*))=
\CK\otimes\Lambda^r\CE,$ where $\CK$ is the 
canonical sheaf (sheaf of sections of the canonical bundle) on $X$.
Next one can show that there
is a short exact sequence of $\CO_{\rm red}$-modules
on $X$
\eqn{eq:JetSe-ch1}{0\ \to\ \O^1(X)\otimes \CF
   \ \to\ {\rm Jet}^1\CF\ \to\ \CF
   \ \to\ 0,}
where ${\rm Jet}^1\CF$ is the sheaf of first-order jets for $\CF$.\footnote{For
a proof, see, e.g., Ref. \cite{Manin}.}
Let now $X$ be some complex three-dimensional manifold which admits 
a spin structure. 
Then $X$ can be extended to a formal Calabi-Yau supermanifold of dimension
$3|4$ by setting $\CE={\rm Jet}^1\CK^{-1/2}$ since from the above sequence
we obtain 
$$\begin{aligned} 
    \Lambda^4\CE\ &\cong\ 
                 \CK^{-1/2}\otimes\Lambda^3(\O^1(X)\otimes\CK^{-1/2})\\
                &\cong\ \CK^{-1/2}\otimes\Lambda^3\O^1(X)\otimes\CK^{-3/2}\\
                &\cong\ \CK^{-1/2}\otimes\CK\otimes\CK^{-3/2}\
                \cong\ \CK^{-1}
  \end{aligned}
$$
and hence ${\rm Ber}\,(X,\CO_{\rm red}(\Lambda^\bullet\CE^*))\!
=\!\CK\otimes\Lambda^4\CE\cong\CK\otimes\CK^{-1}$ is trivial. 
Therefore, we can conclude that 
$(X,\CO_{\rm red}(\Lambda^\bullet\CE^*)$ is a formal Calabi-Yau
supermanifold. For instance, the supertwistor space $\IP^{3|4}$ fits into 
this construction scheme, since $\CK=\CO_{\IP^3}(-4)$ and hence
${\rm Jet}^1\CK^{-1/2}\cong\CO_{\IP^3}(1)\otimes\IC^4$, as can 
most easily be deduced from the Euler sequence \eqref{Euler}.

\section{Real structures}

Up to now, we have only dealt with complex (super)twistor spaces. In order
to discuss real gauge theories, that is, gauge theories living on
either Euclidean or Minkowski spaces (or conformal compactifications
thereof) with some unitary group as gauge group, 
we need to put certain real structures on the supertwistor space $\IT$
and its dual $\IT^*$. This in turn induces real structures on all the 
supermanifolds appearing in our double fibrations. Therefore, our first 
goal is to choose proper coordinates on $\IT$ and $\IT^*$ as well as on 
the supermanifolds participating
in \eqref{DF6}. Recall that in \ref{STS}\kern10pt and
\ref{ATTSII}\kern10pt we have 
already given partial results on this matter.

\paragraph{Local coordinates.}\label{pa:LC-ch1}
In \ref{ATTSII}\kern6pt, we have denoted the coordinates on $\IT$ by
$(z^\a,\pi_\ad,\eta_i)$ and on $\IT^*$ by $(\rho_\a,w^\ad,\te^i)$, 
respectively.\footnote{In fact, they are just homogeneous coordinates
for the projectivized versions.}
Furthermore,
in the same paragraph we also discussed the canonical bilinear form 
\eqn{blf}{\langle(z^\a,\pi_\ad,\eta_i),(\rho_\a,w^\ad,\te^i)\rangle\ =\ 
          z^\a\rho_\a-w^\ad\pi_\ad+2\te^i\eta_i}
as an element of $\IT\otimes\IT^*$. Consider the double fibration 
\eqref{DF6},
\eqna{DF7}{
 \begin{picture}(50,40)
  \put(0.0,0.0){\makebox(0,0)[c]{$\IM_R^{4|2\CN}$}}
  \put(64.0,0.0){\makebox(0,0)[c]{$\IM_L^{4|2\CN}$}}
  \put(34.0,33.0){\makebox(0,0)[c]{$\IM^{4|4\CN}$}}
  \put(7.0,18.0){\makebox(0,0)[c]{$\pi_1$}}
  \put(55.0,18.0){\makebox(0,0)[c]{$\pi_2$}}
  \put(25.0,25.0){\vector(-1,-1){18}}
  \put(37.0,25.0){\vector(1,-1){18}}
 \end{picture}
}
and recall that $\IM_R^{4|2\CN}=F_{2|0}(\IT)$,
$\IM^{4|4\CN}=F_{2|0,2|\CN}(\IT)$ and 
$\IM_L^{4|2\CN}=F_{2|\CN}(\IT)\cong F_{2|0}(\IT^*)$.
According to \ref{STS}\kern6pt, local coordinates on $\IM_R^{4|2\CN}$
are $(x^{\a\ad}_R,\eta^\ad_i)$. In a similar fashion, one may
take $(x^{\a\ad}_L,\te^{i\a})$ as local coordinates on 
$\IM_L^{4|2\CN}$. A $(2|0,2|\CN)$-flag in $\IT$ is the same
thing as a pair of $2|0$-dimensional subspaces in $\IT$ and
$\IT^*$, orthogonal with respect to the bilinear form
\eqref{blf}. Making use of the identifications  
\eqn{IR2}{(z^\a=x_R^{\a\ad}\l_\ad,\pi_\ad=\l_\ad,\eta_i=\eta^\ad_i\l_\ad)
        \quad{\rm and}\quad
       (w^\ad=x_L^{\a\ad}\mu_\ad,\rho_\a=\mu_\a,\te^i=\te^{i\a}\mu_\a),}
as discussed in \ref{STS}\kern10pt in the case of the supertwistor
space together with the orthogonality relation induced by
\eqref{blf}, we find
\eqn{eq:RLcoords-ch1}{x_R^{\a\ad}-x_L^{\a\ad}+2\te^{i\a}\eta^\ad_i\ =\ 0,}
where summation over repeated indices is implied.
This equation can generically be solved by, e.g., putting
\eqn{CACCO}{x_{R,L}^{\a\ad}\ =\ x^{\a\ad}\mp\te^{i\a}\eta^\ad_i.}
We may thus take $(x^{\a\ad},\te^{i\a},\eta^\ad_i)$ as local
coordinates on $\IM^{4|4\CN}$, with the obvious
projections $\pi_{1,2}$
\eqna{}{\pi_1\,:\,(x^{\a\ad},\te^{i\a},\eta^\ad_i)\in\IM^{4|4\CN}\ 
           &\mapsto\ (x_R^{\a\ad},\eta^\ad_i)\in\IM^{4|2\CN}_R,\\
        \pi_2\,:\,(x^{\a\ad},\te^{i\a},\eta^\ad_i)\in\IM^{4|4\CN}\ 
         &\mapsto\ (x_L^{\a\ad},\te^{i\a})\in\IM^{4|2\CN}_L.
}
Clearly, by virtue of the discussion given in \ref{STS}\kern10pt and of
the double fibration \eqref{DF7}, $(x^{\a\ad},\te^{i\a},\eta^\ad_i)$
are defined on
$\pi_1^{-1}(\CM_R^{4|2\CN})\cap\pi_2^{-1}
(\CM_L^{4|2\CN})=:\CM^{4|4\CN}\cong\IC^{4|4\CN}$.

\paragraph{Euclidean signature.}\label{EuSi} 
As we shall see in the next chapter, the supertwistor space will
play a key role in discussing self-dual SYM theories. As the
corresponding equations of motion are natural extensions of
the self-dual YM equations, we are interested in their formulation in 
an Euclidean setting. Though not being subject of the present
discussion, it is also possible to formulate them in the case of 
split signature. For details see, e.g., \cite{Popov:2004rb}.

Let us introduce the $\epsilon$-tensors according to
\eqn{}{(\epsilon^{\a\b})\ =\ (\epsilon^{\ad\bd})\ =\
         \begin{pmatrix}
            0 & 1 \\ -1 & 0 
         \end{pmatrix}
        \qquad{\rm and}\qquad
        (\epsilon_{\a\b})\ =\ (\epsilon_{\ad\bd})\ =\
          \begin{pmatrix}
           0 & -1\\ 1 & 0         
          \end{pmatrix},}
which satisfy $\epsilon_{\a\b}\epsilon^{\b\g}=\d^\g_\a$ and
$\epsilon_{\ad\bd}\epsilon^{\bd\gd}=\d^\gd_\ad$. Furthermore,
define
\eqn{defofT}{({T_i}^j)\ :=\ \begin{pmatrix}
                       0 & 1 & 0 & 0\\
                       -1 & 0 & 0 & 0\\
                       0 & 0 & 0 & 1\\
                       0 & 0& -1 & 0
                      \end{pmatrix}.
}
For the choice $\CN=4$ (the most interesting one for our purposes), 
Euclidean signature will be induced by the antiholomorphic involution
\eqn{E-INV}{\tau_E(z^\a,\pi_\ad,\eta_i)\ :=\ 
         (\epsilon_{\a\b}\bar{z}^\b,\epsilon_{\ad\bd}\bar{\pi}_\bd,
         -{T_i}^j\bar{\eta}_j)}
for the coordinates $(z^\a,\pi_\ad,\eta_i)$ on $\IT$. Here,
summation over repeated indices is implied and bar denotes complex
conjugation. Note that reality of the odd coordinates for Euclidean 
signature can only be imposed if $\CN$ is even
\cite{KotrlaKY,LUKIERSKI}; the $\CN=0$ and $\CN=2$ cases are obtained 
by suitable
truncations of the $\CN=4$ case. Furthermore, we adopt the convention
\eqn{}{\tau_E(ab)\ =\ \tau_E(a)\tau_E(b),}
where $a,b$ are any of the coordinates $(z^\a,\pi_\ad,\eta_i)$. The 
extension of the 
involution $\tau_E$ to any holomorphic function $f$ is defined to be
\eqn{}{\tau_E(f(\cdots))\ :=\ \overline{f(\tau_E(\cdots))}.}

Using \eqref{E-INV} together with 
the incidence relations \eqref{IR2}, we find
\eqn{EI-xandeta}{\tau_E(x^{\a\ad}_R)\ =\ 
             \epsilon^{\a\b}\epsilon^{\ad\bd}\bar{x}^{\b\bd}_R
              \qquad{\rm and}\qquad
             \tau_E(\eta^\ad_i)\ =\ 
             \epsilon^{\ad\bd}{T_i}^j{\bar\eta}^\bd_j} 
for the local coordinates $(x_R^{\a\ad},\eta^\ad_i)$ on 
$\CM_R^{4|8}\subset\IM^{4|8}_R$. The fixed point set of these involutions
then defines antichiral Euclidean superspace 
$\CM_{R,\tau_E}^{4|8}\subset\CM_R^{4|8}$
with $\CM_{R,\tau_E}^{4|8}\cong\IR^{4|8}$. One may choose the
following parametrization:
\eqn{xdefeuc-1}{x^{2{\dot2}}_R\ =\ {\bar x}^{1{\dot1}}_R\ =:\ 
                x^4-\i x^3
                 \qquad{\rm and}\qquad
                 x^{2{\dot1}}_R\ =\ -{\bar x}^{1{\dot2}}_R\ =:\ 
               -x^2+\i x^1,
}
with real $x^\mu$. Hence, the metric is of Euclidean type. Note 
that later on,
we shall also choose parametrizations different from \eqref{xdefeuc-1}.

\paragraph{Minkowski signature.}\label{pa:MS-ch4}
Besides self-dual SYM theories, we are also interested in full
SYM theories which are most interesting when considered on Minkowski 
space-time.
In this situation, the superambitwistor space plays the central 
role in the discussion. 

Minkowski signature is induced by the antiholomorphic involution
\eqn{}{\tau_M(z^\a,\pi_\ad,\eta_i,\rho_\a,w^\ad,\te^i)\ :=\
         (-\bar{w}^\ad,\bar{\rho}_\a,\bar{\te}^i,\bar{\pi}_\ad,
           -\bar{z}^\a,
         \bar{\eta}_i)}
on $\IT\times\IT^*$, where bar denotes, as before, complex conjugation.
This time, however, we choose
\eqn{}{\tau_M(ab)\ =\ \tau_M(b)\tau_M(a),}
where $a,b$ are any of the coordinates on $\IT\times\IT^*$. The 
extension
to holomorphic functions is then given analogously as for $\tau_E$.
 
Next using Eqs. \eqref{IR2} together with the conditions \eqref{CACCO},
we immediately arrive at
\eqn{}{\tau_M(x^{\a\bd})\ =\ -\bar{x}^{\b\ad},\qquad
       \tau_M(\eta^\ad_i)\ =\ \bar{\te}^{i\a}\qquad{\rm and}\qquad
       \tau_M(\te^{i\a})\ =\ \bar{\eta}^\ad_i.
}
The fixed point set 
\eqn{}{\tau_M(x^{\a\bd})\ =\ -\bar{x}^{\b\ad}\ =\ x^{\a\bd}
\qquad{\rm and}\qquad
       \tau_M(\eta^\ad_i)\ =\ \bar{\te}^{i\a}\ =\ \eta^\ad_i
}
together with the parametrization
\eqn{eq:xpar-ch1}{(x^{\a\ad})\ =:\
         \begin{pmatrix}
            -\i(x^0-x^3) & x^2+\i x^1\\
             -x^2+\i x^1 & -\i(x^0+x^3)
         \end{pmatrix}
}
for real $x^\mu$, yields a metric of Minkowski signature, that is,
 $({-}{+}{+}{+})$.

\clearemptydoublepage
\chapter{Self-dual super gauge theory}\label{ch:SDSYM-ch2}
\HRule\\

{\Large A}fter these introductory words on (super)twistor
spaces, we shall now discuss a first application. 
In the introduction we have already seen that 
the twistor approach to gauge theory involves certain holomorphic
vector bundles over spaces appearing in a double
fibration like \eqref{DF3}. As is well known,
holomorphic vector bundles can be described within two
different approaches: the \v Cech and the Dolbeault pictures.
Both pictures, however, turn out to be equivalent -- and
each of it has its own advantages and disadvantages.
In the sequel, we shall be using both on equal footing. 
Therefore, we first describe the equivalence of
both pictures in a more general setting and
then discuss as a first example self-dual SYM theory\footnote{See
Ref. \cite{Semikhatov:1982ig,Volovich:1983ii,Volovich:1984kr,Volovich:1983aa}
for an earlier account on self-dual SYM theory.} and
some related models.     

\section{\v Cech-Dolbeault correspondence}\label{CD-COR-2}

\paragraph{\v Cech cochains, cocycles and cohomology.}
Let us recall some basic definitions already adopted to 
complex supermanifolds.  Consider a complex 
supermanifold $X$ with an open covering $\fU=\{\CU_a\}$. 
Furthermore, we are interested in smooth maps from open subsets of 
$X$ into some Lie (super)group $G$  
as well as in a sheaf $\fS$ of such $G$-valued functions.  
A $q$-cochain of the covering $\fU$ with values in $\fS$ is a 
collection $\psi=\{\psi_{a_0\cdots a_q}\}$ of sections of the sheaf 
$\fS$
over nonempty intersections $\CU_{a_0}\cap\cdots\cap\CU_{a_q}$. We will 
denote
the set of such $q$-cochains by $C^q(\fU,\fS)$. We stress that it has a
group structure, where the multiplication is just pointwise multiplication.

We may define the subsets of cocycles $Z^q(\fU,\fS)\subset C^q(\fU,\fS)$. 
For example, for $q=0,1$ they are given by
\eqna{}{Z^0(\fU,\fS)\ &:=\ \{\psi\in C^0(\fU,\fS)\ |\ \psi_a=\psi_b
           ~~{\rm on}~~\CU_a\cap\CU_b\neq\emptyset\},\\
           Z^1(\fU,\fS)\ &:=\ \{\psi\in C^1(\fU,\fS)\ |\ \psi_{ab}=
           \psi_{ba}^{-1}~~{\rm on}~~\CU_a\cap\CU_b\neq\emptyset\\
           &\kern2.5cm{\rm and}~~\psi_{ab}\psi_{bc}\psi_{ca}=1~~
           {\rm on}~~\CU_a\cap\CU_b\cap\CU_c\neq\emptyset\}.
}
These sets will be of particular interest. We remark that from the 
first of these two definitions it follows that $Z^0(\fU,\fS)$ coincides with 
the group 
$$H^0(X,\fS)\ \equiv\ \fS(X)\ =\ \Gamma(X,\fS),$$
which is the group of global sections of the sheaf $\fS$. Note that in general
the subset $Z^1(\fU,\fS)\subset C^1(\fU,\fS)$ is not a subgroup of the group 
$C^1(\fU,\fS)$.

We say that two cocycles $f,\tilde{f}\in Z^1(\fU,\fS)$ are equivalent if 
$\tilde{f}_{ab}=\psi_a^{-1}f_{ab}\psi_b$ for some $\psi\in C^0(\fU,\fS)$. The set
of equivalence classes induced by this equivalence relation is the first (pointed)
\v Cech cohomology set and denoted by $H^1(\fU,\fS)$. If the $\CU_a$s are all Stein
-- in case of supermanifolds $X$ we need $X_{\rm red}$ to
be covered by Stein manifolds -- we have the bijection
\eqn{}{H^1(\fU,\fS)\ \cong\ H^1(X,\fS),}
otherwise one takes the inductive limit.
To sum up, we see that, for instance, the elements of $H^1(X,\fH)$ with
$\fH:=GL(r|s,\CO_X)$ classify rank $r|s$ locally free sheaves of $\CO_X$-modules
up to isomorphism. Hence,
$H^1(X,\fH)$ is the moduli space of
holomorphic vector bundles over $X$ with complex rank $r|s$.

\paragraph{Dolbeault cohomology.}
Let $X$ be a complex supermanifold and consider a rank $r|s$ complex
vector bundle $\CE\to X$. Furthermore, we let
\eqn{}{\O^{p,q}(X)\ :=\ \Scr Hom_{\CS_X}(\Lambda^p(T^{1,0}_\IC X)
        \otimes\Lambda^q(T^{0,1}_\IC X),\CS_X)}
be the differential $(p,q)$-forms on $X$. In spirit of our discussion 
given in \ref{CCAAT}\kern6pt, we have a natural antiholomorphic
exterior derivative $\pbar\,:\,\O^{p,q}(X)\to\O^{p,q+1}(X)$. A $(0,1)$-connection
on $\CE$ is defined by a covariant differential 
\eqn{}{\nabla^{0,1}\,:\,\CE\ \to\ \O^{0,1}(X)\otimes\CE}
satisfying the Leibniz formula; see also Eq. \eqref{Leibniz}. Locally, it is of the form
\eqn{}{\nabla^{0,1}\ =\ \pbar+\CA^{0,1},}
where $\CA^{0,1}\in\Gamma(X,\O^{0,1}(X)\otimes{\rm End}\,\CE)$. The complex
vector bundle
$\CE$ is said to be holomorphic if the $(0,1)$-connection is flat, that is, if the
corresponding curvature vanishes,
\eqn{flat}{\CF^{0,2}\ =\ (\nabla^{0,1})^2\ =\ \pbar\CA^{0,1}+
      \CA^{0,1}\wedge\CA^{0,1}\ =\ 0.}
In other words, $\nabla^{0,1}$ defines a holomorphic structure on $\CE$.
Note that \eqref{flat} is also called equation of motion of holomorphic
Chern-Simons (hCS) theory. 

Let $\fA^{0,1}$ be the sheaf of solutions to \eqref{flat}.
The group $\Gamma(X,\fS)$, where $\fS:=GL(r|s,\CS_X)$, 
is acting on $\Gamma(X,\fA^{0,1})$ by
\eqn{}{\CA^{0,1}\ \mapsto\ g^{-1}\CA^{0,1}g+g^{-1}\pbar g,}
where $g\in\Gamma(X,\fS)$, and without changing the holomorphic structure on
$\CE$. Therefore, the Dolbeault cohomology set
\eqn{}{H^1_{\nabla^{0,1}}(X,\CE)\ :=\ \Gamma(X,\fA^{0,1})/\Gamma(X,\fS)}
parametrizes all different holomorphic structures on the complex vector bundle
$\CE$. 

\paragraph{Equivalence of \v Cech and Dolbeault pictures.}
Let us now show that the above approaches to holomorphic vector bundles
are actually equivalent. This fact may be 
understood as a non-Abelian generalization of Dolbeault's theorem.
{\Thm\label{C-D-Thm} Let $X$ be a complex supermanifold with an open Stein covering $\fU=\{\CU_a\}$
      and $\CE\to X$ be a rank $r|s$ 
      complex vector bundle over $X$. Then there is a one-to-one
      correspondence between $H^1_{\nabla^{0,1}}(X,\CE)$ and the subset 
      of $H^1(X,\fH)$ consisting of those elements of
      $H^1(X,\fH)$ representing vector bundles which are smoothly
      equivalent to $\CE$, i.e.,
      $$ (\CE,f=\{f_{ab}\},\nabla^{0,1})\ \sim\ (\tilde{\CE},\tilde{f}=\{\tilde{f}_{ab}\},\pbar),$$
      where $\tilde{f}_{ab}=\psi_a^{-1}f_{ab}\psi_b$ for some
      $\psi=\{\psi_a\}\in C^0(\fU,\fS)$.
}\vskip 4mm

\noindent
{\small {\it Proof:} Let $\CE\to X$ be a rank $r|s$ complex vector bundle
 represented by $f=\{f_{ab}\}\in H^1(X,\fS)$. Furthermore,
 consider the subset of $C^0(\fU,\fS)$ consisting of those elements 
 $\psi=\{\psi_a\}$ obeying
 $$\psi_b\pbar\psi^{-1}_b\ =\ f_{ab}^{-1}\psi_a\pbar\psi_a^{-1}f_{ab}+f_{ab}^{-1}\pbar f_{ab} $$
 on $\CU_a\cap\CU_b\neq\emptyset$.
 Due to Eq. \eqref{flat}, elements $\CA^{0,1}=\{\CA^{0,1}_a\}$ of $H^1_{\nabla^{0,1}}(X,\CE)$ are locally
 of the form $\CA^{0,1}_a=\psi_a\pbar\psi_a^{-1}$ and glued together according to the
 above formula. Hence, $\CA^{0,1}\in H^1_{\nabla^{0,1}}(X,\CE)$ determines a zero-cochain
 $\psi=\{\psi_a\}$ with the above property. 
 This $\psi$ can in turn be used to define the transition 
 functions of a rank $r|s$ holomorphic vector bundle $\tilde{\CE}\to X$ by setting
 $$ \tilde{f}_{ab}\ :=\ \psi_a^{-1}f_{ab}\psi_b,$$
 that is, $\pbar\tilde{f}_{ab}=0$. Clearly, the bundle $\tilde{\CE}$
 defined by this $\tilde{f}=\{\tilde{f}_{ab}\}\in H^1(X,\fH)$ is smoothly
 equivalent to $\CE$.
 Conversely, given $\tilde{f}=\{\tilde{f}_{ab}\}\in H^1(X,\fH)$ as transition functions
 of a holomorphic
 vector bundle $\tilde{\CE}$ which is smoothly equivalent to $\CE$, the 
 $\tilde{f}_{ab}$s can always be written in the above form and hence, one can
 reconstruct a differential $(0,1)$-form $\CA^{0,1}$ such that $\CF^{0,2}=0$.

 Bijectivity is shown by virtue of a short exact sequence of sheaves
 $$ 0\ \to\ \fH\ \hookrightarrow\ \fS\ \overset{\d^0}{\to}\ \fA^{0,1}\ \overset{\d^1}{\to}\ 0, $$
 where $\d^0\,:\,\fS\to\fA^{0,1}$ is defined on any open subset $\CU$ on $X$ by
 $\d^0\,:\,\psi_\CU\mapsto\psi_\CU\pbar\psi_\CU^{-1},$ 
 with $\psi_\CU\in\Gamma(\CU,\fS)$. The map $\d^1$ sends $\CA^{0,1}\in\fA^{0,1}$ to
 $\CF^{0,2}$ which by construction vanishes. The above sequence induces an exact
 sequence of cohomology sets
 $$ 0\ \to\ H^0(X,\fH)\ \to\ H^0(X,\fS)\ \to\ H^0(X,\fA^{0,1})\ \to\ H^1(X,\fH)\ \overset{\rho}{\to}\ 
    H^1(X,\fS).$$  
 By definition, $H^1(X,\fH)$ (respectively, $H^1(X,\fS)$) parametrizes holomorphic
 (respectively, smooth) vector bundles over $X$. The kernel of $\rho$ 
 coincides with the subset of $H^1(X,\fH)$ whose elements 
 are mapped into the class of $H^1(X,\fS)$ representing 
 holomorphic vector bundles which are smoothly equivalent to $\CE$.
 By virtue of the exactness of the cohomology sequence, we find
 $$ H^1_{\nabla^{0,1}}(X,\CE)\ =\ H^0(X,\fA^{0,1})/H^0(X,\fS)\ \cong\ \ker\rho. $$

\hfill $\blacksquare$
}\vskip 4mm

\paragraph{Remark.}
In the following, we shall mostly be interested in complex 
vector bundles which are trivial as smooth bundles. Furthermore, we also restrict 
our discussion to rank $r|0\equiv r$ complex vector bundles,
although it straightforwardly generalizes to rank $r|s$.

\section{Self-dual super Yang-Mills theory}

Subject of this section is the discussion of $\CN$-extended self-dual SYM
theory. We first present the \v Cech approach to holomorphic vector
bundles over supertwistor space and derive in this setting the
field equations of self-dual SYM theory on four-dimensional space-time.
Second, we reconsider the whole discussion in the Dolbeault picture. 
The latter then also allows us to formulate appropriate action principles for both,
hCS theory on supertwistor space and self-dual SYM theory on Euclidean
four-dimensional space in  the case of maximal supersymmetry, that is, for $\CN=4$.  

\paragraph{Penrose-Ward transform.}\label{PWT-2}
Let us consider the double fibration \eqref{DF4} and recall
that $\CM^{4|2\CN}_R\cong\IC^{4|2\CN}$, i.e.,
\eqna{DF-21}{
 \begin{picture}(50,40)
  \put(0.0,0.0){\makebox(0,0)[c]{$\CP^{3|\CN}$}}
  \put(64.0,0.0){\makebox(0,0)[c]{$\IC^{4|2\CN}$}}
  \put(34.0,33.0){\makebox(0,0)[c]{$\CF^{5|2\CN}_R$}}
  \put(7.0,18.0){\makebox(0,0)[c]{$\pi_1$}}
  \put(55.0,18.0){\makebox(0,0)[c]{$\pi_2$}}
  \put(25.0,25.0){\vector(-1,-1){18}}
  \put(37.0,25.0){\vector(1,-1){18}}
 \end{picture}
}
where $\CP^{3|\CN}$ is the supertwistor space given 
by \eqref{holofibI} and $\CF_R^{5|2\CN}\cong\IC^{4|2\CN}\times\IC P^1$. 
Recall also the form of the two
projections $\pi_{1,2}$,
\eqna{}{  \pi_1\,:\,(x^{\a\ad}_R,\l^\pm_\ad,\eta^\ad_i)&\ \mapsto\
             (z^\a_\pm=x^{\a\ad}_R\l^\pm_\ad,\pi_\ad^\pm=
           \l^\pm_\ad,\eta_i^\pm=\eta^\ad_i\l^\pm_\ad),\\
            \pi_2\,:\,(x^{\a\ad}_R,\l^\pm_\ad,\eta^\ad_i)&\ \mapsto\
             (x^{\a\ad}_R,\eta^\ad_i).
}
We denote the coverings of $\CP^{3|\CN}$ and  
$\CF^{5|2\CN}$ by $\fU=\{\CU_+,\CU_-\}$ and 
$\hfU=\{\hCU_+,\hCU_-\}$, respectively. Consider a rank $r$ 
holomorphic vector bundle $\CE\to\CP^{3|\CN}$ and its pull-pack 
$\pi^*_1\CE\to\CF^{5|2\CN}$ as defined in \ref{VectBund-1}\kern6pt.
 These bundles are characterized by the transition 
functions $f=\{f_{+-}\}$ on the intersection $\CU_+\cap\CU_-$ and $\pi^*_1f$
on $\hCU_+\cap\hCU_-$. For notational simplicity, we shall use
the same letter, $f$, for the transition functions of both bundles in the 
following course of discussion. By definition of a pull-back, $f$ is constant 
along $\pi_1\,:\,\CF^{5|2\CN}_R\to\CP^{3|\CN}$.
The relative tangent sheaf\footnote{If one has a fibration $\pi\,:\,Z\to X$, the
relative tangent sheaf $TZ/X$ (sheaf of vertical vector fields) is defined
by the following short exact sequence:
$ 0\to TZ/X\to TZ\overset{\pi_*}{\to}\pi^*TX\to0.$} 
$T\CF_R/\CP:=(\O^1(\CF^{5|2\CN}_R)/\pi_1^*\O^1(T\CP^{3|\CN}))^*$ is of rank
$2|\CN$ and freely generated by
\eqn{HVF-2}{D^\pm_\a\ =\ \l^\ad_\pm\partial_{\a\ad}^R
             \qquad{\rm and}\qquad
             D^i_\pm\ =\ \l^\ad_\pm\partial^i_\ad.}
Here, we have again abbreviated
$\partial_{\a\ad}^R:=\partial/\partial x^{\a\ad}_R$ and
$\partial^i_\ad:=\partial/\partial\eta^\ad_i$. In the sequel, we shall
write $\cT:=T\CF_R/\CP$.\footnote{Here, we have just introduced
the letter $\cT$ as abbreviation for the relative tangent sheaf. 
However, it can be put in a general context of integrable distributions
as subsheaves of the tangent sheaf. See \ref{PFC-3}\kern10pt for
details.}
Therefore, the transition
functions of $\pi^*_1\CE$ are annihilated by the vector fields \eqref{HVF-2}.
Letting $\pbar_\CP$ and $\pbar_\CF$ be the anti-holomorphic parts of the 
exterior derivatives on the supertwistor space and its correspondence 
space, respectively, we have $\pi^*_1\pbar_\CP=\pbar_\CF\circ\pi_1^*$. 
Hence, the
transition functions of $\pi^*_1\CE$ are also annihilated by $\pbar_\CF$.

As indicated, we assume that the bundle $\CE$ is smoothly trivial
and moreover $\IC^{4|2\CN}$-trivial, that is, holomorphically trivial 
when restricted to any projective line 
$\IC P^1_{x_R,\eta}\hookrightarrow\CP^{3|\CN}$. These conditions imply that 
there exists some smooth $GL(r,\IC)$-valued functions 
$\psi=\{\psi_+,\psi_-\}\in C^0(\hfU,\fS)$,
 which define trivializations of $\pi_1^*\CE$ over $\hCU_\pm$, such that
$f_{+-}$ can be decomposed as
\eqn{eqforf}{f_{+-}\ =\ \psi^{-1}_+\psi_-}
and
\eqn{}{\pbar_\CF\psi_\pm\ =\ 0.}
Note that in particular this formula implies that the 
$\psi_\pm$s depend holomorphically on $\l_\pm$.
Applying the vector fields \eqref{HVF-2} to \eqref{eqforf}, we realize that
$$ \psi_+D^+_\a\psi_+^{-1}\ =\ 
   \psi_-D^+_\a\psi_-^{-1} 
   \qquad{\rm and}\qquad
   \psi_+D^i_+\psi_+^{-1}\ =\ 
   \psi_-D^i_+\psi_-^{-1}
$$
must be -- by an extension of Liouville's theorem -- at most linear in $\l_+$.
Therefore, 
we may introduce a Lie-algebra valued one-form $\CA_\cT$ which has
components only along $\cT$, such that
\eqna{defofa-2}{D_\a\lrcorner\CA_\cT|_{\hCU_\pm}\ &:=\ 
             \CA^\pm_\a\ = \ 
             \psi_\pm D^\pm_\a\psi_\pm^{-1}\ =\  \l_\pm^\ad\CA_{\a\ad}\,\\
             D^i\lrcorner\CA_\cT|_{\hCU_\pm}\  &:=\ \CA^i_\pm\ 
           =\ \psi_\pm D^i_\pm\psi_\pm^{-1}\ =\ 
             \l_\pm^\ad\CA^i_\ad.
}
In fact, $\CA_\cT$ defines a relative connection 
\eqn{}{\nabla_\cT\,:\,\pi_1^*\CE
       \ \to\ \O^1_\cT(\CF^{5|2\CN}_R)\otimes\pi_1^*\CE,}
which is flat. 
Here, $\Omega^1_\cT(\CF^{5|2\CN}_R):=\cT^*$
are the relative differential one-forms
on the correspondence space.

Eqs. \eqref{defofa-2} can be rewritten as
\eqna{linsys-2}{
  (D^\pm_\a+\CA^\pm_\a)\psi_\pm\ &=\ 0, \\
  (D^i_\pm+\CA^i_\pm)\psi_\pm  \ &=\ 0,\\
   \pbar_\CF\psi_\pm\ &=\ 0.
}
The compatibility conditions for this linear system read as
\eqna{CE-2}{{[\nabla_{\a\ad}^R,\nabla_{\b\bd}^R]}+
              [\nabla_{\a\bd}^R,\nabla_{\b\ad}^R]
             \ &=\ 0,\qquad
             [\nabla^i_\ad,\nabla_{\b\bd}^R]+[\nabla^i_\bd,\nabla_{\b\ad}^R]
             \ =\ 0,\\
             &\kern-.8cm
             \{\nabla^i_\ad,\nabla^j_\bd\}+\{\nabla^i_\bd,\nabla^j_\ad\}\ =\ 0,
}
where we have defined the covariant derivatives 
\eqn{}{\nabla_{\a\ad}^R\ :=\ \partial_{\a\ad}^R+\CA_{\a\ad}
                 \qquad{\rm and}\qquad
                 \nabla^i_\ad\ :=\ \partial^i_\ad+\CA^i_\ad.}
Eqs. \eqref{CE-2} are the constraint equations of $\CN$-extended self-dual
SYM theory. Note that the first of those equations represents the self-duality
equation for the gauge potential $\CA_{\a\ad}$, since
\eqn{}{[\nabla_{\a\ad}^R,\nabla^R_{\b\bd}]\ =\ \epsilon_{\a\b}f_{\ad\bd}+
              \epsilon_{\ad\bd}f_{\a\b},}
where $f_{\ad\bd}$ (respectively, $f_{\a\b}$) is symmetric in its indices.
Furthermore, $f_{\ad\bd}$ (respectively, $f_{\a\b}$) represents the
anti-self-dual (respectively, the self-dual) part of the field strength.  
By virtue of \eqref{CE-2}, the anti-self-dual part is put to zero.

Next let us briefly discuss how to obtain the
functions $\psi_\pm$ in \eqref{linsys-2} from a given gauge potential. Formally,
a solution is given by
\eqn{}{\psi_\pm\ =\ P\exp\left(-\int_{\cC_\pm}\CA\right).
}
Here, ``$P$'' denotes the path-ordering symbol and
\eqn{}{\CA\ =\ \dt x^{\a\ad}\CA_{\a\ad}+\dt \eta^\ad_i\CA^i_\ad.}
The contour $\cC_\pm$ is any real curve within an isotropic two-plane $\IC^{2|\CN}$ 
from a point $(\hat{x},\hat{\eta})$ to a point $(x,\eta)$, with
\eqna{}{x^{\a\ad}(s)\ &=\ \hat{x}^{\a\ad}+s\varepsilon^\a\l^\ad_\pm,\\
       \eta^\ad_i(s)\ &=\ \hat{\eta}^\ad_i+s\varepsilon_i\l^\ad_\pm,}
for $s\in[0,1]$; the choice of the contour plays no role, since the curvature
is zero when restricted to the two-plane.
Furthermore, $(\varepsilon^\a,\varepsilon_i)$ are some free
parameters. 

In \ref{EuSi}\kern6pt, we introduced an antiholomorphic involution
$\tau_E$ corresponding to Euclidean signature. Upon extending this
involution to $\pi_1^*\CE\to\CF^{5|2\CN}$, that is, upon requiring
\eqn{AHI-2}{f_{+-}(\cdots)\ =\ [f_{+-}(\tau_E(\cdots))]^\dagger,}
where dagger denotes Hermitian conjugation, one ends up with real
self-dual SYM fields. In particular, one finds
\eqn{}{\tau_E(\CA_{\a\ad})\ =\ -\epsilon_{\a\b}\epsilon_{\ad\bd}
        \CA_{\b\bd}^\dagger\ =\
       \CA_{\a\ad}
       \quad{\rm and}\quad
       \tau_E(\CA^i_\ad)\ =\ -\epsilon_{\ad\bd}
      {T_j}^i(\CA^j_\bd)^\dagger\ =\ \CA^i_\ad,}
where ${T_i}^j$ has been defined in \eqref{defofT}.
This reduces the gauge group $GL(r,\IC)$ to the unitary group
$U(r)$. Unless otherwise stated, we shall implicitly assume that
\eqref{AHI-2} has been imposed. If one in addition requires that
$\det(f_{+-})=1$, the structure group is reduced further to $SU(r)$.

Before we are discussing the field expansions of the
superfields $\CA_{\a\ad}$ and $\CA^i_\ad$, let us give some
integral formulas
\eqn{pwtrans}{
             \CA_{\a\ad}\ =\ \frac{1}{2\pi\i}\oint_\cC{\rm d}\l_+\,
 \frac{\CA_\a^+}{\l_+\l^\ad_+}
             \qquad{\rm and}\qquad
             \CA^i_\ad\ =\ \frac{1}{2\pi\i}\oint_\cC{\rm d}\l_+\,
            \frac{\CA^i_+}{\l_+\l^\ad_+} 
}
where the contour $\cC=\{\l_+\in\IC P^1\,|\,|\l_+|=1\}$ encircles $\l_+=0$.
In fact, these contour integrals give the explicit form of the 
Penrose-Ward transform.

\paragraph{Field expansions, field equations and action functional.}\label{FE-2}
Let us stick to the $\CN=4$ case. The others are then obtained by
suitable truncations. Recall that the field content of $\CN=4$ 
self-dual SYM theory consists of a (self-dual) 
gauge potential $\Ac~\!\!_{\a\ad}$, four positive chirality spinors 
$\cc\ \!\!^i_\a$, six scalars $\Wc~\!\!^{ij}=-\Wc~\!\!^{ji}$, four negative 
chirality spinors $\cc_{i\ad}$ and an 
anti-self-dual two-form $\Gc_{\ad\bd}$, 
all in the adjoint representation of $SU(r)$. The circle refers to the
lowest component in the superfield expansions of the corresponding superfields
$\CA_{\a\ad}$, $\chi^i_\a$, $W^{ij}$, $\chi_{i\ad}$ and $G_{\ad\bd}$, 
respectively. The constraint equations \eqref{CE-2} can formally be solved
by setting
\eqn{}{[\nabla_{\a\ad}^R,\nabla_{\b\bd}^R]\ =\ \epsilon_{\ad\bd}f_{\a\b},
       \quad
       [\nabla^i_\ad,\nabla_{\b\bd}^R]\ =\ \epsilon_{\ad\bd}\chi^i_\b
       \quad{\rm and}\quad
       \{\nabla^i_\ad,\nabla^j_\bd\}\ =\ \epsilon_{\ad\bd}W^{ij}.}
Using Bianchi identities, we find the remaining fields
\eqn{}{\chi_{i\ad}\ :=\ \tfrac{2}{3}\nabla^j_\ad W_{ij}
       \qquad{\rm and}\qquad
       G_{\ad\bd}\ :=\ -\tfrac{1}{4}\nabla^i_{(\ad}\chi_{i\bd)},}
respectively. Here, we introduced the common abbreviation
$W_{ij}:=\frac{1}{2!}\epsilon_{ijkl}W^{kl}$ and parentheses mean
normalized symmetrization.
Next we follow the literature \cite{HarnadVK,HarnadBC,DevchandGV} and
impose the transversal gauge,
\eqn{tg}{\eta^\ad_i\CA^i_\ad\ =\ 0,}
in order to remove the superfluous gauge degrees of freedom associated with 
the
odd coordinates $\eta^\ad_i$.\footnote{See also Ref. \cite{Arefeva:1986zx}
for a more general setting.} 
Putting it differently, this reduces the 
allowed gauge transformations to ordinary gauge transformations. 
Furthermore, this leads to the recursion operator
$\cD$ given by
\eqn{ro}{\cD\ :=\ \eta^\ad_i\nabla^i_\ad\ =\ \eta^\ad_i\partial^i_\ad.}
Again by virtue of Bianchi identities, one arrives after a somewhat
lengthy calculation at the following set of recursion relations:
\eqna{eq:RR-2}{\cD\CA_{\a\ad}\ &=\ -\epsilon_{\ad\bd}\eta^\bd_i\chi^i_\a,\\
        (1+\cD)\CA^i_\ad\ &=\ \epsilon_{\ad\bd}\eta^\bd_jW^{ij},\\
        \cD W_{ij}\ &=\ -\eta^\ad_{[i}\chi_{j]\ad},\\
        \cD\chi^i_\a\ &=\ -\eta^\ad_j\nabla_{\a\ad}^RW^{ij},\\
        \cD\chi_{i\ad}\ &=\ \eta^\bd_i G_{\ad\bd}+\epsilon_{\ad\bd}\eta^\bd_j
       [W^{jk},W_{ki}],\\
        \cD G_{\ad\bd}\ &=\ \eta^\gd_i\epsilon_{\gd(\ad}[\chi_{j\bd)},W^{ij}],}
where, as before, parentheses mean normalized symmetrization while the brackets
denote normalized antisymmetrization of the enclosed indices.\footnote{Note that
these equations resemble the supersymmetry transformations, but nevertheless
they should not be confused with them.} 
These equations determine all 
superfields to $(n+1)$-st order, provided one knows them to $n$-th
order in the odd coordinates. At this point, it is helpful  
to present some formulas which simplify this argumentation 
a lot.
Consider some generic superfield $f$. Its explicit
$\eta$-expansion looks as
\eqn{}{f\ =\ \fc+\sum_{k\geq1}\eta^{\gd_1}_{j_1}\cdots
\eta^{\gd_k}_{j_k}\,  f_{\gd_1\cdots\gd_k}^{j_1\cdots j_k}.}
Furthermore, we have $\cD f=\eta^{\gd_1}_{j_1}[\ \
]_{\gd_1}^{j_1}$, where the bracket $[\ \ ]_{\gd_1}^{j_1}$ is a
composite expression of some superfields. For example, we have
$\cD\CA_{\a\ad}=\eta^{\gd_1}_{j_1} [\a\ad]_{\gd_1}^{j_1}$,
with $[\a\ad]_{\gd_1}^{j_1}=-
\epsilon_{\ad\gd_1}\chi^{j_1}_\ad$.  Let now
\eqn{}{
\cD[\ \ ]_{\gd_1\cdots\gd_k}^{j_1\cdots j_k}\ =\
\eta^{\gd_{k+1}}_{j_{k+1}}[\ \
]_{\gd_1\cdots\gd_{k+1}}^{j_1\cdots
       j_{k+1}}.}
Then we find by induction
\eqn{eq:rec1-ch2}{f\ =\ \fc+\sum_{k\geq1}\frac{1}{k!}\,
\eta^{\gd_1}_{j_1}\cdots\eta^{\gd_k}_{j_k}\, \overset{\circ}{[\ \ ]}\
\!\!_{\gd_1\cdots\gd_k}^{j_1\cdots j_k}.}
In case the recursion relation of
$f$ was given by $(1+\cD)f=\eta^{\gd_1}_{j_1}[\ \
]_{\gd_1}^{j_1}$, as it happens to be for $\CA^i_\ad$, then
$\fc=0$ and the superfield expansion is of the form
\eqn{eq:rec2-ch2}{f\ =\ \sum_{k\geq1}\frac{k}{(k+1)!}\,
\eta^{\gd_1}_{j_1}\cdots\eta^{\gd_k}_{j_k}\, \overset{\circ}{[\ \ ]}\
\!\!_{\gd_1\cdots\gd_k}^{j_1\cdots j_k}.}
Using these expressions, one obtains the following
results for the superfields $\CA_{\a\ad}$ and $\CA^i_\ad$:
\eqna{componentexp-2}{
     \CA_{\a\ad}\ &=\ \Ac~\!\!_{\a\ad}+\epsilon_{\ad\bd}\cc\ 
                      \!\!^i_\a\eta^\bd_i+\cdots,\\
     \CA^i_\ad  \ &=\ \tfrac{1}{2!}\epsilon_{\ad\bd}\Wc~\!^{ij}
              \eta^\bd_j-\tfrac{1}{3!}
 \epsilon^{ijkl}\epsilon_{\ad\bd}\cc_{k\gd}\eta^\gd_l
                      \eta^\bd_j\ +\\
                  &~~~~~~~\ +\tfrac{3}{2\cdot 4!}\epsilon^{ijkl}\epsilon_{\ad\bd}
                      (\Gc_{\gd\dd}\d^m_{l}+\epsilon_{\gd\dd}[\Wc~\!^{mn},
                      \Wc_{nl}])\eta^\gd_k\eta^\dd_m\eta^\bd_j+\cdots.
}

Upon substituting the superfield expansions \eqref{componentexp-2}
into the constraint equations \eqref{CE-2}, we obtain
\eqna{fieldeqn-2}{
  \fc_{\ad\bd}\ &=\ 0,\cr
  \epsilon^{\a\b}\cnab\ \!\!^R_{\a\ad}\cc\ \!\!^i_\b\ &=\ 0,\\
  \overset{\circ}{\square}\ \!\!^R\Wc~\!^{ij}\ &=\ -
  \epsilon^{\a\b}\{\cc\ \!\!^i_\a,\cc\ \!\!^j_\b\},\\
   \epsilon^{\ad\bd}\cnab\ \!\!^R_{\a\ad}\cc_{i\bd}\ &=\ 2
  [\Wc_{ij},\cc\ \!\!^j_\a],\\
  \epsilon^{\ad\bd}\cnab\ \!\!^R_{\a\ad}\Gc_{\bd\gd}\ &=\ 
  \{\cc\ \!\!^i_\a,\cc_{i\gd}\}
  - \tfrac{1}{2}[\Wc_{ij},\cnab\ \!\!^R_{\a\gd}\Wc~\!^{ij}],
}
which are the $\CN=4$ self-dual SYM equations. The equations for less
supersymmetry are obtained from those by suitable truncations.
We have also introduced the abbreviation
\eqn{boxop}{\overset{\circ}{\square}\ \!\!^R\ :=\ 
    \tfrac{1}{2}\epsilon^{\a\b}\epsilon^{\ad\bd}
\cnab\ \!\!^R_{\a\ad}\cnab\ \!\!^R_{\b\bd}.}
We stress that Eqs. \eqref{fieldeqn-2} represent the field equations
to lowest order in the superfield expansions. With the help of the
recursion operator \eqref{ro}, one may verify that they are in one-to-one
correspondence with the constraint equations \eqref{CE-2}; see also Ref.
\cite{DevchandGV} for a detailed discussion.

Furthermore, one easily checks that the above field equations can
be derived from the following action functional:
\eqn{action-Siegel}{S\ =\ \int{\rm d}^4x_R\,{\rm tr}
 \left\{\Gc\ \!\!^{\ad\bd}\fc_{\ad\bd}+
         \cc\ \!\!^{i\a}\cnab\ \!\!^R_{\a\ad}\cc\ 
         \!\!^\ad_i+\tfrac{1}{2}\Wc_{ij}
                \overset{\circ}{\square}\ \!\!^R\Wc\ \!\!^{ij}
             +\Wc_{ij}\{\cc\ \!\!_\a^i,\cc\ \!\!^{j\a}\} \right\},}
which has first been introduced by Siegel \cite{SiegelZA}.

\paragraph{HCS theory on $\CP^{3|4}$.}\label{HCS-2}
Let us now try to understand the twistor analog of the
action functional \eqref{action-Siegel}. So far, we have worked
in the \v Cech approach to holomorphic vector bundles. In particular,
we started with a smoothly trivial holomorphic vector bundle 
$(\CE,f=\{f_{+-}\},\pbar_\CP)$ over 
supertwistor space and additionally required holomorphic triviality
along any $\IC P^1_{x_R,\eta}\hookrightarrow\CP^{3|4}$;
for the sake of concreteness, we again stick to the $\CN=4$ case.  
As we have learned in the previous section, there is another equivalent
approach -- the Dolbeault approach. To switch to this picture, 
we need to find
$(\tilde{\CE},\tilde{f}=\{\mathbbm{1}_r\},\pbar_\CP+\CA^{0,1})$, since
$\CE$ is assumed to be smoothly trivial. Moreover, within this approach 
we shall be
able to write down an action functional on supertwistor space yielding
the functional \eqref{action-Siegel} after performing suitable integrations.
Up to now, it is not known how to formulate
an appropriate action principle within the \v Cech approach. This is mainly 
due to the fact that
the \v Cech approach makes the construction manifestly on-shell: certain
holomorphic functions (the transition functions) on supertwistor space yield 
solutions to the equations of motion of $\CN=4$ SYM theory via contour
integrals of the form \eqref{pwtrans}.

First, let us make a more careful analysis of the real structure $\tau_E$ 
as introduced in \ref{EuSi}\kern6pt. There we have seen that the
fixed point set $\tau_E(x^{\a\ad}_R,\eta^\ad_i)=(x^{\a\ad}_R,\eta^\ad_i)$ -- 
cf. also Eqs. \eqref{EI-xandeta} -- 
is a real slice $\IR^{4|8}\subset\IC^{4|8}\frac{}{}$ corresponding to 
Euclidean
superspace. When restricting to $\IR^{4|8}$, we have the
diffeomorphisms
\eqn{}{\IR^{4|8}\times S^2\ \cong\ \IC^{2|4}\times\IC P^1\ \cong\ \CP^{3|4}.}
In fact, the map from $\CP^{3|4}$ with coordinates 
$(z^\a_\pm,z^3_\pm,\eta_i^\pm)$ to 
$\IC^{2|4}\times\IC P^1$ with coordinates 
$(x^{\a\dot1},\eta^{\dot1}_i,\l_\pm)$ is
explicitly given by  
\eqna{}{x^{1{\dot1}}\ =\ \frac{z^1_++z_+^3\bz^2_+}{1+z_+^3\bz_+^3}\ =\
\frac{\bz_-^3z_-^1+\bz_-^2}{1+z_-^3\bz_-^3},~~~ &x^{2\dot1}\ =\
\frac{z^2_+-z_+^3\bz^1_+}{1+z_+^3\bz_+^3}\ =\
\frac{\bz_-^3z^2_--\bz_-^1}{1+z_-^3\bz_-^3}~,\\\lambda_\pm\ =\
&z_\pm^3
}
and
\eqna{}{\eta_1^{\dot1}\ &=\
\frac{\eta_1^++z_+^3\bar{\eta}_2^+}{1+z_+^3\bz_+^3}\ =\
\frac{\bz_-^3\eta_1^-+\bar{\eta}_2^-}{1+z_-^3\bz_-^3},&
\eta_2^{\dot1}\ &=\
\frac{\eta_2^+-z_+^3\bar{\eta}_1^+}{1+z_+^3\bz_+^3}\ =\
\frac{\bz_-^3\eta_2^--\bar{\eta}_1^-}{1+z_-^3\bz_-^3},\\
\eta_3^{\dot1}\ &=\
\frac{\eta_3^++z_+^3\bar{\eta}_4^+}{1+z_+^3\bz_+^3}\ =\
\frac{\bz_-^3\eta_3^-+\bar{\eta}_4^-}{1+z_-^3\bz_-^3},&
\eta_4^{\dot1}\ &=\
\frac{\eta_4^+-z_+^3\bar{\eta}_3^+}{1+z_+^3\bz_+^3}\ =\
\frac{\bz_-^3\eta_4^--\bar{\eta}_3^-}{1+z_-^3\bz_-^3}.
}
These relations define a (smooth) projection
\eqn{F-21}{\CP^{3|4}\ \to\ \IR^{4|8}.}
Therefore, we may conclude that in the Euclidean setting no
double fibration like \eqref{DF-21} is needed. It is
rather enough to restrict the discussion to the
nonholomorphic fibration \eqref{F-21}. This, however,
is a very special feature of the present setting and we shall
find other examples where double fibrations -- even
if reality conditions are imposed -- are 
inevitable. 

Let us continue with the fibration \eqref{F-21}. The vector fields 
\eqref{HVF-2} which
generate the relative tangent sheaf 
$\cT=T\CF_R/\CP$ do
now, together with $\partial_{\bl_\pm}$, generate the antiholomorphic 
tangent 
sheaf $T^{0,1}_\IC\CP^{3|4}$, since $\partial_{\bz^\a_\pm}$,
$\partial_{\bz^3_\pm}$ and $\partial_{\bar{\eta}^\pm_i}$ can be 
rewritten
according to
\eqna{}{\der{\bz^\a_\pm}\ &=\ \epsilon_{\a\b}\g_\pm\bV_\b^\pm,\\
        \der{\bz_+^3}\ &=\
        \bV_3^+-\g_+x^{\a\dot1}\bV_\a^+-\g_+\eta_i^{\dot1}
        \bV^i_+,\quad\der{\bz_-^3}\ =\
        \bV_3^--\g_-x^{\a\dot2}\bV_\a^-+\g_-\eta_i^{\dot2}\bV^i_-,\\
        \der{\bar{\eta}_i^\pm}\ &=\ \g_\pm {T_j}^i\bV_\pm^j,
}
where we have defined 
\eqn{AHVF-2}{\bV^\pm_\a\ :=\ \l^\ad_\pm\partial^R_{\a\ad},\qquad
       \bV^\pm_3\ := \partial_{\bl_\pm}\qquad{\rm and}\qquad
       \bV^i_\pm\ :=\ \l^\ad_\pm\partial^i_\ad}         
and
\eqn{gam}{\g_\pm\ :=\ \frac{1}{1+\l_\pm\bl_\pm}.}
This makes it obvious that 
$T^{0,1}_\IC\CP^{3|4}=\langle\bV^\pm_\a,\bV^\pm_3,\bV^i_\pm\rangle$.
The matrix $({T_i}^j)$ has been defined in \eqref{defofT}.
A short calculation reveals that the sheaf of differential $(0,1)$-forms
on $\CP^{3|4}$ is, for instance, freely generated by the sections
\eqn{dual-2}{\bE^\a_\pm\ =\ -\g_\pm\hl^\pm_\ad\dt x^{\a\ad}_R,\qquad
       \bE^3_\pm\ =\ \dt\bl_\pm,\qquad{\rm and}\qquad
       \bE^\pm_i\ =\ -\g_\pm\hl^\pm_\ad\dt\eta^\ad_i,}
which, in fact, are dual to $(\bV^\pm_\a,\bV^\pm_3,\bV^i_\pm)$. 
Here, we have introduced
\eqna{}{(\hl_\ad^+)&\ :=\ \left(\begin{array}{cc} 0 & -1 \\ 1 & 0
\end{array}\right)\left(\begin{array}{c} 1\\ \bl_+
\end{array}\right)\ =\ \left(\begin{array}{c}-\bl_+\\1
\end{array}\right),\\
(\hl_\ad^-)&\ :=\ \left(\begin{array}{cc} 0 & -1 \\ 1 & 0
\end{array}\right)\left(\begin{array}{c} \bl_-\\ 1
\end{array}\right)\ =\ \left(\begin{array}{c}-1\\\bl_-
\end{array}\right).}
Thus, $\pbar_\CP$ is given by
\eqn{}{\pbar_\CP|_{\CU_\pm}\ =\ \dt \bz_\pm^\a\der{\bz^\a_\pm}+\dt \bz_\pm^3\der{\bz^3_\pm}+\dt\bar{\eta}^\pm_i\der{\bar{\eta}_i^\pm}\ =\
\bE^\a_\pm\bV_\a^\pm+\bE^3_\pm\bV^\pm_3+\bE^i_\pm\bV^\pm_i,}
where, as before, $\CU_\pm$ are the two sets covering the supertwistor 
space.

After this digression, we can now proceed as in \ref{PWT-2}\kern10pt
and discuss holomorphic vector bundles $(\CE,f=\{f_{+-}\},\pbar_\CP)$ 
over $\CP^{3|4}$ which are
smoothly trivial and in addition $\IR^{4|8}$-trivial. Eventually, one
again finds a linear system of the form \eqref{linsys-2}, that is,
\eqna{linsys-22}{
  (\bV^\pm_\a+\CA^\pm_\a)\psi_\pm\ &=\ 0, \\
   \bV^\pm_3\psi_\pm\ &=\ 0,\\
  (\bV^i_\pm+\CA^i_\pm)\psi_\pm  \ &=\ 0,   
}
where $f_{+-}=\psi_+^{-1}\psi_-$.
But this time, $\CA_\a^\pm$ and $\CA^i_\pm$ (and, of course, $\CA_3^\pm$, 
which is absent in the present gauge)
are interpreted as components of a differential $(0,1)$-form 
$\CA^{0,1}\in\Gamma(\CP^{3|4},\O^{0,1}(\CP^{3|4})\otimes{\rm End}\,
\tilde{\CE})$,
where the bundle $\tilde{\CE}\to\CP^{3|4}$ is smoothly equivalent to 
$\CE\to\CP^{3|4}$, i.e.,
$$(\CE,f=\{f_{+-}\},\pbar_\CP)\ \sim\ (\tilde{\CE},\tilde{f}=
\{\mathbbm{1}_r\},\pbar_\CP+\CA^{0,1}).$$
This is nothing but a special case of Thm. \ref{C-D-Thm}
Note that
\eqn{}{\CA^{0,1}|_{\CU_\pm}\ =\ \psi_\pm\pbar_\CP\psi_\pm^{-1},
\qquad{\rm with}\qquad
       \CA^{0,1}|_{\CU_+}\ =\ \CA^{0,1}|_{\CU_-}}
by smooth triviality of $\tilde{\CE}$.
By following the analysis of \ref{FE-2}\kern6pt, one again reproduces the
equations of motion of $\CN=4$ self-dual SYM theory. So, we do not
repeat the argumentation at this point.

Instead, we shall now change the trivialization of $\CE$.
In fact, there exist 
matrix-valued functions $\hat{\psi}=\{\hat{\psi}_+,
\hat{\psi}_-\}\in C^0(\fU,\fS)$ such
that\footnote{Recall that $\CP^{3|4}=\CO_{\IC P^1}(1)
\otimes\IC^2\oplus\Pi\CO_{\IC P^1}(1)\otimes\IC^3$.}
\eqn{strafo}{
  f_{+-}\ =\ \psi_+^{-1}\psi_-\ =\ \hat{\psi}_+^{-1}\hat{\psi}_-,
  \qquad{\rm with}\qquad\bV^i_\pm\hat{\psi}_\pm\ =\ 0.}
From \eqref{strafo} it then follows that
\eqn{}{g\ :=\ \psi_+\hat{\psi}_+^{-1}\ =\ \psi_-\hat{\psi}_-^{-1}}
is a globally well-defined matrix-valued function generating a gauge
transformation
\eqna{5.24}{
\psi_\pm\ &\mapsto\ \hat{\psi}_\pm\!\!&&=\ g^{-1}\psi_\pm,\\
\CA_\alpha^\pm\ &\mapsto\ \hat{\CA}_\alpha^\pm\!\!&&=\
g^{-1}\CA_\alpha^\pm g+g^{-1}\bV^\pm_\alpha g\ =\
\hat{\psi}_\pm\bV^\pm_\alpha\hat{\psi}^{-1}_\pm,\\
0\ =\ \CA_3^\pm\ &\mapsto\ \hat{\CA}_3^\pm\!\!&&=\ 
g^{-1}\bV^\pm_3g\ =\ \hat{\psi}_\pm\bV^\pm_3
\hat{\psi}_\pm^{-1},\\
\CA^i_\pm\ &\mapsto\ \hat{\CA}^i_\pm\!\!&&=\ g^{-1}\CA^i_\pm g+
g^{-1}\bV_\pm^ig\ =\
\hat{\psi}_\pm\bV_\pm^i\hat{\psi}^{-1}_\pm\ =\ 0.
}
Thus, we end up with
\eqna{linsysb-2.1}{
(\bV_\a^\pm+\hat{\CA}_\a^\pm)\hat{\psi}_\pm\ &=\ 0,\\
(\bV^\pm_3+\hat{\CA}_3^\pm)\hat{\psi}_\pm\ &=\ 0,\\
 \bV^i_\pm\hat{\psi}_\pm\ &=\ 0,\\
}
which is gauge equivalent to 
\eqref{linsys-22}.

The compatibility conditions of the linear system
\eqref{linsysb-2.1} are, of course, the field equations of hCS
theory on the supertwistor space $\CP^{3|4}$. On $\CU_\pm$, they read as
\eqna{shCS1}{
\bV_\a^\pm\hat{\CA}_\b^\pm-\bV_\b^\pm\hat{\CA}_\a^\pm+
[\hat{\CA}_\a^\pm,\hat{\CA}_\b^\pm]\ &=\ 0,\\
\bV^\pm_3\hat{\CA}_\a^\pm-\bV_\a^\pm\hat{\CA}_3^\pm+
[\hat{\CA}_3^\pm,\hat{\CA}_\a^\pm]\ &=\ 0.
}
As in \ref{FE-2}\kern6pt, we now have to find the explicit superfield
expansions of the components $\hat{\CA}_\a^\pm$ and $\hat{\CA}_3^\pm$,
respectively. However, their form is fixed by the geometry of  
supertwistor space.
Recall that $\hat{\CA}_\a^\pm$ and $\hat{\CA}_3^\pm$ are 
$\CO_{\IC P^1}(1)$- and $\overline{\CO}_{\IC P^1}(-2)$-valued. 
Together with the fact that the $\eta^\pm_i$s take values
in the bundle $\Pi\CO_{\IC P^1}(1)$, this determines the dependence of
$\hat{\CA}^\pm_\a$ and $\hat{\CA}_3^\pm$ on
$\l_\pm$ and $\bl_\pm$ \cite{Popov:2004rb}, 
\eqna{somefields}{
\hat{\CA}_\a^\pm\ &=\ \l_\pm^\ad\,\Ac~\!\!_{\a\ad}+\eta_i^\pm\cc\ \!\!^i_\a+
\g_\pm\,\tfrac{1}{2!}\,\eta^\pm_i\eta^\pm_j\,\hl^\ad_\pm\,
 \Wc\ \!\!^{ij}_{\a\ad}+\g_\pm^2\,\,\tfrac{1}{3!}\,\eta^\pm_i\eta^\pm_j\eta^\pm_k\,
 \hl_\pm^\ad\,\hl_\pm^\bd\,\cc\ \!\!^{ijk}_{\a\ad\bd}\ +\\
  &\kern3cm+\ \g_\pm^3\,
 \tfrac{1}{4!}\,\eta^\pm_i\eta^\pm_j\eta^\pm_k\eta^\pm_l\,
 \hl_\pm^\ad\,\hl_\pm^\bd\,\hl_\pm^\gd\,
 \Gc\ \!\!^{ijkl}_{\a\ad\bd\gd},\\
\hat{\CA}_3^\pm\ &=\ \pm\g_\pm^2\,\tfrac{1}{2!}\,\eta^\pm_i\eta^\pm_j\,
 \Wc\ \!\!^{ij}\pm\g_\pm^3\,\tfrac{1}{3!}\,\eta^\pm_i\eta^\pm_j\eta^\pm_k\,\hl_\pm^\ad\,
 \cc\ \!\!^{ijk}_\ad\ \pm\\
  &\kern3cm\pm\ \g_\pm^4\,\tfrac{1}{4!}\,\eta^\pm_i\eta^\pm_j\eta^\pm_k
 \eta^\pm_l\,\hl_+^\ad\,\hl_\pm^\bd \Gc\ \!\!^{ijkl}_{\ad\bd}.
}
Here,
$\Ac~\!\!_{\a\ad}$, $\cc\ \!\!_\a^i$, $\Wc\ \!\!^{ij}$, $\cc_{i\ad}
:=\frac{1}{3!}\epsilon_{ijkl}\cc\ \!\!^{jkl}_\ad$ and
$\Gc_{\ad\bd}:=\frac{1}{4!}\epsilon_{ijkl}\Gc\ \!\!^{ijkl}_{\ad\bd}$ is again the 
field content of $\CN=4$ self-dual SYM theory.
Note that, of course, the above expansions are unique only up to gauge 
transformations generated by group-valued functions which may depend on
$\l_\pm$ and $\bl_\pm$. Also the absence of terms to
zeroth and first order in $\eta_i^\pm$ in the expansion of $\hat{\CA}^\pm_3$ 
is due to the existence of a gauge in which $\hat{\CA}^\pm_3$ vanishes
identically.
Moreover, not all coefficient fields are independent degrees of freedom. Some of
them are composite expressions, 
\eqn{expA2}{
 \Wc\ \!\!^{ij}_{\a\ad}=-\cnab\ \!\!_{\a\ad}^R\Wc\ \!\!^{ij},\quad
 \cc\ \!\!^{ijk}_{\a\ad\bd}=-\tfrac{1}{2}\cnab\ \!\!_{\a(\ad}^R
 \cc\ \!\!^{ijk}_{\bd)}\quad{\rm and}\quad
 \Gc\ \!\!^{ijkl}_{\a\ad\bd\gd}=-\tfrac{1}{3}\cnab\ \!\!_{\a(\ad}^R
 \Gc\ \!\!^{ijkl}_{\bd\gd)},
}
which follow upon substituting the field expansions \eqref{somefields}
into the second equation of \eqref{shCS1}. Again we have
abbreviated $\cnab\ \!\!_{\a\ad}^R:=\partial^R_{\a\ad}+\Ac~\!\!_{\a\ad}$. 
The field expansions
\eqref{somefields} together with the first equation  of \eqref{shCS1}
eventually reproduce \eqref{fieldeqn-2}.
 
Now we have all ingredients to give the twistor analog of the
action functional \eqref{action-Siegel}. In fact, we have just seen that
the equations of motion of hCS theory,
$$\pbar_\CP\CA^{0,1}+\CA^{0,1}\wedge\CA^{0,1}\ =\ 0,$$
reproduce the equations of motion of $\CN=4$ self-dual SYM theory.
Luckily, as was discussed in \ref{FCYS}\kern10pt of the previous chapter,
the supertwistor space $\CP^{3|4}$ is a formal Calabi-Yau supermanifold.
In particular, this means that it admits a globally defined nowhere
vanishing holomorphic volume form $\O$. On the patches $\CU_\pm$ of 
$\CP^{3|4}$, it is given by
\eqn{}{\O|_{\CU_\pm}\ =\ \pm
       \dt z^1_\pm\wedge\dt z^2_\pm\wedge\dt z^3_\pm
       \dt\eta^\pm_1\dt\eta^\pm_2\dt\eta^\pm_3\dt\eta^\pm_4.}
Assuming that $\CA^{0,1}$ contains no antiholomorphic odd
components and does not depend on $\bar{\eta}^\pm_i$ (see our
above discussion), we may write down \cite{Witten:1992fb,Witten:2003nn}
\eqn{bla}{S\ =\ \int_\CY\O\wedge{\rm tr}\left\{\CA^{0,1}\wedge
       \pbar_\CP\CA^{0,1}+\tfrac{2}{3}\CA^{0,1}\wedge\CA^{0,1}\wedge
       \CA^{0,1}\right\},}
where $\CY$ is the submanifold of $\CP^{3|4}$ constrained by 
$\bar{\eta}^\pm_i=0$. The action functional \eqref{bla} of hCS theory
represents the twistor analog of \eqref{action-Siegel} we were looking 
for. Upon substituting the field expansions \eqref{somefields} into
\eqref{bla}, integrating over the odd coordinates and over
the Riemann sphere, we eventually end up with \eqref{action-Siegel}.

\paragraph{Summary.} 
Even though we have restricted our above discussion to the
$\CN=4$ case, one may equally well talk about the cases with
less supersymmetry, i.e., the cases with $\CN<4$. Of course, then the 
whole story is restricted to the level of the equations of motion, as there
are no appropriate action principles. Nevertheless, we may collect all the 
things said above and summarize as follows:
{\Thm\label{SDYM-HCS} There is a one-to-one correspondence between gauge 
equivalence classes of local solutions to the $\CN$-extended self-dual 
SYM equations on  four-dimensional space-time and equivalence classes of
holomorphic vector bundles $\CE$ over supertwistor space $\CP^{3|\CN}$ 
which are smoothly trivial and holomorphically trivial on any projective line 
$\IC P^1_{x_R,\eta}\hookrightarrow\CP^{3|\CN}$.}\vskip 4mm

Putting it differently, by Thm. \ref{C-D-Thm} we let
 $H^1_{\nabla^{0,1}}(\CP^{3|\CN},\tilde{\CE})$ be the moduli space of
hCS theory on $\CP^{3|\CN}$ for vector bundles $\tilde{\CE}$ smoothly equivalent 
to $\CE$. Furthermore, we denote by $\CM_{\rm SDYM}^\CN$ the moduli space
of $\CN$-extended self-dual SYM theory obtained from the solution space 
by quotiening with respect to the group of gauge transformations. Then we
have a bijection
\eqn{}{H^1_{\nabla^{0,1}}(\CP^{3|\CN},\tilde{\CE})\ \cong\ \CM^\CN_{\rm SDYM}.}

\section{Other self-dual models in four dimensions}\label{sec:OSDM-ch2}

Above we have related $\CN$-extended self-dual SYM theory to
hCS theory on supertwistor space $\CP^{3|\CN}$. The $\CN=4$ case
turned out to be very special in the sense of allowing to write down
action functionals for self-dual SYM and hCS theories. Clearly, the reason 
is the formal
Calabi-Yau property of $\CP^{3|4}$. The natural question one may now pose 
is that of extending the above approach to other geometries but at 
the same time 
keeping the formal Calabi-Yau property. One such class of geometries is 
weighted projective superspaces \cite{Witten:2003nn,Popov:2004nk}. 
In fact, they are natural
extensions of the projective superspace $\IC P^{m|n}$. The following
discussion is based on the work done together with Alexander Popov 
\cite{Popov:2004nk}.

\paragraph{Weighted projective superspaces.}\label{WPS-2}
First, consider ordinary weighted projective spaces. They
are defined by some $\IC^*$-action
on the complex space $\IC^{m+1}$. By letting $(z^1,\ldots,z^{m+1})$ 
be coordinates on $\IC^{m+1}$, we define the weighted projective space
$W\IC P^m[k_1,\ldots,k_{m+1}]$ for $k_i\in\IZ$ according to
\eqn{WPS}{ W\IC P^m[k_1,\ldots,k_{m+1}]\ :=\ (\IC^{m+1}\setminus\{0\})/\IC^*,}
where the $\IC^*$-action is given by
\eqn{}{(z^1,\ldots,z^{m+1})\ \mapsto\ (t^{k_1}z^1,\ldots,t^{k_{m+1}}z^{m+1})
       \qquad{\rm for}\qquad t\in\IC^*.}
Clearly, what we have just defined is a toric variety and as such it need
not be a manifold. In general, there may be nontrivial fixed points
under coordinate identifications leading to singularities. However, we 
shall ignore this subtlety at this point and assume that
the generic expression \eqref{WPS} is a complex manifold. Anyhow,
our later discussions will be unaffected by this issue since, 
in spirit of our above discussion, we shall be considering only certain 
subsets which are always manifolds.

In analogy to $\IC P^{m|n}=(\CO_{\IC P^m},\CO_{\rm red}
(\Lambda^\bullet(\IC^n\otimes\CO_{\IC P^m}(-1)))$ from  
Chap. \ref{STG-Chapter}, we define the weighted projective superspace
$W\IC P^{m|n}[k_1,\ldots,k_{m+1}|l_1,\ldots,l_n]$ for $k_i,l_i\in\IZ$
by
\eqna{}{W\IC P^{m|n}[k_1,\ldots,k_{m+1}|l_1,\ldots,l_n]\ &:=\
       (W\IC P^m[k_1,\ldots,k_{m+1}],\CO_{W\IC P^{m|n}}),\\
        &\kern-5cm\CO_{W\IC P^{m|n}}\ :=\  
        \CO_{\rm red}(\Lambda^\bullet(\CO_{W\IC P^m}(-l_1)
          \oplus\cdots\oplus\CO_{W\IC P^m}(-l_n))).}
Furthermore, by extending the Euler sequence \eqref{Euler} to the present
setting (see, e.g., \cite{Huebsch} for the purely even case), 
one may readily deduce that the first Chern class is given by
\eqn{}{c_1(W\IC P^{m|n})\ =\ \left(\sum_{i=1}^{m+1}k_i-\sum_{i=1}^nl_i\right)x,}
where $x:=c_1(\CO_{W\IC P^{m|n}}(1))$. Hence,
for appropriate numbers $k_i$ and $l_i$, the weighted projective
superspace $W\IC P^{m|n}$ becomes a formal Calabi-Yau supermanifold.
Note also that with this definition, we have
$$ W\IC P^{m|n}[1,\ldots,1|1,\ldots,1]\ \equiv\ \IC P^{m|n}.$$

\paragraph{HCS theory on $\CP^{3|2}_{p,q}$.}
For the sake of concreteness, let us now consider an open subset of  
$W\IC P^{3|2}[1,1,1,1|p,q]$ defined by
\eqn{}{\CP^{3|2}_{p,q}\ := \ W\IC P^{3|2}[1,1,1,1|p,q]
\setminus W\IC P^{1|2}[1,1|p,q].}
This space can be identified with 
\eqn{}{\CO_{\IC P^1}(1)\otimes\IC^2\oplus\Pi\CO_{\IC P^1}(p)
\oplus\Pi\CO_{\IC P^1}(q)\ \to\ \IC P^1}
and as such, it can be covered by two patches, which we denote by 
$\CU_\pm$. Obviously, for the particular combination
$p+q=4$, it becomes a formal Calabi-Yau supermanifold. In the following, 
we shall 
only be interested in this case. Let 
$[z^\a,\pi_\ad,\eta_i]$ be homogeneous coordinates on 
$W\IC P^{3|2}[1,1,1,1|p,q]$. Since the
body of $\CP^{3|2}_{p,q}$ is the twistor space $\CP^3$, 
we may take as local coordinates \eqref{LC-1} together with
\eqna{}{\eta_1^+\ &:=\ \frac{\eta_1}{\pi_{\dot1}^p}\qquad{\rm and}\qquad
            \eta_2^+\ :=\ \frac{\eta_2}{\pi_{\dot1}^q}
           \qquad {\rm on}\qquad\CU_+,\\
           \eta_1^-\ &:=\ \frac{\eta_1}{\pi_{\dot2}^p}\qquad{\rm and}\qquad
           \eta_2^-\ :=\ \frac{\eta_2}{\pi_{\dot2}^q}
           \qquad{\rm on}\qquad\CU_-,}
which are related by $\eta_1^+=(z_+^3)^p\eta_1^-$ and 
$\eta_2^+=(z_+^3)^q\eta_2^-$ 
on the intersection $\CU_+\cap\CU_-$. Note that as even coordinates on
$\CP^{3|2}_{p,q}$ either $(z^\a_\pm,z_\pm^3)$ or 
$(x^{\a\ad}_R,\l_\pm)$ can be used if proper reality conditions as
those discussed in \ref{EuSi}\kern10pt and in \ref{HCS-2}\kern6pt, respectively,
have been imposed. Therefore, we can again take $\bV^\pm_\a$ and $\bV_3^\pm$ of
\eqref{AHVF-2} 
as even generators of the antiholomorphic tangent sheaf 
$T_\IC^{0,1}\CP^{3|2}_{p,q}$.

Having given all the ingredients, we may now consider hCS theory on 
$\CP^{3|2}_{p,q}$. 
Let $\CE$ be a smoothly trivial rank $r$ complex vector bundle
over  $\CP^{3|2}_{p,q}$ equipped with a holomorphic structure 
$\CA^{0,1}\in H^1_{\nabla^{0,1}}(\CP^{3|2}_{p,q},\CE)$, that is, 
$\CA^{0,1}$ is
subject to \eqref{flat}. Furthermore, by virtue of the
twistor approach, we shall assume 
that there exists a gauge in which the component $ \CA^\pm_3$ is 
zero. This corresponds to the holomorphic triviality on any
$\IC P^1_{x_R,\eta}\hookrightarrow\CP^{3|2}_{p,q}$
of the holomorphic vector bundle which is  associated
to any solution of hCS theory.
The equations of motion \eqref{flat} of hCS theory on the patches 
$\CU_\pm$ of $\CP^{3|2}_{p,q}$ are again given by \eqref{shCS1},
since there exists a gauge in which $\CA^{0,1}$ does neither contain
antiholomorphic odd components nor depend on $\bar{\eta}_i$. Let
us now discuss particular examples.

\paragraph{HCS theory on $\CP^{3|2}_{1,3}$.}
Consider the case $p=1$ and $q=3$, where the fermionic coordinates 
$\eta_1^\pm$ and $\eta_2^\pm$ are $\Pi\CO_{\IC P^1}(1)$- and 
$\Pi\CO_{\IC P^1}(3)$-valued, 
respectively. Therefore, the components $\CA_\a^\pm$ and 
$\CA_3^\pm$ of the $(0,1)$-form $\CA^{0,1}$ are again
$\CO_{\IC P^1}(1)$- and $\overline{\CO}_{\IC P^1}(-2)$-valued. 
As before, this fixes the dependence of $\CA_\a^\pm$ and $\CA_3^\pm$ 
on $\l_\pm$ and $\bl_\pm$ up to gauge transformations. In particular, 
we obtain 
\eqna{}{
      \CA_\a^\pm\ &=\ \l^\ad_\pm\Ac_{\a\ad}+\eta^\pm_1\cc_\a+\tfrac{1}{2!{\sqrt3}}
           \eta^\pm_2\g^2_\pm\hl^\ad_\pm\hl^\bd_\pm
      \overset{\circ}{\psi}_{\a\ad\bd}+\tfrac{1}{3!}\eta^\pm_1\eta^\pm_2\g^3_\pm
            \hl^\ad_\pm\hl^\bd_\pm\hl^\gd_\pm
      \Gc_{\a\ad\bd\gd},\\
      \CA_3^\pm\ &=\ \pm\tfrac{1}{{\sqrt3}}
              \eta^\pm_2\g^3_\pm\hl_\pm^\ad\overset{\circ}{\psi}_\ad\pm\tfrac{1}{2!}
      \eta_1^\pm\eta_2^\pm\g^4_\pm\hl^\ad_\pm\hl^\bd_\pm\Gc_{\ad\bd},
}
where $\g_\pm$ has been defined in \eqref{gam}. As before, not all
component fields are independent degrees of freedom. Upon substituting
the above expansions into \eqref{shCS1}, we find
\eqn{}{\overset{\circ}{\psi}_{\a\ad\bd}\ =\ -\cnab\ \!\!_{\a(\ad}^R
            \overset{\circ}{\psi}_{\bd)}\qquad{\rm and}\qquad
            \Gc_{\a\ad\bd\gd}\ =\ -\cnab\ \!\!^R_{\a(\ad}\Gc_{\bd\gd)}.}
The field equations resulting from \eqref{shCS1} are then easily
obtained to be
\eqna{}{\fc_{\ad\bd}\ &=\ 0,\\
          \epsilon^{\a\b}\cnab\ \!\!^R_{\a\ad}\cc_\b\ &=\ 0,\\
          \epsilon^{\ad\bd}\cnab\ \!\!^R_{\a\ad}\overset{\circ}\psi_\bd\ &=\ 0,\\
          \epsilon^{\ad\bd}\cnab\ \!\!^R_{\a\ad}\Gc_{\bd\gd}\ &=\ \{\cc_\a,
          \overset{\circ}\psi_\gd\}.
}

As one may check, these equations follow also from the action functional 
\eqn{actionI}{S\ =\ \int\,\dt^4x_R\ {\rm tr}\,\left\{
                    \Gc\ \!\!^{\ad\bd}\fc_{\ad\bd}+
                    \overset{\circ}{\psi}\ \!\!^\ad\cnab\ \!\!^R_{\a\ad}\cc\ \!\!^\a
                 \right\},}
which can be obtained from \eqref{bla} by integration over the odd coordinates 
and over the sphere $\IC P^1$. Note that this action has an obvious 
supersymmetry the transformation laws being
\eqna{}{ \d_\xi \Ac_{\a\ad}\ =\ \xi_\ad\cc_\a\qquad&{\rm and}\qquad
         \d_\xi \Gc_{\ad\bd}\ =\ -\epsilon^{\a\b}\xi_{(\ad}\cnab\ \!\!^R_{\a\bd)}\cc_\b,\\
         \d_\xi \cc_\a\ =\ 0\qquad&{\rm and}\qquad
           \d_\xi\overset{\circ}{\psi}_\ad\ =\ -\xi^\bd(\Gc_{\ad\bd}+\fc_{\ad\bd}),\
}
where $\xi_\ad$ is a constant (anticommuting) spinor.
The action describes a truncation of $\CN=4$ self-dual
SYM theory for which all the scalars and three of 
the dotted and three of the undotted fermions are put to zero. 

\paragraph{HCS theory on $\CP^{3|2}_{2,2}$.}
Now we consider the case $p=q=2$, i.e., the odd coordinates $\eta_i^\pm$
take values in $\Pi\CO_{\IC P^1}(2)$. The  equations of motion of hCS theory
on  $\CP^{3|2}_{2,2}$ have the same form \eqref{shCS1}. Again, the functional
dependence on $\l_\pm$ and $\bl_\pm$ is fixed up to gauge transformations by the 
geometry of $\CP^{3|2}_{2,2}$. That is, this dependence has the form
\eqna{fieldexpaII}{
      \CA_\a^\pm\ &=\ \l^\ad_\pm \Ac_{\a\ad}+\tfrac{1}{\sqrt{3}}
                  \eta_i^\pm\g_\pm\hl^\ad_\pm\overset{\circ}{\phi}\ \!\!^i_{\a\ad}+
                  \tfrac{1}{3!}\eta^\pm_1\eta^\pm_2\g^3_\pm
                 \hl^\ad_\pm\hl^\bd_\pm\hl^\gd_\pm
                  \Gc_{\a\ad\bd\gd},\\
      \CA_3^\pm\ &=\ \pm\tfrac{1}{\sqrt{3}}\eta^\pm_i\g^2_\pm
                   \overset{\circ}{\phi}\ \!\!^i\pm
                   \tfrac{1}{2!}\eta_1^\pm\eta^\pm_2\g^4_\pm
                    \hl^\ad_\pm\hl^\bd_\pm\Gc_{\ad\bd}
}
together with
\eqn{condiII}{\overset{\circ}{\phi}\ \!\!^i_{\a\ad}\ =\ -\cnab\ \!\!^R_{\a\ad}
       \overset{\circ}{\phi}\ \!\!^i
            \qquad{\rm and}\qquad
            \Gc_{\a\ad\bd\gd}\ =\ -\cnab\ \!\!^R_{\a(\ad}\Gc_{\bd\gd)}.}
The remaining nontrivial equations read 
\eqna{eq:hier-ch2}{\fc_{\ad\bd}\ &=\ 0,\\
        \overset{\circ}{\square}\ \!\!^R\overset{\circ}{\phi}\ \!\!^i\ &=\ 0,\\
          \epsilon^{\ad\bd}\cnab\ \!\!^R_{\a\ad}\Gc_{\bd\gd}\ &=\ 
          \epsilon_{ij}\{\overset{\circ}{\phi}\ \!\!^i,
          \cnab\ \!\!^R_{\a\gd}\overset{\circ}{\phi}\ \!\!^j\},
}
where $\overset{\circ}{\square}\ \!\!^R$ has been introduced in \eqref{boxop}.
The associated action functional is given by
\eqn{actionII}{S\ =\ \int\,\dt^4x_R\ {\rm tr}\,\left\{
                    \Gc\ \!\!^{\ad\bd}\fc_{\ad\bd}+
                     \epsilon_{ij}\overset{\circ}{\phi}\ \!\!^i
                     \overset{\circ}{\square}\ \!\!^R
                   \overset{\circ}{\phi}\ \!\!^j\right\},}
and can be obtained from \eqref{bla}.
Note that formally \eqref{actionII} looks as the 
bosonic truncation of the self-dual $\CN=4$ SYM theory, i.e., all the
spinors and four of the six scalars of self-dual $\CN=4$ SYM theory
are put to zero. However, in \eqref{actionII} the parity of the scalars 
$\overset{\circ}{\phi}\ \!\!^i$ is 
different, as they are Gra{\ss}mann odd. To understand their nature, 
note that in the expansions \eqref{fieldexpaII} we have two 
Gra{\ss}mann odd vectors $\overset{\circ}{\phi}\ \!\!^i_{\a\ad}$ which satisfy 
the equations
\eqna{eqforsca}{\epsilon^{\a\b}\cnab\ \!\!^R_{\a\ad}
             \overset{\circ}{\phi}\ \!\!^i_{\b\bd}\ &=\ 0,\\
        \epsilon^{\a\b}\epsilon^{\ad\bd}\cnab\ \!\!^R_{\a\ad}
\overset{\circ}{\phi}\ \!\!^i_{\b\bd}\ &=\ 0}
following from \eqref{shCS1}. Solutions to these equations describe tangent vectors
$\d\Ac_{\a\ad}$ (with assigned odd parity) to the solution space of the 
self-duality equations $\fc_{\ad\bd}=0$ \cite{WittenZE,VafaTF}.
However, due to the
first equation of \eqref{condiII} (which solves the first equation of
\eqref{eqforsca} and reduces the second to  
$\overset{\circ}{\square}\ \!\!^R\overset{\circ}{\phi}\ \!\!^i=0$), 
the $\overset{\circ}{\phi}\ \!\!^i_{\a\ad}$s are projected to zero in the
moduli space of solutions to the equations $\fc_{\ad\bd}=0$. By choosing 
$\overset{\circ}{\phi}\ \!\!^i=0$, we remain with the equations 
\eqn{}{\fc_{\ad\bd}\ =\ 0\qquad{\rm and}\qquad
            \epsilon^{\ad\bd}\cnab\ \!\!^R_{\a\ad}\Gc_{\bd\gd}\ =\ 0,} 
which can be obtained from the Lorentz-invariant Siegel action \cite{SiegelZA} 
\eqn{actionIV}{S\ =\ \int\,\dt^4x_R\ {\rm tr}\ \left\{\Gc\ \!\!^{\ad\bd}\fc_{\ad\bd}\right\},}
describing self-dual YM theory.

\paragraph{HCS theory on $\CP^{3|2}_{4,0}$.}
Finally, we want to discuss the case in which the odd coordinate $\eta^\pm_1$
has weight four and $\eta^\pm_2$ weight zero, i.e., we consider  
$\CP^{3|2}_{4,0}$. Proceeding as in the 
previous two paragraphs, we obtain the following field expansions:
\eqna{}{
      \CA_\a^\pm\ &=\ \l^\ad_\pm\Ac_{\a\ad}+
                  \tfrac{1}{3!}\eta^\pm_1\g_\pm^3\hl^\ad_\pm\hl^\bd_\pm\hl^\gd_\pm
                  \cc_{\a\ad\bd\gd}+\eta^\pm_2\l^\ad_\pm\overset{\circ}{\psi}_{\a\ad}+
                  \tfrac{1}{3!}\eta^\pm_1\eta^\pm_2
                  \g^3_\pm\hl^\ad_\pm\hl^\bd_\pm\hl^\gd_\pm\Gc_{\a\ad\bd\gd},\\
      \CA_3^\pm\ &=\ \pm\tfrac{1}{2!}\eta^\pm_1\g^4_\pm\hl_\pm^\ad\hl_\pm^\bd 
                   \cc_{\ad\bd}\pm\tfrac{1}{2!}
             \eta^\pm_1\eta^\pm_2\g^4_\pm\hl^\ad_\pm\hl^\bd_\pm 
                  \Gc_{\ad\bd}
}
together with the conditions
\eqn{}{\cc_{\a\ad\bd\gd}\ =\ -\cnab\ \!\!^R_{\a(\ad}\cc_{\bd\gd)}
              \qquad{\rm and}\qquad
        \Gc_{\a\ad\bd\gd}\ =\ -\cnab\ \!\!^R_{\a(\ad}\Gc_{\bd\gd)}
              -\{\overset{\circ}{\psi}_{\a(\ad},\cc_{\bd\gd)}\}.}
The field equations of this theory read as
\eqna{eomIII}{ \fc_{\ad\bd}\ &=\ 0,\\
         \epsilon^{\a\b}\cnab\ \!\!^R_{\a(\ad}\overset{\circ}{\psi}_{\b\bd)}\ &=\ 0,\\        
          \epsilon^{\ad\bd}\cnab\ \!\!^R_{\a\ad}\cc_{\bd\gd}\ &=\ 0,\\
  \epsilon^{\ad\bd}\cnab\ \!\!^R_{\a\ad}\Gc_{\bd\gd}\ &=\ 
         -\epsilon^{\ad\bd}\{\overset{\circ}{\psi}_{\a\ad},\cc_{\bd\gd}\}.
}
In this case, the action functional from which these equations arise is
\eqn{actionIII}{S\ =\ \int\,\dt^4x_R\ {\rm tr}\,\left\{
                    \Gc\ \!\!^{\ad\bd}\fc_{\ad\bd}+\epsilon^{\a\b}
                     (\cnab\ \!\!^R_{\a\ad}
             \overset{\circ}{\psi}_{\b\bd})\cc\ \!\!^{\ad\bd}\right\}.}
This time, the multiplet contains a space-time vector 
$\overset{\circ}{\psi}_{\a\ad}$ and an anti-self-dual two-form $\cc_{\ad\bd}$ which
are both Gra{\ss}mann odd. Such fields are well known from topological 
YM theories \cite{WittenZE,VafaTF}. In this respect, the model
\eqref{eomIII}, \eqref{actionIII} can be understood as a truncated 
self-dual sector of these theories. One may, of course, also consider more than 
just two fermionic coordinates in order to enlarge the multiplet. 
This may lead to other truncations of topological YM theories. We will
come to related issues when dealing with (truncated) self-dual SYM 
hierarchies in Chap. \ref{HS-CHAPTER}. Note that the above constructions 
have been formalized in \cite{Saemann:2004tt} in the context of exotic
supermanifolds.

\clearemptydoublepage  
\chapter{Supersymmetric Bogomolny monopole equations}\label{ch:SBME-ch3}
\HRule\\

{\Large A}pproximately two decades ago, it has been conjectured by Ward 
\cite{WardGZ,Ward86,Ward90} 
that  all integrable models in less than four space-time 
dimensions can be obtained from the self-dual
YM equations in four dimensions.
Typical examples of such systems are the nonlinear Schr\"odinger equation, the
Korteweg-de Vries equation, the sine-Gordon model, etc. All of these models follow 
from the self-dual YM equations by incorporating suitable algebraic 
ans\"atze for the self-dual gauge potential followed by a dimensional 
reduction. Also the Bogomolny 
monopole equations on $\IR^3$, describing static Yang-Mills-Higgs monopoles 
in the Prasad-Sommerfield limit, may be added to this list. In fact, Hitchin 
showed \cite{Hitchin:1982gh} that the Bogomolny monopole equations
can be described by
using twistorial methods. He constructed a twistor space -- the so-called
mini-twistor space -- corresponding to $\IR^3$. Geometrically, it is the
space of oriented lines in $\IR^3$. Furthermore, he then gave the 
construction of a Penrose-Ward transform relating equivalence
classes of certain holomorphic vector bundles over mini-twistor space to
gauge equivalence classes of solutions to
the Bogomolny monopole equations. Our subsequent discussion is devoted
to an extension of Hitchin's approach to a supersymmetric setting. It is based
on the work done together with Alexander Popov and Christian S\"amann 
\cite{Popov:2005uv}. We will obtain the mini-supertwistor space,
which leads us to a twistorial description of the supersymmetrized Bogomolny
model. To jump ahead of our story
a bit the mini-supertwistor space can be considered as an open subset
of the weighted projective superspace $W\IC P^{2|4}[2,1,1|1,1,1,1]$ and as
such it is a formally Calabi-Yau; cf. also \ref{WPS-2}\kern6pt. 
Furthermore, on the way of our discussion 
we will meet with the notion of Cauchy-Riemann structures
(see, e.g., Ref. \cite{LeBrun:1984} for the purely even case) which naturally 
generalize the notion of complex structures. This allows us to use 
tools familiar from complex geometry. 

\section{Cauchy-Riemann supermanifolds}\label{sec:CRSM-ch3} 

The supersymmetric Bogomolny monopole equations we are interested
in are obtained from the four-dimensional $\CN=4$ self-dual SYM 
equations by a dimensional reduction $\IR^4\to\IR^3$. In this
section, we study in detail the meaning of this reduction on the
level of the supertwistor space. Note that we shall always be
working in the real setting as discussed in \ref{HCS-2}\kern6pt. 
In particular, we find that the supertwistor
space $\CP^{3|4}$, when interpreted as the real manifold
$\IR^{4|8}\times S^2$, reduces to the space $\IR^{3|8}\times S^2$.
As a complex manifold, however, $\CP^{3|4}$ reduces to the rank
$1|4$ holomorphic vector bundle
$\CP^{2|4}:=\CO_{\IC P^1}(2)\oplus\Pi\CO_{\IC P^1}(1)\otimes\IC^4$. Due
to this difference, the twistor correspondence gets more involved.
For instance, we need a double fibration. We also show that 
$\IR^{3|8}\times S^2$ can be equipped with so-called
Cauchy-Riemann structures.

\paragraph{Dimensional reduction $\IR^{4|8}\times
S^2\to\IR^{3|8}\times S^2$.}
It is well known
that the Bogomolny equations on $\IR^3$ describing BPS monopoles
\cite{Bogomolny:1975de,Prasad:1975kr} can be obtained from the
self-dual YM equations on $\IR^4$ by demanding the components of a gauge 
potential to be independent of $x^4$
and by putting the four-component of the gauge potential
to be the Higgs field
\cite{Manton:1977ht,Hitchin:1982gh,Atiyah:1988jp}. Obviously, one
can similarly reduce the $\CN=4$ self-dual SYM equations \eqref{fieldeqn-2} on
$\IR^4$ by imposing the $\der{x^4}$-invariance condition on all
the fields 
$$(\fc_{\a\b},\cc\ \!\!^i_\a,\Wc\ \!\!^{ij},\cc_{i\ad},
\Gc_{\ad\bd})$$ in the supermultiplet and obtain supersymmetric
Bogomolny equations on $\IR^3$. Recall that both $\CN=4$ self-dual SYM
theory and $\CN=4$ SYM theory have an $SU(4)\cong Spin(6)$
R-symmetry group. In the case of the full $\CN=4$ SYM
theory, the R-symmetry group and supersymmetry get enlarged to
$Spin(7)$ and $\CN=8$ supersymmetry by a reduction from four to
three dimensions, cf. Ref. \cite{Chiou:2005jn}. However, the situation
in the dimensionally reduced $\CN=4$ self-dual SYM theory is more involved
since there is no parity symmetry interchanging left-handed and
right-handed fields, and only the $SU(4)$ subgroup of
$Spin(7)$ is manifest as an R-symmetry of the Bogomolny model.

Recall that on $\IR^4\cong\IC^2$ we may use the complex coordinates
$x^{\a\ad}_R$ satisfying the reality conditions induced by
\eqref{EI-xandeta} or the real coordinates $x^\mu$ defined by
\eqn{xdefeuc-2}{x^{2{\dot2}}_R\ =\ {\bar x}^{1{\dot1}}_R\ =:\ -\i (x^1-\i x^2)
                 \qquad{\rm and}\qquad
                 x^{2{\dot1}}_R\ =\ -{\bar x}^{1{\dot2}}_R\ =:\ -\i (x^3-\i x^4).
}
This choice differs from the one given in \eqref{xdefeuc-1}
by a change of coordinates. However, it is most convenient
for our subsequent discussion. It will also yield a better comparability
with related literature.

Translations generated by the vector field $\cT_4:=\der{x^4}$ are
isometries of $\IR^{4|8}$ and by taking the quotient with respect
to the action of the Abelian group 
\eqn{}{\cG\ :=\ \{\exp(a\cT_4)\,|\,x^4\mapsto x^4+a,a\in \IR\}} 
generated by $\cT_4$, we
obtain the superspace $\IR^{3|8}\cong\IR^{4|8}/\cG$. Recall that
the eight odd complex coordinates $\eta_i^\ad$ satisfy certain
reality conditions induced by
\eqref{EI-xandeta}. The vector field $\cT_4$ is
trivially lifted to $\IR^{4|8}\times S^2$ 
and therefore the supertwistor space,
considered as the smooth supermanifold $\IR^{4|8}\times S^2$, is
reduced to $\IR^{3|8}\times S^2\cong (\IR^{4|8}\times S^2)/\cG$. In
other words, smooth $\cT_4$-invariant functions on
$\CP^{3|4}\cong\IR^{4|8}\times S^2$ can be considered as ``free''
smooth functions on the supermanifold $\IR^{3|8}\times S^2$.

Recall that the
rotation group $SO(4)$ of Euclidean four-dimensional space is locally
isomorphic to $SU(2)_L\times SU(2)_R\cong Spin(4)$. Upon
dimensional reduction to three dimensions, the rotation group
$SO(3)$ of $(\IR^3,\delta_{rs})$ with $r,s=1,2,3$ is locally
$SU(2)\cong Spin(3)$, which is the diagonal group
${\rm diag}(SU(2)_L\times SU(2)_R)$. Therefore, the distinction
between undotted, i.e., $SU(2)_L$, and dotted, i.e.,
$SU(2)_R$, indices disappears. This implies that one can
relabel the bosonic coordinates $x^{\a\bd}_R$ by 
$x^{\ad\bd}_R$ and split them as
\eqn{}{x^{\ad\bd}_R\ =\ x^{(\ad\bd)}_R+x^{[\ad\bd]}_R\ :=\
\tfrac{1}{2}(x^{\ad\bd}_R+x^{\bd\ad}_R)+
\tfrac{1}{2}(x^{\ad\bd}_R-x^{\bd\ad}_R),}
into symmetric
\eqn{ycoords}{y^{\ad\bd}\ :=\  -\i x^{(\ad\bd)}_R,\ \
       y^{\dot1\dot1}\ =\ -\bar{y}^{\dot2\dot2}\ =\ (x^1+\i x^2)\ =:\ y,\ \
       y^{\dot1\dot2}\ =\ \bar{y}^{\dot1\dot2}\ =\ -x^3}
and antisymmetric
\eqn{}{x^{[\ad\bd]}_R\ =\  \epsilon^{\ad\bd}x^4}
parts. 
More abstractly, this splitting corresponds to the
decomposition ${\bf 4}\cong{\bf 3}\oplus{\bf 1}$ of the irreducible real
vector representation ${\bf 4}$ of the group
$Spin(4)\cong SU(2)_L\times SU(2)_R$ into two irreducible
real representations ${\bf 3}$ and ${\bf 1}$ of the group
$Spin(3)\cong SU(2)={\rm diag}(SU(2)_L\times SU(2)_R)$. For
future use, we also introduce the derivations
\eqn{eq:someder_ch3}{\partial_{(\ad\bd)}\ :=\
 \frac{\i}{2}\left(\der{x^{\ad\bd}_R}+\der{x^{\bd\ad}_R}\right),}
which read explicitly as
\eqn{}{\partial_{(\dot1\dot1)}\ =\ \der{y^{\dot1\dot1}},
\qquad\partial_{(\dot1\dot2)}\ =\
\frac{1}{2}\der{y^{\dot1\dot2}}\qquad{\rm and}\qquad 
\dpar_{(\dot2\dot2)}\ =\ \der{y^{\dot2\dot2}}.}
Altogether, we thus have
\eqn{}{\der{x^{\ad\bd}_R}\ =\
-\i\partial_{(\ad\bd)}-\tfrac{1}{2}\epsilon_{\ad\bd}\der{x^4}.}

\paragraph{Holomorphic reduction $\CP^{3|4}\to\CP^{2|4}$.}
The vector field
$\cT_4=\der{x^4}$ yields a free twistor space action of the
Abelian group $\cG\cong \IR$ which is the real part of the
holomorphic action of the complex group $\cG_\IC\cong \IC$. In
other words, we have
\eqna{}{\cT_4\ &=\ \der{x^4}\ =\
\derr{z_+^a}{x^4}\der{z_+^a}+\derr{\bz_+^a}{x^4}\der{\bz_+^a}\\
 &=\
\left(-\der{z^2_+}+z_+^3\der{z^1_+}\right)+\left(-\der{\bz^2_+}+
\bz_+^3\der{\bz^1_+}\right)\ =:\ \cT_+'+\bar{\cT}_+'}
in the coordinates $(z_+^a,\eta_i^+)$ for $a=1,2,3$ on $\CU_+$, where
\eqn{}{\cT_+'\ :=\ \cT'|_{\CU_+}\ =\ 
      -\der{z^2_+}+z_+^3\der{z_+^1}}
is a holomorphic vector field on $\CU_+$. Similarly, we
obtain
\eqn{}{\cT_4\ =\ \cT'_-+\bar{\cT}'_-,
       \qquad{\rm with}\qquad
        \cT'_-\ :=\ \cT'|_{\CU_-}\ =\ -z^3_-\der{z_-^2}+\der{z^1_-}}
on $\CU_-$ and $\cT_+'=\cT_-'$ on
$\CU_+\cap\CU_-$. For holomorphic functions $f$ on
$\CP^{3|4}$ we clearly have $\cT_4f=\cT' f$,
and therefore $\cT'$-invariant holomorphic functions on
$\CP^{3|4}$ can be considered as ``free'' holomorphic functions on
a reduced space $\CP^{2|4}\cong \CP^{3|4}/\cG_\IC$ obtained as
the quotient space of $\CP^{3|4}$ by the action of the complex
Abelian group $\cG_\IC$ generated by $\cT'$.

In the purely even case,
the space $\CP^2\cong \CP^3/\cG_\IC$ was called mini-twistor
space \cite{Hitchin:1982gh} and we shall refer to $\CP^{2|4}$ as
the mini-supertwistor space. To sum up, the reduction of the
supertwistor correspondence induced by the $\cT_4$-action is
described by the following diagram:
\eqna{mrsuperdblfibration2}{
\begin{picture}(150,100)
\put(0.0,0.0){\makebox(0,0)[c]{$\CP^{2|4}$}}
\put(140.0,0.0){\makebox(0,0)[c]{$\IR^{3|8}$}}
\put(0,90.0){\makebox(0,0)[c]{$\CP^{3|4}$}}
\put(30.0,88.0){\makebox(0,0)[c]{$\cong$}}
\put(73.0,90.0){\makebox(0,0)[c]{$\IR^{4|8}\times S^2$}}
\put(102.0,90.0){\vector(1,0){20}}
\put(140,90.0){\makebox(0,0)[c]{$\IR^{4|8}$}}
\put(73.0,50.0){\makebox(0,0)[c]{$\IR^{3|8}\times S^2$}}
\put(26.0,27.0){\makebox(0,0)[c]{$\pi_1$}}
\put(110.0,27.0){\makebox(0,0)[c]{$\pi_2$}}
\put(59.0,37.0){\vector(-4,-3){40}}
\put(82.0,37.0){\vector(4,-3){40}}
\put(0.0,80.0){\vector(0,-1){70}}
\put(72.0,80.0){\vector(0,-1){20}}
\put(140.0,80.0){\vector(0,-1){70}}
\end{picture}
}
Here, ``$\downarrow$" symbolizes projections generated by the action
of the groups $\cG$ or $\cG_\IC$ and $\pi_2$ is the canonical
projection. The projection $\pi_1$ will be described momentarily.

\paragraph{Geometry of mini-supertwistor space.}
It is not difficult to see that the
functions
\eqna{}{w_+^1\ &:=\ -\i(z_+^1+z_+^3z_+^2),\quad
        w_+^2\ :=\ z_+^3\quad{\rm and}\quad
       \eta_i^+\quad{\rm on}\quad\CU_+,\\
       w_-^1\ &:=\ -\i(z_-^2+z_-^3z_-^1),\quad
       w_-^2\ :=\ z_-^3\quad{\rm and}\quad\eta_i^-
       \quad{\rm on}\quad\CU_-}
are constant along the $\cG_\IC$-orbits in $\CP^{3|4}$ and thus
descend to the patches $\CV_\pm:=\CU_\pm\cap\CP^{2|4}$
covering the (orbit) space $\CP^{2|4}\cong \CP^{3|4}/\cG_\IC$. On
the overlap $\CV_+\cap\CV_-$, we have
\eqn{}{w_+^1\ =\ \frac{1}{(w_-^2)^2}w_-^1,\qquad
       w_+^2\ =\ \frac{1}{w_-^2}\qquad{\rm and}\qquad
       \eta_i^+\ =\ \frac{1}{w_-^2}\eta_i^-}
which coincides with the transformation laws of canonical coordinates
on the total space
\eqn{}{\CO_{\IC P^1}(2)\oplus\Pi\CO_{\IC P^1}(1)\otimes\IC^4\ =\ \CP^{2|4}}
of the holomorphic vector bundle
\eqn{}{\CP^{2|4}\ \to\ \IC P^1.}
Clearly, this space is a formal Calabi-Yau supermanifold. Hence,
it comes with a globally well-defined nowhere vanishing holomorphic volume form
\eqn{volumemini}{\O|_{\CV_\pm}\ =\ \pm\dt w_\pm^1\wedge \dt w_\pm^2
         \dt\eta_1^\pm\cdots\dt\eta_4^\pm.}
The body of this supermanifold is the mini-twistor
space \cite{Hitchin:1982gh}
$$\CP^2\ \cong\ \CO_{\IC P^1}(2).$$
Note that the space $\CP^{2|4}$ can be considered as an
open subset of the weighted projective superspace
$W\IC P^{2|4}[2,1,1|1,1,1,1]$.

The real structure $\tau_E$ (cf. our discussion given in \ref{EuSi}\kern6pt) on
$\CP^{3|4}$ induces a real structure on $\CP^{2|4}$ acting on
local coordinates by the formula
\eqn{eq:3.16}{
\tau_E(w_\pm^1,w_\pm^2,\eta_i^\pm)\ =\
\left(-\frac{\bar{w}_\pm^1}{(\bar{w}_\pm^2)^2},
-\frac{1}{\bar{w}_\pm^2},\pm\frac{1}{\bar{w}^2_\pm}
{T_i}^j\bar{\eta}_j^\pm\right),}
where the matrix $({T_i}^j)$ has been defined in
\eqref{defofT}. From \eqref{eq:3.16}, one sees that $\tau_E$ has no
fixed points in $\CP^{2|4}$ but leaves invariant projective lines
$\IC P^1_{x,\eta}\hookrightarrow\CP^{2|4}$ defined by the equations
\eqna{eq:3.17}{w_+^1\ &=\ y-2\l_+x^3-\l_+^2\bar{y},&\eta_i^+&\ =\
\eta_i^{\dot1}+\l_+\eta_i^{\dot2},&&\mbox{with}~~\l_+\ =\ w_+^2\
\in\ U_+,\\
w_-^1\ &=\ \l_-^2y-2\l_-x^3-\bar{y}~,&\eta_i^-&\ =\
\l_-\eta_i^{\dot1}+\eta_i^{\dot2},&&\mbox{with}~~\l_-\ =\ w_-^2\
\in\ U_-}
for fixed $(x,\eta)\in\IR^{3|8}$. Here, $y=x^1+\i x^2$,
$\bar{y}=x^1-\i x^2$ and $x^3$ are coordinates on $\IR^3$ and $U_\pm$ 
denote again the canonical patches covering $\IC P^1$.

By using the coordinates \eqref{ycoords}, we can rewrite
\eqref{eq:3.17} as
\eqn{eq:3.18}{
w_\pm^1\ =\ \l_\ad^\pm\l_\bd^\pm
y^{\ad\bd},~~~w_\pm^2\ =\ \l_\pm\qquad{\rm and}\qquad \eta_i^\pm\ =\
\eta_i^\ad\l_\ad^\pm,}
where the explicit form of $\l_\ad^\pm$ has been given in
\ref{someLC}\kern6pt. In fact, Eqs. \eqref{eq:3.18} are the
incidence relations which lead to the double fibration
\eqna{eq:3.19}{
\begin{picture}(50,40)
\put(0.0,0.0){\makebox(0,0)[c]{$\CP^{2|4}$}}
\put(64.0,0.0){\makebox(0,0)[c]{$\IR^{3|8}$}}
\put(34.0,33.0){\makebox(0,0)[c]{$\CF^{5|8}$}}
\put(7.0,18.0){\makebox(0,0)[c]{$\pi_1$}}
\put(55.0,18.0){\makebox(0,0)[c]{$\pi_2$}}
\put(25.0,25.0){\vector(-1,-1){18}}
\put(37.0,25.0){\vector(1,-1){18}}
\end{picture}}
where $\CF^{5|8}\cong \IR^{3|8}\times S^2$, $\pi_2$ is again the
canonical projection onto $\IR^{3|8}$ and the projection $\pi_1$
is defined by the formula
\eqn{eq:3.20}{
\pi_1(x^r,\l_\pm,\eta_i^\ad)\ =\
\pi_1(y^{\ad\bd},\l_\ad^\pm,\eta_i^\ad)\ =\
(w^1_\pm,w_\pm^2,\eta_i^\pm),}
where $r=1,2,3$, and $w_\pm^{1,2}$ and $\eta_i^\pm$ are given in
\eqref{eq:3.18}. The diagram \eqref{eq:3.19}, which is a part of
\eqref{mrsuperdblfibration2}, describes the following proposition:
{\Pro There exist the following geometric correspondences:
\vspace*{-2mm}
\begin{center} 
 \begin{tabular}{cccc}
   {\rm (i)} & point $p$ in $\CP^{2|4}$ & $\quad\longleftrightarrow\quad$ & 
                                    oriented $\IR^{1|0}_p$ in $\IR^{3|8}$\\
   {\rm (ii)} & $\tau_E$ -invariant 
 $\IC P^1_{x,\eta}\hookrightarrow\CP^{2|4}$&$\quad\longleftrightarrow\quad$ & 
                                    point $(x,\eta)$ in $\IR^{3|8}$
 \end{tabular}
\end{center}}\vskip 4mm
\paragraph{Cauchy-Riemann supermanifolds.}
Consider the double fibration \eqref{eq:3.19}. 
The correspondence space $\IR^{3|8}\times S^2$ is the smooth 
$5|8$-dimensional supermanifold. As it is of ``wrong" dimensionality,
it cannot be a complex supermanifold but it can be 
understood as a so-called
Cauchy-Riemann (CR) supermanifold, i.e., as a 
partially complex supermanifold. Recall that a CR structure on a
smooth supermanifold $X$ of real dimension $m|n$ is a locally
direct subsheaf
$\cDb$ of rank $r|s$ of the complexified tangent sheaf $T_\IC X$
such that $\cD\cap\cDb=\{0\}$ and $\cDb$ is involutive
(integrable), i.e., $\cDb$ is
closed with respect to the Lie superbracket. Of course, the
distribution $\cD$ is
integrable if $\cDb$ is integrable. The pair $(X,\cDb)$ is
called a CR supermanifold of dimension $m|n=\dim_\IR X$, of rank
$r|s=\dim_\IC \cDb$ and of codimension $m-2r|n-2s$. In particular, a CR
structure on $X$ in the special case $m|n=2r|2s$ is a complex structure
on $X$. Thus, the notion of CR supermanifolds generalizes that of
complex supermanifolds.

Given a CR supermanifold $(X,\cDb)$, we let $\O^k(X):=\Lambda^k T_\IC^*X$
be the sheaf of complex-valued smooth differential $k$-forms. Then we define
locally free subsheaves of $\CS_X$-modules by
\eqn{}{\hat{\O}^{p,q}_{\rm CR}(X)\ :=\ \{\o\in\O^{p+q}(X)\,|\,\bV_0\wedge\bV_1\wedge
       \cdots\wedge\bV_q\lrcorner\,\o = 0,\ \forall\,\bV_i\in\cDb\}.}
Furthermore, we set $\hat{\O}^{p,-1}_{\rm CR}(X):=\{0\in\O^0\}(X)\}$. 
Clearly, we have
$$\dt\,:\, \hat{\O}^{p,q}_{\rm CR}(X)\ \to\ \hat{\O}^{p,q+1}_{\rm CR}(X).$$ 
We now define
complex-valued differential $(p,q)$-forms $\O^{p,q}_{\rm CR}(X)$ on $X$
according to
\eqn{}{\O^{p,q}_{\rm CR}(X)\ :=\ \hat{\O}^{p,q}_{\rm CR}(X)/\hat{\O}^{p+1,q-1}_{\rm CR}(X).}
Then we can introduce a natural family of $\pbar$-operators, i.e.,
$\pbar\,:\, \O^{p,q}_{\rm CR}(X)\to \O^{p,q+1}_{\rm CR}(X)$ by requiring
that the diagrams
\begin{center}
\begin{tabular}{ccccccc}
     && $0$ && $0$ && \\
     && $\uparrow$ && $\uparrow$ && \\
 $\cdots$ & $\overset{\pbar}{\to}$ & $\O^{p,q}_{\rm CR}(X)$ &
 $\overset{\pbar}{\to}$ & $\O^{p,q+1}_{\rm CR}(X)$ &
 $\overset{\pbar}{\to}$ & $\cdots$ \\
     && $\uparrow$ && $\uparrow$ && \\
 $\cdots$ & $\overset{\rm d}{\to}$ & ${\hat\O}^{p,q}_{\rm CR}(X)$ &
 $\overset{\rm d}{\to}$ & ${\hat\O}^{p,q+1}_{\rm CR}(X)$ &
 $\overset{\rm d}{\to}$ & $\cdots$ \\
     && $\uparrow$ && $\uparrow$ && \\
 $\cdots$ & $\overset{\rm d}{\to}$ & ${\hat\O}^{p+1,q-1}_{\rm CR}(X)$ &
 $\overset{\rm d}{\to}$ & ${\hat\O}^{p+1,q}_{\rm CR}(X)$ &
 $\overset{\rm d}{\to}$ & $\cdots$ \\
     && $\uparrow$ && $\uparrow$ && \\
     && $0$ && $0$ &&
\end{tabular}
\end{center}
should commute. 

\paragraph{Cauchy-Riemann supertwistors.}\label{CRST-ex}
Let us now come back to our example \eqref{eq:3.19}.
On the manifold $\IR^{3|8}\times S^2$, one can introduce several CR
structures. For instance, we may choose
\eqn{}{\CF_0^{5|8}\ :=\ (\IR^{3|8}\times S^2,\cDb_0)
      \ \cong\ \IR^{1|0}\times\IC^{1|4}\times\IC P^1}
for the distribution
\eqn{}{\cDb_0\ =\ \left\langle\der{\bar{y}},\der{\bl_\pm},
\der{\bar{\eta}_i^{\dot1}}\right\rangle.}
Another one is obtained by setting
\eqn{}{\CF^{5|8}\ :=\ (\IR^{3|8}\times S^2,\cDb=\pi_1^*T^{0,1}_\IC\CP^{2|4}),}
i.e.,
\eqn{eq:3.23}{\cDb\ =\
\left\langle\pi_1^*\der{\bw^1_\pm},\pi_1^*\der{\bw^2_\pm},
\pi_1^*\der{\bar{\eta}_i^\pm}\right\rangle.}
In the following, we shall suppress the explicit appearance of
$\pi_1^*$.
In spirit of LeBrun's \cite{LeBrun:1984}, we call $\CF^{5|8}$ the CR supertwistor
space.\footnote{In the purely even case, a CR
five-dimensional manifold can be
constructed as a sphere bundle over an arbitrary three-dimensional manifold
with conformal metric \cite{LeBrun:1984}.}
Clearly, all the criteria for a CR structure are satisfied for
our above two choices and moreover, in both cases the CR structures
have rank $2|4$.

Let us denote the covering of $\CF^{5|8}$ by $\hfV=\{\hCV_+,\hCV_-\}$.
Up to now, we have used the coordinates
$(y,\bar{y},x^3,\l_\pm,\bl_\pm,\eta_i^\ad)$ or
$(y^{\ad\bd},\l_\ad^\pm,\hl_\ad^\pm,\eta_i^\ad)$ on the
two patches $\hCV_\pm$. More appropriate for the distribution
\eqref{eq:3.23} are, however, the coordinates \eqref{eq:3.17}
together with
\eqna{eq:3.24}{
  w_+^3\ &:=\ \tfrac{1}{1+\l_+\bl_+}
  \left[\bl_+y+(1-\l_+\bl_+)x^3+\l_+\bar{y}\right]
  \qquad{\rm on}\qquad\hCV_+,\\
  w_-^3\ &:=\ \tfrac{1}{1+\l_-\bl_-}
   \left[\l_-y+(\l_-\bl_--1)x^3+\bl_-\bar{y}\right]
\qquad{\rm on}\qquad\hCV_-.}
In terms of the coordinates \eqref{ycoords} and $\l_\ad^\pm$, 
we can rewrite \eqref{eq:3.17} and
\eqref{eq:3.24} concisely as
\eqn{eq:3.25}{
  w_\pm^1\ =\ \lambda_\ad^\pm\lambda_\bd^\pm y^{\ad\bd},~~~
  w_\pm^2\ =\ \lambda_\pm,~~~ w_\pm^3\ =\
 -\gamma_\pm\lambda_\ad^\pm\hat{\lambda}^\pm_\bd
y^{\ad\bd}\quad{\rm and}\quad\eta_i^\pm\ =\ \eta_i^\ad\lambda_\ad^\pm,
}
where the factors $\gamma_\pm$ have been given in \eqref{gam}.
Note that $w_\pm^3$ is real and all the other
coordinates in \eqref{eq:3.25} are complex. The relations between
the coordinates on $\hCV_+\cap\hCV_-$ follow
directly from their definitions \eqref{eq:3.25}.

The coordinates
$w_\pm^{1,2}$ and $\eta_i^\pm$ have already appeared in
\eqref{eq:3.18} since $\CP^{2|4}$ is a complex subsupermanifold of
$\CF^{5|8}$. Recall that formulas \eqref{eq:3.20} together
with \eqref{eq:3.25} define a projection
\eqn{eq:3.26}{
\pi_1\ :\ \CF^{5|8}\ \to\ \CP^{2|4}}
onto mini-supertwistor space $\CP^{2|4}$. The fibers over
points $p\in\CP^{2|4}$ of this projection are real one-dimensional
spaces $\ell_p\cong \IR$ parametrized by $w_\pm^3$.
Note that the pull-back to $\CF^{5|8}$ of the real structure
$\tau_E$ on $\CP^{2|4}$ given in \eqref{eq:3.16} reverses the
orientation of each line $\ell_p$, since $\tau_E(w_\pm^3)=-w_\pm^3$.

In order to clarify the geometry of the fibration \eqref{eq:3.26},
we note that the body $\CF^5\cong\IR^3\times S^2$ of the
supermanifold $\CF^{5|8}$ can be considered as the sphere bundle
\eqn{}{
S(T\IR^3)\ =\ \{(x,u)\in T\IR^3\,|\,\delta_{rs}u^r u^s=1\}\ \cong\
\IR^3\times S^2 }
whose fibers at points $x\in\IR^3$ are spheres of unit vectors in
$T_x\IR^3$ \cite{Hitchin:1982gh}. Since this bundle is trivial,
its projection onto $\IR^3$ is obviously $\pi_2(x,u)=x$. Moreover,
the complex two-dimensional mini-twistor space $\CP^2$ can be
described as the space of all oriented lines in $\IR^3$. That is,
any such line $\ell$ is defined by a unit vector $u^r$ in the
direction of $\ell$ and a shortest vector $v^r$ from the origin in
$\IR^3$ to $\ell$, and one can show \cite{Hitchin:1982gh} that
\eqn{}{
\CP^2\ =\ \{(v,u)\in
 T\IR^3\,|\,\delta_{rs}u^rv^s=0\,,~\delta_{rs}u^ru^s=1\}\ \cong\
 T\IC P^1\ \cong\ \CO_{\IC P^1}(2).}

The fibers of the projection $\pi_1\,:\,\IR^3\times S^2\to
\CP^2$ are the orbits of the action of the group $\cG'\cong \IR$
on $\IR^3\times S^2$ given by 
$(v^r,u^s)\mapsto(v^r+tu^r,u^s)$ for $t\in\IR$ and
\eqn{}{
\CP^2\ \cong\ \IR^3\times S^2/\cG'.}

Recall that $\CF^5\cong \IR^3\times S^2$ is a (real) hypersurface
in the twistor space $\CP^3$. On the other hand, $\CP^2$ is a
complex two-dimensional submanifold of $\CF^5$ and therefore
$$\CP^2\ \subset\ \CF^5\ \subset\ \CP^3.$$
Similarly, we have in the supertwistor case
$$\CP^{2|4}\ \subset\ \CF^{5|8}\ \subset\ \CP^{3|4}.$$

The formulas given in
\eqref{eq:3.17} and \eqref{eq:3.24}, respectively, define a coordinate
transformation
$(y,\bar{y},x^3,\lambda_\pm,\bl_\pm,\eta_i^\ad)\mapsto(w_\pm^a,\eta_i^\pm)$
on $\CF^{5|8}$. From corresponding inverse formulas defining
the transformation
$(w_\pm^a,\eta_i^\pm)\mapsto(y,\bar{y},x^3,\lambda_\pm,\bl_\pm,\eta_i^\ad)$,
we obtain
\eqna{}{
\der{w_+^1}\ &=\
\gamma_+^2\left(\der{y}-\bl_+\der{x^3}-\bl_+^2\der{\bar{y}}\right)\ =:\ \gamma_+^2
W_1^+,\\
\der{w_+^2}\ &=\
W_2^++2\gamma_+^2(x^3+\lambda_+\bar{y})W_1^+-\gamma_+^2(\bar{y}-2\bl_+
x^3-\bl_+^2 y) W_3^+-\gamma_+\bar{\eta}_i^{\dot1} V_+^i,\\
\der{w_+^3}\ &=\
2\gamma_+\left(\lambda_+\der{y}+\bl_+\der{\bar{y}}+\frac{1}{2}
(1-\lambda_+\bl_+)\der{x^3}\right)\
=:\ W_3^+,
}
as well as
\eqna{}{
\der{w_-^1}\ &=\ \gamma_-^2\left(\bl_-^2\der{y}-\bl_-\der{x^3}
-\der{\bar{y}}\right)\ =:\ \gamma_-^2W_1^-,\\
\der{w_-^2}\ &=\
W_2^-+2\gamma_-^2(x^3-\lambda_-y)W_1^-+\gamma_-^2(\bl_-^2\bar{y}
-2\bl_-x^3-y)W_3^-+\gamma_-\bar{\eta}^{\dot2}_i
V_-^i,\\
\der{w_-^3}\ &=\
2\gamma_-\left(\bl_-\der{y}+\lambda_-\der{\bar{y}}+\frac{1}{2}
(\lambda_-\bl_--1)\der{x^3}\right)\
=:\ W_3^-,
}
where $W_2^\pm:=\der{\lambda_\pm}$.
Thus, when working in the
coordinates $(y,\bar{y},x^3,\lambda_\pm,\bl_\pm,\eta_i^\ad)$ on
$\hCV_\pm\subset \CF^{5|8}$, we will use the even vector
fields $W_a^\pm$ with $a=1,2,3$ and the odd vector fields
$V_\pm^i$ together with their complex conjugates $\bW_{1,2}^\pm$ and
$\bV_\pm^i$, respectively. Note
that the vector field $W_3^\pm$ is real!

Hence, we learn that when using the coordinates
$(y^{\ad\bd},\l^\pm_\ad,\hl^\pm_\ad,\eta^\ad_i)$,
the CR tangent sheaf 
$T^{1,0}_{\rm CR}\CF^{5|8}$ is freely generated by 
\eqna{eq:3.34}{
 W_1^\pm\ =\
 \hl^\ad_\pm\hl^\bd_\pm\partial_{(\ad\bd)},~~~W_2^\pm\ =\
\partial_{\lambda_\pm},~~~ W_3^\pm\ =\
2\gamma_\pm\hl^\ad_\pm\lambda^\bd_\pm\partial_{(\ad\bd)},\\
V_\pm^i\ =\ -\hl_\pm^\ad T_j{}^i\der{\eta_j^\ad}~\hspace{4.0cm}~
} 
while $T^{0,1}_{\rm CR}\CF^{5|8}$ is generated by
\eqna{eq:3.35}{
 \bW_1^\pm\ =\
 -\lambda^\ad_\pm\lambda_\pm^\bd\partial_{(\ad\bd)},~~~\bW_2^\pm\
=\ \partial_{\bl_\pm},~~~ 
\bV_\pm^i\ =\ \lambda_\pm^\ad \der{\eta_i^\ad}.}

It is not too difficult to see that
forms dual to the vector fields \eqref{eq:3.34} and
\eqref{eq:3.35} are
\eqna{eq:3.37}{
 \T_\pm^1\ :=\  \gamma^2_\pm\lambda_\ad^\pm\lambda_\bd^\pm\dt
 y^{\ad\bd},~~~ \Theta_\pm^2\ :=\  \dt \lambda_\pm,~~~
 \T_\pm^3\ :=\  -\gamma_\pm\lambda_\ad^\pm\hl_\bd^\pm\dt
 y^{\ad\bd},\\ E_i^\pm\ :=\ \gamma_\pm\lambda_\ad^\pm
 T_i{}^j\dt \eta_j^\ad\hspace{4cm}
}
and
\eqn{eq:3.37b}{
\bT_\pm^1\ =\ -\gamma^2_\pm\hl_\ad^\pm\hl_\bd^\pm\dt
y^{\ad\bd},~~~\bT_\pm^2\ =\ \dt \bl_\pm,~~~
 \bE_i^\pm\ =\
-\gamma_\pm\hl^\pm_\ad\dt \eta^\ad_i,}
where $T_i{}^j$ has been given in \eqref{defofT}. The exterior
derivative $\pbar$ on $\CF^{5|8}$ reads as
\eqn{}{
        \pbar|_{\hCV_\pm}\ =\ \dt \bw_\pm^1\der{\bw_\pm^1}+\dt
         \bw_\pm^2\der{\bw_\pm^2} +\dt
        \bar{\eta}_i^\pm\der{\bar{\eta}_i^\pm}   
        \ =\ \bT_\pm^1\bW_1^\pm+\bT_\pm^2\bW_2^\pm+\bE_i^\pm \bV^i_\pm.}
Note again that $\T_\pm^3$ and $W_3^\pm$ are both
real. To homogenize the notation later on, we shall also
use $\bW_3^\pm$ and $\partial_{\bw^\pm_3}$ instead of $W_3^\pm$ and
$\partial_{w^\pm_3}$, respectively.

\section{Partially holomorphic Chern-Simons theory}\label{sec:phCSt-3}

We have discussed how mini-supertwistor and CR supertwistor
spaces arise via dimensional reductions from supertwistor
space of four-dimensional space-time. Subject of this section is
the discussion of a generalization of Chern-Simons theory and its
relatives to this setup. We call the theory we are about to
introduce partially holomorphic Chern-Simons theory or phCS
theory for short. Roughly speaking, this theory is a mixture of
Chern-Simons and holomorphic Chern-Simons (hCS) theory on the CR
supertwistor space $\CF^{5|8}$ which has one real and two complex
even dimensions. This theory is a reduction of hCS theory on
$\CP^{3|4}$. As we will show below, there is a one-to-one
correspondence between the moduli space of solutions to the
equations of motion of phCS theory on $\CF^{5|8}$ and the moduli
space of solutions to the supersymmetrized Bogomolny equations
on $\IR^3$, quite similar to Thm. \ref{SDYM-HCS} for $\CN$-extended
self-dual SYM theory.

\paragraph{Partially flat connections.}\label{PFC-3}
Let $X$ be a smooth supermanifold of real dimension 
$m|n$ and $T_\IC X$ the complexified
tangent sheaf of $X$. A locally direct subsheaf $\cT\subset T_\IC X$ is
integrable if i) $\cT\cap\overline{\cT}$ has constant rank $r|s$ and
ii) $\cT$ and $\cT\cap\overline{\cT}$ are closed under the Lie
superbracket. Note that a CR structure is the special case of
an integrable distribution $\cT$ with $r|s=0|0$.

For any smooth function $f$ on $X$, let $\dt_\cT f$ denote the
restriction of $\dt f$ to $\cT$, i.e., $\dt_\cT$ is the
composition
\eqn{}{\CS_X\ \overset{\dt}{\to}\ T_\IC^*X\
        \to\ \cT^*,}
where $\cT^*$ denotes the sheaf
of smooth complex-valued differential one-forms dual to $\cT$;
cf., e.g., Rawnsley's discussion for the purely even case 
\cite{Rawnsley}. The operator
$\dt_\cT$ can be extended to act on relative differential $k$-forms 
denoted by $\Omega^k_\cT(X):=\Lambda^k\cT^*$,
\eqn{}{\dt_\cT\,:\, \Omega^k_\cT(X)\ \rightarrow\
\Omega_\cT^{k+1}(X).
}

Let $\CE$ be a complex vector bundle over $X$. A connection
on $\CE$ along the distribution $\cT$ -- a $\cT$-connection
-- is an even morphism of sheaves
\eqn{}{
 \nabla_\cT\, :\, \CE\ \to\ \O^1_\cT(X)\otimes \CE}
satisfying the Leibniz formula
\eqn{}{\nabla_\cT(f\sigma)\ =\ f\nabla_\cT \sigma+\dt_\cT f\otimes \sigma,}
where $\sigma$ is a local section of $\CE$ and $f$
is a local smooth function. This $\cT$-connection extends to 
\eqn{}{
 \nabla_\cT\, :\, \Omega^k_\cT(X,\CE)\ \to\ \Omega_\cT^{k+1}(X,\CE),}
where $\Omega^k_\cT(X,\CE):=\O^k_\cT(X)\otimes \CE$.
Locally, $\nabla_\cT$ has the form
\eqn{}{\nabla_\cT\ =\ \dt_\cT+\CA_\cT,}
where the standard ${\rm End}\,\CE$-valued $\cT$-connection one-form
$\CA_\cT$ has components only along the distribution $\cT$. As
usual, $\nabla^2_\cT$ naturally induces
\eqn{}{\CF_\cT\ \in\ \Gamma(X,\O^2_\cT(X)\otimes{\rm End}\,\CE)}
which is the curvature of $\CA_\cT$. We say that $\nabla_\cT$
is flat, if $\CF_\cT=0$. For a flat $\nabla_\cT$, the pair
$(\CE,\nabla_\cT)$ is called a $\cT$-flat vector bundle. 
In particular, if $\cT$ is a CR structure then
$(\CE,\nabla_\cT)$ is a CR vector bundle. Moreover, if $\cT$ is the
integrable sheaf $T^{0,1}_\IC X$ on some complex supermanifold $X$ then the
$\cT$-flat complex vector bundle $(\CE,\nabla_\cT)$ is a holomorphic
bundle.

\paragraph{Field equations on the CR supertwistor space.}\label{FECRCTS-3}
Consider the
CR supertwistor space $\CF^{5|8}$ and a distribution $\cT$
generated by the vector fields $\bW_1^\pm$, $\bW_2^\pm$,
$\bV^i_\pm$ from the CR structure $\cDb$ and $\bW_3^\pm$.
This distribution is integrable since all conditions described in
\ref{PFC-3}\kern10pt are satisfied, e.g., the only nonzero commutator is
\eqn{}{[\bW_2^\pm,\bW_3^\pm]\ =\ \pm2 \gamma_\pm^2\bW_1^\pm}
and therefore $\cT$ is closed under the Lie superbracket. Also,
\eqn{}{\cV\ :=\ \cT\cap\overline{\cT}}
is of (real) rank $1|0$ and hence integrable. The
vector field $\bW_3^\pm$ is a basis section for $\cV$ over the patches
$\hCV_\pm\subset \CF^{5|8}$. Note that 
mini-supertwistor space $\CP^{2|4}$ is a subsupermanifold of
$\CF^{5|8}$ transversal to the leaves of $\cV=\cT\cap\overline{\cT}$
and furthermore, $\cT|_{\CP^{2|4}}=\cDb$. Thus, we have an
integrable distribution $\cT$ defined by
\eqn{defofct}{0\ \to\ \pi^*_1 T^{0,1}_\IC\CP^{2|4}\ \to\ 
 \cT\ \to\ 
(\O^{1,0}_{\rm CR}(\CF^{5|8})/\pi_1^*\O^{1,0}(\CP^{2|4}))^*\ \to\ 0}
on the CR supertwistor space $\CF^{5|8}$ and we will denote by $\cT_b$ its
part generated by the even vector fields $\bW_1^\pm$,
$\bW_2^\pm$ and $\bW_3^\pm$,
\eqn{}{\cT_b\ :=\ \langle\bW_1^\pm,\bW_2^\pm,\bW_3^\pm\rangle.}

Let $\CE$ be a trivial
rank $r$ complex vector bundle over $\CF^{5|8}$ and $\CA_\cT$ a
$\cT$-connection one-form on $\CE$ with
$\cT$ given by \eqref{defofct}. Consider now the subspace $\CX$ of
$\CF^{5|8}$ which is parametrized by the same even coordinates
but only the holomorphic odd coordinates of $\CF^{5|8}$,
i.e., on $\CX$, all objects are holomorphic in $\eta_i^\pm$. As
it was already noted, the 
mini-supertwistor space is a formal Calabi-Yau supermanifold. In
particular, this ensures the existence of a holomorphic volume
form on $\CP^{2|4}$. Moreover, $\CP^{2|4}\subset\CF^{5|8}$ and the
pull-back $\hat{\O}$ of this form is globally defined
on $\CF^{5|8}$. Locally, on the patches
$\hCV_\pm\subset \CF^{5|8}$, one obtains
\eqn{eq:4.12}{
\hat{\O}|_{\hCV_\pm}\ =\ \pm\dt w_\pm^1\wedge \dt
w_\pm^2\dt \eta_1^\pm\cdots\dt\eta_4^\pm.}
This well-defined integral form allows us to integrate on $\CX$
by pairing it with elements from $\Omega^3_{\cT_b}(\CX)$.
Assume
that $\CA_\cT$ neither contains antiholomorphic odd components
nor depends on $\bar{\eta}_i^\pm$,
\eqn{eq:4.13}{
\bV_\pm^i\lrcorner \CA_\cT\ =\ 0\qquad{\rm and}\qquad\bV_\pm^i(\CA^\pm_a)\ =\
0,~~~\mbox{with}~~~ \CA^\pm_a\ :=\  \bW_a^\pm\lrcorner\CA_\cT,}
that is, $\CA_\cT\in\Omega^1_{\cT_b}(\CX,{\rm End}\,\CE)$. Now, we
introduce a CS-type action functional
\eqn{eq:4.14}{
 S\ =\ \int_\CX\hat{\O}\wedge{\rm tr}\,\left\{\CA_\cT\wedge
\dt_\cT\CA_\cT+\tfrac{2}{3}
\CA_\cT\wedge\CA_\cT\wedge\CA_\cT\right\},}
where
\eqn{}{\dt_\cT|_{\hCV_\pm}\ =\ \dt \bw_\pm^a\der{\bw_\pm^a}+\dt
\bar{\eta}_i^\pm\der{\bar{\eta}_i^\pm}}
is the $\cT$-part of the exterior derivative $\dt$ on $\CF^{5|8}$.

The action \eqref{eq:4.14} leads to the CS-type field equations
\eqn{eq:4.16}{
\dt_\cT\CA_\cT+\CA_\cT\wedge\CA_\cT\ =\ 0,}
which are the equations of motion of partially holomorphic Chern-Simons (phCS) theory. 
In the nonholonomic basis $(\bW_a^\pm, \bV_\pm^i)$ of the distribution
$\cT$ over $\hCV_\pm\subset\CF^{5|8}$, these equations read
as
\eqna{eq:4.17}{
\bW_1^\pm\CA_2^\pm-\bW^\pm_2\CA_1^\pm+[\CA_1^\pm,\CA_2^\pm]\ &=\
0~,\\
\bW_2^\pm\CA_3^\pm-\bW^\pm_3\CA_2^\pm+[\CA_2^\pm,\CA_3^\pm]\mp2
\gamma_\pm^2\CA_1^\pm\ &=\ 0,\\
\bW_1^\pm\CA_3^\pm-\bW^\pm_3\CA_1^\pm+[\CA_1^\pm,\CA_3^\pm]\ &=\
0,}
where the components $\CA_a^\pm$ have been defined in
\eqref{eq:4.13}.

\paragraph{Equivalence to supersymmetric Bogomolny equations.}\label{par:esbe-3}
Note that from
\eqref{eq:3.34} and \eqref{eq:3.35}, it follows that
\eqn{}{
\bW_1^+\ =\ \lambda_+^2\bW_1^-,\quad\bW_2^+\ =\
-\bl_+^{-2}\bW_2^-\quad{\rm and}\quad \gamma_+^{-1}\bW_3^+\ =\
\lambda_+\bl_+\left(\gamma_-^{-1}\bW_3^-\right)}
and therefore $\CA_1^\pm$, $\CA_2^\pm$ and
$\gamma_\pm^{-1}\CA_3^\pm$ take values in the bundles $\CO_{\IC P^1}(2)$,
$\bCO_{\IC P^1}(-2)$ and 
$\CO_{\IC P^1}(1)\otimes\bCO_{\IC P^1}(1)$, respectively. Together with
the definitions \eqref{eq:4.13} of $\CA^\pm_a$ and the fact that the
$\eta_i^\pm$s are $\Pi\CO_{\IC P^1}(1)$-valued, this determines
the dependence of $\CA_a^\pm$ on $\eta_i^\pm$, $\lambda_\pm$ and
$\bl_\pm$ to be
\eqn{}{\CA_1^\pm\ =\ -\lambda_\pm^\ad \CB_\ad^\pm
       \qquad{\rm and}\qquad \CA_3^\pm\ =\ 2\gamma_\pm\hl^\ad_\pm\CB_\ad^\pm}
with the abbreviation
\eqna{eq:4.19}{
   \CB_\ad^\pm\ :=\  &\lambda^\bd_\pm
    \Bc_{\ad\bd}+\eta_i^\pm\cc\ \!\!^i_\ad+\tfrac{1}{2!}
    \gamma_\pm\eta_i^\pm\eta_j^\pm\hl^\bd_\pm\Wc\ \!\!^{ij}_{\ad\bd} 
    +\tfrac{1}{3!}\gamma_\pm^2\eta_i^\pm\eta_j^\pm\eta_k^\pm
    \hl^\bd_\pm\hl^\gd_\pm\cc\ \!\!^{ijk}_{\ad\bd\gd}\ +
   \\&+\ \tfrac{1}{4!}\gamma_\pm^3\eta_i^\pm\eta_j^\pm\eta_k^\pm
     \eta_l^\pm\hl_\pm^\bd\hl_\pm^\gd\hl_\pm^\dd
       \Gc\ \!\!^{ijkl}_{\ad\bd\gd\dd}}
and
\eqna{eq:4.19c}{
  &\CA_2^\pm\ =\
  \pm\tfrac{1}{2!}\gamma_\pm^2\eta_i^\pm\eta_j^\pm\Wc\ \!\!^{ij}\pm
  \tfrac{1}{3!}\gamma_\pm^3\eta_i^\pm\eta_j^\pm\eta_k^\pm
   \hl^\ad_\pm\cc\ \!\!_\ad^{ijk}\pm\tfrac{1}{4!}\gamma_\pm^4
   \eta_i^\pm\eta_j^\pm\eta_k^\pm\eta_l^\pm
  \hl^\ad_\pm\hl^\bd_\pm\Gc\ \!\!^{ijkl}_{\ad\bd}.}
The expansions
\eqref{eq:4.19} and \eqref{eq:4.19c} are defined up to gauge 
transformations generated
by group-valued functions which may depend on $\lambda_\pm$ and
$\bl_\pm$. In particular, it is assumed in this twistor
correspondence that for solutions to \eqref{eq:4.17}, there exists
a gauge in which terms of zeroth and first order in $\eta_i^\pm$
are absent in $\CA_2^\pm$. Recall that in the \v Cech approach, this
corresponds to the holomorphic triviality of the bundle
defined by such solutions when restricted to
projective lines. Putting it differently, we consider a subset in the set
of all solutions of phCS theory on $\CF^{5|8}$, and we will always
mean this subset when speaking about solutions to phCS theory. 

Note that in
\eqref{eq:4.19} and \eqref{eq:4.19c}, all coefficient
fields $\Bc_{\ad\bd},\cc\ \!\!_\ad^i,\ldots$
depend only on $y^{\ad\bd}$.
Furthermore, not all of them represent independent degrees
of freedom. Upon substituting the superfield expansions 
\eqref{eq:4.19} and \eqref{eq:4.19c} 
into the first two equations of \eqref{eq:4.17}, we find the relations
\eqna{eq:4.constr}{
 \Wc\ \!\!^{ij}_{\ad\bd}\ &=\
-\left(\partial_{(\ad\bd)}\Wc\ \!\!^{ij}+[\Bc_{\ad\bd},\Wc\ \!\!^{ij}]\right),\\
\cc\ \!\!^{ijk}_{\ad(\bd\gd)}\ &=\ -\tfrac{1}{2}
\left(\partial_{(\ad(\bd)}\cc\ \!\!^{ijk}_{\gd)}+
[\Bc_{\ad(\bd},\cc\ \!\!^{ijk}_{\gd)}]\right),\\
\Gc\ \!\!^{ijkl}_{\ad(\bd\gd\dd)}\ &=\
-\tfrac{1}{3}\left(\partial_{(\ad(\bd)}\Gc\ \!\!^{ijkl}_{\gd\dd)}+
[\Bc_{\ad(\bd},\Gc\ \!\!^{ijkl}_{\gd\dd)}]\right)}
showing that $\Wc\ \!\!_{\ad\bd}^{ij}$,
$\cc\ \!\!_{\ad\bd\gd}^{ijk}$ and
$\Gc\ \!\!^{ijkl}_{\ad\bd\gd\dd}$ are composite fields. 
Furthermore, the field
$\Bc_{\ad\bd}$ can be decomposed into its symmetric part, denoted
by $\Ac_{\ad\bd}=\Ac_{(\ad\bd)}$, and its antisymmetric part,
proportional to $\Pc$, such that
\eqn{eq:4.B}{
\Bc_{\ad\bd}\ =\ \Ac_{\ad\bd}-\tfrac{\i}{2}\epsilon_{\ad\bd}\Pc.}
Hence, we have recovered the covariant derivative $\cnab_{\ad\bd}
=\partial_{(\ad\bd)}+\Ac_{\ad\bd}$ and the (scalar) Higgs field
$\Pc$. Defining
\eqn{eq:4.chi}{
\cc_{i\ad}\ :=\
\tfrac{1}{3!}\epsilon_{ijkl}\cc\ \!\!_\ad^{jkl}
\qquad{\rm and}\qquad\Gc_{\ad\bd}\
:=\  \tfrac{1}{4!}\epsilon_{ijkl}\Gc\ \!\!^{ijkl}_{\ad\bd},}
we have thus obtained the supermultiplet in three dimensions:
$$\Ac_{\ad\bd},\cc\ \!\!^i_\ad,\Pc,\Wc\ \!\!^{ij},\cc_{i\ad},
\Gc_{\ad\bd}.$$
We shall again abbreviate $\Wc_{ij}:=\frac{1}{2!}\epsilon_{ijkl}\Wc\ \!\!^{kl}$.
Eqs. \eqref{eq:4.17} together with the field expansions
\eqref{eq:4.19} and \eqref{eq:4.19c}, the constraints \eqref{eq:4.constr} and the
definitions \eqref{eq:4.B} and \eqref{eq:4.chi} yield the
maximally supersymmetrically extended Bogomolny monopole equations:
\eqna{eq:4.24}{
 \fc_{\ad\bd}\ &=\ -\tfrac{\i}{2}\cnab_{\ad\bd}\Pc,\\
 \epsilon^{\bd\gd}\cnab_{\ad\bd}\cc\ \!\!^i_\gd\ &=\
 -\tfrac{\i}{2}[\Pc,\cc\ \!\!^i_\ad],\\
 \overset{\circ}{\triangle}\Wc\ \!\!^{ij}\ &=\ -\tfrac{1}{4}[\Pc,[\Wc\ \!\!^{ij},\Pc]]-
 \epsilon^{\ad\bd}\{\cc\ \!\!^i_\ad,\cc\ \!\!^j_\bd\},\\
 \epsilon^{\bd\gd}\cnab_{\ad\bd}\cc_{i\gd}\ &=\
 -\tfrac{\i}{2}[\cc_{i\ad},\Pc]+2[\Wc\ \!\!_{ij},\cc\ \!\!^j_\ad],\\
 \epsilon^{\bd\gd}\cnab_{\ad\bd}\Gc_{\gd\dd}\ &=\ -\tfrac{\i}{2}[\Gc_{\ad\dd},\Pc]+
 \{\cc\ \!\!^i_\ad,\cc_{i\dd}\}-\tfrac{1}{2}[\Wc_{ij},\cnab_{\ad\dd}\Wc\ \!\!^{ij}]+
 \tfrac{\i}{4}\epsilon_{\ad\dd}[\Wc_{ij},[\Pc,\Wc\ \!\!^{ij}]],}
which can also be derived from Eqs. \eqref{fieldeqn-2} by
demanding that all the fields in \eqref{fieldeqn-2} are independent
of the coordinate $x^4$. Here, we have introduced the
abbreviation
\eqn{}{\overset{\circ}{\triangle}\ :=\ \tfrac{1}{2}\epsilon^{\ad\bd}
 \epsilon^{\gd\dd}\cnab_{\ad\gd}
\cnab_{\bd\dd}.}

Note that
\eqref{eq:4.12} can be rewritten as
$ \hat{\O}|_{\hCV_\pm}=\pm\Theta_\pm^1\wedge
\Theta_\pm^2\dt\eta_1^\pm\cdots\dt\eta_4^\pm$,
where the differential one-forms $\Theta_\pm^{1,2}$ have been given in
\eqref{eq:3.37}. Substituting this expression and the expansions
\eqref{eq:4.19} and \eqref{eq:4.19} into the action \eqref{eq:4.14}, we arrive after a
straightforward calculation at the action
\eqna{eq:4.27}{
 S\ =\ \int\dt^3x\,{\rm tr}\Big\{\Big.&\Gc\ \!\!^{\ad\bd}
\left(\fc_{\ad\bd}+\tfrac{\i}{2}\cnab_{\ad\bd}\Pc\right)+
 \cc\ \!\!^{i\ad}
\cnab_{\ad\bd}\cc\ \!\!^\bd_i+
\tfrac{1}{2}\Wc_{ij}\overset{\circ}{\triangle}\Wc\ \!\!^{ij}\ +\\&+\ \tfrac{\i}{2}
 \cc\ \!\!^i_\ad[\cc\ \!\!^\ad_i,\Pc]+
 \Wc_{ij}\{\cc\ \!\!^i_\ad,\cc\ \!\!^{j\ad}\}+\tfrac{1}{8}
[\Wc_{ij},\Pc][\Wc\ \!\!^{ij},\Pc]\Big.\Big\},}
producing \eqref{eq:4.24}.

\paragraph{PhCS theory in the \v Cech description.}
Our starting point in \ref{FECRCTS-3}\kern10pt 
was to consider a trivial complex vector bundle $\CE$
over $\CF^{5|8}$ endowed with a $\cT$-connection. Such a
$\cT$-connection $\nabla_\cT=\dt_\cT+\CA_\cT$ on $\CE$ is flat if
$\CA_\cT$ solves Eqs. \eqref{eq:4.16}, and then
$(\CE,f=\{\mathbbm{1}_r\},\nabla_\cT)$ is a $\cT$-flat bundle in the
Dolbeault description. As for holomorphic vector bundles
(cf. Sec. \ref{CD-COR-2}), one may turn to the \v Cech description
of $\cT$-flat bundles in which the connection one-form $\CA_\cT$
disappears and all the information is hidden in a transition
function. In fact, let $\CE\to X$ be a rank $r|s$ complex
vector bundle over some smooth supermanifold $X$ and denote by 
\eqn{}{H^1_{\nabla_\cT}(X,\CE)\ =\ \Gamma(X,\fA_\cT)/\Gamma(X,\fS)}
the moduli space of phCS theory, 
where $\fA_\cT$ is the sheaf of solutions to \eqref{eq:4.16}
and $\fS=GL(r|s,\CS_X)$, as before. Furthermore, we shall need the subsheaf
$\CC_X\subset\CS_X$ of $\cT$-functions on $X$, that is, ${\rm d}_\cT f=0$ for 
$f\in\Gamma(U,\CC_X)$ as well as the sheaf $\fC:=GL(r|s,\CC_X)$. 
Quite generically, we may then state the following theorem:
{\Thm\label{C-D-Tflat-Thm} 
      Let $X$ be a smooth supermanifold with an open Stein covering $\fU=\{\CU_a\}$
      and $\CE\to X$ be a rank $r|s$ 
      complex vector bundle over $X$. Then there is a one-to-one
      correspondence between $H^1_{\nabla_\cT}(X,\CE)$ and the subset 
      of $H^1(X,\fC)$ consisting of those elements of
      $H^1(X,\fC)$ representing vector bundles which are smoothly
      equivalent to $\CE$, i.e.,
      $$ (\CE,f=\{f_{ab}\},\nabla_\cT)\ \sim\ (\tilde{\CE},
       \tilde{f}=\{\tilde{f}_{ab}\},\dt_\cT),$$
      where $\tilde{f}_{ab}=\psi_a^{-1}f_{ab}\psi_b$ for some
      $\psi=\{\psi_a\}\in C^0(\fU,\fS)$.
}\vskip 4mm
 
The proof is similar to the one of Thm. \ref{C-D-Thm} For that reason, we shall
omit it here and instead continue with our example.

Consider our smoothly trivial 
vector bundle $\CE\to\CF^{5|8}$ from above. 
Since the $\cT$-connection one-form $\CA_\cT$ is flat, it
is given as a pure gauge configuration on each patch and we have
\eqn{eq:4.28}{
\CA_\cT|_{\hCV_\pm}\ =\ {\psi}_\pm\dt_\cT{\psi}_\pm^{-1},}
together with the gluing condition
\eqn{eq:4.29}{
\psi_+\dt_\cT\psi_+^{-1}\ =\ {\psi}_-\dt_\cT{\psi}_-^{-1}}
for the trivial bundle $\CE$. Therefore, we can define a $\cT$-flat complex vector bundle
$\tilde{\CE}\to\CF^{5|8}$ with the canonical flat
$\cT$-connection $\dt_\cT$ and the transition function
\eqn{eq:4.31}{
\tilde{f}_{+-}\ :=\  {\psi}_+^{-1}{\psi}_-.}
The condition $\dt_\cT\tilde{f}_{+-}=0$ reads explicitly as
\eqna{eq:4.33}{
\bW_1^+\tilde{f}_{+-}\ =\ 0,~~~\bW_2^+\tilde{f}_{+-}&\ =\ 0,~~~
\bV_+^i\tilde{f}_{+-}\ =\ 0,\\
\bW_3^+\tilde{f}_{+-}&\ =\ 0.
}
Recall that the vector fields appearing in the first line generate
the antiholomorphic distribution $\cDb$, which is the chosen CR 
structure. In other
words, the bundle $\tilde{\CE}$ is holomorphic along the
mini-supertwistor space $\CP^{2|4}\subset\CF^{5|8}$ and flat along
the fibers of the projection $\pi_1:\CF^{5|8}\to\CP^{2|4}$
as follows directly from second line of \eqref{eq:4.33}. 
Let us now additionally assume
that $\tilde{\CE}$ is $\IR^{3|8}$-trivial, that is,
holomorphically trivial when restricted to any projective line
$\IC P^1_{x,\eta}\hookrightarrow\CF^{5|8}$ given by \eqref{eq:3.18}.
Recall that this
assumption was already used in \eqref{eq:4.19c}.
As before, this
extra ingredient guarantees the existence of a gauge in which the
component $\CA_2^\pm$ of $\CA_\cT$ vanishes. Hence, there exist
$GL(r,\IC)$-valued functions 
$\hat{\psi}=\{\hat{\psi}_+,\hat{\psi}_-\}\in C^0(\hfV,\fS)$ 
such that
\eqn{}{\tilde{f}_{+-}\ =\ {\psi}_+^{-1}{\psi}_-\ =\
 \hat{\psi}_+^{-1}\hat{\psi}_-,\qquad{\rm with}\qquad
\bW_2^\pm\hat{\psi}_\pm\ =\ 0}
and
\eqn{}{g\ :=\  \psi_+\hat{\psi}^{-1}_+\ =\ {\psi}_-\hat{\psi}_-^{-1}
}
is a matrix-valued function generating a gauge transformation
\eqn{}{{\psi}_\pm\ \mapsto\ \hat{\psi}_\pm\ =\ g^{-1}{\psi}_\pm,}
which acts on the gauge potential according to
\eqna{eq:4.38}{
\CA_1^\pm&\ \mapsto\  \hat{\CA}_1^\pm\ =\ g^{-1}\CA_1^\pm
g+g^{-1}\bW_1^\pm
g\ =\ \hat{\psi}_\pm\bW_1^\pm\hat{\psi}_\pm^{-1},\\
\CA_2^\pm&\ \mapsto\  \hat{\CA}_2^\pm\ =\ g^{-1}\CA_2^\pm
g+g^{-1}\bW_2^\pm
g\ =\ \hat{\psi}_\pm\bW_2^\pm\hat{\psi}_\pm^{-1}\ =\ 0,\\
\CA^\pm_3&\ \mapsto\  \hat{\CA}_3^\pm\ =\ g^{-1}\CA_3^\pm
g+g^{-1}\bW_3^\pm g\ =\ \hat{\psi}_\pm\bW_3^\pm
\hat{\psi}_\pm^{-1},\\
0\ =\ \CA_\pm^i\ :=\ \psi_\pm\bV^i_\pm\psi_\pm^{-1}&\ \mapsto\
\hat{\CA}_\pm^i\ =\ g^{-1}\bV^i_\pm g\ =\
\hat{\psi}_\pm\bV^i_\pm\hat{\psi}_\pm^{-1}.}
In this new gauge, one generically has $\hat{\CA}^i_\pm\neq 0$.

Note that \eqref{eq:4.28} can be
rewritten as the following linear system of differential
equations:
\eqna{eq:4.39}{
 (\bW_a^\pm+\CA_a^\pm)\psi_\pm&\ =\ 0,\\ 
 \bV_\pm^i\psi_\pm&\ =\ 0.}
The compatibility conditions of this linear system are 
Eqs. \eqref{eq:4.17}. This means that for any solution
$\CA^\pm_a$ to \eqref{eq:4.17}, one can construct solutions
$\psi_\pm$ to \eqref{eq:4.39} and, conversely, for any given
$\psi_\pm$ obtained via a splitting \eqref{eq:4.31} of a
transition function $\tilde{f}_{+-}$, one can construct a solution
\eqref{eq:4.28} to \eqref{eq:4.17}.

Similarly, Eqs. \eqref{eq:4.38} can be rewritten as the
gauge equivalent linear system
\eqna{eq:4.40}{
 (\bW_1^\pm+\hat{\CA}_1^\pm)\hat{\psi}_\pm&\ =\ 0,\\
 \bW_2^\pm\hat{\psi}_\pm&\ =\ 0,\\
 (\bW_3^\pm+\hat{\CA}_3^\pm)\hat{\psi}_\pm&\ =\ 0,\\
 (\bV_\pm^i+\hat{\CA}_\pm^i)\hat{\psi}_\pm&\ =\ 0.}
Note that due to the holomorphicity of $\hat{\psi}_\pm$ in
$\lambda_\pm$ and the condition $\hat{\CA}_\cT^+=\hat{\CA}_\cT^-$
on $\hCV_+\cap\hCV_-$, the components
$\hat{\CA}_1^\pm$, $\gamma_\pm^{-1}\hat{\CA}_3^\pm$ and
$\hat{\CA}_\pm^i$ must take the form
\eqn{eq:4.41}{
 \hat{\CA}^\pm_1\ =\ -\lambda_\pm^\ad\lambda_\pm^\bd
\CB_{\ad\bd},~~~ \gamma_\pm^{-1}\hat{\CA}_3^\pm\ =\
2\hl^\ad_\pm\lambda_\pm^\bd \CB_{\ad\bd}\qquad{\rm and}\qquad\hat{\CA}^i_\pm\
=\ \lambda^\ad_\pm\CA^i_\ad,}
with $\lambda$-independent superfields
$\CB_{\ad\bd}:=\CA_{\ad\bd}-\frac{\i}{2}\epsilon_{\ad\bd}\Phi$
and $\CA_\ad^i$. Introducing the first-order differential operators
$D_{\ad\bd}=\partial_{(\ad\bd)}+\CB_{\ad\bd}$ and
$\nabla_\ad^i=\partial_\ad^i+\CA_\ad^i$,
we arrive at the following compatibility conditions of the linear
system \eqref{eq:4.40}:
\eqna{eq:4.42}{
{}&[D_{\ad\gd},D_{\bd\dd}]+[D_{\ad\dd},
D_{\bd\gd}]\ =\ 0,
~~~~[\nabla^i_\ad,D_{\bd\gd}]+[\nabla_\gd^i,D_{\bd\ad}]\ =\
0,\\&\hspace{3cm} \{\nabla_\ad^i,\nabla_\bd^j\}+\{\nabla_\bd^i,
\nabla_\ad^j\}\ =\ 0.}
These equations also follow from \eqref{eq:4.16} after
substituting the expansions \eqref{eq:4.41}. These equations
can be understood as the constraint equations of our supersymmetric
Bogomolny model. In fact, proceeding as in \ref{FE-2}\kern6pt, we can
derive all the superfields together with their field expansions
and the equations of motion they are subject to. Eventually, one finds
\eqref{eq:4.24}. So, let us only sketch this way.

Eqs. \eqref{eq:4.42} can equivalently be rewritten as
\eqn{constraint}{
       [D_{\ad\gd},D_{\bd\dd}]\ =\ \epsilon_{\gd\dd}
       \Sigma_{\ad\bd},
       ~~~
       {[\nabla_\ad^i,D_{\bd\gd}]}\ =\ \epsilon_{\ad\gd}
       \Sigma^i_\bd
       \quad{\rm and}\quad
       \{\nabla^i_\ad,\nabla^j_\bd\}\ =\ \epsilon_{\ad\bd}\Sigma^{ij},}
where $\Sigma_{\ad\bd}=\Sigma_{\bd\ad}$ and
$\Sigma^{ij}=-\Sigma^{ji}$. Note that the first equation in
\eqref{eq:4.42} immediately shows that
$f_{\ad\bd}=-\frac{\i}{2}\nabla_{\ad\bd}\Phi$ and thus the
contraction of the first equation of \eqref{constraint} with
$\epsilon^{\gd\dd}$ gives
$\Sigma_{\ad\bd}=f_{\ad\bd}-\frac{\i}{2}\nabla_{\ad\bd}\Phi=
2f_{\ad\bd}$. The Bianchi identities for the differential
operators $D_{\ad\bd}$ and $\nabla^i_\ad$ yield in a
straightforward manner further field equations, which allow us to
identify the superfields $\Sigma^i_\ad$ and $\Sigma^{ij}$ with
the spinors $\chi^i_\ad$ and the scalars $W^{ij}$,
respectively. Moreover, $\chi_{i\ad}$ is given by
$\chi_{i\ad}:=\frac{1}{3}\epsilon_{ijkl}\nabla^j_\ad W^{kl}$
and $G_{\ad\bd}$ is defined by
$G_{\ad\bd}:=-\frac{1}{4}\nabla^i_{(\ad}\chi_{i\bd)}$.
Collecting the above information, one obtains the equations
of motion for the superfields
$\CA_{\ad\bd}$, $\chi^i_\ad$, $\Phi$,
$W^{ij}$, $\chi_{i\ad}$ and $G_{\ad\bd}$ which take
the same form as \eqref{eq:4.24} but with all the fields now being
superfields. Thus, the projection of the superfields onto the
zeroth order components of their $\eta$-expansions gives
\eqref{eq:4.24}.  

To extract the physical field content from the superfields, we
need their explicit expansions in powers of $\eta_i^\ad$. For
this, one again imposes the transversal gauge condition
\eqref{tg}. The constraint equations \eqref{constraint} together 
with the Bianchi identities yield the recursion relations
analogously to Eqs. \eqref{eq:RR-2}. In fact, one simply
needs to dimensionally reduce \eqref{eq:RR-2} to obtain the 
recursion relations for the present setting. Then by iterating
the equations one may compute the superfield expansions. Furthermore,
as in \ref{FE-2}\kern10pt one deduces the one-to-one correspondence between
the constraint equations \eqref{constraint} and the field
equations \eqref{eq:4.24}. We shall come back to this issue when
dealing with explicit solution construction algorithms for 
Eqs. \eqref{eq:4.24} in Sec. \ref{sec:solution}.

\section{Holomorphic BF theory}\label{sec:hBF-2}

In the preceding section, we have defined a theory on the CR
supertwistor space $\CF^{5|8}$ entering into the double fibration
\eqref{eq:3.19} which we called partially holomorphic Chern-Simons
theory. We have shown that this theory is equivalent to a
supersymmetric Bogomolny-type Yang-Mills-Higgs theory in three
Euclidean dimensions. The purpose of this section is to show, that
one can also introduce a theory (including an action functional)
on mini-supertwistor space $\CP^{2|4}$, which is equivalent to
phCS theory on $\CF^{5|8}$. Thus, one can define at each level of
the double fibration \eqref{eq:3.19} a theory accompanied by a
proper action functional and, moreover, these three theories are
all equivalent.

\paragraph{Field equations of hBF theory on $\CP^{2|4}$.}
Consider the mini-supertwistor space $\CP^{2|4}$. Let $E$ be a
trivial rank $r$ complex vector bundle over $\CP^{2|4}$ with a
connection one-form $\CA$. Assume that its $(0,1)$-part
$\CA^{0,1}$ contains no antiholomorphic odd components and
does not depend on $\bar{\eta}_i^\pm$, that is,
$\bV_\pm^i\lrcorner\CA^{0,1}=0$ and
$\bV_\pm^i(\partial_{\bw_\pm^{1,2}}\lrcorner\CA^{0,1})=0$. Recall
that on $\CP^{2|4}$ we have a holomorphic volume form $\Omega$
which is locally given by \eqref{volumemini}. Hence, we can define a
holomorphic BF (hBF) type theory -- as introduced and discussed in Refs.
\cite{Popov:1999cq,Ivanova:2000xr,Ivanova:2000af,Baulieu:2004pv} -- on $\CP^{2|4}$
with the action
\eqn{eq:5.1}{
 S\ =\ \int_{\CY}\Omega\wedge{\rm tr}\{B(\pbar_\CP
\CA^{0,1}+\CA^{0,1}\wedge\CA^{0,1})\}\ =\ \int_{\CY}\Omega\wedge
{\rm tr}\{B \CF^{0,2}\}.}
Here, $B$ is a scalar field in the adjoint representation of the
group $GL(r,\IC)$, $\pbar_\CP$ is the antiholomorphic part of the
exterior derivative on $\CP^{2|4}$ and $\CF^{0,2}$ the 
$(0,2)$-part of the curvature two-form. The space $\CY$ is the
subsupermanifold of $\CP^{2|4}$ constrained by $\bar{\eta}_i^\pm=0$. 

The equations of motion following from the action functional
\eqref{eq:5.1} are
\eqna{eq:5.2}{
\pbar_\CP\CA^{0,1}+\CA^{0,1}\wedge\CA^{0,1}&\ =\ 0,\\
\pbar_\CP B+[\CA^{0,1},B]&\ =\ 0.}
These equations as well as the Lagrangian in \eqref{eq:5.1} can be
obtained from Eqs. \eqref{eq:4.16} and the Lagrangian in
\eqref{eq:4.14}, respectively, by imposing the condition
$\partial_{\bw_\pm^3}\CA_{\bw_\pm^a}=0$ and identifying
\eqn{}{\CA^{0,1}|_{\CV_\pm}\ =\ \dt \bw_\pm^1\CA_{\bw_\pm^1}+\dt
\bw_\pm^2\CA_{\bw_\pm^2}\qquad{\rm and}\qquad B^\pm\ :=\  B|_{\CV_\pm}\ =\
\CA_{\bw_\pm^3}.}
Note that $\CA_{\bw_\pm^3}$ behaves on $\CP^{2|4}$ as a scalar.
Thus, \eqref{eq:5.2} can be obtained from \eqref{eq:4.16} by
demanding invariance of all fields under the action of the group
$\cG'$ from \ref{CRST-ex}\kern10pt such that
$\CP^{2|4}\cong\CF^{5|8}/\cG'$. 

\paragraph{\v Cech description.}
When restricted to the patches $\CV_\pm$, Eqs.
\eqref{eq:5.2} can be solved by
\eqn{eq:5.4}{
\CA^{0,1}|_{\CV_\pm}\ =\
\tilde{\psi}_\pm\pbar_\CP\tilde{\psi}_\pm^{-1}\qquad{\rm and}\qquad B^\pm\ =\
\tilde{\psi}_\pm B^\pm_0\tilde{\psi}_\pm^{-1},}
where $B_0^\pm$ is a holomorphic $\mathfrak{gl}(r,\IC)$-valued function on
$\CV_\pm$,
\eqn{}{\pbar_\CP B_0^\pm\ =\ 0.}
On the intersection $\CV_+\cap\CV_-$, we have the
gluing conditions
\eqn{eq:5.6}{
\tilde{\psi}_+\pbar_\CP\tilde{\psi}_+^{-1}\ =\
\tilde{\psi}_-\pbar_\CP\tilde{\psi}_-^{-1}\qquad{\rm and}\qquad
\tilde{\psi}_+B_0^+\tilde{\psi}_+^{-1}\ =\
\tilde{\psi}_-B_0^-\tilde{\psi}_-^{-1}}
as $E$ is a trivial bundle. From \eqref{eq:5.6}, we learn that
\eqn{eq:5.7}{
\tilde{f}_{+-}\ :=\  \tilde{\psi}_+^{-1}\tilde{\psi}_-}
can be identified with the holomorphic transition function of a
bundle $\tilde{E}$ with the canonical holomorphic structure
$\pbar_\CP$, and
\eqn{}{B_0^+\ =\ \tilde{f}_{+-}B_0^-\tilde{f}_{+-}^{-1},}
i.e., $B_0\in H^0(\CP^{2|4},{\rm End}\,\tilde{E})$ and $B\in
H^0(\CP^{2|4},{\rm End}\, E)$. Note that the pull-back
$\pi_1^*\tilde{E}$ of the bundle $\tilde{E}$ to the space
$\CF^{5|8}$ can be identified with the bundle $\tilde{\CE}$,
\eqn{}{\tilde{\CE}\ =\ \pi_1^*\tilde{E},}
with the transition function
$\tilde{f}_{+-}=\psi_+^{-1}\psi_-=\tilde{\psi}_+^{-1}\tilde{\psi}_-$
if one additionally assumes that 
$\tilde{E}$ is $\IR^{3|8}$-trivial, that is, 
holomorphically trivial on any $\IC P^1_{x,\eta}\hookrightarrow\CP^{2|4}$.
Recall that the transition functions of the bundle $\tilde{\CE}$
do not depend on $w_\pm^3$ and therefore they can always be
considered as the pull-backs of transition functions of a bundle
$\tilde{E}$ over $\CP^{2|4}$.

\paragraph{Moduli space.}
By construction, $B=\{B^\pm\}$ is a $\mathfrak{gl}(r,\IC)$-valued function
generating trivial infinitesimal gauge transformations of
$\CA^{0,1}$ and therefore it does not contain any physical degrees
of freedom. Remember that solutions to the first equation of
\eqref{eq:5.2} are defined up to gauge transformations
\eqn{eq:5.10}{
\CA^{0,1}\ \mapsto\ \tilde{\CA}^{0,1}\ =\ g\CA^{0,1}g^{-1}+g\pbar_\CP
g^{-1}}
generated by smooth $GL(r,\IC)$-valued functions $g$ on
$\CP^{2|4}$. Clearly, the transformations \eqref{eq:5.10} do not change the
holomorphic structure $\nabla^{0,1}$ on the bundle $E$.
On infinitesimal level, the transformations
\eqref{eq:5.10} take the form
\eqn{}{\delta\CA^{0,1}\ =\ \pbar_\CP B+[\CA^{0,1},B],}
with $B\in H^0(\CP^{2|4},{\rm End}\, E)$ and such a field $B$, solving
the second equation of
\eqref{eq:5.2}, generates holomorphic transformations such that
$\delta \CA^{0,1}=0$. Their finite version is
\eqn{}{\tilde{\CA}^{0,1}\ =\ g\CA^{0,1}g^{-1}+g\pbar_\CP g^{-1}\ =\
\CA^{0,1},}
and for a gauge potential $\CA^{0,1}$ given by \eqref{eq:5.4},
such a $g$ takes the form
\eqn{}{g_\pm\ =\
\tilde{\psi}_\pm\e^{B_0^\pm}\tilde{\psi}_\pm^{-1},\quad{\rm with}\quad
g_+\ =\ g_-\quad{\rm on}\quad\CV_+\cap\CV_-.}

\paragraph{Summary.}
Collecting all the things derived in the preceding sections, we may
now summarize our discussion as follows: 

\pagebreak[4]
{\Thm\label{thm:summmary-3} The are one-to-one correspondences between 
equivalence classes
of $\IR^{3|8}$-trivial holomorphic vector bundles $E$ over mini-supertwistor
space $\CP^{3|8}$, equivalence classes of $\IR^{3|8}$-trivial $\cT$-flat
vector bundles $\CE$ over CR supertwistor space $\CF^{5|8}$ -- all
being smoothly trivial -- and gauge
equivalence classes of local solutions to the maximally supersymmetrized
Bogomolny monopole equations on $\IR^3$.}\vskip 4mm 

By virtue of Thms.
\ref{C-D-Thm} and \ref{C-D-Tflat-Thm}, we let i)
 $H^1_{\nabla^{0,1}}(\CP^{2|4},\tilde{E})$ be the moduli space of
hBF theory on $\CP^{2|4}$ for vector bundles $\tilde{E}$ smoothly equivalent 
to $E$ and ii) $H^1_{\nabla_\cT}(\CF^{5|8},\tilde{\CE})$ be the
moduli space of phCS theory on $\CF^{5|8}$ for vector bundles $\tilde{\CE}$ 
smoothly equivalent to $\CE$, respectively. Then we have the bijections
\eqn{}{H^1_{\nabla^{0,1}}(\CP^{2|4},\tilde{E})\ \cong\ 
  H^1_{\nabla_\cT}(\CF^{5|8},\tilde{\CE})\ \cong\ \CM_{\rm sB},}
where $\CM_{\rm sB}$ denotes the moduli space
of the supersymmetric Bogomolny
monopole equations on $\IR^3$ obtained from the solution space 
by quotiening with respect to the group of gauge transformations. 

Pictorially, we have established the following diagram:
$$\begin{aligned}
\begin{picture}(180,70)(0,-5)
\put(0.0,0.0){\makebox(0,0)[c]{hBF theory on $\CP^{2|4}$}}
\put(160.0,14.0){\makebox(0,0)[c]{supersymmetric}}
\put(160.0,0.0){\makebox(0,0)[c]{ Bogomolny model on $\IR^3$}}
\put(80.0,50.0){\makebox(0,0)[c]{phCS theory on $\CF^{5|8}$}}
\put(40.0,40.0){\vector(-1,-1){30}}
\put(10.0,10.0){\vector(1,1){30}}
\put(100.0,40.0){\vector(1,-1){18}}
\put(118.0,22.0){\vector(-1,1){18}} \put(56,0){\vector(1,0){37}}
\put(86,0){\vector(-1,0){30}}
\end{picture}
\end{aligned}$$

\section{Massive fields}

In \cite{Chiou:2005jn} (see also the 
subsequent work \cite{Chiou:2005pu}), Chiou et al. developed
a twistor string theory corresponding to a certain massive SYM
theory in three dimensions. It was argued, that the
mass terms in this theory arise from coupling the R-symmetry
current to a constant background field when performing the
dimensional reduction. In this section, we want to study the
analogous construction for the supersymmetric Bogomolny model
which we discussed in the previous sections. We focus on the
geometric origin of the additional mass terms by discussing the
associated twistor description. More explicitly, we establish a
correspondence between holomorphic bundles over the deformed
mini-supertwistor space introduced in \cite{Chiou:2005jn} and
solutions to massive supersymmetric Bogomolny equations in three
dimensions.

\paragraph{Mini-supertwistor and CR supertwistor spaces as vector
bundles.}
We start from
the observation that mini-supertwistor space $\CP^{2|4}$ can
be considered as the total space of a rank $0|4$ holomorphic
vector bundle over mini-twistor space $\CP^2$, that is,
\eqn{eq:6.2}{\CP^{2|4}\ \to\ \CP^2.}
The mini-twistor space $\CP^2$ is covered by two
patches, say $\CW_\pm$, with coordinates $w_\pm^1$ and
$w_\pm^2$. The additional fiber coordinates in the
vector bundle $\CP^{2|4}$ over $\CP^2$ are the Gra{\ss}mann
variables $\eta_i^\pm$. For later convenience, we rearrange them
into the vector $\eta^\pm=(\eta_i^\pm)$. On
$\CW_+\cap\CW_-$, we have the relation
\eqn{eq:6.3}{
\eta^+\ =\ \varphi_{+-}\eta^-,}
with the transition function
\eqn{eq:6.4}{
\varphi_{+-}\ =\ w_+^2 (\delta_i{}^j)\ =\ w_+^2\mathbbm{1}_4.}

The CR supertwistor
space $\CF^{5|8}$ is a CR vector bundle, i.e., it has
a transition function annihilated by the vector fields
$\partial_{\bw_\pm^1}$, $\partial_{\bw_\pm^2}$ from the distribution
$\cDb$ on $\CF^5$, over the CR twistor space
$\CF^5\cong\IR^3\times S^2$,
\eqn{eq:6.5}{
\CF^{5|8}\ \rightarrow\ \CF^5,}
with complex fiber coordinates $\eta_i^\pm$ 
over the patches $\hCW_\pm$ covering $\CF^5$. Recall that we have
the double fibration
\eqna{eq:6.5.2}{
\begin{picture}(50,40)
\put(0.0,0.0){\makebox(0,0)[c]{$\CP^{2}$}}
\put(64.0,0.0){\makebox(0,0)[c]{$\IR^{3}$}}
\put(34.0,33.0){\makebox(0,0)[c]{$\CF^{5}$}}
\put(7.0,18.0){\makebox(0,0)[c]{$\nu_1$}}
\put(55.0,18.0){\makebox(0,0)[c]{$\nu_2$}}
\put(25.0,25.0){\vector(-1,-1){18}}
\put(37.0,25.0){\vector(1,-1){18}}
\end{picture}
}
in the purely even case and the transition function of the
vector bundle \eqref{eq:6.5} can be identified with
\eqn{}{\nu_1^*\varphi_{+-}\ =\ \lambda_+(\delta_i{}^j)\ =\
\lambda_+\mathbbm{1}_4,}
i.e., we have the same transformation \eqref{eq:6.3}
relating $\eta^+$ to $\eta^-$ on $\hCW_+\cap\hCW_-$.
Note that our notation often does not
distinguish between objects on $\CP^2$ and their pull-backs to
$\CF^5$.

Thus, we arrive at the diagram
\eqna{}{
\begin{picture}(100,100)
\put(0.0,0.0){\makebox(0,0)[c]{$\CP^{2}$}}
\put(0.0,53.0){\makebox(0,0)[c]{$\CP^{2|4}$}}
\put(96.0,0.0){\makebox(0,0)[c]{$\IR^{3}$}}
\put(96.0,53.0){\makebox(0,0)[c]{$\IR^{3|8}$}}
\put(51.0,33.0){\makebox(0,0)[c]{$\CF^{5}$}}
\put(51.0,85.0){\makebox(0,0)[c]{$\CF^{5|8}$}}
\put(16.5,23.0){\makebox(0,0)[c]{$\nu_1$}}
\put(76.5,23.0){\makebox(0,0)[c]{$\nu_2$}}
\put(16.5,75.0){\makebox(0,0)[c]{$\pi_1$}}
\put(76.5,75.0){\makebox(0,0)[c]{$\pi_2$}}
\put(37.5,25.0){\vector(-3,-2){25}}
\put(55.5,25.0){\vector(3,-2){25}}
\put(37.5,78.0){\vector(-3,-2){25}}
\put(55.5,78.0){\vector(3,-2){25}}
\put(0.0,45.0){\vector(0,-1){37}}
\put(90.0,45.0){\vector(0,-1){37}}
\put(45.0,78.0){\vector(0,-1){37}}
\end{picture}}
combining the double fibrations \eqref{eq:3.19} and
\eqref{eq:6.5.2}.

\paragraph{Deformed mini-supertwistor and CR supertwistor
spaces.}
Let us define a holomorphic vector bundle
\eqn{eq:6.9}{
\CP^{2|4}_M\ \to\ \CP^2}
with complex coordinates $\tilde{\eta}^\pm\ =\
(\tilde{\eta}_i^\pm)$ on the fibers over
$\CW_\pm\subset \CP^2$ which are related by the transition
function
\eqn{eq:6.10}{
\tilde{\varphi}_{+-}\ =\ w_+^2\e^{\frac{w^1_+}{w^2_+}M}}
on the intersection $\CW_+\cap\CW_-$, i.e.,
\eqn{eq:6.11}{
\tilde{\eta}^+\ =\ \tilde{\varphi}_{+-}\tilde{\eta}^-.}
For reasons which will become more transparent in the later
discussion, we demand that $M$ is traceless and Hermitian. It
is also assumed that $M$ is constant. 

This supermanifold $\CP^{2|4}_M$ was introduced in
\cite{Chiou:2005jn} as the target space of twistor string
theory\footnote{For this, $\CP^{2|4}_M$ has to be a formal Calabi-Yau
supermanifold, which is the reason underlying the above
restriction to $\tr M=0$.} which corresponds
by our subsequent discussion to a SYM theory
in three dimensions with massive spinors and both massive and
massless scalar fields. In the
following, we provide a twistorial derivation of analogous mass
terms in our supersymmetric Bogomolny model and explain their
geometric origin.

Consider the rank $0|4$ holomorphic vector bundle
\eqref{eq:6.9} and its pull-back
\eqn{eq:6.12}{
\CF^{5|8}_M\ :=\  \nu_1^*\CP^{2|4}_M\ \to\ \CF^5}
to the space $\CF^5$ from the double fibration \eqref{eq:6.5.2}.
Note that the supervector bundle $\CF_M^{5|8}\to\CF^5$ is
smoothly equivalent to the supervector bundle
$\CF^{5|8}\to\CF^5$ since in the coordinates
$(y^{\ad\bd},\lambda_\pm,\bl_\pm)=(y,\bar{y},x^3,\lambda_\pm,\bl_\pm)$
on $\CF^5$, the pulled-back transition function
$\nu_1^*\tilde{\varphi}_{+-}$ can be split
\eqn{eq:6.13}{
\nu_1^*\tilde{\varphi}_{+-}\ =\
\lambda_+\e^{\frac{1}{\lambda_+}y^{\ad\bd}\lambda^+_\ad
\lambda^+_\bd M}\ =\ \varphi_+(\lambda_+\mathbbm{1}_4)\varphi_-^{-1}\
\sim\ \lambda_+\mathbbm{1}_4.}
Here,
\eqn{}{
\varphi_+\ =\ \e^{-(x^3+\lambda_+\bar{y})M}\ =\ \e^{\lambda^+_\ad
y^{\ad\dot2}M}\quad{\rm and}\quad \varphi_-\ =\ \e^{(x^3-\lambda_-y)M}\ =\
\e^{-\lambda^-_\ad y^{\ad\dot1}M}}
are matrix-valued functions well-defined on the patches $\hCW_+$
and $\hCW_-$, respectively. Remember that $\tilde{\eta}_i^+$ and
$\tilde{\eta}_i^-$ are related by \eqref{eq:6.11} and their pull-backs to
$\CF^5$ (which we denote again by the same letter) are related by
the transition function \eqref{eq:6.13}. Therefore, we have
\eqn{eq:6.15}{
(\varphi_+^{-1}\tilde{\eta}^+)\ =\
\lambda_+(\varphi_-^{-1}\tilde{\eta}^-).}
From this we conclude that
\eqn{eq:6.16}{
\tilde{\eta}^+\ =\ \varphi_+\eta^+\ =\
\e^{\lambda_\ad^+y^{\ad\dot2}M}\eta^+\qquad{\rm and}\qquad \tilde{\eta}^-\ =\
\varphi_-\eta^-\ =\ \e^{-\lambda_\ad^-y^{\ad\dot1}M}\eta^-,}
where $\eta^\pm=(\eta_i^\pm)$ are the fiber coordinates of the
bundle \eqref{eq:6.5} related by \eqref{eq:6.3} on the
intersection $\hCW_+\cap\hCW_-$.

The fibration $\CF_M^{5|8}\to\CF^5$ can also be understood in the 
Dolbeault picture. It follows from \eqref{eq:6.16} that
\eqn{eq:6.17}{
\bW_1^\pm\tilde{\eta}^\pm_i\ =\ 0,\qquad\bW^\pm_2\tilde{\eta}_i^\pm\ =\ 0
\qquad{\rm and}\qquad
\bW_3^\pm\tilde{\eta}_i^\pm+M_i{}^j\tilde{\eta}^\pm_j\ =\ 0.}
Recall that the vector fields $\bW_a^\pm$ generate an integrable
distribution $\cT_b=\langle\bW_a^\pm\rangle$
together with the operator
\eqn{}{\dt_{\cT_b}|_{\hCW_\pm}\ =\ \dt \bw_\pm^a\der{\bw_\pm^a},}
which annihilates the transition function \eqref{eq:6.13} of the
bundle \eqref{eq:6.12}. Due to formulas \eqref{eq:6.13} and
\eqref{eq:6.17}, the vector bundle $\CF_M^{5|8}$ with
canonical $\cT_b$-flat connection $\dt_{\cT_b}$ is diffeomorphic
to the vector bundle $\CF^{5|8}$ with the $\cT_b$-flat
connection $\nabla_{\cT_b}=\dt_{\cT_b}+\CA_{\cT_b}$ the components
$\CA_a^\pm=\bW_a^\pm\lrcorner\CA_{\cT_b}$ of which are
given by
\eqn{eq:6.19}{
 \CA^\pm_1\ =\ 0,\qquad\CA^\pm_2\ =\
0\qquad{\rm and}\qquad\CA^\pm_3\ =\ M.}
In other words, we have an equivalence of the following data:
\eqn{eq:6.20}{
(\CF^{5|8}_M,\tilde{\varphi},\dt_{\cT_b})\ \sim\
(\CF^{5|8},\varphi=
\{\lambda_+\mathbbm{1}_4\},\nabla_{\cT_b}).}

By construction, the connection one-form $\CA_{\cT_b}$,
given explicitly in \eqref{eq:6.19}, is a solution to the field
equations
\eqn{}{\dt_{\cT_b}\CA_{\cT_b}+\CA_{\cT_b}\wedge\CA_{\cT_b}\ =\ 0}
of phCS theory on $\CF^5$, which are equivalent via the arguments
of \ref{par:esbe-3}\kern10pt 
to the Bogomolny equations on $\IR^3$. Due to this
correspondence, \eqref{eq:6.19} is equivalent to a solution of the
Bogomolny equations with vanishing YM gauge potential
$a_{\ad\bd}$ and constant Higgs field
\eqn{eq:6.22}{
\phi\ =\ (\phi_i{}^j)\ =\ -\i(M_i{}^j),}
which takes values in the Lie algebra $\mathfrak{su}(4)$ of the R-symmetry
group $SU(4)$. Thus, the data \eqref{eq:6.20} are equivalent to
the trivial vector bundle $\IR^{3|8}\to\IR^3$ together with the
differential operator
\eqn{}{D_{\ad\bd}\ =\
\partial_{(\ad\bd)}-\tfrac{1}{2}\epsilon_{\ad\bd}M}
encoding the information about the matrix $M$, i.e.,
\eqn{eq:6.24}{
(\CF_M^{5|8},\tilde{\varphi},\dt_{\cT_b})\ \sim \
(\CF^{5|8},\varphi,\nabla_{\cT_b})\ \sim\
(\IR^{3|8},D_{\ad\bd}).}

Note that the gauge potential $A_{\ad\bd}$ corresponding
to\footnote{Here, $A_s$ with $s=1,2,3$ are the components of the
ordinary gauge potential in three dimensions.} $A_s\in\mathfrak{u}(r)$ in
a different basis and the Higgs fields $\Phi\in\mathfrak{u}(r)$ considered
in Sec. \ref{sec:phCSt-3} can be combined with $a_{\ad\bd}$ and $\phi$ into
the fields
\eqn{}{A_{\ad\bd}\otimes\mathbbm{1}_4+\mathbbm{1}_r\otimes a_{\ad\bd}
\qquad{\rm and}\qquad
\Phi\otimes\mathbbm{1}_4+\mathbbm{1}_r\otimes\phi}
acting on the tensor product $V_{U(r)}\otimes V_{SU(4)}$ of
the (adjoint) representation space $V_{U(r)}$ of the gauge group
and the representation space $V_{SU(4)}$ of the R-symmetry
group.

For the sake of completeness, we note that the deformed complex 
vector bundle $\CP^{2|4}_M\to \CP^2$ with the transition function
$\tilde{\varphi}_{+-}$ from \eqref{eq:6.10} and the holomorphic
structure
\eqn{}{\pbar_b|_{\CW_\pm}\ =\
\dt\bw_\pm^1\der{\bw_\pm^1}+\dt\bw_\pm^2\der{\bw_\pm^2}}
is smoothly equivalent to the bundle $\CP^{2|4}\to\CP^2$
with the transition function $\varphi_{+-}$ from \eqref{eq:6.4}
and the holomorphic structure defined by the fields
$\CA^{0,1}$ and $B$ with the components
\eqn{}{\CA_{\bw^1_\pm}\ =\ 0,\qquad\CA_{\bw^2_\pm}\ =\
\mp\frac{w_\pm^1}{(1+w_\pm^2\bw_\pm^2)^2}M\qquad{\rm and}
\qquad B_\pm\ =\ \CA_{\bw^3_\pm}\ =\ M.}
The fields $\CA^{0,1}$ and $B$ obviously satisfy the
field equations
\eqn{}{\pbar_b\CA^{0,1}+\CA^{0,1}\wedge\CA^{0,1}\ =\
0\qquad{\rm and}\qquad \pbar_bB+[\CA^{0,1},B]\ =\ 0}
of hBF theory on $\CP^2$. By repeating the discussion of Sec.
\ref{sec:hBF-2}, one can now show the equivalence of the data
\eqn{}{(\CP^{2|4}_M,\tilde{\varphi}_,\pbar_b)\ \sim\
(\CP^{2|4},\varphi=\{\lambda_+\mathbbm{1}_4\},\nabla^{0,1}_b)\ \sim\
(\CF^{5|8}_M,\tilde{\varphi},\dt_{\cT_b}),}
which extends the equivalences described in \eqref{eq:6.24}.

\paragraph{The deformed CR supertwistor space as a
supermanifold.}
For developing a twistor correspondence involving the deformed CR
supertwistor space $\CF_M^{5|8}$, the description of $\CF_M^{5|8}$
as a rank $0|4$ complex vector bundle with a constant gauge
potential \eqref{eq:6.19} which twists the direct product of even
and odd spaces is not sufficient. We rather have to interpret the
total space of $\CF_M^{5|8}$ as a supermanifold with deformed CR
structure and deformed distribution $\cT_M$.

Let us begin with the vector fields on $\CF_M^{5|8}$.
Remember that a covariant derivative along a vector field on the
base space of a bundle can be lifted to a vector field on the
total space of the bundle. In our case of the bundle
\eqref{eq:6.12}, the lift of \eqref{eq:6.17} reads as
\eqn{}{\bW_1^\pm\tilde{\eta}_i^\pm\ =\ 0,\qquad
       \bW_2^\pm\tilde{\eta}_i^\pm\ =\ 0\qquad{\rm and}\qquad
      \left(\bW_3^\pm+M_k{}^j\tilde{\eta}_j^\pm\der{\tilde{\eta}_k^\pm}\right)
      \tilde{\eta}_i^\pm\ =\ 0.}
To see the explicit form of the vector fields corresponding to the
integrable distribution
\eqn{eq:6.30}{
 \cT_M\ =\ \left\langle\der{\bw_\pm^a},\der{\bar{\tilde{\eta}}_i^\pm}\right\rangle}
on $\CF^{5|8}_M$, it is convenient to switch to the coordinates
$(y^{\ad\bd},\lambda_\pm,\bl_\pm,\eta_i^\ad)$ by the
formulas
\eqna{eq:6.31}{
w_\pm^1\ =\ \lambda_\ad^\pm\lambda_\bd^\pm y^{\ad\bd},\quad
w_\pm^2\ =\ \lambda_\pm\quad{\rm and}\quad
w_\pm^3\ =\ -\gamma_\pm\lambda_\ad^\pm\hl_\bd^\pm y^{\ad\bd},\\
\tilde{\eta}_i^+\ =\ \left(\e^{\lambda^+_\ad y^{\ad\dot2}
M}\right)_i^{~j}\eta_j^\bd\lambda_\bd^+\quad{\rm and}\quad\tilde{\eta}_i^-\ =\
\left(\e^{-\lambda^-_\ad y^{\ad\dot1}
M}\right)_i^{~j}\eta_j^\bd\lambda_\bd^-.}
By a straightforward calculation, we obtain
\eqna{eq:6.32}{
\dt_{\cT_M}|_{\hCV_\pm}&\ =\ \dt
\bw_\pm^a\der{\bw_\pm^a}+\dt
\bar{\tilde{\eta}}_i^\pm\der{\bar{\tilde{\eta}}_i^\pm}\\
&\ =\ \bT^1_\pm \bar{\CW}_1^\pm+\bT^2_\pm
\bar{\CW}_2^\pm+(\bT^3_\pm
\mp\gamma_\pm^2\hl^\pm_\ad\hl^\pm_\bd
y^{\ad\bd}\T^2_\pm)\bar{\CW}_3^\pm+
\bar{\CE}_i^\pm\bV_\pm^i,}
where
\eqna{eq:6.33}{
\bar{\CW}_1^\pm&\ :=\ \bW_1^\pm\mp\lambda_\pm(T\bar{M}T)_i{}^j
\hl_\ad^\pm\eta_j^\ad\bV_\pm^i,\qquad\bar{\CW}_2^\pm\ :=\
\bW_2^\pm,\\
\bar{\CW}_3^\pm&\ :=\
\bW_3^\pm+\gamma_\pm(TM)_i{}^j\lambda_\ad^\pm\eta_j^\ad
V_\pm^i+\gamma_\pm(T\bar{M}T)_i{}^j\hl_\ad^\pm\eta_j^\ad
\bV_\pm^i,\\
\bar{\CE}^+_i&\ :=\
\bar{E}_i^++\gamma_+\hl_\ad^+\hl^+_\bd\eta^\bd_j
(T\bar{M}T)_i{}^j\dt y^{\dot1\ad},\\
\bar{\CE}^-_i&\ :=\
\bar{E}_i^-+\gamma_-\hl_\ad^-\hl^-_\bd\eta^\bd_j
(T\bar{M}T)_i{}^j\dt y^{\dot2\ad}}
and $\bW_a^\pm,V_\pm^i,\bV_\pm^i$ and $\bT_\pm^a,\T_\pm^a$
were given in \eqref{eq:3.34}--\eqref{eq:3.37b} and
$(AB)_i{}^j:=A_i{}^kB_k{}^j$. Actually, the formulas
\eqref{eq:6.31} and their inverses define a diffeomorphism between
the supermanifolds $\CF_M^{5|8}=(\IR^{3|8}\times S^2,\cT_M)$ and
$\CF^{5|8}=(\IR^{3|8}\times S^2,\cT)$ which have different
integrable distributions $\cT_M$ and $\cT$ (and different CR
structures).

Next we need the vector fields on $\CP_M^{2|4}$.
In the above discussion, we used a transformation from the
coordinates $\tilde{\eta}_i^\pm$ to the coordinates $\eta_i^\pm$
on $\CF^{5|8}_M$, which are (pulled-back) sections of $\Pi\CO_{\IC P^1}(1)$.
The corresponding splitting of the transition function was given
in \eqref{eq:6.13}--\eqref{eq:6.15}. One can find a similar
splitting of the transition function \eqref{eq:6.10} also on the
complex supermanifold $\CP^{2|4}_M$ and obtain new coordinates
$\hat{\eta}_i^\pm$, which are sections of $\Pi \CO_{\IC P^1}(1)$, as well.
Explicitly, we have
$$\e^{\frac{w_+^1}{w_+^2}M}\ =\
\e^{\left(1-\frac{1}{1+w_+^2\bw_+^2}\right)\frac{w_+^1}{w_+^2}
M}\e^{\frac{w_+^1}{w_+^2(1+w_+^2\bw_+^2)} M}\ =\
\e^{\frac{\bw_+^2 w_+^1}{1+w_+^2\bw_+^2} M}\e^{\frac{\bw_-^2
w_-^1}{1+w_-^2\bw_-^2}M},$$
which yields the formulas
\eqn{}{\tilde{\eta}^+\ =\ \e^{\frac{\bw_+^2 w_+^1}{1+w_+^2\bw_+^2}
M}\hat{\eta}^+\qquad{\rm and}\qquad\tilde{\eta}^-\ =\ \e^{-\frac{\bw_-^2
w_-^1}{1+w_-^2\bw_-^2}M}\hat{\eta}^-.}
From this and \eqref{eq:6.11} it follows that
\eqn{}{\hat{\eta}_i^+\ =\ w_+^2\hat{\eta}_i^-}
and these coordinates have the desired property. Furthermore, in
the $(0,1)$-part of the differential
\eqna{}{
\pbar_{\CP_M}|_{\CV_\pm}\ &=\
\dt \bw^1_\pm\der{\bw^1_\pm}+\dt \bw^2_\pm\der{\bw^2_\pm}+
\dt \bar{\tilde{\eta}}_i^\pm\der{\bar{\tilde{\eta}}_i^\pm}\\
 &=\ \dt \bar{\hat{w}}^1_\pm\partial_{\bar{\hat{w}}^1_\pm}+\dt
\bar{\hat{w}}^2_\pm\left(\partial_{\bar{\hat{w}}^2_\pm}\mp
\gamma^2_\pm\hat{w}^1_\pm
M_i{}^j\hat{\eta}_j^\pm\partial_{\hat{\eta}^\pm_i}\right)\ +\\
  &\kern4cm+\ \left(\phantom{\partial_{\bar{\hat{w}}^2_\pm}\mp}\kern-1cm\dt
\bar{\hat{\eta}}^\pm_i\pm\gamma^2_\pm\bar{\hat{w}}^1_\pm
\bar{M}_i{}^j\bar{\hat{\eta}}_j^\pm\dt
\hat{w}^2_\pm\right)\partial_{\bar{\hat{\eta}}^\pm_i},}
where we introduced $\hat{w}^{1,2}_\pm=w_\pm^{1,2}$ for clarity,
we see explicitly the deformation of the complex structure from
$\CP^{2|4}$ to $\CP^{2|4}_M$. Note that the coordinates
$\hat{\eta}_i^\pm$ can be pulled-back to $\CF_M^{5|8}$, and there they are
related to the coordinates $\eta^\pm$ by
\eqn{}{\hat{\eta}^\pm\ =\ \e^{-w_\pm^3 M}\eta^\pm.}

\paragraph{Mass-deformed Bogomolny equations from phCS theory
on $\CF_M^{5|8}$.}
Now we have all ingredients for discussing phCS theory on 
deformed CR supertwistor space $\CF_M^{5|8}$. 
The deformed mini-supertwistor space $\CP^{2|4}_M$ fits into a
double fibration
\eqna{eq:6.36}{
\begin{picture}(50,50)(0,-7)
\put(0.0,0.0){\makebox(0,0)[c]{$\CP_M^{2|4}$}}
\put(64.0,0.0){\makebox(0,0)[c]{$\IR^{3|8}$}}
\put(34.0,37.0){\makebox(0,0)[c]{$\CF_M^{5|8}$}}
\put(7.0,20.0){\makebox(0,0)[c]{$\pi_1$}}
\put(55.0,20.0){\makebox(0,0)[c]{$\pi_2$}}
\put(25.0,27.0){\vector(-1,-1){18}}
\put(37.0,27.0){\vector(1,-1){18}}
\end{picture}}
similarly to the undeformed case $M=0$. Recall that we had a
holomorphic integral form on $\CP^{2|4}$ locally defined by
\eqn{}{\Omega|_{\CV_\pm}\ =\ \pm\dt w_\pm^1\wedge \dt
w_\pm^2\dt\eta_1^\pm\cdots\dt\eta_4^\pm.}
One can extend $\Omega$ to a nonvanishing holomorphic volume form
\eqn{eq:6.37}{
\Omega^M|_{\CV_\pm}\ =\ \pm\dt w_\pm^1\wedge \dt
w_\pm^2\dt\tilde{\eta}_1^\pm\cdots\dt\tilde{\eta}_4^\pm}
on $\CP^{2|4}_M$ if and only if $\tr M=0$ \cite{Chiou:2005jn}.
This is the reason why we imposed this condition from the very
beginning. Similarly to the discussion of phCS theory on
$\CF^{5|8}$ in Sec. \ref{sec:hBF-2}, we consider a subsupermanifold
$\CX_M$ of $\CF^{5|8}_M$ which is defined by the constraints
$\bar{\tilde{\eta}}_i^\pm=0$. Clearly, the latter equations are
equivalent to $\bar{\eta}_i^\pm=0$ and therefore $\CX_M$ is
diffeomorphic to $\CX$. Note that the pull-back of the
holomorphic integral form \eqref{eq:6.37} to $\CF_M^{5|8}$
coincides with $\pi_1^*\Omega$,
\eqn{}{\tilde{\Omega}^M|_{\hCV_\pm}\ :=\
\pi_1^*\Omega^M|_{\hCV_\pm}\ =\
\pm\Theta^1_\pm\wedge\Theta^2_\pm\dt \tilde{\eta}_1^\pm\cdots\dt
\tilde{\eta}_4^\pm\ =\  \pm\Theta^1_\pm\wedge \Theta^2_\pm\dt
\eta_1^\pm\cdots\dt \eta_4^\pm}
which is due to \eqref{eq:6.16} and the tracelessness of $M$.

From here on, we proceed as in Sec. \ref{sec:hBF-2} and consider a trivial
rank $r$ complex vector bundle over the CR supertwistor space
$\CF_M^{5|8}$ with a connection $\CA_{\cT_M}$ along the integrable
distribution $\cT_M$ defined in \eqref{eq:6.30} and
\eqref{eq:6.32}. By assuming that
$\bV_\pm^i\lrcorner\CA_{\cT_M}=0$ and
$\bV_\pm^i(\bar{\CW}_a^\pm\lrcorner \CA_{\cT_M})=0$, we may define
the action functional
\eqn{eq:6.39}{
  S\ =\
  \int_{\CX_M}\tilde{\Omega}^M\wedge{\rm tr}\left\{\CA_{\cT_M}\wedge
\dt_{\cT_M}\CA_{\cT_M}+\tfrac{2}{3}\CA_{\cT_M}\wedge\CA_{\cT_M}
\wedge\CA_{\cT_M}\right\}}
of deformed phCS theory. The equations of motion keep the form
\eqref{eq:4.16} up to relabeling $\cT$ by $\cT_M$ and in
components $\CA_a^\pm:=\bar{\CW}_a^\pm\lrcorner\CA_{\cT_M}$, we
have
\eqna{eq:6.40}{
  \bar{W}_1^\pm\CA^\pm_2-\bar{W}^\pm_2\CA^\pm_1+
  [\CA^\pm_1,\CA^\pm_2]&\ =\ 0,\\[5pt]
  \bar{W}_2^\pm\CA_3^\pm-\bar{W}^\pm_3\CA^\pm_2+
  [\CA^\pm_2,\CA_3^\pm]\mp2\gamma_\pm^2\CA^\pm_1
  &\ =\ M_j{}^i\eta_i^\pm\der{\eta_j^\pm}\CA^\pm_2,\\
  \bar{W}_1^\pm\CA_3^\pm-\bar{W}_3^\pm\CA^\pm_1+
  [\CA^\pm_1,\CA^\pm_3] &\ =\
  M_j{}^i
      \eta_i^\pm\der{\eta_j^\pm}\CA^\pm_1,}
where the vector fields \eqref{eq:6.33} have already been
substituted. The dependence of the components $\CA_a^\pm$ on
$\lambda_\pm,\bl_\pm$ and $\eta_i^\pm$ is of the same form as the
one given in \eqref{eq:4.19} but with coefficient
functions obeying $M$-deformed equations. 

Substituting the expansions of the form \eqref{eq:4.19} for
$\CA_a^\pm$ and our vector fields $\bar{W}_a^\pm$ into
\eqref{eq:6.40}, we obtain mass-deformed supersymmetric Bogomolny
equations:
\eqna{eq:6.41}{
  \fc_{\ad\bd}\ &=\ -\tfrac{\i}{2}\cnab_{\ad\bd}\Pc,\\
    \epsilon^{\bd\gd}\cnab_{\ad\bd}\cc\ \!\!^i_\gd
             -\tfrac{1}{2}M_j{}^i\cc\ \!\!^j_\ad\ &=\
             -\tfrac{\i}{2}[\Pc,\cc\ \!\!^i_\ad],\\
 \overset{\circ}{\triangle}\Wc\ \!\!^{ij}+M_k{}^{[i}M_l{}^{[j]}\Wc\ \!\!^{k]l}\ &=\
-\tfrac{1}{4}[\Pc,[\Wc\ \!\!^{ij},\Pc]]
 -\i M_k{}^{[i}[\Pc,\Wc\ \!\!^{j]k}]-\epsilon^{\ad\bd}
 \{\cc\ \!\!^i_\ad,\cc\ \!\!^j_\bd\},\\
\epsilon^{\bd\gd}\cnab\ \!\!_{\ad\bd}\cc_{i\gd}-\tfrac{1}{2}
           M_i{}^j\cc_{j\ad}\ &=\ -\tfrac{\i}{2}
           [\cc_{i\ad},\Pc]+2[\Wc_{ij},\cc\ \!\!^j_\ad],\\
  \epsilon^{\bd\gd}\cnab_{\ad\bd}\Gc_{\gd\dd}\ &=\
  -\tfrac{\i}{2}[\Gc_{\ad\dd},\Pc]+
  \{\cc\ \!\!^i_\ad,\cc_{i\dd}\}
  -\tfrac{1}{2}[\Wc_{ij},\cnab_{\ad\dd}\Wc\ \!\!^{ij}]\ +\ \\
 &\kern1cm +\tfrac{\i}{4}\epsilon_{\ad\dd}[\Wc\ \!\!_{ij},
 [\Pc,\Wc\ \!\!^{ij}]]+\tfrac{1}{2}
   \epsilon_{\ad\dd}M_m{}^k[\Wc_{kl},\Wc\ \!\!^{lm}].}

Eqs. \eqref{eq:6.40} show that, as in the undeformed case
\eqref{eq:4.17}, some of the fields appearing in the expansions of
$\CA_a^\pm$ are not independent degrees of freedom but composite
fields. In fact, we find
\eqna{eq:6.42}{
       \Wc\ \!\!^{ij}_{\ad\bd}\ &=\ -\left(\partial_{(\ad\bd)}\Wc\ \!\!^{ij}+
       [\Bc_{\ad\bd},\Wc\ \!\!^{ij}]
         -\epsilon_{\ad\bd}M_k{}^{[i}\Wc\ \!\!^{j]k}\right),\\
       \cc\ \!\!^{ijk}_{\ad(\bd\gd)}\ &=\
       -\tfrac{1}{2}\left(\partial_{(\ad(\bd)}
              \cc\ \!\!^{ijk}_{\gd)}+
              [\Bc_{\ad(\bd},\cc\ \!\!^{ijk}_{\gd)}]
               +\tfrac{3}{2}\epsilon_{\ad(\bd}
               M_l{}^{[i}\cc\ \!\!^{jk]l}_{\gd)}\right),\\
       \Gc\ \!\!^{ijkl}_{\ad(\bd\gd\dd)}\ &=\
       -\tfrac{1}{3}\left(\partial_{(\ad(\bd)}
          \Gc\ \!\!^{ijkl}_{\gd\dd)}+[\Bc_{\ad(\bd},\Gc\ \!\!^{ijkl}_{\gd\dd)}]-
         2\epsilon_{\ad(\bd}M_m{}^{[i}\Gc\ \!\!^{jkl]m}_{\gd\dd)}\right).}

Finally, upon substituting our superfield expansions for $\CA_a^\pm$
into the action \eqref{eq:6.39} and integrating over the odd
coordinates and over the Riemann sphere, we end up with
\eqn{eq:6.43}{
 S\ =\ S_0-\tfrac{1}{2}\int\dt^3 x\
{\rm tr}\left\{
      \cc_{i\ad}M_j{}^i\cc\ \!\!^{j\ad}-\Wc_{ij}
      M_k{}^iM_l{}^{[j}\Wc\ \!\!^{k]l}+
       \i\Pc M_k{}^i[\Wc_{ij},\Wc\ \!\!^{jk}]\right\},}
where $S_0$ is the action
functional for the massless supersymmetric Bogomolny equations as
given by \eqref{eq:4.27}.

\paragraph{Summary.}
We have described a one-to-one correspondence between
gauge equivalence classes of local solutions to the supersymmetric
Bogomolny equations with massive fermions and scalar fields and
equivalence classes of $\cT_M$-flat bundles over the CR
supertwistor space $\CF_M^{5|8}$ which are holomorphically trivial
on each $\IC P^1_{x,\eta}\hookrightarrow\CF^{5|8}_M$. We have also 
described
a one-to-one correspondence between the equivalence classes of
$\cT_M$-flat complex vector bundles over $\CF_M^{5|8}$ and of
holomorphic vector bundles over the deformed mini-supertwistor
space $\CP^{2|4}_M$. The assumption that these bundles become
holomorphically trivial on projective lines translates in the
Dolbeault description into a one-to-one correspondence between
gauge equivalence classes of solutions to the field equations of
i) hBF theory on the deformed mini-supertwistor space
$\CP_M^{2|4}$, ii) phCS theory on the CR supertwistor space
$\CF_M^{5|8}$ and iii) massive supersymmetric Bogomolny model on
Euclidean three-dimensional space $\IR^3$.

In fact, this section's discussion can be understood as 
a corollary of Thm. \ref{thm:summmary-3}: complex structure
deformations on the mini-supertwistor space induce, for appropriately
chosen CR structures, CR structure deformations on the CR 
supertwistor space and, of course, vice versa. Our above discussion
is the translation of Thm. \ref{thm:summmary-3} to this deformed
setting.

\section{Solution generating techniques}\label{sec:solution} 

In the preceding sections, we have presented in detail the
relations between supersymmetric Bogomolny monopole equations on the
Euclidean space $\IR^3$ and field equations of phCS theory on
the CR supertwistor space $\CF^{5|8}$ as well as hBF theory on the
mini-supertwistor space $\CP^{2|4}$. We have shown that the moduli
spaces of solutions to the field equations of these three theories
are bijective. Furthermore, we introduced mass-deformed versions
of these field theories. In this section, we want to show how the
twistor correspondences described in the previous sections can be
used for constructing explicit solutions to the supersymmetric
Bogomolny equations. In fact, any solution to the standard
Bogomolny equations given as a pair $(\Ac_{\ad\bd},\Pc)$ of a
gauge potential and a Higgs field can be extended to a solution
including the remaining fields of the supersymmetrically extended
Bogomolny equations in a nontrivial fashion. The subsequent 
discussion is devoted to this issue. However, we are not
considering this task in full generality but merely give
some flavor of how the algorithms work. For simplicity, we 
also restrict ourselves to the case
when only the fields $\Ac_{\ad\bd},\Pc$ and $\Gc_{\ad\bd}$ are
non-zero. In this case, the supersymmetric Bogomolny equations
\eqref{eq:4.24} simplify to
\eqna{eq:7.1}{
-\tfrac{1}{2}\epsilon^{\gd\dd}\left(\partial_{(\ad\gd)}
\Ac_{\bd\dd}-\partial_{(\bd\dd)}\Ac_{\ad\gd}+
[\Ac_{\ad\gd},\Ac_{\bd\dd}]\right)&\ =\ -\tfrac{\i}{2}
\left(\partial_{(\ad\bd)}\Pc+[\Ac_{\ad\bd},\Pc]\right),\\
\epsilon^{\gd\dd}\left(\partial_{(\ad\gd)}\Gc_{\bd\dd}+
[\Ac_{\ad\gd},\Gc_{\bd\dd}]\right)
&\ =\ -\tfrac{\i}{2}[\Gc_{\ad\bd},\Pc].}
First, we discuss Abelian solutions to these equations, which
correspond to Dirac monopole-antimonopole systems. After this,
we present two algorithms which generate non-Abelian solutions.

\paragraph{Abelian solutions.}
In the Abelian case, the system
\eqref{eq:7.1} simplifies further to
\eqna{eq:7.2}{
\epsilon^{\gd\dd}\left(\partial_{(\ad\gd)}\Ac_{\bd\dd}-
\partial_{(\bd\dd)}\Ac_{\ad\gd}\right)&\ =\ \i\partial_{(\ad\bd)}\Pc,\\
\epsilon^{\gd\dd}\partial_{(\ad\gd)}\Gc_{\bd\dd}&\ =\ 0.}
It is convenient to rewrite these equations in terms of the real
coordinates $x^r$ on $\IR^3$ with $r=1,2,3$ as
\eqna{eq:7.4}{
\tfrac{1}{2}\epsilon_{rst}(\partial_s\Ac_t-\partial_t\Ac_s)&\ =\
\partial_r\Pc,\\ \partial_r\Gc_r&\ =\ 0,\\
\epsilon_{rst}\partial_s\Gc_t&\ =\ 0.}
From the second equation of \eqref{eq:7.4}, it follows that
\eqn{}{\Gc_r\ =\
\tfrac{1}{2}\epsilon_{rst}(\partial_s\Ac\ \!\!'_t-\partial_t\Ac\ \!\!'_s),}
and from third one, we obtain
\eqn{eq:7.8}{
 \Gc_r\ =\ -\partial_r\Pc\ \!\!',}
where the sign in \eqref{eq:7.8} was chosen to match the fact that
in four dimensions, $\Gc_r$ corresponds to an anti-self-dual
two-form with components $\Gc_{\mu\nu}=\bar{\eta}^r_{\mu\nu}\Gc_r$ and
helicity $-1$, where $\bar{\eta}^r_{\mu\nu}$ are the 't Hooft tensors.
Here, $\Ac\ \!\!'_r$ and $\Pc\ \!\!'$ are a vector and a scalar,
respectively. Therefore, Eqs. \eqref{eq:7.4} can be
rewritten as
\eqna{eq:7.9}{
\tfrac{1}{2}\epsilon_{rst}(\partial_s\Ac_t-\partial_t\Ac_s)&\ =\
\partial_r\Pc,\\
\tfrac{1}{2}\epsilon_{rst}(\partial_s\Ac\ \!\!'_t-\partial_t\Ac\ \!\!'_s)&\ =\
-\partial_r\Pc\ \!\!'.}
It is well known that the first equation describes Dirac
monopoles while the second one Dirac antimonopoles (see,
e.g., Atiyah et al. \cite{Atiyah:1988jp} and references therein). Thus, the
action \eqref{eq:4.27} with only the fields $\fc_{\ad\bd}$, $\Pc$
and $\Gc_{\ad\bd}$ being non-zero can be considered as a proper
action for the description of monopole-antimonopole systems.

Let us consider a configuration of $m_1$ Dirac monopoles and $m_2$
antimonopoles located at points $a_i=(a_i^1,a_i^2,a_i^3)$ with
$i=1,\ldots, m_1$ and $i=m_1+1,\ldots, m_1+m_2$, respectively.
Moreover, we assume for simplicity that $a_i^{1,2}\neq a_j^{1,2}$
for $i\neq j$. Such a configuration is then described by the
fields
\eqna{eq:7.11}{
 \Ac\ \!\!^{N}&\ =\ \sum_{j=1}^{m_1}\Ac\ \!\!^{N,j}~,&\Ac\ \!\!^{S}&\ =\
\sum_{j=1}^{m_1}\Ac\ \!\!^{S_,j},
&\Pc\ \!\!^N&\ =\ \Pc\ \!\!^S\ =\ \sum_{j=1}^{m_1}\frac{\i}{2r_j},\\
\Ac\ \!\!'_{N}&\ =\
\sum_{j=m_1+1}^{m_1+m_2}\overset{\circ}{\bar{\CA}}\ \!\!^{N,j},&\Ac\ \!\!'^{S}&\ =\
\sum_{j=m_1+1}^{m_1+m_2}\overset{\circ}{\bar{\CA}}\ \!\!^{S_,j},&\Pc\ \!\!'^N&\ =\
\Pc\ \!\!^S\ =\ \sum_{j=m_1+1}^{m_1+m_2}\frac{\i}{2r_j},}
where $\Ac\ \!\!^{N,j}= \Ac\ \!\!^{N,j}_m\dt x^m$ and $\Ac\ \!\!^{S,j}=\Ac\ \!\!^{S,j}_m\dt x^m$
with  
\eqna{eq:7.13}{
 \Ac\ \!\!_1^{N,j}&\ =\ \frac{\i x_j^2}{2r_j(r_j+x_j^3)},&\Ac\ \!\!_2^{N,j}&\ =\
\frac{-\i
x_j^1}{2r_j(r_j+x_j^3)},&\Ac\ \!\!_3^{N,j}&\ =\ 0,\\
\Ac\ \!\!_1^{S,j}&\ =\ -\frac{\i x_j^2}{2r_j(r_j-x_j^3)},&\Ac\ \!\!_2^{S,j}&\ =\
\frac{\i x_j^1}{2r_j(r_j-x_j^3)},&\Ac\ \!\!_3^{S,j}&\ =\ 0,}
\eqn{}{
x_j^s\ =\ x^s-a_j^s~,~~~r_j^2\ =\ \delta_{rs}x_j^rx_j^s~.}
Here, $N$ and $S$ denote the following two regions in $\IR^3:$
\eqna{}{\IR^3_{N,m_1+m_2}&\ :=\
\IR^3\ \backslash\bigcup_{i=1}^{m_1+m_2}
\left\{x^1=a_i^1,\,x^2=a_i^2,\,x^3\leq
a_i^3 \right\},\\
\IR^3_{S,m_1+m_2}&\ :=\
\IR^3\ \backslash\bigcup_{i=1}^{m_1+m_2}
\left\{x^1=a_i^1,\,x^2=a_i^2,\,x^3\geq
a_i^3 \right\}}
and bar stands for complex conjugation. Note that
\eqn{}{
\IR^3_{N,m_1+m_2}\cup\IR^3_{S,m_1+m_2}\ =\
\IR^3\backslash\{a_1,\ldots,a_{m_1+m_2}\}}
and the configuration \eqref{eq:7.11}, \eqref{eq:7.13} has
delta-function sources at the points $a_i$ with
$i=1,\ldots,m_1+m_2$.

\paragraph{Non-Abelian solutions via a contour integral.}\label{par:NASCI-ch3}
For the gauge group $SU(2)$, one can consider the Wu-Yang point
monopole \cite{wuyang} and its generalizations to configurations
describing $m_1$ monopoles and $m_2$ antimonopoles
\cite{Popov:2004rt}. This solution, which is singular at points
$a_i$, $i=1,\ldots,m_1+m_2$, is a solution to Eqs.
\eqref{eq:7.1} for $\mathfrak{su}(2)$-valued fields. However, it is just an
Abelian configuration in disguise, as it is equivalent to the
multi-monopole configuration \eqref{eq:7.11}, \eqref{eq:7.13}
\cite{Popov:2004rt}.

One can construct true non-Abelian solutions to \eqref{eq:7.1} as
follows. Let us first consider a configuration $\Ac_{\ad\bd}=0=\Pc$ and
$\Gc_{\ad\bd}\neq 0$. Then from \eqref{eq:7.1} one obtains the
equation
\eqn{eq:7.9b}{
 \epsilon^{\bd\gd}\partial_{(\ad\bd)}\Gc_{\gd\dd}\ =\ 0.}
All solutions to this equation can be described in
the twistor approach \cite{Penrose:in} via a contour integral
\eqn{eq:7.10b}{
 \Gc_{\ad\bd}\ =\ \frac{1}{2\pi\i}\oint_{\cC}\dt\lambda_+\lambda^+_\ad\lambda^+_\bd
\overset{\circ}{\Upsilon}_{+-}(w_+^1,w_+^2),}
where $\overset{\circ}{\Upsilon}_{+-}(w_+^1,w_+^2)$ is a Lie-algebra valued meromorphic
function of $w_+^1=\lambda^+_\ad\lambda^+_\bd y^{\ad\bd}$ and
$w_+^2=\lambda_+$ holomorphic in the vicinity of the curve
$\cC\cong S^1\subset\IC P^1$. From \eqref{eq:7.10b} it follows
that nontrivial contributions to $\Gc_{\ad\bd}$ are only given
by those $\overset{\circ}{\Upsilon}_{+-}$ which define elements of the cohomology group
$H^1(\CP^2,\mathfrak{gl}(r,\CO(-4))$. It is easy to see that
\eqref{eq:7.10b} satisfies \eqref{eq:7.9b} due to the identity
 $$\lambda_+^\bd\partial_{(\ad\bd)}\overset{\circ}{\Upsilon}_{+-}\ =\
\derr{\overset{\circ}{\Upsilon}_{+-}}{w_+^1}\lambda_+^\bd\lambda_\bd^+\lambda_\ad^+\
=\ 0, $$
which appears after pulling the derivatives $\partial_{(\ad\bd)}$
under the integral.

Consider now a fixed solution $(\Ac_{\ad\bd},\Pc)$ of the
Bogomolny equations given by the first line of \eqref{eq:7.1}.
One may take, e.g., the $SU(2)$ BPS
monopole \cite{Prasad:1975kr,Bogomolny:1975de}. In the twistor
approach, we can find functions $\hat{\psi}_\pm$ solving the
linear system
\eqn{eq:7.13a}{
 \lambda_\pm^\bd(\partial_{(\ad\bd)}+\Ac_{\ad\bd}-
\tfrac{\i}{2}\epsilon_{\ad\bd}\Pc)\hat{\psi}_\pm\
=\ 0\qquad{\rm and}\qquad \partial_{\bl_\pm}\hat{\psi}_\pm\ =\ 0,}
which is equivalent to the linear system of phCS theory. These
$\hat{\psi}_\pm$ are known explicitly for many cases, e.g., for
our chosen example of the $SU(2)$ BPS monopole, they have been given by Ward in
\cite{Ward:1981jb,Ward:1981jb-2}. Using $\hat{\psi}_\pm$, we can introduce
dressed fields $\Gc_{\ad\bd}$ by the formula
\eqn{eq:7.13b}{
 \Gc_{\ad\bd}\ =\ \frac{1}{2\pi\i}\oint_{\cC}\dt \lambda_+
\lambda^+_\ad\lambda^+_\bd\hat{\psi}_+\overset{\circ}{\Upsilon}_{+-}
\hat{\psi}_-^{-1},}
where $\overset{\circ}{\Upsilon}$ is chosen as above.
One can straightforwardly check that with this choice,
\eqn{eq:7.16}{
\epsilon^{\bd\gd}\left(\partial_{(\ad\bd)}\Gc_{\gd\dd}+[\Ac_{\ad\bd}-
\tfrac{\i}{2}\epsilon_{\ad\bd}\Pc,\Gc_{\gd\dd}]\right)\ =\ 0,}
and therefore the configuration $(\Ac_{\ad\bd},\Pc,\Gc_{\ad\bd})$
satisfies \eqref{eq:7.1}. The explicit form of a $\Gc_{\ad\bd}$
for a given $\Ac_{\ad\bd}$ and $\Pc$ is obtained by performing
the contour integral \eqref{eq:7.13b} along $\cC$ after a
proper choice of the Lie-algebra valued function $\overset{\circ}{\Upsilon}$.
Recall that the configuration $(\Ac_s,\Pc)$ will be real, i.e., the
fields will take values in the Lie algebra $\mathfrak{su}(r)$, if the
matrix-valued functions $\hat{\psi}_\pm$ in \eqref{eq:7.13a}
satisfy the reality condition as the one induced by \eqref{AHI-2} and $\det
(\hat{\psi}_+^{-1}\hat{\psi}_-)=1$. Imposing a proper reality
condition on the function $\overset{\circ}{\Upsilon}_{+-}$ will ensure the
skew-Hermiticity of $\Gc_s$.

\paragraph{Solutions via nilpotent dressing 
transformations in the \v Cech approach.}\label{pa:soldres-ch3}
In this section, we will present a novel algorithm for
constructing solutions to Eqs. \eqref{eq:7.1} based on
the twistor description of hidden symmetry algebras in the
self-dual SYM theory in four dimensions
\cite{Wolf:2004hp} -- cf. our discussion presented in
Chap. \ref{HS-CHAPTER}. Recall that we have
described a one-to-one correspondence between equivalence classes 
of transition functions of $\cT$-flat vector bundles
over the CR supertwistor space $\CF^{5|8}$ obeying
certain triviality conditions and gauge equivalence classes
of solutions to the supersymmetric Bogomolny
equations on $\IR^3$. We can, however, associate with any open
subset $\hCV_+\cap\hCV_-\subset\CF^{5|8}$ an
infinite number of classes of transition functions, which
in turn yield an infinite number of gauge equivalence classes
of solutions to the supersymmetric Bogomolny equations. Therefore, 
one naturally meets with a possibility of constructing new solutions 
from a given one, that is, with dressing transformations. In the remainder of
this chapter, we  
discuss a particular example of such a construction but first we briefly 
introduce some necessary background material. At this stage, we
present the latter in a more applied formulation and without precise
mathematical terminology. For a more thorough exposition, we refer 
to Chap. \ref{HS-CHAPTER}, where also a \v Cech cohomological interpretation 
of the underlying structure is given. 

 We consider the linear
system \eqref{eq:4.40}, which can be rewritten as
\eqn{eq:7.15}{
(\bV_\ad^\pm+\hat{\CA}_\ad^\pm)\hat{\psi}_\pm\ =\ 0,\quad
\partial_{\bl_\pm}\hat{\psi}_\pm\ =\ 0\quad{\rm and}\quad
(\bV_\pm^i+\hat{\CA}_\pm^i)\hat{\psi}_\pm\ =\ 0,}
where we have defined
\eqn{}{\bV_\ad^\pm\ :=\  \lambda_\pm^\bd\partial_{(\ad\bd)}
\qquad{\rm and}\qquad
\hat{\CA}_\ad^\pm\ :=\  \bV_\ad^\pm\lrcorner\hat{\CA}_{\cT},}
as before.\footnote{Note
that the first and third equations of \eqref{eq:4.40} are equivalent to
$$\lambda_\pm^\ad(\bV_\ad^\pm+\hat{\CA}_\ad^\pm)\hat{\psi}_\pm\ =\ 0
\qquad{\rm and}\qquad 
\hl_\pm^\ad(\bV_\ad^\pm+\hat{\CA}_\ad^\pm)\hat{\psi}_\pm\ =\ 0,$$
respectively, which together imply
$(\bV_\ad^\pm+\hat{\CA}_\ad^\pm)\hat{\psi}_\pm=0$.}
From arguments similar to those used subsequent to
\eqref{eq:4.40}, we have
$\hat{\CA}_\ad^\pm=\lambda_\pm^\bd\CB_{\ad\bd}$ and
$\hat{\CA}_\pm^i=\lambda_\pm^\ad\CA_\ad^i$ with
$\lambda$-independent superfields $\CB_{\ad\bd}$ and
$\CA^i_\ad$. The compatibility conditions for the linear system
\eqref{eq:7.15} are Eqs. \eqref{eq:4.42}. From this
linear system one also derives that
$\tilde{f}_{+-}=\hat{\psi}_+^{-1}\hat{\psi}_-$ is $\cT$-flat,
i.e.,
\eqn{eq:7.17}{
\bV_\ad^\pm\tilde{f}_{+-}\ =\ 0,\qquad\partial_{\bl_\pm}\tilde{f}_{+-}\
=\ 0\qquad{\rm and}\qquad \bV_\pm^i\tilde{f}_{+-}\ =\ 0.}

The key idea is to study infinitesimal deformations of the
transition function $\tilde{f}_{+-}$ of the $\cT$-flat vector
bundle preserving \eqref{eq:7.17} and the triviality properties
discussed above. More explicitly, given such a function
$\tilde{f}_{+-}=\hat{\psi}_+^{-1}\hat{\psi}_-$ (with
$\partial_{\bl_\pm}\hat{\psi}_\pm=0$), we consider
\eqn{eq:7.18}{
\tilde{f}_{+-}+\delta \tilde{f}_{+-}\ =\
(\hat{\psi}_++\delta\hat{\psi}_+)^{-1}
(\hat{\psi}_-+\delta\hat{\psi}_-),}
where $\delta$ represents some generic infinitesimal deformation.
Note that any infinitesimal $\cT$-flat deformation (that is,
preserving \eqref{eq:7.17}) is allowed since for small
perturbations, the trivializability property of the bundle
$\tilde{\CE}$ on the holomorphic curves $\IC P^1_{x,\eta}\hookrightarrow
\CF^{5|8}$ is preserved (cf. Chap. \ref{HS-CHAPTER}). 
Upon introducing the Lie-algebra valued function
\eqn{M:5.2}{
\phi_{+-}\ := \ \hat{\psi}_+(\delta
\tilde{f}_{+-})\hat{\psi}_-^{-1}}
and linearizing \eqref{eq:7.18}, we have to find a splitting
\eqn{M:5.3}{
\phi_{+-}\ =\ \phi_+-\phi_-,}
where the Lie-algebra valued functions $\phi_\pm$ can be extended
to holomorphic functions in $\lambda_\pm$, which yields
\eqn{}{\delta\hat{\psi}_\pm\ =\ -\phi_\pm\hat{\psi}_\pm.}
To find these $\phi_\pm$ from $\phi_{+-}$ means to solve the
infinitesimal Riemann-Hilbert problem. Clearly, such solutions are
not unique, as we have the freedom
\eqn{M:5.5}{
\phi_{+-}\ =\ \phi_+-\phi_-\ =\ (\phi_++\omega)-(\phi_-+\omega)\
    =:\ \tilde{\phi}_+-\tilde{\phi}_-,}
with $\tilde{\phi}_\pm:=\phi_\pm+\omega$, where the function 
$\omega$ is independent of $\l_\pm$.
This freedom can be used to
preserve the transversal gauge condition \eqref{tg}. The preservation
of this gauge is
in fact needed in order to compare the deformed superfields and 
the ones one has started with as one wishes to derive the induced
transformations of the component fields.

Linearizing \eqref{eq:7.15}, we get
\eqn{M:5.6}{
\delta\hat{\CA}_\ad^\pm\ =\ \bar{\nabla}_\ad^\pm\phi_\pm \qquad{\rm
and}\qquad
       \delta\hat{\CA}^i_\pm\ =\ \bar{\nabla}^i_\pm\phi_\pm,}
where we have introduced the operators
$\bar{\nabla}_\ad^\pm:=\bV^\pm_\ad+\hat{\CA}^\pm_\ad$ and
$\bar{\nabla}^i_\pm:=\bV^i_\pm+\hat{\CA}^i_\pm$. From \eqref{eq:7.15},
\eqref{eq:7.17} and \eqref{eq:7.18} it follows that
\eqn{}{\bar{\nabla}^\pm_\ad\phi_{+-}\ =\ 0\ =\ \bar{\nabla}^i_\pm\phi_{+-},}
and we eventually arrive at the formulas
\eqn{M:5.8}{
\delta\CB_{\ad\bd}\ =\ \frac{1}{2\pi\i}
\oint_{\cC}\dt\l_+\frac{\bar{\nabla}^+_\ad\phi_+}{\lambda_+\lambda^\bd_+}\,
       \qquad{\rm and}\qquad
       \delta\CA_\ad^i\ =\ \frac{1}{2\pi\i}\oint_{\cC}\dt\lambda_+
       \frac{\bar{\nabla}^i_+\phi_+}{\lambda_+\lambda^\ad_+},}
where the contour is $\cC=\{\lambda_+\in
\IC P^1\,|\,|\lambda_+|=1\}$. Thus, the consideration of
infinitesimal deformations of the transition function of some
$\cT$-flat vector bundle over the CR supertwistor space
$\CF^{5|8}$ obeying certain triviality conditions gives by virtue
of the integral formulas \eqref{M:5.8} infinitesimal
deformations of the components $\CB_{\ad\bd}$ and $\CA^i_\ad$,
which satisfy -- by construction -- the linearized  constraint
equations \eqref{constraint} and thus the supersymmetric
Bogomolny equations \eqref{eq:4.24}. 
Once again, we have a
one-to-one correspondence between equivalence classes of local
solutions, with equivalence induced on the gauge theory side by
infinitesimal gauge transformations and on the twistor side by
transformations of the form
$\phi_\pm=\psi_\pm\chi_\pm\psi_\pm^{-1}$, where the $\chi_\pm$s are
functions globally defined on $\hCV_\pm\subset\CF^{5|8}$
and annihilated by all vector fields from the distribution $\cT$.
Putting it differently, we have just presented the ``infinitesimal"
version of Thm. \ref{thm:summmary-3}.

Let us now exemplify our discussion by describing how to construct
explicit solutions to \eqref{eq:7.1}. Consider a $\cT$-flat vector
bundle $\tilde{\CE}\to\CF^{5|8}$ of rank $r$ which is
holomorphically trivial when restricted to any projective line
$\IC P^1_{x,\eta}\hookrightarrow\CF^{5|8}$. Assume further that a
transition function $\tilde{f}_{+-}$ of $\tilde{\CE}$ is chosen
such that all the fields $\cc\ \!\!^i_\ad$, $\Wc\
\!\!^{ij}$, $\cc_{i\ad}$ and $\Gc_{\ad\bd}$
vanish identically, i.e., we start with the field equation
\eqn{}{\fc_{\ad\bd}\ =\ -\tfrac{\i}{2}\cnab_{\ad\bd}\Pc.}
Without loss of generality, we may assume that the transition
function of $\tilde{\CE}$ can be split as
$\tilde{f}_{+-}=\hat{\psi}_+^{-1}\hat{\psi}_-$, where the
$\hat{\psi}_\pm$s do not depend on the fermionic coordinates
$\eta^\pm_i$.

Suppose now that
\eqn{M:5.13}{
   \delta \tilde{f}_{+-}\ :=\ -\tfrac{1}{4!}
                  \epsilon^{j_1\cdots j_4}\eta^+_{j_1}\cdots\eta^+_{j_4}
                      [X,\tilde{f}_{+-}],}
where $X\in\mathfrak{gl}(r,\IC)$. Considering vector bundles subject to 
the reality conditions induced by \eqref{AHI-2}, one restricts the
perturbations to those preserving these conditions. In our
example, they read explicitly as
\eqn{}{\delta \tilde{f}_{+-}\ =\ -\tfrac{1}{4!}
                  \epsilon^{j_1\cdots j_4}\left(\eta^+_{j_1}\cdots\eta^+_{j_4}
                      +\eta^-_{j_1}\cdots\eta^-_{j_4}
                      \right)[X,\tilde{f}_{+-}],}
with $X\in\mathfrak{su}(r)$. For illustrating reasons and to simplify
equations, we shall be continuing with \eqref{M:5.13}.
Then a short calculation
reveals that any splitting \eqref{M:5.3} is of the form
\eqn{M:5.14}{
\phi_{+-}\ =\ \phi_+-\phi_-\ =\ -\tfrac{1}{4!}
       \epsilon^{j_1\cdots j_4}\eta^+_{j_1}\cdots\eta^+_{j_4}(\pc_+-
       \pc_-),}
with $\pc_\pm:= -[X,\hat{\psi}_\pm]\hat{\psi}_\pm^{-1}$.
Introducing the shorthand notation 
\eqn{}{\eta^{\gd_1\cdots\gd_4}\ :=\ 
-\tfrac{1}{4!}\epsilon^{j_1\cdots j_4}
        \eta^{\gd_1}_{j_1}\cdots\eta^{\gd_4}_{j_4},}
we find
\eqn{M:5.15}{
\phi_{+-}\ =\
    \eta^{\dot{2}\dot{2}\dot{2}\dot{2}}\pc\ \!\!^4_{+-}+
    4\eta^{\dot{2}\dot{2}\dot{2}\dot{1}}\pc\ \!\!^3_{+-}+
    6\eta^{\dot{2}\dot{2}\dot{1}\dot{1}}\pc\ \!\!^2_{+-}+
    4\eta^{\dot{2}\dot{1}\dot{1}\dot{1}}\pc\ \!\!^1_{+-}+
    \eta^{\dot{1}\dot{1}\dot{1}\dot{1}}\pc\ \!\!^0_{+-},}
where we have used the fact that $\eta^{\gd_1\cdots\gd_4}$ is
totally symmetric. In addition, we defined
\eqn{}{\pc\ \!\!^m_{+-}\ :=\
\lambda_+^m\pc_+-\lambda_+^m\pc_-\ :=\
       \pc\ \!\!^m_+-\pc\ \!\!^m_-.}
The functions $\pc\ \!\!^m_\pm$ can be Laurent-expanded as
($m\geq0$)
\eqn{}{\pc\ \!\!^m_\pm\ =\ \sum_{n=0}^\infty\lambda_+^{\pm
n}\pc\
       \!\!^{m(n)}_\pm}
with
\eqn{M:5.18}{
\pc\ \!\!^{m(n)}_+\ =\
        \begin{cases}
         \delta_{m,0}\pc\ \!\!^{0(0)}_+ & n=0\\
        \pc\ \!\!^{0(n-m)}_+-\pc\ \!\!^{0(m-n)}_- & n>0
        \end{cases}\qquad{\rm and}\qquad
       \pc\ \!\!^{m(n)}_-\ =\ \pc\ \!\!^{0(m+n)}_-.}
Combining the expansion
\eqn{eq:exp-2}{\phi_\pm\ =\ \sum_{n=0}^\infty\lambda_+^{\pm n}\phi_\pm^{(n)}}
with \eqref{M:5.14}--\eqref{M:5.18}, we therefore find
\eqn{M:5.19}{
\phi_-^{(n)}\ =\
    \eta^{\dot{2}\dot{2}\dot{2}\dot{2}}\pc\ \!\!^{0(4+n)}_-+
    4\eta^{\dot{2}\dot{2}\dot{2}\dot{1}}\pc\ \!\!^{0(3+n)}_-+
    6\eta^{\dot{2}\dot{2}\dot{1}\dot{1}}\pc\ \!\!^{0(2+n)}_-+
    4\eta^{\dot{2}\dot{1}\dot{1}\dot{1}}\pc\ \!\!^{0(1+n)}_-+
    \eta^{\dot{1}\dot{1}\dot{1}\dot{1}}\pc\
    \!\!^{0(n)}_-,}
and a similar expression for $\phi_+^{(n)}$. 

At this point, we
have to choose an $\omega$ to preserve 
the transversal gauge. Explicitly, a possible $\omega$ is
given by
\eqn{eq:om}{\omega\ =\
        -\eta^{\dot{2}\dot{2}\dot{2}\dot{1}}\pc\ \!\!^{0(3)}_--
        3\eta^{\dot{2}\dot{2}\dot{1}\dot{1}}\pc\ \!\!^{0(2)}_--
        3\eta^{\dot{2}\dot{1}\dot{1}\dot{1}}\pc\ \!\!^{0(1)}_--
         \eta^{\dot{1}\dot{1}\dot{1}\dot{1}}\pc\ \!\!^{0(0)}_-.}
Let us justify this result. In \eqref{M:5.5}, we
noticed a freedom in splitting the Lie-algebra valued function
$\phi_{+-}$ and we claimed that it can be used to guarantee
the transversal gauge condition \eqref{tg} which, of course, 
translates to the requirement
$\eta^\ad_i\d\CA^i_\ad=0$. In fact, from the second equation
of \eqref{M:5.6}, we obtain
\eqna{B:4}{
\delta\CA^i_{\dot{1}}\ &=\ \nabla_{\dot{1}}^i\phi_+^{(0)}-
          \nabla_{\dot{2}}^i\phi_+^{(1)}\ =\ \nabla_{\dot{1}}^i\phi_-^{(0)},\\
        \delta\CA^i_{\dot{2}}\ &=\ \nabla_{\dot{2}}^i\phi_+^{(0)}\ =\
        -\nabla_{\dot{1}}^i\phi_-^{(1)}+\nabla_{\dot{2}}^i\phi_-^{(0)},}
where we have inserted \eqref{eq:exp-2}.
The contraction of these equations with $\eta^\ad_i$ yields the
constraints
\eqn{B:5}{
    \cD\phi^{(0)}_-+\cD\omega\ =\ \eta^{\dot{2}}_i
    \nabla^i_{\dot{1}}\phi_-^{(1)}
    \qquad{\rm and}\qquad
    \cD\phi^{(0)}_++\cD\omega\ =\ \eta^{\dot{1}}_i
    \nabla^i_{\dot{2}}\phi_+^{(1)}.}
Here, we used the fact that $\tilde{\phi}^{(0)}_\pm=\phi_\pm^{(0)}+\omega$ and
$\tilde{\phi}^{(1)}_\pm=\phi_\pm^{(1)}$, respectively, and recalled
the definition $\cD=\eta^\ad_i\nabla^i_\ad$. Thus, a splitting
\eqref{M:5.5} with an $\omega$ satisfying \eqref{B:5} yields a
deformation of the gauge potential which respects the
transversal gauge condition. In our present example, Eqs.
\eqref{B:5} simplify to
\eqn{M:5.20}{
   \cD\phi^{(0)}_-+\cD\omega\ =\
   \eta^{\dot{2}}_i\partial^i_{\dot{1}}\phi_-^{(1)}
   \qquad{\rm and}\qquad
   \cD\phi^{(0)}_++\cD\omega\ =\
   \eta^{\dot{1}}_i\partial^i_{\dot{2}}\phi_+^{(1)}.}
Since our particular deformation \eqref{M:5.13} is of fourth order
in the odd coordinates, we may assume that
$\omega=\eta^{\gd_1\cdots\gd_4}\omega_{\gd_1\cdots\gd_4}$.
Then, after some algebraic manipulations, the expansions of
$\phi_\pm^{(n)}$ given by \eqref{M:5.19} together with
\eqref{M:5.20} and the ansatz for $\omega$ lead to \eqref{eq:om}.

The perturbations $\delta\CB_{\ad\bd}$ and
$\delta\CA^i_\ad$ are obtained from Eqs. \eqref{M:5.8}
according to
\eqna{M:5.22}{
\delta\CB_{\ad\dot{1}}\ &=\ \overset{\circ}{D}_{\ad\dot{1}}
        (\phi^{(0)}_-+\omega)\ =\ \overset{\circ}{D}_{\ad\dot{1}}
        (\phi^{(0)}_++\omega)-\overset{\circ}{D}_{\ad\dot{2}}\phi^{(1)}_+,\\
      \delta\CB_{\ad\dot{2}}\ &=\ -\overset{\circ}{D}_{\ad\dot{1}}
       \phi_-^{(1)}+\overset{\circ}{D}_{\ad\dot{2}}(\phi^{(0)}_-+\omega)
          \ =\ \overset{\circ}{D}_{\ad\dot{2}}(\phi^{(0)}_++\omega)}
and
\eqna{}{
\delta\CA^i_{\dot{1}}\ &=\ \partial_{\dot{1}}^i
    (\phi_-^{(0)}+\omega)\ =\ \partial_{\dot{1}}^i(\phi_+^{(0)}+\omega)-
    \partial_{\dot{2}}^i\phi_+^{(1)}~,\\
    \delta\CA^i_{\dot{2}}\ &=\
    -\partial_{\dot{1}}^i\phi_-^{(1)}+
    \partial_{\dot{2}}^i(\phi_-^{(0)}+\omega)
    \ =\ \partial_{\dot{2}}^i(\phi_+^{(0)}+\omega).}
Remember that $\overset{\circ}{D}_{\ad\bd}=\partial_{(\ad\bd)}+\Bc_{\ad\bd}$.
Consider now the expansions
\eqna{M:5.23}{
\delta\CB_{\ad\bd}\ &=\ \d\Bc_{\ad\bd}+
    \sum_{k\geq1}\frac{1}{k!}\,
    \eta^{\gd_1}_{j_1}\cdots\eta^{\gd_k}_{j_k}\,
    \delta\overset{\circ}{[\ad\bd]}\ \!\!_{\gd_1\cdots\gd_k}^{j_1\cdots j_k},\\
    \delta\CA^i_\ad\ &=\
    \sum_{k\geq1}\frac{k}{(k+1)!}\,
    \eta^{\gd_1}_{j_1}\cdots\eta^{\gd_k}_{j_k}\,
    \delta\overset{\circ}{ [\,{}^{\,i}_\ad]}\
    \!\!_{\gd_1\cdots\gd_k}^{j_1\cdots j_k},}
where the brackets $[\ \ ]_{\gd_1\cdots\gd_k}^{j_1\cdots j_k}$
are composite expressions of some superfields, cf.\ also 
our discussion given in \ref{FE-2}\kern6pt. Since our particular deformation of the transition function
implies that $\phi_\pm+\omega=\CO(\eta^4)$, the resulting
deformations of $\CB_{\ad\bd}$ and $\CA^i_\ad$ are of the form
$\CB_{\ad\bd}=\CO(\eta^4)$ and $\CA^i_\ad=\CO(\eta^3)$,
respectively. In transversal gauge, the explicit superfield
expansions of $\CB_{\ad\bd}$ and $\CA^i_\ad$ show that
$\delta\Bc_{\ad\bd}=\delta\cc\ \!\!^i_\ad=
 \delta\Wc\ \!\!^{ij}=\delta\cc_{i\ad}=0$.
Together with the recursion relations \eqref{eq:RR-2} (of course,
applied to the present setting obtained via dimensional
reduction), they
moreover imply that the variation of all higher order terms than
of fourth (respectively, of third) order of $\CA_{\a\ad}$
(respectively, of $\CA^i_\ad$) in the $\eta$-expansions vanish.
Hence, from \eqref{M:5.23} we find
\eqna{}{\delta\CB_{\ad\bd}\ &=\ \tfrac{1}{2\cdot4!}\epsilon^{j_1j_2j_3j_4}
\eta^{\gd_1}_{j_1}\eta^{\gd_2}_{j_2}\eta^{\gd_3}_{j_3}
\eta^{\gd_4}_{j_4}
\epsilon_{\bd\gd_1}\overset{\circ}{D}_{\ad\gd_2}
\delta\Gc_{\gd_3\gd_4},\\
\delta\CA^i_\ad\ &=\
\tfrac{3}{4}\epsilon^{ij_1j_2j_3}\eta^{\gd_1}_{j_1}
\eta^{\gd_2}_{j_2}\eta^{\gd_3}_{j_3}\epsilon_{\ad\gd_1}
\delta\Gc_{\gd_2\gd_3}.}
 Comparing these equations with \eqref{M:5.22}
and the $\eta$-expansions of $\phi_\pm^{(0)}$, $\phi_\pm^{(1)}$
and $\omega$ given earlier, we arrive at
\eqn{}{\delta\Gc_{\dot{1}\dot{1}}\ =\ 2\pc\
\!\!^{0(1)}_-~,\qquad
       \delta\Gc_{\dot{1}\dot{2}}\ =\ 2\pc\
       \!\!^{0(2)}_-\qquad{\rm and}\qquad
       \delta\Gc_{\dot{2}\dot{2}}\ =\ 2\pc\
       \!\!^{0(3)}_-}
together with the field equations
\eqn{}{\fc_{\ad\bd}\ =\
-\tfrac{\i}{2}\cnab_{\ad\bd}\Pc\qquad{\rm and}\qquad
       \epsilon^{\bd\gd}\cnab_{\ad\bd}\delta\Gc_{\gd\dd}\
       =\ -\tfrac{\i}{2}
       [\delta\Gc_{\ad\dd},\Pc].}
Since Eqs. \eqref{eq:7.1} are linear in $\Gc_{\ad\bd}$,
we hence have generated a solution
to \eqref{eq:7.1} starting from a solution to the first equation
of \eqref{eq:7.1}, that is, we may identify $\d\Gc_{\ad\bd}$ with
$\Gc_{\ad\bd}$. Thus, knowing the explicit splitting
$\tilde{f}_{+-}=\hat{\psi}_+^{-1}\hat{\psi}_-$, we can define
functions $\pc_\pm=-[X,\hat{\psi}_\pm]\hat{\psi}_\pm^{-1}$
which then in turn yield $\Gc_{\ad\bd}$.

\clearemptydoublepage
\chapter{Super Yang-Mills theory}\label{ch:SYMT}
\HRule\\

{\Large M}ain subject of this chapter is the twistorial description of
the maximally supersymmetric Yang-Mills theory in four space-time 
dimensions, that is, of $\CN=4$ (respectively, $\CN=3$) SYM theory 
\cite{Gliozzi:1976qd,Brink:1976bc}. 
Notice that $\CN=4$ and $\CN=3$ SYM theories represent
the same physical theory.
The only difference between both formulations lies 
in what part of the R-symmetry group is made manifest: for
$\CN=4$ it is $SU(4)$ while for $\CN=3$ it is 
just the subgroup $U(1) \times SU(3)$.
The twistor description of $\CN=3$ SYM theory dates back to Witten's
work \cite{Witten:1978xx} done in the late 70ies of the last century.
The idea is roughly to discuss equivalence classes of
certain holomorphic vector bundles over superambitwistor space 
$\IL^{5|6}$ which in turn are in one-to-one correspondence with
gauge equivalence classes of solutions to the $\CN=3$ SYM equations.
The superambitwistor description 
involving $\IL^{5|8}$ does not yield the $\CN=4$ SYM equations directly. 
That is why we want to focus on the $\CN=3$ formulation
of $\CN=4$ SYM theory. Moreover, in this respect
it is worth mentioning that Manin \cite{Manin} generalized the theorems of
Witten \cite{Witten:1978xx} and Isenberg et al. \cite{Isenberg:1978kk} 
by showing that there is a one-to-one correspondence
between equivalence classes of certain holomorphic vector bundles 
over superambitwistor space $\IL^{5|2\CN}$ -- for $\CN\leq3$ -- which 
admit an extension 
to a $(3-\CN)$-th formal neighborhood of $\IL^{5|2\CN}$ in 
$\IP^{3|\CN}\times\IP^{3|\CN}_*$ and gauge equivalence classes
of solutions to the $\CN$-extended SYM equations in four dimensions.   
We first discuss
$\CN=3$ SYM theory and then briefly comment on SYM theories with
less supersymmetry.

\section{$\CN=3$ super Yang-Mills theory}

Let us begin our discussion by recalling that in \ref{ATTSII}\kern10pt
we have defined superambitwistor space $\IL^{5|6}$ in terms
of flag supermanifolds. In particular, $\IL^{5|6}$ is participating in the 
following double fibration:
\eqna{eq:DF-ch4}{
\begin{picture}(50,40)
  \put(0.0,0.0){\makebox(0,0)[c]{$\IL^{5|6}$}}
  \put(64.0,0.0){\makebox(0,0)[c]{$\IM^{4|12}$}}
  \put(34.0,33.0){\makebox(0,0)[c]{$\IF^{6|12}$}}
  \put(7.0,18.0){\makebox(0,0)[c]{$\pi_1$}}
  \put(55.0,18.0){\makebox(0,0)[c]{$\pi_2$}}
  \put(25.0,25.0){\vector(-1,-1){18}}
  \put(37.0,25.0){\vector(1,-1){18}}
 \end{picture}
}
Moreover, we have shown that superambitwistor space can be viewed
as a hypersurface in $\IP^{3|3}\times\IP^{3|3}_*$ determined by the
zero locus
\eqn{eq:QE-ch4}{z^\a\rho_\a-w^\ad\pi_\ad+2\theta^i\eta_i\ =\ 0,}
where $[z^\a,\pi_\ad,\eta_i,\rho_\a,w^\ad,\theta^i]$ are
homogeneous coordinates on $\IP^{3|3}\times\IP^{3|3}_*$. In order to be able
to continuing as in the preceding two chapters, we first need to give local
coordinates at each stage of \eqref{eq:DF-ch4}.

\paragraph{Local coordinates.}\label{par:LC-ch4}
In \ref{pa:LC-ch1}\kern6pt, we have already introduced local coordinates
on $\IM^{4|12}$. In particular, we had $(x^{\a\ad},\eta_i^\ad,\theta^{i\a})$
on $\CM^{4|12}\cong \IC^{4|12}$. As in Chap. \ref{STG-Chapter}, we are now 
making use of the two projections $\pi_{1,2}$ to introduce the affine parts of
$\IF^{6|12}$ and $\IL^{5|6}$ according to
\eqn{}{\CF^{6|12}\ :=\ \pi_2^{-1}(\IC^{4|12})\qquad{\rm and}\qquad
       \CL^{5|6}\ :=\ \pi_1(\pi_2^{-1}(\IC^{4|12})).} 
Analogously to \eqref{le:LL-ch1}, one may show that
\eqn{}{\CF^{6|12}\ \cong\ \IC^{4|12}\times Y,\qquad{\rm with}\qquad Y\ :=\ \IC P^1\times\IC P^1_*.}
This makes it obvious that $\CF^{6|12}$ can be covered 
by four coordinate patches which we denote by
$\hfW=\{\hCW_a\}$, with $a,b,\ldots=1,\ldots,4$.
Letting $\l_\pm\in U_\pm$ and $\mu_\pm\in V_\pm$ be local coordinates on 
\eqn{}{\IC P^1\times\IC P^1_*\ =\ \underbrace{(U_+\times V_+)}_{=:\,W_1}\cup
                             \underbrace{(U_+\times V_-)}_{=:\,W_2}\cup
                             \underbrace{(U_-\times V_+)}_{=:\,W_3}\cup
                             \underbrace{(U_-\times V_-)}_{=:\,W_4},}
where $U_\pm$ (respectively, $V_\pm$) are the canonical patches covering $\IC P^1$ 
(respectively, $\IC P^1_*$), we thus may take
\eqna{}{(x^{\a\ad},\l_+,\mu_+,\eta^\ad_i,\theta^{i\a})\qquad&{\rm on}\qquad\hCW_1,\\
        (x^{\a\ad},\l_+,\mu_-,\eta^\ad_i,\theta^{i\a})\qquad&{\rm on}\qquad\hCW_2,\\
        (x^{\a\ad},\l_-,\mu_+,\eta^\ad_i,\theta^{i\a})\qquad&{\rm on}\qquad\hCW_3,\\
        (x^{\a\ad},\l_-,\mu_-,\eta^\ad_i,\theta^{i\a})\qquad&{\rm on}\qquad\hCW_4.}
In the sequel, we shall collectively denote them by $(x^{\a\ad},\l_{(a)},\mu_{(a)},\eta^\ad_i,\theta^{i\a})$
on the patch $\hCW_a=\IC^{4|12}\times W_a\subset\CF^{6|12}$. 

Next we need coordinates on $\CL^{5|6}$. By our above construction, $\CL^{5|6}$ is an open 
subset of $\IL^{5|6}$ 
and as such it can be viewed as a degree two
hypersurface in $\CP^{3|3}\times\CP^{3|3}_*$
-- in fact, $\CL^{5|6}=\IL^{5|6}\cap(\CP^{3|3}\times\CP^{3|3}_*)$. Here,
$\CP^{3|3}$ represents the
supertwistor space as given in \eqref{holofibI} and $\CP^{3|3}_*$ is the dual supertwistor
space
\eqn{}{\CO_{\IC P^1_*}(1)\otimes\IC^2\oplus\Pi\CO_{\IC P^1_*}(1)\otimes\IC^3\ \to\ \IC P^1_*.} 
 As before,
we denote the covering of $\CP^{3|3}$ by $\fU=\{\CU_+,\CU_-\}$ and moreover, that of
$\CP^{3|3}_*$ by $\fV=\{\CV_+,\CV_-\}$. Then
the product $\CP^{3|3}\times\CP^{3|3}_*$ is covered by four patches according to 
\eqn{}{\fU\times\fV\ =\ \{\CU_+\times\CV_+,
\CU_+\times \CV_-,\CU_-\times \CV_+,\CU_-\times \CV_-\},}
and we may set 
\eqn{}{\CL^{5|6}\ =\ \bigcup_{a=1}^4\CW_a,\quad{\rm with}\quad
      \fW\ =\ (\fU\times\fV)\cap\IL^{5|6} \quad{\rm and}\quad    \fW\ =\ \{\CW_a\}.}
Let now $(z^\a_\pm,z^3_\pm,\eta^\pm_i)$ (respectively, $(w^\ad_\pm,w^3_\pm,\te^i_\pm)$)
be local coordinates on $\CP^{3|3}$ (respectively, on $\CP^{3|3}_*$), that is, we take 
\eqna{}{z^\a_+\ &:=\ \frac{z^\a}{\pi_{\dot1}},\qquad
    z^3_+\ :=\ \frac{\pi_{\dot2}}{\pi_{\dot1}}\qquad{\rm and}\qquad
     \eta_i^+\ :=\  \frac{\eta_i}{\pi_{\dot1}}\qquad
      {\rm on}\qquad\CU_+, \\
       z^\a_-\ &:=\ \frac{z^\a}{\pi_{\dot2}},\qquad
    z^3_-\ :=\ \frac{\pi_{\dot1}}{\pi_{\dot2}}\qquad{\rm and}\qquad 
       \eta_i^-\ :=\  \frac{\eta_i}{\pi_{\dot2}}\qquad
      {\rm on}\qquad\CU_-,
}
and
\eqna{}{w^\ad_+\ &:=\ \frac{w^\ad}{\rho_1},\qquad
    w^3_+\ :=\ \frac{\rho_2}{\rho_1}\qquad{\rm and}\qquad
     \te^i_+\ :=\  \frac{\te^i}{\rho_1}\qquad
      {\rm on}\qquad\CV_+, \\
       w^\ad_-\ &:=\ \frac{w^\ad}{\rho_2},\qquad
    w^3_-\ :=\ \frac{\rho_1}{\rho_2}\qquad{\rm and}\qquad 
       \te^i_-\ :=\  \frac{\te^i}{\rho_2}\qquad
      {\rm on}\qquad\CV_-.
}
The induced coordinates on $\CL^{5|6}$ are then given by
\eqna{}{
 (z^\a_{(1)},z^3_{(1)},\eta_i^{(1)},w_{(1)}^\ad,w^3_{(1)},\te_{(1)}^i)|_{\CL^{5|6}}\ &=\
 (z^\a_+,z^3_+,\eta_i^+,w_+^\ad,w^3_+,\te_+^i)|_{\CL^{5|6}}\quad&&{\rm on}\quad\CW_1,\\
 (z^\a_{(2)},z^3_{(2)},\eta_i^{(2)},w_{(2)}^\ad,w^3_{(2)},\te_{(2)}^i)|_{\CL^{5|6}}\ &=\
 (z^\a_+,z^3_+,\eta_i^+,w_-^\ad,w^3_-,\te_-^i)|_{\CL^{5|6}}\quad&&{\rm on}\quad\CW_2,\\
 (z^\a_{(3)},z^3_{(3)},\eta_i^{(3)},w_{(3)}^\ad,w^3_{(3)},\te_{(3)}^i)|_{\CL^{5|6}}\ &=\
 (z^\a_-,z^3_-,\eta_i^-,w_+^\ad,w^3_+,\te_+^i)|_{\CL^{5|6}}\quad&&{\rm on}\quad\CW_3,\\
 (z^\a_{(4)},z^3_{(4)},\eta_i^{(4)},w_{(4)}^\ad,w^3_{(4)},\te_{(4)}^i)|_{\CL^{5|6}}\ &=\
 (z^\a_-,z^3_-,\eta_i^-,w_-^\ad,w^3_-,\te_-^i)|_{\CL^{5|6}}\quad&&{\rm on}\quad\CW_4,
}
which are transformed on nonempty intersections $\CW_a\cap\CW_b$  
in an obvious way which is obtained from the transformation laws on
$\CP^{3|3}$ and $\CP^{3|3}_*$, respectively.
Of course, these coordinates are not independent as they are 
subject to (cf. Eq. \eqref{eq:QE-ch4})
\eqn{eq:QElocal-ch4}{z^\a_{(a)}\rho_\a^{(a)}-w^\ad_{(a)}\pi_\ad^{(a)}+2\theta^i_{(a)}\eta_i^{(a)}\ =\ 0,}
where $\pi^{(1)}_\ad=\pi^{(2)}_\ad=\pi^+_\ad$, $\pi^{(3)}_\ad=\pi^{(4)}_\ad=\pi^-_\ad$ and
$\rho^{(1)}_\a=\rho^{(3)}_\a=\rho^+_\a$, $\rho^{(2)}_\a=\rho^{(4)}_\a=\rho^-_\a$ together with
$(\pi^+_\ad)=\ \!^t(1,z^3_+)$, $(\pi^-_\ad)=\ \!^t(z^3_-,1)$ and
$(\rho^+_\a)=\ \!^t(1,w^3_+)$, $(\rho^-_\a)=\ \!^t(w^3_-,1)$. 
Note that there is no summation over $a$.

Altogether, we therefore obtain the following double fibration 
\eqna{eq:DF2-ch4}{
\begin{picture}(50,40)
  \put(0.0,0.0){\makebox(0,0)[c]{$\CL^{5|6}$}}
  \put(64.0,0.0){\makebox(0,0)[c]{$\IC^{4|12}$}}
  \put(34.0,33.0){\makebox(0,0)[c]{$\CF^{6|12}$}}
  \put(7.0,18.0){\makebox(0,0)[c]{$\pi_1$}}
  \put(55.0,18.0){\makebox(0,0)[c]{$\pi_2$}}
  \put(25.0,25.0){\vector(-1,-1){18}}
  \put(37.0,25.0){\vector(1,-1){18}}
 \end{picture}
}
together with the two holomorphic projections
\eqna{eq:HP-ch4}{\pi_1\,:\,(x^{\a\ad},\l^{(a)}_\ad,\mu^{(a)}_\a,\eta^\ad_i,\te^{i\a})\ 
          &\mapsto\ (z^\a_{(a)}=(x^{\a\ad}-\te^{i\a}\eta_i^\ad)\l^{(a)}_\ad,z^3_{(a)}=\l_{(a)},\\
       &\kern-3cm w^\ad_{(a)}=(x^{\a\ad}+\te^{i\a}\eta_i^\ad)\mu^{(a)}_\a,w^3_{(a)}=\mu_{(a)},
                   \eta^{(a)}_i=\eta^\ad_i\l^{(a)}_\ad,
                  \te_{(a)}^i=\te^{i\a}\mu^{(a)}_\a),\\
        \pi_2\,:\,(x^{\a\ad},\l^{(a)}_\ad,\mu^{(a)}_\ad,\eta^\ad_i,\te^{i\a})\ 
          &\mapsto\ (x^{\a\ad},\eta^\ad_i,\te^{i\a});
}
cf. also our discussion presented in 
\ref{someLC}\kern6pt, \ref{STS}\kern10pt and \ref{pa:LC-ch1}\kern6pt.
In these equations, we have used the notation
$\l^{(1)}_\ad=\l^{(2)}_\ad=\l^+_\ad$, $\l^{(3)}_\ad=\l^{(4)}_\ad=\l^-_\ad$ and
$\mu^{(1)}_\a=\mu^{(3)}_\a=\mu^+_\a$, $\mu^{(2)}_\a=\mu^{(4)}_\a=\mu^-_\a$.

In fact, Eqs. \eqref{eq:HP-ch4} illustrate Prop.~\ref{GTC3}:
a fixed point $p\in\CL^{5|6}$ corresponds to a super null line 
$\IC_p^{1|6}\subset\IC^{4|12}$ and furthermore, a fixed point 
$(x,\eta,\te)\in\IC^{4|12}$ 
corresponds to a holomorphic embedding of 
$Y_{x,\eta,\te}\hookrightarrow\CL^{5|6}$. 
To see that $\IC^{1|6}_p$ is null, we solve 
\eqna{eq:IR-ch4}{z^\a_{(a)}\ =\ (x^{\a\ad}-\te^{i\a}\eta^\ad_i)\l^{(a)}_\ad,\qquad
       w^\ad_{(a)}\ =\ (x^{\a\ad}+\te^{i\a}\eta^\ad_i)\mu^{(a)}_\a,\\
       \eta^{(a)}_i\ =\ \eta^\ad_i\l_\ad^{(a)},\qquad\te^i_{(a)}\ =\ \te^{i\a}\mu_\a^{(a)}\kern2cm}
for a generic $p\in\CL^{5|6}$,
\eqna{}{x^{\a\ad}\ =\ \hat{x}^{\a\ad}+\varepsilon\mu^\a_{(a)}\l^\ad_{(a)}+
                       \varepsilon_i\l^\ad_{(a)}
     \hat{\te}^{i\a}+\varepsilon^i\mu_{(a)}^\a\hat{\eta}^\ad_i,\\
        \eta^\ad_i\ =\ \hat{\eta}^\ad_i+\varepsilon_i\l^\ad_{(a)},\qquad
         \te^{i\a}\ =\ \hat{\te}^{i\a}+\varepsilon^i\mu^\a_{(a)},\kern.2cm}
where $\varepsilon$, $\varepsilon_i$ and $\varepsilon^i$ are arbitrary.
Here, $(\hat{x}^{\a\ad},\hat{\eta}^\ad_i,\hat{\te}^{i\a})$ denotes a
particular solution to \eqref{eq:IR-ch4}. This shows that
$\IC^{1|6}_p$ is null. Hence, $\CL^{5|6}$ is the space of all super
null lines in $\IC^{4|12}$.

\paragraph{Remark.}\label{par:R-Chap4} 
Though we are mainly interested in $\CN=3$, let us again 
stick to generic values of $\CN$.
In Secs. \ref{TS} and \ref{sec:STS-ch1}, we have seen that
the supertwistor space $\CP^{3|\CN}$ is an open subset
in $\IC P^{3|\CN}$ describing small deformations of the 
Riemann sphere $\IC P^1$ inside $\IC P^{3|\CN}$. In fact, a
similar argument can be given for $\CL^{5|2\CN}$. Let us consider the
following sequence of embeddings:
\eqn{}{Y\ \overset{\varphi}{\hookrightarrow}\ 
    \IL^{5|2\CN}\ \hookrightarrow\ \IP^{3|\CN}\times\IP^{3|\CN}_*.} 
Then we have an exact sequence 
\eqn{eq:ES-ch4}{0\ \to\ N\ \to\ N_Y\ \to\ 
     \varphi^*N_{\IL^{5|2\CN}}\ \to\ 0,} 
where $N_Y$ is the normal sheaf of
$Y$ in $\IP^{3|\CN}\times\IP^{3|\CN}_*$ and
$N_{\IL^{5|2\CN}}$ that of $\IL^{5|2\CN}$ 
in $\IP^{3|\CN}\times\IP^{3|\CN}_*$. In particular, we have
$$ 
\begin{aligned} 
N_Y\ &\cong\ {\rm pr}_1^*\CP^{3|\CN}\oplus{\rm pr}_2^*\CP^{3|\CN}_*,\\
  \varphi^*N_{\IL^{5|2\CN}}\ &\cong\ \CO_Y(1,1).
\end{aligned}$$
Here, pr$_{1,2}$ are again the two projections from $Y=\IC P^1\times\IC P^1_*$ to
the first and second factors and $\CO_Y(1,1)$ is the
sheaf of sections of the divisor line bundle of the diagonal $\IC P^1\subset Y$. Thus, 
$N$ denotes the normal sheaf of $Y$ in
$\IL^{5|2\CN}$. In fact, $\CL^{5|2\CN}$ is a rank $3|2\CN$ holomorphic vector
bundle over $Y$ and moreover, $N\cong\CL^{5|2\CN}$. Hence,
\eqn{eq:DESATS-ch4}{0\ \to\ \CL^{5|2\CN}\ \to\ {\rm pr}_1^*\CP^{3|\CN}\oplus{\rm pr}_2^*\CP^{3|\CN}_*\ 
         \to\ \CO_Y(1,1)\ \to\ 0.}
The sequence \eqref{eq:DESATS-ch4} induces a long exact cohomology sequence
\eqna{eq:LECS-ch4}{
 0\ &\to\ H^0(Y,\CL^{5|2\CN})\ \to\
         H^0(Y,N_Y)\ \overset{\kappa}{\to}\ 
          H^0(Y,\CO_Y(1,1))\ \to\ \\ 
        &\kern.5cm\to\  H^1(Y,\CL^{5|2\CN})\ \to\ 
          H^1(Y,N_Y)\ \to\ 
          H^1(Y,\CO_Y(1,1))\ \to\ \cdots.}
As before, we let $\pi_\ad$ and $\rho_\a$ be homogeneous coordinates on 
$Y$. An element of 
$H^0(Y,N_Y)$ is described by a $4|2\CN$-tuple
$(z^\a,w^\ad,\eta_i,\theta^i)$, where $z^\a$ and $w^\ad$
are even homogeneous bidegree $(1,0)$ respectively, 
bidegree $(0,1)$
polynomials in $(\pi_\ad,\rho_\a)$ while $\eta_i$ and $\te^i$ are
odd polynomials of bidegree $(1,0)$ and $(0,1)$, respectively.
The map $$\kappa\,:\,H^0(Y,N_Y)\ \to\ 
H^0(Y,\CO_Y(1,1))$$ is then given by
\eqn{}{ \kappa\,:\,(z^\a,w^\ad,\eta_i,\theta^i)\ \mapsto\
              z^\a\rho_\a-w^\ad\pi_\ad+2\te^i\eta_i}
Clearly, this mapping is surjective. Hence, the sequence
\eqref{eq:LECS-ch4} splits to give a short exact sequence
\eqn{eq:SECS-ch4}{
 0\ \to\ H^0(Y,\CL^{5|2\CN})\ \to\
          H^0(Y,N_Y)\
    \to\ H^0(Y,\CO_Y(1,1))\ \to\ 0.}
Using K\"unneth's formula\footnote{Recall the following fact.
Let $\CE_i$ be a holomorphic
vector bundle over $X_i$ for $i=1,2$. Furthermore, denote the
two projections from $X=X_1\times X_2$ to $X_i$ by pr$_i$ and consider
the bundle $\CE={\rm pr}_1^*\CE_1\otimes{\rm pr}_2^*\CE_2$. 
K\"unneth's formula then says:
$$ H^k(X,\CE)\ \cong\ \bigoplus_{p+q=k} H^p(X_1,\CE_1)\otimes
   H^q(X_2,\CE_2).$$
}, one readily verifies that
\eqna{eq:SECS2-ch4}{H^0(Y,\CO_Y(m,n))\ &\cong\ \IC^{(m+1)(n+1)}\\
       H^1(Y,\CO_Y(m,n))\ &\cong\ 0}
for $m,n\geq0$. By virtue of \eqref{eq:SECS-ch4}, we 
therefore find
\eqn{}{
     H^0(Y,\CL^{5|2\CN})\ \cong\ \IC^{4|4\CN}.}
Similarly, one may show that $H^1(Y,\CL^{5|2\CN})\cong0$
(cf. Eqs \eqref{eq:SECS2-ch4}).     

\paragraph{Penrose-Ward transform.}\label{par:PWTN3-ch4}
After having presented the setup, we shall
now come to the description of $\CN=3$ SYM theory. For this,
we consider the double fibration \eqref{eq:DF2-ch4} together
with a rank $r$ 
holomorphic vector bundle $\CE\to\CL^{5|6}$ which is characterized
by the transition functions $f=\{f_{ab}\}$ and its pull-back
$\pi_1^*\CE$ to the supermanifold $\CF^{6|12}$. We denote again
the pulled-back transition functions by the same letter $f$.
By definition of a pull-back, the transition functions $f$ are constant along
$\pi_1\,:\CF^{6|12}\to\CL^{5|6}$. The relative tangent sheaf
$\cT:=(\O^1(\CF^{5|6})/\pi_1^*\O^1(\CL^{5|6}))^*$ is of rank $1|6$ and freely generated 
by
\eqn{eq:HVF-ch4}{D_{(a)}\ =\ \mu^\a_{(a)}\l^\ad_{(a)}\partial_{\a\ad},\qquad
       D^i_{(a)}\ =\ \l^\ad_{(a)} D^i_\ad\qquad{\rm and}\qquad
       D_i^{(a)}\ =\ \mu^\a_{(a)} D_{i\a},}
where $\partial_{\a\ad}:=\partial/\partial x^{\a\ad}$ and 
\eqn{eq:DefOfDer-ch4}{D^i_\ad\ =\ \partial^i_\ad+\te^{i\a}\partial_{\a\ad}
        \qquad{\rm and}\qquad
       D_{i\a}\ =\ \partial_{i\a}+\eta^\ad_i\partial_{\a\ad},}
with $\partial^i_\ad:=\partial/\partial\eta^\ad_i$ and 
$\partial_{i\a}:=\partial/\partial\te^{i\a}$.
Hence, the transition
functions of $\pi^*_1\CE$ are annihilated by the vector fields \eqref{eq:HVF-ch4}.
Clearly, they are also annihilated by $\pbar_\CF$.

Next we want to assume that the vector bundle $\CE\to\CL^{5|6}$ is smoothly 
trivial and in addition $\IC^{4|12}$-trivial, i.e.,
holomorphically trivial when restricted to any submanifold\footnote{See also
Prop. \ref{GTC3}}
$Y_{x,\eta,\te}\hookrightarrow\CL^{5|6}$.
Together, these conditions imply that there exist some
$\psi=\{\psi_a\}\in C^0(\fW,\fS)$, which define 
trivializations of $\pi_1^*\CE$, such that $f=\{f_{ab}\}$ can be decomposed as
\eqn{eq:SF-ch4}{f_{ab}\ =\ \psi^{-1}_a\psi_b}
and 
\eqn{}{\pbar_\CF\psi_a\ =\ 0.}
In particular, $\psi_a$ depends
holomorphically on $\l_a$ and $\mu_a$.
Applying the vector fields \eqref{eq:HVF-ch4} to \eqref{eq:SF-ch4}, we realize that
by virtue of an extension of Liouville's theorem to $\IC P^1\times\IC P^1_*$,
the expressions
$$\psi_a D_{(a)}\psi_a^{-1}\ =\ \psi_b D_{(a)}\psi_b^{-1},\quad
       \psi_a D_{(a)}^i\psi_a^{-1}\ =\ \psi_b D_{(a)}^i\psi_b^{-1},\quad
       \psi_a D^{(a)}_i\psi_a^{-1}\ =\ \psi_b D^{(a)}_i\psi_b^{-1}
$$
must be at most linear in $\l_{(a)}$ and $\mu_{(a)}$. Therefore, we may introduce
a relative connection one-form 
$\CA_\cT\in\Gamma(\CF^{6|12},\O^1_\cT(\CF^{6|12})\otimes{\rm End}\,\pi_1^*\CE)$ such that
\eqna{eq:DefofA-ch4}{D\lrcorner\CA_\cT|_{\hCW_a}\ &:=\ \CA_{(a)}\ 
            =\ \psi_a D_{(a)}\psi_a^{-1}\ =\ \mu^\a_{(a)}\l^\ad_{(a)}\CA_{\a\ad},\\
       D^i\lrcorner\CA_\cT|_{\hCW_a}\ &=:\ \CA_{(a)}^i\ 
             =\ \psi_a D_{(a)}^i\psi_a^{-1}\ =\ \l^\ad_{(a)}\CA_\ad^i,\\
        D_i\lrcorner\CA_\cT|_{\hCW_a}\ &:=\ \CA^{(a)}_i\ 
            =\ \psi_a D^{(a)}_i\psi_a^{-1}=\ \mu^\a_{(a)}\CA_{i\a},
}
and hence
\eqna{eq:LS1-ch4}{\mu^\a_{(a)}\l^\ad_{(a)}(\partial_{\a\ad}+\CA_{\a\ad})\psi_a\ &=\ 0,\\
         \l^\ad_{(a)}(D^i_\ad+\CA_\ad^i)\psi_a\ &=\ 0,\\
          \mu^\a_{(a)}(D_{i\a}+\CA_{i\a})\psi_a\ &=\ 0,\\
           \pbar_\CF\psi_a\ &=\ 0.}
The compatibility conditions for the linear system \eqref{eq:LS1-ch4} read as
\eqna{eq:CC1-ch4}{\{\nabla^i_{(\ad},\nabla^j_{\bd)}\}\ &=\ 0,\\
          \{\nabla_{i(\a},\nabla_{j\b)}\}\ &=\ 0,\\
           \{\nabla_{i\a},\nabla^j_\bd\}-2\d^j_i\nabla_{\a\bd}\ &=\ 0,
}
where we have introduced
\eqn{eq:DEFofCD-ch4}{\nabla^i_\ad\ :=\ D^i_\ad+\CA^i_\ad,\quad
       \nabla_{i\a}\ :=\ D_{i\a}+\CA_{i\a}\quad{\rm and}\quad
       \nabla_{\a\ad}\ :=\ \partial_{\a\ad}+\CA_{\a\ad}.}
Eqs. \eqref{eq:CC1-ch4} are the constraint equations of $\CN=3$ SYM theory.

Next let us discuss how to obtain the
functions $\psi_a$ in \eqref{eq:LS1-ch4} from a given gauge potential. Formally,
a solution is given by
\eqn{eq:SOL-ch4}{\psi_a\ =\ P\exp\left(-\int_{\cC_a}\CA\right).
}
Here, ``$P$" denotes the path-ordering symbol and
\eqn{}{\CA\ =\ E^{\a\ad}\CA_{\a\ad}+E^\ad_i\CA^i_\ad+E^{i\a}\CA_{i\a},}
where the basis $(E^{\a\ad},E^\ad_i,E^{i\a})$ is dual to 
$(\partial_{\a\ad},D^i_\ad,D_{i\a})$. It is explicitly given by
\eqn{}{E^{\a\ad}\ =\ \dt x^{\a\ad}+\eta^\ad_i\dt\te^{i\a}+
       \te^{i\a}\dt\eta^\ad_i,\qquad
       E^\ad_i\ =\ \dt\eta^\ad_i\qquad{\rm and}\qquad E^{i\a}\ =\ \dt\te^{i\a}.
}
The contour $\cC_a$ is a any real curve within a super light ray $\IC^{1|6}$ 
from a point $(\hat{x},\hat{\eta},
\hat{\te})$ to a point $(x,\eta,\te)$, with
\eqna{curve}{x^{\a\ad}(s)\ &=\ \hat{x}^{\a\ad}+s(\varepsilon\mu^\a_{(a)}\l^\ad_{(a)}
          +\varepsilon_i\l^\ad_{(a)}\te^{i\a}+\varepsilon^i\mu^\a_{(a)}\eta^\ad_i),\\
       \eta^\ad_i(s)\ &=\ \hat{\eta}^\ad_i+s\varepsilon_i\l^\ad_{(a)},\qquad
       \te^{i\a}(s)\ =\ \hat{\te}^{i\a}+s\varepsilon^i\mu^\a_{(a)},}
for $s\in[0,1]$; the choice of the contour plays no role, since the curvature
is zero when restricted to the super light ray.
Furthermore, $(\varepsilon,\varepsilon_i,\varepsilon^i)$ are some free
parameters. 

Before discussing the superfield expansions of the above gauge potentials,
let us say a few words about reality conditions. So far, we have
entirely worked in a complex setting. However, we eventually want to
discuss real $\CN=3$ SYM theory on Minkowski space. In order to achieve this,
we need to extend our in \ref{pa:MS-ch4}\kern11pt introduced
antiholomorphic involution $\tau_M$ to $\pi_1^*\CE\to\CF^{6|12}$, that is,
we have to require
\eqn{eq:realf-ch4}{f_{12}^\dagger\ =\ f_{31},\qquad f_{14}^\dagger\ =\ f_{41}
       \qquad{\rm and}\qquad
       f_{23}^\dagger\ =\ f_{23}}
on appropriate intersections. In these equations,
we have used the shorthand notation $f_{ab}^\dagger:=[f_{ab}(\tau_M(\cdots))]^\dagger$.
Upon imposing \eqref{eq:realf-ch4}, we find
\eqn{eq:RCPSI-ch4}{\psi_1^\dagger\ =\ \psi_1^{-1},\qquad
       \psi_2^\dagger\ =\ \psi_3^{-1}\qquad{\rm and}\qquad
       \psi_4^\dagger\ =\ \psi_4^{-1}}
and hence,
\eqna{}{\tau_M(\CA_{\a\bd})\ &=\ \CA_{\b\ad}^\dagger\ =\ \CA_{\a\bd},\\
       \tau_M(\CA_\ad^i)\ &=\ \CA_{i\a}^\dagger\ =\ \CA_\ad^i,\\
       \tau_M(\CA_{i\a})\ &=\ \CA^{i\dagger}_\ad\ =\ \CA_{i\a},
}
as desired. In this case, the gauge group $GL(r,\IC)$ 
is reduced to the 
unitary group $U(r)$ and as we have seen in \ref{PWT-2}\kern6pt, the
additional requirement $\det(f_{ab})=1$ yields $SU(r)$.

\paragraph{Remark.}
If $f=\{f_{ab}\}$ is independent of say $\mu_{(a)}$ then one
may assume without loss of generality that 
$f_{12}=f_{34}=\mathbbm{1}_r$ 
and $f_{13}=f_{24}=f_{+-}(x_R^{\a\ad}\l_\ad^+,\l_+,\eta^\ad_i\l^+_\ad)$
\cite{Tafel:1985qk}.
All other transition functions are obtained from those by virtue of
the cocycle conditions.
In this case, the linear system \eqref{eq:LS1-ch4} reduces to that of
self-dual SYM theory given in \eqref{linsys-2}. Hence, by assuming that
all transition function are independent of $\mu_{(a)}$, one ends up
with self-dual SYM theory. Similarly, by imposing independence of
$\l_{(a)}$, one eventually obtains the anti-self-dual SYM equations.

\paragraph{Field expansions, field equations and action functional.}
Let us now show that the constraint equations \eqref{eq:CC1-ch4} imply
the equations of motion of $\CN=3$ SYM theory (and vice versa).
The analysis is, however, basically the same as the one given in 
\ref{FE-2}\kern6pt, for instance.
Therefore, we shall be rather brief and merely quote results. More details
about the derivation can be found in Harnad et al. 
\cite{HarnadVK,HarnadBC,Harnad:1988rs}.

Eqs. \eqref{eq:CC1-ch4} can be formally solved according to
\eqna{eq:CC2-ch4}{\{\nabla^i_\ad,\nabla^j_\bd\}\ &=\ 
         2\epsilon_{\ad\bd}\epsilon^{ijk}W_k,\\
       \{\nabla_{i\a},\nabla_{j\b}\}\ &=\ 2\epsilon_{\a\b}\epsilon_{ijk}W^k,\\
       \{\nabla_{i\a},\nabla^j_\ad\}\ &=\ 2\d^j_i\nabla_{\a\ad}.}
It follows then from Bianchi identities that the odd spinor superfields
are given by
\eqna{}{\psi_\a\ &=\ \tfrac{1}{3}\nabla_{i\a}W^i,\\
       \psi_\ad\ &=\ \tfrac{1}{3}\nabla^i_\ad W_i,\\
       \chi_{i\ad}\ &=\ -\tfrac{1}{2}\epsilon^{\a\b}[\nabla_{\a\ad},\nabla_{i\b}],\\
       \chi^i_\a\ &=\ -\tfrac{1}{2}\epsilon^{\ad\bd}[\nabla_{\a\ad},\nabla^i_\bd].} 
As before, one imposes the transversal gauge condition
\eqn{eq:TGC-ch4}{\eta^\ad_i\CA^i_\ad+\te^{i\a}\CA_{i\a}\ =\ 0}
to remove superfluous gauge degrees of freedom associated with the odd coordinates.
This time, the recursion operator takes the form
\eqn{}{\cD\ =\ \eta^\ad_i\nabla^i_\ad+\te^{i\a}\nabla_{i\a}\ =\ 
              \eta^\ad_i\partial^i_\ad+\te^{i\a}\partial_{i\a}.}
With the help of Bianchi identities one may prove the following recursion relations:
\eqna{}{\cD\CA_{\a\ad}\ &=\ -\epsilon_{\a\b}\te^{i\b}\chi_{i\ad}
             +\epsilon_{\ad\bd}\eta^\bd_i\chi^i_\a,\\
        (1+\cD)\CA^i_\ad\ &=\ 2\epsilon_{\ad\bd}\epsilon^{ijk}
             \eta^\bd_jW_k+2\te^{i\a}\CA_{\a\ad},\\
        (1+\cD)\CA_{i\a}\ &=\ 2\epsilon_{\a\b}\epsilon_{ijk}
                \te^{j\b}W^k+2\eta^\ad_i\CA_{\a\ad},\\
        \cD W_i\ &=\ \eta^\ad_i\psi_\ad+\epsilon_{ijk}\te^{j\a}\chi^k_\a,\\
        \cD W^i\ &=\ \te^{i\a}\psi_\a+\epsilon^{ijk}\eta^\ad_j\chi_{k\ad},\\
         \cD\psi_\a\ &=\ \epsilon_{\a\b}\epsilon_{ijk}
                 \te^{i\b}[W^j,W^k]+2\eta^\ad_i\nabla_{\a\ad}W^i,\\
        \cD\psi_\ad\ &=\ \epsilon_{\ad\bd}\epsilon^{ijk}
             \eta^\bd_i[W_j,W_k]+2\te^{i\a}\nabla_{\a\ad}W_i,\\
        \cD\chi^i_\a\ &=\ -2\epsilon^{ijk}\eta^\ad_j
             \nabla_{\a\ad}W_k+2\te^{i\b}f_{\a\b}+
                            \epsilon_{\a\b}\te^{j\b}(\d_j^i[W^k,W_k]-2[W_j,W^i]),\\
          \cD\chi_{i\ad}\ &=\ 
          -2\epsilon_{ijk}\te^{j\a}\nabla_{\a\ad}W^k+2\eta^\bd_if_{\ad\bd}+
                            \epsilon_{\ad\bd}\eta^\bd_j(\d^j_i[W_k,W^k]-2[W^j,W_i]),
 }
where again $f_{\a\b}$ (respectively, $f_{\ad\bd}$) represents the 
self-dual (respectively, anti-self-dual)
part of the field strength. Note that
these equations resemble the supersymmetry transformations, but nevertheless
they should not be confused with them.  
 Now one can start to iterate these equations to obtain 
all the superfields order by order in the odd coordinates. Using formulas 
similar to Eqs. \eqref{eq:rec1-ch2} and
\eqref{eq:rec2-ch2}, one eventually finds:
\eqna{}{\CA^i_\ad\ &=\ \epsilon_{\ad\bd}\epsilon^{ijk}\eta^\bd_j\Wc_k+
                       \te^{i\a}\Ac_{\a\ad}+
             \tfrac{2}{3}\epsilon_{\ad\bd}\epsilon^{ijk}\eta^\bd_j
                        \eta^\gd_k\psc_\gd\ -\\
                 &\kern2cm-\ \tfrac{4}{3}\epsilon_{\ad\bd}
                \te^{i\a}\eta^\bd_j\cc\ \!\!^j_\a+
              \tfrac{2}{3}\epsilon_{\ad\bd}\te^{j\a}\eta^\bd_j\cc\ 
                      \!\!^i_\a-\tfrac{2}{3}\epsilon_{\a\b}
                \te^{i\a}\te^{j\b}\cc_{j\ad}+\ \cdots,\\
         \CA_{i\a}\ &=\ \epsilon_{\a\b}\epsilon_{ijk}\te^{j\b}\Wc\ 
                   \!\!^k+\eta_i^\ad\Ac_{\a\ad}+
             \tfrac{2}{3}\epsilon_{\a\b}\epsilon_{ijk}\te^{j\b}\te^{k\g}\psc_\g\ +\\
                 &\kern2cm+\ \tfrac{4}{3}\epsilon_{\a\b}\te^{j\b}\eta^\ad_i\cc_{j\ad}-
              \tfrac{2}{3}\epsilon_{\a\b}\te^{j\b}\eta^\ad_j\cc_{i\bd}
                       -\tfrac{2}{3}\epsilon_{\ad\bd}
                \eta^\ad_i\eta^\bd_j\cc\ \!\!^j_\a+\ \cdots.
}
Upon substituting these expansions into the constraint 
equations \eqref{eq:CC2-ch4}, we obtain
the equations of motion of $\CN=3$ SYM theory
\eqna{eq:eomN3-ch4}{\epsilon^{\a\b}\cnab_{\a\ad}\psc_\b+
                      [\cc_{i\ad},\Wc\ \!\!^i]\ &=\ 0,\\
        \epsilon^{\ad\bd}\cnab_{\a\ad}\psc_\bd+[\cc\ \!\!^i_\a,\Wc_i]\ &=\ 0,\\ 
        \epsilon^{\a\b}\cnab_{\a\ad}\cc\ \!\!^j_\b+[\cc_{i\ad},\Wc_k]
                      \epsilon^{ijk}-[\psc_\ad,\Wc\ \!\!^j]\ &=\ 0,\\
         \epsilon^{\ad\bd}\cnab_{\a\ad}\cc_{j\bd}+[\cc\ \!\!^i_\a,\Wc\ \!\!^k]
                      \epsilon_{ijk}-[\psc_\a,\Wc_j]\ &=\ 0,\\
           \overset{\circ}{\square}\Wc_j+[[\Wc\ \!\!^i,\Wc_j],\Wc_i]-
          \tfrac{1}{2}[[\Wc\ \!\!^i,\Wc_i],\Wc_j]+
         \tfrac{1}{2}\epsilon^{\ad\bd}
        \{\cc_{j\ad},\psc_\bd\}-\tfrac{1}{4}
         \epsilon_{ijk}\epsilon^{\a\b}\{\cc\ \!\!^i_\a,\cc\ \!\!^k_\b\}\ &=\ 0,\\
        \overset{\circ}{\square}\Wc\ \!\!^j+[[\Wc_i,\Wc\ \!\!^j],\Wc\ \!\!^i]-
          \tfrac{1}{2}[[\Wc_i,\Wc\ \!\!^i],\Wc\ \!\!^j]+
         \tfrac{1}{2}\epsilon^{\a\b}
        \{\cc\ \!\!^j_\a,\psc_\b\}-\tfrac{1}{4}\epsilon^{ijk}
            \epsilon^{\ad\bd}\{\cc_{i\ad},\cc_{k\bd}\}\ &=\ 0,\\
          \epsilon^{\a\b}\cnab_{\a\gd}\fc_{\b\g}+
               \epsilon^{\ad\bd}\cnab_{\g\ad}\fc_{\bd\gd}
          +\{\cc\ \!\!^k_\g,\cc_{k\gd}\}+\{\psc_\g,\psc_\gd\}+
            [\Wc\ \!\!^i,\cnab_{\g\gd}\Wc_i]
           +[\Wc_i,\cnab_{\g\gd}\Wc\ \!\!^i]\ &=\ 0,
}
where we have defined
\eqn{}{\overset{\circ}{\square}\ :=\ \tfrac{1}{2}\epsilon^{\a\b}
            \epsilon^{\ad\bd}\cnab_{\a\ad}\cnab_{\b\bd}.}
It can now be shown by induction and with the help
of the recursion operator $\cD$ that Eqs. \eqref{eq:eomN3-ch4} 
are in one-to-one correspondence
with the constraint equations \eqref{eq:CC2-ch4}. 
For details, see \cite{HarnadVK,HarnadBC,Harnad:1988rs}.

Finally, it remains to give an appropriate
action functional producing the equations of motion \eqref{eq:eomN3-ch4}. 
A straightforward calculation reveals that they can be obtained by varying
\eqna{eq:AN3-ch4}{S\ &=\ \int\,\dt^4 x\,{\rm tr}\left\{\fc_{\a\b}\fc\ 
                \!\!^{\a\b}+\fc_{\ad\bd}\fc\ \!\!^{\ad\bd}+
                   \cc\ \!\!^{i\a}\cnab_{\a\ad}\cc\ \!\!^\ad_i+2\Wc\ \!\!^i
                    \overset{\circ}{\square}\Wc_i
            +\Wc_i\{\cc\ \!\!^i_\a,\psc\ \!\!^\a\}\ +\right.\\
              &\kern3cm\left.+\ \Wc\ \!\!^i\{\cc_{i\ad},\psc\ 
               \!\!^\ad\}+\tfrac{1}{2}\epsilon_{ijk}\{\cc\ \!\!^i_\a,
          \cc\ \!\!^{j\a}\}\Wc\ \!\!^k+\tfrac{1}{2}\epsilon^{ijk}
                 \{\cc_{i\ad},\cc_{j\bd}\}\Wc_k\ - \right.\\[5pt]
           &\kern4cm\left. -\ [\Wc_i,\Wc\ \!\!^j][\Wc_j,\Wc\ \!\!^i]+
             \tfrac{1}{2}[\Wc_i,\Wc\ \!\!^i][\Wc_j,\Wc\ \!\!^j]\right\}.} 

\paragraph{HCS theory on $\CL^{5|6}$.}
Using the Dolbeault approach to holomorphic vector bundles,
we gave the twistor interpretation of 
Siegel's action functional of $\CN=4$ self-dual SYM theory 
in \ref{HCS-2}\kern5pt. It
turned out to be the action functional of hCS theory on supertwistor
space $\CP^{3|4}$. The existence of an appropriate action principle on
$\CP^{3|4}$ was due to the formal Calabi-Yau property
of the latter. As we have seen, also superambitwistor space
$\CL^{5|6}$ is a formal Calabi-Yau supermanifold. Thus, there
is a globally well-defined nowhere vanishing holomorphic volume form.
On the patch $\CW_a\subset\CL^{5|6}$, it is given by
\eqn{eq:HM-ch4}{\O|_{\CW_a}\ =\ \frac{(-)^{\lfloor\frac{a}{2}\rfloor}}{2\pi\i}
         \oint_\cC\frac{\dt^3 z_{(a)}\wedge\dt^3 w_{(a)}\dt^3\eta_{(a)}\dt^3\te_{(a)}}{
                z^\a_{(a)}\rho_\a^{(a)}
          -w^\ad_{(a)}\pi_\ad^{(a)}+2\te^i_{(a)}\eta_i^{(a)}},}
where $\cC$ is any contour encircling 
$\CL^{5|6}\hookrightarrow\CP^{3|3}\times\CP^{3|3}_*$ and
$\dt^3 z_{(a)}:=\dt z^1_{(a)}\wedge\dt z^2_{(a)}\wedge\dt z^3_{(a)}$, etc.
The problem which prevents us from writing down an action functional
like \eqref{bla} is the ``wrong" dimensionality of superambitwistor
space: the holomorphic volume form \eqref{eq:HVF-ch4} is a form of
type $(5|6,0)$ while the Chern-Simons form is of type $(0,3|0)$, as one
is again interested in a subsupermanifold $\CY\subset\CL^{5|6}$ determined
by $\bar{\eta}_i^{(a)}=0=\bar{\te}^i_{(a)}$. Thus, in total one obtains a form of
type $(5|6,3|0)$ which, of course, cannot be integrated over $\CY$. 

In order to circumvent this problem, Mason et al. \cite{Mason:2005kn} (cf. also
Ref. \cite{Mason:2005zm}) proposed to
instead consider a real codimension $2|0$ CR supermanifold\footnote{For a 
definition of a CR supermanifold, see Sec. \ref{sec:CRSM-ch3}.} in 
superambitwistor space. Their construction is based on an 
Euclidean signature as for Minkowski signature the CR supermanifold
under consideration will no longer be smooth but have singularities. In 
this setting it is, however, possible to use the holomorphic 
volume form \eqref{eq:HM-ch4}
to give an action functional for phCS theory (see also our discussion
given in Sec. \ref{sec:phCSt-3}) reproducing the action functional of
$\CN=3$ SYM theory. Though their construction works for an Euclidean
setting, it remains an open question to find an appropriate twistor 
interpretation of \eqref{eq:AN3-ch4} in the case of Minkowski signature. 

\paragraph{Summary.}\label{par:Sum-ch4} Let us now summarize:
{\Thm There is a one-to-one correspondence between gauge equivalence classes
of local solutions to the $\CN=3$ SYM equations on four-dimensional Minkowski
space and equivalence classes of
holomorphic vector bundles $\CE$ over superambitwistor space $\CL^{5|6}$ which
are smoothly trivial and holomorphically trivial on any submanifold 
$(\IC P^1\times\IC P^1_*)_{x,\eta,\te}\hookrightarrow\CL^{5|6}$.}\vskip 4mm

If we let $H^1_{\nabla^{0,1}}(\CL^{5|6},\tilde{\CE})$ be the moduli space
of hCS theory on $\CL^{5|6}$ for vector bundles $\tilde{\CE}$ smoothly
equivalent to $\CE$, we may equivalently write
\eqn{}{H^1_{\nabla^{0,1}}(\CL^{5|6},\tilde{\CE})\ \cong\ \CM^{\CN=3}_{\rm SYM},}
where $\CM^{\CN=3}_{\rm SYM}$ denotes the moduli space
of $\CN=3$ SYM theory. The latter is obtained from the solution space
by quotiening with respect to the group of gauge transformations.

\paragraph{Remark.}
In Chap. \ref{ch:SBME-ch3}, we have described the dimensional
 reduction of the supertwistor space and obtained
the mini-supertwistor space. In this setting, we established
a correspondence between hBF theory on mini-supertwistor space 
and a supersymmetric Bogomolny model in three dimensions.
As for this model, which was obtained
by a dimensional reduction of $\CN=4$ self-dual SYM theory, 
one can establish a twistor correspondence for the full 
$\CN=6$ (respectively, $\CN=8$) SYM theory in 
three dimensions by using a dimensional 
reduction of $\CL^{5|6}$. In fact, one can establish:
\begin{equation}
\begin{aligned}
\begin{picture}(100,95)(0,-5)
\put(0.0,0.0){\makebox(0,0)[c]{$\CL^{4|6}$}}
\put(0.0,52.0){\makebox(0,0)[c]{$\CL^{5|6}$}}
\put(96.0,0.0){\makebox(0,0)[c]{$\IC^{3|12}$}}
\put(96.0,52.0){\makebox(0,0)[c]{$\IC^{4|12}$}}
\put(51.0,33.0){\makebox(0,0)[c]{$\CF^{5|12}$}}
\put(51.0,85.0){\makebox(0,0)[c]{$\CF^{6|12}$}}
\put(37.5,25.0){\vector(-3,-2){25}}
\put(55.5,25.0){\vector(3,-2){25}}
\put(37.5,77.0){\vector(-3,-2){25}}
\put(55.5,77.0){\vector(3,-2){25}}
\put(0.0,45.0){\vector(0,-1){37}}
\put(90.0,45.0){\vector(0,-1){37}}
\put(45.0,78.0){\vector(0,-1){37}}
\end{picture}
\end{aligned}
\end{equation}
The details about this correspondence, the construction of
the field equations, etc. can be found in 
\cite{Saemann:2005ji,Saemann:2005xq}.

\section{Thickenings and $\CN<3$ super Yang-Mills theory}

For the sake of completeness, we shall now briefly talk about SYM theories
with less supersymmetry, that is, for $\CN<3$. In order to discuss
them, we first need the notion of formal neighborhoods.
This will then allow us to provide a twistor formulation of these
theories. However, we stress in advance that we will not prove any
of the subsequent statements but rather quote results.
\paragraph{Formal neighborhoods.}\label{pa:FN-ch4}
Let $X$ be a complex supermanifold and $Y$ a sub(super)manifold. Furthermore,
let $\CI$ be the ideal subsheaf of all holomorphic functions in $\CO_X$
which vanish on $Y$. Then we have a short exact sequence of sheaves
\eqn{}{0\ \to\ \CI\ \to\ \CO_X\ \to\ \CO_X/\CI\ \to\ 0}
on $X$. The sheaf $\CO_X/\CI$ can then be identified with $\CO_Y$. Generally
speaking, there is an isomorphism
\eqn{}{\CO_X|_Y\ \cong\ \CO_X/\CI\oplus\CI/\CI^2\oplus\CI^2/\CI^3\oplus\cdots}
given by the Taylor expansion of a germ of $f$ at any point of $Y$. Note that
$\CI/\CI^2$ can be identified with the conormal sheaf of $Y$ in $X$.

Next we define
\eqn{}{\CO^{(k)}_Y\ :=\ \CO_X/\CI^{k+1}.}
Then $(Y,\CO^{(k)}_Y)$ is called the $k$-th formal neighborhood of $Y$ in $X$.
Note that $\CO_Y^{(k)}$ can be expanded according to
\eqn{}{\CO_Y^{(k)}\ \cong\ \CO_X/\CI\oplus\CI/\CI^2\oplus
           \CI^2/\CI^3\oplus\cdots\oplus\CI^k/\CI^{k+1}.} 
Putting it differently, the $k$-th formal neighborhood contains formal
Taylor expansions in the normal sheaf direction up to order $k$.

\paragraph{Manin's theorem for $\CN\leq3$ SYM theory.}
Let us consider the double fibration
\eqna{eq:DF3-ch4}{
\begin{picture}(50,40)
  \put(0.0,0.0){\makebox(0,0)[c]{$\IL^{5|2\CN}$}}
  \put(64.0,0.0){\makebox(0,0)[c]{$\IM^{4|4\CN}$}}
  \put(34.0,33.0){\makebox(0,0)[c]{$\IF^{6|4\CN}$}}
  \put(7.0,18.0){\makebox(0,0)[c]{$\pi_1$}}
  \put(55.0,18.0){\makebox(0,0)[c]{$\pi_2$}}
  \put(25.0,25.0){\vector(-1,-1){18}}
  \put(37.0,25.0){\vector(1,-1){18}}
 \end{picture}
}
as introduced in \ref{ATTSII}\kern6pt. Then we may state the following
theorem \cite{Manin}:
{\Thm Let $U$ be an open subset of $\IM^{4|4\CN}$ such that any null line
intersects $U$ in a simply connected set. Then there is a one-to-one
correspondence between equivalence classes of $U$-trivial holomorphic vector 
bundles which admit an extension to a $(3-\CN)$-th formal neighborhood 
of $\IL^{5|2\CN}$ in $\IP^{3|\CN}\times\IP^{3|\CN}_*$ and gauge
equivalence classes of solutions to the $\CN$-extended SYM equations
on $U$.}\vskip 4mm

Details about the proof can be found in \cite{Manin}. 
Furthermore, related aspects such as superfield expansions, etc., 
are discussed in Refs. \cite{Eastwood,Harnad:1988rs}.

\clearemptydoublepage
\chapter{Hidden symmetries in self-dual super Yang-Mills theory}\label{HS-CHAPTER}
\HRule\\

{\Large T}he purpose of this chapter is the discussion of hidden 
infinite-dimensional symmetry algebras in $\CN$-extended self-dual 
SYM theory. The main tools we shall be using 
are built upon the twistor correspondence, which we established
in Chap. \ref{ch:SDSYM-ch2}. 

Since Pohlmeyer's
work \cite{PohlmeyerYA}, it has been known that self-dual 
YM theory possess infinitely many hidden nonlocal symmetries 
and hence infinitely many conserved nonlocal charges. 
As was shown in 
\cite{ChauGI1,ChauGI2,ChauGI3,UENO,Dolan,Crane},
these symmetries are affine extensions of internal symmetries
with an underlying Kac-Moody structure. For a review on that 
matter, we refer also to
\cite{Dolan2}. Furthermore, in \cite{Popov:1995qb} 
affine extensions of conformal symmetries have been found. 
As a result, certain Kac-Moody-Virasoro-type algebras 
associated with space-time symmetries were obtained. 
A systematic investigation of symmetries based on 
twistor theory and on \v Cech and Dolbeault cohomology methods
was performed in \cite{Popov:1998} (see also Refs. 
\cite{Popov:1998fb,Ivanova:1997cu} and the book \cite{MasonRF}). Therein,
all symmetries of the self-dual YM equations were derived.

One of the goals of this chapter is a generalization of the
above symmetries to the self-dual SYM equations. 
The subsequent discussion is based on \cite{Wolf:2004hp,Wolf:2005sd}. 
In particular, we will consider perturbations 
of transition functions of holomorphic vector bundles 
over supertwistor space. Using the Penrose-Ward transform, 
we relate these perturbations to 
symmetries of $\CN$-extended self-dual SYM theory.
After some general words on hidden symmetry algebras, we exemplify
our discussion by constructing Kac-Moody symmetries which come 
from affine extensions of internal symmetries. 
Furthermore, we also
consider affine extensions of the superconformal algebra resulting 
in super Kac-Moody-Virasoro-type symmetries. Moreover, 
by focussing on certain Abelian subalgebras of the affinely 
extended superconformal algebra, we introduce 
supermanifolds which we call generalized supertwistor spaces. 
These spaces then allow us to introduce 
so-called hierarchies which describe sets of graded 
Abelian symmetries. This generalizes the results known for 
 self-dual YM theory 
\cite{Mason,Takasaki,Mason:1992vd,Ablowitz,MasonRF,IvanovaZT}. We remark that such 
symmetries of the latter theory are intimately connected 
with one-loop maximally helicity violating amplitudes 
\cite{BardeenGK,Cangemi1,Cangemi2,Rosly:1996vr,Gorsky:2005sf}.

\section{Holomorphicity and symmetries}\label{sec:HandS-ch5}

As indicated, the key idea for studying symmetries within
the supertwistor framework is, for a given 
Stein covering $\fU=\{\CU_a\}$ of supertwistor space 
$\CP^{3|\CN}$, to consider
infinitesimal deformations of the transition functions
$f=\{f_{ab}\}\in H^1(\CP^{3|\CN},\fH)$
of some smoothly trivial rank $r$ holomorphic vector bundle 
$\CE$ over $\CP^{3|\CN}$ which in addition is
$\IC^{4|2\CN}$- (respectively, $\IR^{4|2\CN}$-)
trivial.\footnote{In \ref{pa:soldres-ch3}\kern6pt,
we have already used the idea of infinitesimal 
deformations for the construction of solutions to the equations
of motion of the supersymmetrized Bogomolny model. Here,
we are going to formalize the things used in that paragraph.}
Recall that $\fH=GL(r,\CO_{\CP^{3|\CN}})$. 
Generically, an infinitesimal deformation looks as
\eqn{}{f\ \mapsto\ f'\ =\ f+ \d f\qquad\longleftrightarrow\qquad
       f_{ab}\ \mapsto\ f'_{ab}\ =\ f_{ab}+\d f_{ab}.}
Clearly, such perturbations have to obey certain criteria:
\begin{itemize}\setlength{\itemsep}{-1mm}
 \item They must be holomorphic and moreover preserve the cocycle
       conditions, that is, $f'=\{f'_{ab}\}\in H^1(\CP^{3|\CN},\fH)$.
 \item They must preserve the holomorphic
       triviality on any $\IC P^1_{x_R,\eta}\hookrightarrow
       \CP^{3|\CN}$.
 \item In case one is interested in real YM fields, they must 
       in addition respect 
       the reality conditions induced by the antiholomorphic
       involution $\tau_E$ introduced in \ref{EuSi}\kern6pt. 
\end{itemize}
\noindent
The first point is not really a restriction. Remember that in Chaps.
\ref{STG-Chapter} and \ref{ch:SDSYM-ch2} we have seen that there exists 
a two-set Stein covering of $\CP^{3|\CN}$. Therefore, 
no cocycle conditions appear when working with this covering.\footnote{In
principle, there are transformations which can change the
covering. Then one needs to consider the common refinement. For details,
see \cite{Popov:1998}. In
the sequel, we shall only be interested in deformations which preserve
the chosen covering.}
The second point  
deserves some discussion. Besides $H^1(\CP^{3|\CN},\fH)$, consider
the Abelian group (by addition) $H^1(\CP^{3|\CN},{\rm Lie}\,\fH)$,
where ${\rm Lie}\,\fH:=\mathfrak{gl}(r,\CO_{\CP^{3|\CN}})$. This group
parametrizes infinitesimal deformations of the trivial bundle
$\CE_0$.\footnote{Note that infinitesimal
deformations of an arbitrary holomorphic vector bundle
$\CE\to X$ are parametrized by $H^1(X,{\rm End}\,\CE)$.} 
Furthermore, 
$\dim_\IC H^1(\CP^{3|\CN},{\rm Lie}\,\fH)=\infty$, that is, in
an arbitrarily small neighborhood of the trivial bundle $\CE_0$
there exists an infinite number of holomorphically nontrivial
bundles $\CE$. If we let 
$H^1(\IC P^1_{x_R,\eta},{\rm Lie}\,\fH_{\IC P^1_{x_R,\eta}})$
be the restriction of the cohomology group
$H^1(\CP^{3|\CN},{\rm Lie}\,\fH)$
to $\IC P^1_{x_R,\eta}\hookrightarrow\CP^{3|\CN}$ for some fixed
$(x_R,\eta)\in\IC^{4|2\CN}$, then 
$H^1(\IC P^1_{x_R,\eta},{\rm Lie}\,\fH_{\IC P^1_{x_R,\eta}})$ 
parametrizes infinitesimal deformations of $\CE_0$ restricted
to ${\IC P^1_{x_R,\eta}}$. 
However, 
$$H^1(\IC P^1_{x_R,\eta},{\rm Lie}\,\fH_{\IC P^1_{x_R,\eta}})\ \cong\ 0$$
which follows from the vanishing of $H^1(\IC P^1,\CO_{\IC P^1})$.
This in turn implies that small enough deformations do
not change the trivializability of $\CE$ over $\IC P^1_{x_R,\eta}
\hookrightarrow\CP^{3|\CN}$. In fact, this is a version of the  
Kodaira-Spencer-Nirenberg theorem \cite{Kodaira}. 
Therefore, point two from the
above list is likewise no restriction on the allowed perturbations.
Altogether, any infinitesimal holomorphic deformation 
is allowed. The only restriction one has 
to implement is point three from the above list. If not 
otherwise stated, we shall always assume that our deformations
are compatible with the reality conditions.
   
\paragraph{Kac-Moody symmetries.}\label{pa-KM-ch5}
Let us consider the nonholomorphic fibration $\CP^{3|\CN}\to\IR^{4|2\CN}$
given by \eqref{F-21}. Choose again the canonical covering $\fU=\{\CU_+,\CU_-\}$
of the supertwistor space. Given a smoothly and $\IR^{4|2\CN}$-trivial 
rank $r$ holomorphic
vector bundle $\CE\to\CP^{3|\CN}$, we define an action 
of the one-cochain group $C^1(\fU,\fH)$ on $Z^1(\fU,\fH)$ by
\eqn{defcz}{h\,:\, f_{+-}\ \mapsto\ h_{+-}f_{+-}h_{-+}^{-1},}
where $h=\{h_{+-},h_{-+}\}\in C^1(\fU,\fH)$ and 
$f=\{f_{+-}\}\in Z^1(\fU,\fH)$. Obviously, the group 
$C^1(\fU,\fH)$ acts transitively on
$Z^1(\fU,\fH)$, since for an arbitrary \v Cech one-cocycle
$f\in Z^1(\fU,\fH)$ one can find an 
$h\in C^1(\fU,\fH)$ such that $f_{+-}=h_{+-}h_{-+}^{-1}$ and
$f_{-+}=h_{-+}h_{+-}^{-1}$. The stabilizer of the trivial \v Cech 
one-cocycle, given by $f=\{\mathbbm{1}_r\}$, is the subgroup 
\eqn{}{ C_\Delta^1(\fU,\fH)\ :=\ \{h\in C^1(\fU,\fH)
   \,|\,h_{+-}=h_{-+}\}}
implying that $Z^1(\fU,\fH)$ can be identified with the 
coset
\eqn{}{Z^1(\fU,\fH)\ \cong\ C^1(\fU,\fH)/C_\Delta^1(\fU,\fH).}
Then the moduli space $H^1(\CP^{3|\CN},\fH)$ is given by 
a double coset space
\eqn{}{H^1(\CP^{3|\CN},\fH)\ \cong\ 
      C^0(\fU,\fH)\backslash C^1(\fU,\fH)/C_\Delta^1(\fU,\fH),}
where the zero-cochain group acts according to
\eqn{}{h\,:\, f_{+-}\ \mapsto\ h_+f_{+-}h_-^{-1},}
with $h=\{h_+,h_-\}\in C^0(\fU,\fH)$.

Let us now study infinitesimal deformations of the transition function
$f=\{f_{+-}\}$. Recalling the definition $\fS=GL(r,\CS_{\CP^{3|\CN}})$,
we introduce the subsheaves $\fP\subset\fS$ and 
${\rm Lie}\,\fP\subset{\rm Lie}\,\fS$ consisting of those 
$GL(r,\IC)$-valued, respectively, $\mathfrak{gl}(r,\IC)$-valued smooth functions
which are holomorphic in $\l_\pm$. 
We have a
natural infinitesimal action of the group $C^1(\fU,\fH)$
on the space $Z^1(\fU,\fH)$ 
which is induced by the linearization of \eqref{defcz}. That is, 
we get
\eqn{defczi}{\d f_{+-}\ =\ \d h_{+-}f_{+-}-f_{+-}\d h_{-+},}
where $\d h=\{\d h_{+-},\d h_{-+}\}\in C^1(\fU,{\rm Lie}\,\fH)$ and 
$f=\{f_{+-}\}\in Z^1(\fU,\fH)$. Then we introduce a
$\mathfrak{gl}(r,\IC)$-valued function
\eqn{defofphi}{\phi_{+-}\ :=\ \psi_+(\d f_{+-}f_{+-}^{-1})\psi_+^{-1}\ =\ 
     \psi_+(\d f_{+-})\psi_-^{-1},}
where $\psi=\{\psi_+,\psi_-\}\in C^0(\fU,\fP)$ and $f_{+-}=\psi_+^{-1}
\psi_-$, as before. From Eq. \eqref{defczi} it is then immediate that
$$\phi_{+-}\ =\ -\phi_{-+}.$$
Moreover, $\phi_{+-}$ is annihilated by the
vector field $\partial_{{\bar\l}_\pm}$. Thus, it 
defines an element $\phi=\{\phi_{+-}\}\in Z^1(\fU,{\rm Lie}\,\fP)$.
However, any one-cocycle with values in the sheaf
${\rm Lie}\,\fP$ is a one-coboundary since 
$H^1(\CP^{3|\CN},{\rm Lie}\,\fP)\cong0$. This is an immediate consequence
of the vanishing of $H^1(\IC P^1,\CO_{\IC P^1})$ and
of the acyclicity of the sheaf $\CS_{\CP^{3|\CN}}$.
Therefore, we have
\eqn{eqphii}{\phi_{+-}\ =\ \phi_+-\phi_-,}
where 
$\phi=\{\phi_+,\phi_-\}\in C^0(\fU,{\rm Lie}\,\fP)$.
Linearizing $f_{+-}=\psi_+^{-1}\psi_-$, we obtain
$$\d f_{+-}\ =\ f_{+-}\psi_-^{-1}\d\psi_-
           -\psi_+^{-1}\d\psi_+f_{+-}$$
and hence by virtue of Eqs. \eqref{defofphi} and 
\eqref{eqphii} we find
\eqn{eqforhpmi}{\d\psi_\pm\ =\ -\phi_\pm\psi_\pm.}
In summary, given some $\d h=\{\d h_{+-},\d h_{-+}\}
\in C^1(\fU,{\rm Lie}\fH)$ one derives via
\eqref{defczi}--\eqref{eqphii} the perturbations $\d\psi_\pm$.
 Moreover, we point out that finding such
$\phi_\pm$ means to solve the infinitesimal variant of the Riemann-Hilbert 
problem. Obviously, the splitting \eqref{eqphii}, \eqref{eqforhpmi}
and hence solutions to the Riemann-Hilbert problem are not unique, 
as we certainly have the freedom to
consider a new ${\tilde\phi}$ shifted by functions $\o_\pm$, 
${\tilde\phi}_\pm=\phi_\pm+\o_\pm$, such that $\o_+=\o_-$ on the 
intersection $\CU_+\cap\CU_-$, that is, $\o\in Z^0(\fU,{\rm Lie}\,\fP)
\equiv H^0(\CP^{3|\CN},{\rm Lie}\,\fP)$. 

Furthermore, infinitesimal variations of the linear system
\eqref{linsys-22} yield
\eqna{deforeqa}{
   \d\CA^\pm_\a\ &=\ \d\psi_\pm\bV^\pm_\a\psi_\pm^{-1}+
                             \psi_\pm\bV^\pm_\a\d\psi_\pm^{-1},\\
             \d\CA^i_\pm \ &=\ \d\psi_\pm\bV^i_\pm\psi_\pm^{-1}+
                             \psi_\pm\bV^i_\pm\d\psi_\pm^{-1}.
}
Substituting \eqref{eqforhpmi} into these equations, we arrive at
\eqn{deforeqai}{\d\CA^\pm_\a\ =\ \bnab^\pm_\a\phi_\pm
               \qquad{\rm and}\qquad
               \d\CA^i_\pm\ =\ \bnab^i_\pm\phi_\pm.}
Here, we have introduced the definitions
\eqn{defofnabi}{\bnab^\pm_\a\ :=\  \bV^\pm_\a+\CA^\pm_\a
               \qquad{\rm and}\qquad
               \bnab^i_\pm\ :=\ \bV^i_\pm+\CA^i_\pm.}
Note that \eqref{defofnabi} acts adjointly in \eqref{deforeqai}. 
Therefore, \eqref{deforeqai} together with \eqref{eqphii} imply that
\eqn{deforeqaii}{\bnab^\pm_\a\phi_{+-}\ =\ 0\qquad{\rm and}\qquad
               \bnab^i_\pm\phi_{+-}\ =\ 0.}
One may easily check that the choice 
$\phi_\pm=\psi_\pm\chi_\pm\psi_\pm^{-1}$, where 
$\chi=\{\chi_+,\chi_-\}\in C^0(\fU,{\rm Lie}\,\fH)$, 
implies $\d\CA_{\a\ad}=0$ and $\d\CA^i_\ad=0$, respectively. Hence, such
$\phi_\pm$ define trivial perturbations. On the other hand, infinitesimal gauge
transformations of the form
$$\d\CA_{\a\ad}\ =\ \nabla_{\a\ad}^R\o
  \qquad{\rm and}\qquad
  \d\CA^i_\ad\ =\ \nabla^i_\ad\o,$$ 
where $\o$ is some smooth $\mathfrak{su}(r)$-valued function on $\IR^{4|2\CN}$,
imply (for irreducible gauge potentials) that 
$$\phi_\pm\ =\ \o.$$ 
Hence, such $\phi_\pm$ do not depend on $\l_\pm$. In particular, we have 
$\phi_{+-}=0$ and hence $\d f_{+-}=0$. 

Finally, we obtain the formulas
\eqn{infpwtrans}{
             \d\CA_{\a\ad}\ =\ \frac{1}{2\pi\i}\oint_\cC{\rm d}\l_+\,
           \frac{\bnab_\a^+\phi_+}{\l_+\l^\ad_+}
             \qquad{\rm and}\qquad
             \d\CA^i_\ad\ =\ \frac{1}{2\pi\i}\oint_\cC{\rm d}\l_+\,
            \frac{\bnab^i_+\phi_+}{\l_+\l^\ad_+} 
}
where the contour $\cC=\{\l_+\in\IC P^1\,|\,|\l_+|=1\}$ encircles $\l_+=0$.
We see that generically the outcome of the transformation $\d\CA$ is a
highly nonlocal expression depending on $\CA$, i.e., we may write 
$\d\CA=F[\CA,\partial\CA,\ldots]$, where $F$ is a functional whose explicit
form is determined by $\phi_\pm$.

Summarizing, to any infinitesimal deformation, 
$f\mapsto f+\d f$, of the transition function $f$ we have
associated a symmetry transformation $\CA\mapsto\CA+\d\CA$, that is, 
solutions to the linearized  
$\CN$-extended self-dual SYM equations. In fact, we have obtained
a one-to-one correspondence between equivalence classes of deformations
of holomorphic vector bundles over supertwistor space which are 
subject to certain triviality conditions and gauge equivalence classes
of solutions to the linearized field equations of $\CN$-extended SYM
theory. Putting it differently, 
our above discussion represents the ``infinitesimal"
version of Thm. \ref{SDYM-HCS}

\paragraph{Virasoro-type symmetries.}
Above we have introduced Kac-Moody symmetries which were generated by 
the algebra $C^1(\fU,{\rm Lie}\,\fH)$. In this paragraph
we focus on symmetries, which are related to the group of local
biholomorphisms of supertwistor space $\CP^{3|\CN}$. We 
shall only be interested in this subgroup of the
diffeomorphism group of $\CP^{3|\CN}$ as generic local diffeomorphisms
would change the complex structure of the supertwistor space.
However, this in turn would induce a change of the conformal structure
and a metric on $\IR^{4|2\CN}$ as was demonstrated in the purely
even setting by Penrose \cite{Penrose:1976js} and 
Atiyah et al. \cite{Atiyah:wi}. As we
want to discuss symmetries of the self-dual SYM equations on
$\IR^4$, we need to consider those diffeomorphisms which 
preserve the complex structure, that is, biholomorphisms.
Let us denote the group of local biholomorphisms by $\CH_{\CP^{3|\CN}}$.
Furthermore, we are again choosing the canonical covering $\fU=\{\CU_+,\CU_-\}$
of $\CP^{3|\CN}$ together with the coordinates
$(Z^I_\pm)=(z_\pm^\a,z^3_\pm,\eta_i^\pm)$.
On the intersection $\CU_+\cap\CU_-$, they are related by transition 
functions $t=\{t_{+-}^I\}$,
\eqn{}{Z_+^I\ =\ t^I_{+-}(Z_-^J).}  
To $\CH_{\CP^{3|\CN}}$ one associates the algebra $C^0(\fU,T\CP^{3|\CN})$
of zero-cochains on $\CP^{3|\CN}$ with values in the tangent sheaf 
$T\CP^{3|\CN}$ of supertwistor space. In order to
define an appropriate action of $C^0(\fU,T\CP^{3|\CN})$,
let us first consider the
algebra $C^1(\fU,T\CP^{3|\CN})$ whose elements are collections of vector fields 
\eqn{elementsofc}{\chi\ =\ \{\chi_{+-},\chi_{-+}\}\ =\ 
               \left\{\chi^I_{+-}\partial_I^+,\chi^I_{-+}\partial_I^-
                 \right\},}
where we have abbreviated $\partial^\pm_I:=\partial/\partial Z^I_\pm$.
In particular, $\chi_{+-}$ and $\chi_{-+}$ are elements of the algebra
$T\CP^{3|\CN}|_{\CU_+\cap\CU_-}$ of holomorphic vector fields on the intersection
$\CU_+\cap\CU_-$. Thus, $C^1(\fU,T\CP^{3|\CN})$
can be decomposed according to
\eqn{}{C^1(\fU,T\CP^{3|\CN})\ \cong\ T\CP^{3|\CN}|_{\CU_+\cap\CU_-}\oplus
    T\CP^{3|\CN}|_{\CU_+\cap\CU_-}.}
Kodaira-Spencer deformation theory \cite{Kodaira} then tells us that 
the algebra 
$C^1(\fU,T\CP^{3|\CN})$ acts on the transition functions $t_{+-}^I$ 
according to
\eqn{kodspe}{\d t^I_{+-}\ =\ \chi_{+-}^I-\chi^J_{-+}\partial_J^-t^I_{+-}}
which can equivalently be rewritten as
\eqn{kodspetwo}{\d t_{+-}\ :=\ \d t^I_{+-}\partial^+_I\ =\ 
               \chi_{+-}-\chi_{-+}.}
Consider a subalgebra
\eqn{}{C^1_\Delta(\fU,T\CP^{3|\CN})\ :=\ \{\chi\in C^1(\fU,T\CP^{3|\CN})\,|\,
   \chi_{+-}=\chi_{-+}\}}
of the algebra $C^1(\fU,T\CP^{3|\CN})$. Then the space 
\eqn{}{Z^1(\fU,T\CP^{3|\CN})\ =\ \{\chi\in C^1(\fU,T\CP^{3|\CN})\,|\, 
          \chi_{+-}=-\chi_{-+}\}}
is given by the quotient
\eqn{}{Z^1(\fU,T\CP^{3|\CN})\ \cong\ C^1(\fU,T\CP^{3|\CN})/C^1_\Delta(\fU,T\CP^{3|\CN}).}
We stress that the
transformations \eqref{kodspe} change the complex structure of $\CP^{3|\CN}$
if $\chi_{+-}\neq\chi_+-\chi_-$, where 
$\{\chi_+,\chi_-\}\in C^0(\fU,T\CP^{3|\CN})$. Therefore,
\eqn{}{H^1(\CP^{3|\CN},T\CP^{3|\CN})\ \cong\ Z^1(\fU,T\CP^{3|\CN})/C^0(\fU,T\CP^{3|\CN})}
is the tangent space (at a chosen complex structure)
of the moduli space of deformations of the complex structure on $\CP^{3|\CN}$.

According to \eqref{kodspetwo}, we may define an action of
the algebra $C^1(\fU,T\CP^{3|\CN})$ on the transition functions $f=\{f_{+-}\}$
of holomorphic vector bundles over supertwistor space by
\eqn{kodspethree}{\d f_{+-}\ :=\ \chi_{+-}(f_{+-})-\chi_{-+}(f_{+-}).}
However, as we have just seen, such transformations may change the
complex structure on $\CP^{3|\CN}$. Therefore, we let the
algebra $C^0(\fU,T\CP^{3|\CN})$ act on $t=\{t^I_{+-}\}$ and $f=\{f_{+-}\}$ 
via
\eqn{}{C^0(\fU,T\CP^{3|\CN})\ni\{\chi_+,\chi_-\}\ \mapsto\ 
       \{\chi_+-\chi_-,\chi_--\chi_+\}\in C^1(\fU,T\CP^{3|\CN})}
together with Eqs. \eqref{kodspetwo} and \eqref{kodspethree}, i.e.,
\eqna{}{ \d t_{+-}\ &=\ \chi_+-\chi_-,\\
         \d f_{+-}\ &=\ \chi_{+}(f_{+-})-\chi_{-}(f_{+-}).}
These transformations do preserve the complex structure on
supertwistor space.

After this digression, we may now follow the lines presented 
in \ref{pa-KM-ch5}\kern10pt to arrive at the formulas \eqref{infpwtrans} 
for symmetries. We shall not repeat the argumentation at
this point. We refer to symmetries obtained in this 
way as Virasoro-type symmetries since they are associated with the group of 
local biholomorphisms.

\paragraph{All symmetries.}
With the help of \v Cech cohomology and the Penrose-Ward transform,
we have shown how Kac-Moody and Virasoro-type symmetries
in $\CN$-extended self-dual SYM theory arise. In the next
paragraph, we will make the symmetry algebras more transparent
by proving a theorem relating certain deformation algebras on the
twistor side to symmetry algebras on the self-dual SYM side. 
Furthermore, in Sec. \ref{sec:EEHSA-ch5} we then give
explicit examples of symmetry transformations and corresponding
symmetry algebras. Before dealing with these issues, however,
let us summarize our preceding discussion. 
In fact, above we have given all possible continuous symmetry transformations
acting on the solution space  
of the equations of motion of self-dual SYM theory.
If we let $\fU=\{\CU_+,\CU_-\}$ be the canonical covering of the
supertwistor space $\CP^{3|\CN}$, $\fH=GL(r,\CO_{\CP^{3|\CN}})$
and $\CH_{\CP^{3|\CN}}$ be the group of local biholomorphisms of $\CP^{3|\CN}$
associated with the algebra $C^0(\fU,T\CP^{3|\CN})$, 
we may state the following result:
{\Thm The full group of continuous (symmetry) transformations acting
      on the space $Z^1(\fU,\fH)$ of smoothly and
      $\IR^{4|2\CN}$-trivial holomorphic vector bundles
      $\CE\to\CP^{3|\CN}$ of rank $r$
      is a semidirect product $$\CH_{\CP^{3|\CN}}\ltimes C^1(\fU,\fH).$$ By virtue
      of the Penrose-Ward transform, one obtains a corresponding
      group action on the solution space of $\CN$-extended 
      self-dual SYM theory in four dimensions.}\vskip 4mm

\noindent
A proof of this theorem (for $\CN=0$) can be found in Ref. 
\cite{Popov:1998}. The argumentation for $\CN>0$ goes along
similar lines. 

\paragraph{General symmetry algebras.}\label{par:GSA-ch5}
So far, we have worked within a quite abstract scheme. Let us now 
present a more concrete relationship between certain deformation 
algebras and hidden symmetry algebras \cite{Wolf:2005sd}. 
Suppose we are given some indexed set $\{\d_I\}$ of infinitesimal
variations $\d_I f$ of the transition function $f=\{f_{+-}\}$
of our holomorphic
vector bundle $\CE$ over supertwistor space. 
Suppose further that the $\d_I$s satisfy 
a deformation algebra of the form 
\eqn{defalg}{[\d_I,\d_J\}\ =\ {f_{IJ}}^K\d_K,} where 
generically ${f_{IJ}}^K\in C^1(\fU,\CO_{\CP^{3|\CN}})$, with
$${f_{IJ}}^K\ =\ -(-)^{p_Ip_J}{f_{JI}}^K,$$
and $[\cdot,\cdot\}$ denotes the supercommutator.
{\Thm\label{thm:symtrafo-ch5}
       Given a deformation algebra of the form
      \eqref{defalg} with constant ${f_{IJ}}^K\in\IC$, 
      the corresponding symmetry algebra on the gauge theory side 
      has exactly the same form modulo possible gauge 
      transformations.}\vskip 4mm

\noindent
Putting it differently, the linearized Penrose-Ward transform is, 
besides being an isomorphism between the tangent spaces of the moduli
spaces $H^1_{\nabla^{0,1}}(\CP^{3|\CN},\CE)$ and
$\CM^\CN_{\rm SDYM}$ (see also Thm. \ref{SDYM-HCS}),  
a Lie algebra homomorphism. Let us now prove the theorem.\vskip 4mm

\noindent
{\small {\it Proof:}
Let $\CE\to\CP^{3|\CN}$ be a smoothly and $\IR^{4|2\CN}$-trivial holomorphic
vector bundle of rank $r$. Recall that the transition function
$f=\{f_{+-}\}\in H^1(\CP^{3|\CN},\fH)$ 
can be split according to $f_{+-}=\psi^{-1}_+\psi_-$,
with $\psi=\{\psi_+,\psi_-\}\in C^0(\fU,\fP)$. Consider
$$[\d_1,\d_2]\ =\ (-)^{p_Ip_J}\varepsilon^I\varrho^J[\d_I,\d_J\},$$
where $\varepsilon^I$ and $\varrho^J$ are the infinitesimal parameters
of the transformations $\d_1$ and $\d_2$, respectively. By virtue of
\eqref{deforeqai}, we may write
$$[\d_1,\d_2]\CA_\pm\ =\ \d_1(\CA_\pm+\d_2\CA_\pm)-\d_1\CA_\pm
       -\d_2(\CA_\pm+\d_1\CA_\pm)+\d_2\CA_\pm,$$
where $\CA_\pm$ stands symbolically for both components 
$\CA^\pm_\a=\l^\ad_\pm\CA_{\a\ad}$ and $\CA^i_\pm=\l^\ad_\pm\CA_{\ad}$.
Then one easily checks that
$$[\d_1,\d_2]\CA_\pm\ =\ \bnab_\pm\Sigma_{\pm12},\qquad{\rm with}\qquad
        \Sigma_{\pm12}\ :=\ \d_1\phi^2_\pm-\d_2\phi_\pm^1+[\phi^1_\pm,\phi^2_\pm],$$
where $\bnab_\pm$ represents any of the covariant derivatives given in 
\eqref{defofnabi}. Note that we use the notation 
$\phi^1_\pm=\varepsilon^I\phi_{\pm I}$ and
similarly for $\phi^2_\pm$. Hence,
$\Sigma_{\pm 12}=(-)^{p_Ip_J}\varepsilon^I\varrho^J\Sigma_{\pm IJ}$.
Next one considers the commutator
$${[\d_1,\d_2]}f_{+-}\ =\ \d_1(f_{+-}+\d_2f_{+-})-\d_1f_{+-}
       -\d_2(f_{+-}+\d_1f_{+-})+\d_2f_{+-}.$$
Using definition \eqref{defofphi} and the resulting splitting 
\eqref{eqphii} for the deformations $\d_{1,2}f_{+-}$, 
one obtains after some tedious but straightforward algebraic
manipulations 
$${[\d_1,\d_2]f_{+-}}\ =\ \psi_+^{-1}(
     \Sigma_{+12}-\Sigma_{-12})\psi_-,$$
where $\Sigma_{\pm12}$ has been introduced above. By assumption, it
must also be equal to
$${[\d_1,\d_2]f_{+-}}\ =\ \d_3 f_{+-},\qquad{\rm with}\qquad \d_3\ 
              =\ \varepsilon^I\varrho^J {f_{IJ}}^K\d_K,$$
i.e., 
$${[\d_1,\d_2]f_{+-}}\ =\ \psi_+^{-1}(\phi^3_+-\phi^3_-)\psi_-,$$
where $\phi^3_\pm=\varepsilon^I\varrho^J {f_{IJ}}^K\phi_{\pm K}$.
By comparing this equation with the previously obtained result 
for $[\d_1,\d_2]f_{+-}$,
we may conclude that
$$\Sigma_{\pm12}\ =\ \phi^3_\pm + \o^3\ =\ 
     \varepsilon^I\varrho^J ({f_{IJ}}^K\phi_{\pm K}+\o_{IJ})$$
since by assumption ${f_{IJ}}^K\in\IC$. Here,
$\o^3$ (respectively, $\o_{IJ}$) belongs to $H^0(\CP^{3|\CN},{\rm Lie}\,\fP)$.
Thus, by our above discussion, it represents an infinitesimal gauge 
transformation.
\hfill $\blacksquare$
}\vskip 4mm

Hence, any Lie-algebra type deformation algebra on the 
twistor side yields by virtue of the Penrose-Ward transform
a symmetry algebra on the gauge theory side being of the same
form. Furthermore, we also realize that if 
${f_{IJ}}^K\in C^1(\fU,\CO_{\CP^{3|\CN}})$, the algebra on the 
gauge theory side will generically no longer close because of the 
dependence of ${f_{IJ}}^K$ on $(x^{\a\ad}_R,\eta^\ad_i)$ through
$(z^\a_\pm,z^3_\pm,\eta^\pm_i)$. In fact, in order to be
able to compare $\Sigma_{\pm12}$ 
with $\phi^3_\pm$ (which will no longer be given by
$\varepsilon^I\varrho^J {f_{IJ}}^K\phi_{\pm K}$) from the above proof,
one has to Laurent-expand ${f_{IJ}}^K$ in powers of $\l_\pm$.
The resulting coefficient functions will solely depend
on $(x^{\a\ad}_R,\eta^\ad_i)\in\IR^{4|2\CN}$. Therefore,
the space-time derivatives appearing in the transformations 
\eqref{deforeqai} will destroy the
``would be" algebra on the self-dual SYM side. As a matter
of fact, if one allows ${f_{IJ}}^K$ to depend only on 
$z^3_\pm=\l_\pm$, no such problems will occur. That is, the
algebra, though modified, will still close. More details
about these issues can be found in the next section
when explicitly dealing with affine extensions of superconformal
symmetries.

\section{Explicit examples of hidden symmetry 
algebras}\label{sec:EEHSA-ch5} 

By now it is clear that self-dual SYM theory possesses
infinite-dimensional hidden symmetry algebras.
To exemplify our discussion, let us now describe some explicit symmetry 
algebras. We begin with affine extensions of internal
symmetries and afterwards discuss affine extensions
related to space-time symmetries. As before,
we consider a smoothly and $\IR^{4|2\CN}$-trivial
rank $r$ holomorphic vector bundle $\CE$ over 
supertwistor space, the latter being covered by
$\fU=\{\CU_+,\CU_-\}$. The transition function
of $\CE$ is again given by $f=\{f_{+-}\}=\{\psi_+^{-1}\psi_-\}$,
with $\psi=\{\psi_+,\psi_-\}\in C^0(\fU,\fP)$.
\paragraph{Kac-Moody symmetries.}
The prime example is Kac-Moody symmetries associated
with internal symmetries. Let $X_I$ be any generator
of the gauge group $\mathfrak{su}(r)$ with
\eqn{}{[X_I,X_J]\ =\ {f_{IJ}}^KX_K,}
where the ${f_{IJ}}^K$s are the structure constants
of $\mathfrak{su}(r)$. Then we may define the
following infinitesimal deformation
\eqn{}{\d^m_If_{+-}\ :=\ \tfrac{1}{2}(\l_+^m+(-\l_-)^m)[X_I,f_{+-}]
       \qquad{\rm for}\qquad m\in\IN_0.}
It is clearly a holomorphic deformation and moreover,
one may readily check that it preserves the
reality condition \eqref{AHI-2}. Notice that by comparing 
with Eq. \eqref{defczi}, we see that 
$$\d^m_I h_{+-}\ =\ \d^m_Ih_{-+}\ =\ \tfrac{1}{2}(\l_+^m+(-\l_-)^m)X_I$$
define one-cochains $\d^m_I h\in C^1(\fU,{\rm Lie}\,\fH)$.
For $m=0$, the transformations
of the components of the gauge potential turn out to be (after
properly fixing the freedom in solving the infinitesimal Riemann-Hilbert
problem; cf. Eq. \eqref{phisolzeroKMY})
\eqn{}{\d^0_I\CA_{\a\ad}\ =\ [X_I,\CA_{\a\ad}]\qquad{\rm and}\qquad
       \d^0_I\CA^i_\ad\ =\ [X_I,\CA^i_\ad].}
Thus, they represent an internal symmetry transformation.
Then a short calculation reveals that
\eqn{eq:KMA-ch5}{[\d_I^m,\d_J^n]\ =\ \tfrac{1}{2}{f_{IJ}}^K(\d_K^{m+n}
           +(-)^n\d_K^{|m-n|}),}
which is the analytic half of a centerless Kac-Moody algebra
$\widehat{\mathfrak{su}(r)}$. 
One can bring \eqref{eq:KMA-ch5} in
a more familiar form as follows. Define $\Delta^0_I:=\d^0_I$ and 
$\Delta^1_I:=\d^1_I$. Furthermore, let
\eqn{}{\Delta^m_I\ :=\ \sum_{k=0}^m c^m_k\d^k_I,} 
with some (real) coefficients $c^m_k$ to be determined. Next consider
\eqn{}{[\Delta^m_I,\Delta^1_J]\ =:\ {f_{IJ}}^K\Delta^{m+1}_K,}
that is, the $(m+1)$-th generator is recursively defined by
$m$-th one. Then
\eqn{}{[\Delta^m_I,\Delta^n_J]\ =\ {f_{IJ}}^K\Delta^{m+n}_K,}
for all $m,n\geq0$.
In fact, this is a direct consequence of Eq. \eqref{eq:KMA-ch5} and 
can be proven by induction with the help of Jacobi
identity for the generators $\Delta^m_I$. As this is rather
straightforward, we leave the 
explicit verification to the interested reader.

Next we need to find -- by means of the discussion of the previous section 
-- the action of $\d^m_I$ for $m>0$ on the 
components $\CA_{\a\ad}$ and $\CA^i_\ad$ of 
the gauge potential. Consider the function $\phi^m_{+-I}$ as defined 
in \eqref{defofphi}. We obtain
\eqarr{eq:adeform-ch5}{ 
      \phi^m_{+-I}
           &=&\phi^m_{+I}-\phi^m_{-I}\notag\\ 
           &=&\psi_+\tfrac{1}{2}(\l_+^m+(-\l_-)^m)[X_I,f_{+-}]\psi_-^{-1}\ =\ 
               \tfrac{1}{2}(\l_+^m+(-\l_-)^m)\psi_+[X_I,\psi_+^{-1}\psi_-]\psi_-^{-1}\notag\\
           &=&-\tfrac{1}{2}(\l_+^m+(-\l_-)^m)[X_I,\psi_+]\psi_+^{-1}+
                 \tfrac{1}{2}(\l_+^m+(-\l_-)^m)[X_I,\psi_-]\psi_-^{-1}\notag\\
           &=&\tfrac{1}{2}(\l_+^m+(-\l_-)^m)\phi^0_{+I}-
               \tfrac{1}{2}(\l_+^m+(-\l_-)^m)\phi^0_{-I},
}
where in the last step we have introduced the functions $\phi^0_{\pm I}$ 
which are solutions of the Riemann-Hilbert problem for $m=0$, 
\eqn{phisolzeroKMY}{\phi^0_{\pm I}\ =\ -[X_I,\psi_\pm]\psi_\pm^{-1}.} As the 
$\phi^0_{\pm I}$s are holomorphic and nonsingular in $\l_\pm$ on their 
respective domains, we expand them in powers of $\l_+$ on the intersection 
$\CU_+\cap\CU_-$ according to
\eqn{phiexpKMY}{\phi_{\pm I}^0\ =\ \sum_{n=0}^\infty\l_+^{\pm n}
               \phi^{0(n)}_{\pm I}.}
Furthermore, Eqs. \eqref{infpwtrans} imply
\eqna{perturbationAKMYpi}{
     \d^m_I\CA_{\a{\dot1}}\ &=\ \nabla_{\a{\dot1}}\phi^{m(0)}_{+I}-
                                \nabla_{\a{\dot2}}\phi^{m(1)}_{+I}      
                           \ =\ \nabla_{\a{\dot1}}\phi^{m(0)}_{-I},\\
     \d^m_I\CA_{\a{\dot2}}\ &=\ \nabla_{\a{\dot2}}\phi^{m(0)}_{+I}      
                           \ =\ -\nabla_{\a{\dot1}}\phi^{m(1)}_{-I}
                                +\nabla_{\a{\dot2}}\phi^{m(0)}_{-I},
}
and similarly for $\CA^i_\ad$.

Upon substituting the expansions \eqref{phiexpKMY} into \eqref{eq:adeform-ch5},
we find the following splitting of $\phi_{+- I}^m$ into 
$\phi^m_{\pm I}$:
\eqn{phisolKMY}{\phi^m_{\pm I}\ =\ \tfrac{1}{2}(\pm)^m\l_+^{\pm m}\phi^0_{\pm I}
         +\tfrac{1}{2}(\mp)^m\sum_{n=0}^\infty\l_+^{\pm n}\phi^{0(m+n)}_{\pm I}
         -\tfrac {1}{2}(\pm)^m\sum_{n=0}^{m-1}\l_+^{\pm(m-n)}\phi^{0(n)}_{\mp I}.}
Recall that solutions to the Riemann-Hilbert problem are not unique. 
For instance, we could have added to $\phi^m_{+I}$ any smooth function 
which does not depend on $\l_\pm$. But at the same time we had to add the same 
function to $\phi^m_{-I}$, as well. However, we have learned that such 
shifts of ``zero-modes" will eventually result in gauge 
transformations. As we are not interested in such trivial symmetries, 
solution \eqref{phisolKMY} turns out to be the appropriate choice. 
Note that for $m=0$, Eq. \eqref{phisolKMY} is an
identity. Expanding the functions $\phi^m_{\pm a}$ in powers 
of $\l_+$, we obtain for $m>0$
\eqn{eq:c-1-ch5}{ \phi^{m(n)}_{\pm I}\ =\
          \begin{cases}
            \tfrac{1}{2}(\mp)^m\phi^{0(m)}_{\pm I} &{\rm for}\quad n=0\\
            \tfrac{1}{2}\left\{(\pm)^m\phi^{0(n-m)}_{\pm I}+(\mp)^m\phi^{0(m+n)}_{\pm I}
                         -(\pm)^m\phi^{0(m-n)}_{\mp I}\right\} &{\rm for}\quad n>0
          \end{cases}.
}
Note that $\phi_{\pm I}^{0(-n)}=0$ for all $n>0$.
Combining these expressions with the transformation laws
\eqref{perturbationAKMYpi}, we have thus given the action of
$\d^m_I$ on $\CA_{\a\ad}$ and $\CA_\ad^i$, respectively. Furthermore,
by Thm. \ref{thm:symtrafo-ch5} we know that the $\d^m_I$s satisfy
\eqn{eq:KMA2-ch5}{[\d_I^m,\d_J^n]\ =\ \tfrac{1}{2}{f_{IJ}}^K(\d_K^{m+n}
           +(-)^n\d_K^{|m-n|})}
modulo gauge transformations. For an explicit derivation, see
also Ref. \cite{Wolf:2004hp}. Following the discussion subsequent to
\eqref{eq:KMA-ch5}, we eventually arrive at
\eqn{}{[\Delta^m_I,\Delta^n_J]\ =\ {f_{IJ}}^K\Delta^{m+n}_K,}
for all $m,n\geq0$ modulo gauge transformations. Altogether, we
have thus obtained an infinite-dimensional affine symmetry
algebra in $\CN$-extended
self-dual SYM theory, which is the analytic half of the affine
Lie algebra
$\widehat{\mathfrak{su}(r)}$.

\paragraph{Virasoro-type symmetries.}
Let us now discuss affine extensions of superconformal
symmetries. However, we first need 
some preliminaries.
The superconformal group 
for $\CN\neq 4$ is locally isomorphic to a real form of the super 
matrix group $SU(4|\CN)$.
In the case of maximal $\CN=4$ supersymmetries, the supergroup $SU(4|4)$ is not
semi-simple and the superconformal group is considered to be a real form of the
semi-simple part $PSU(4|4)\subset SU(4|4)$.
The generators of the superconformal group  are the translation generators 
$P_{\a\ad}$, $Q_{i\a}$ and $Q^i_\ad$, the dilatation generator $D$, the 
generators of special conformal transformations $K_{\a\ad}$, $K^{i\a}$ and
$K^\ad_i$, the rotation generators $J_{\a\b}$ and $J_{\ad\bd}$, the generators 
$R_i^j$ of the R-symmetry and the generator of the axial symmetry 
$A$. The latter one is absent in the case of maximal $\CN=4$ supersymmetries.
The superconformal algebra then takes the following form:
\eqna{eq:SCA-ch5}{
    {[P_{\a\ad},K_{\b\bd}]}\ &=\ \tfrac{1}{2}(\epsilon_{\a\b}J_{\ad\bd}+\epsilon_{\ad\bd}
                               J_{\a\b})-\tfrac{1}{4}\epsilon_{\a\b}\epsilon_{\ad\bd}D,\\
    \{Q_{i\a},Q^j_\ad\}\ &=\ -2\d^j_iP_{\a\ad},\qquad
    \{K^{i\a},K^\ad_j\}\ =\ -2\d_j^iK^{\a\ad},\\
    \{Q_{i\a},K^{j\b}\}\ &=\ -2\d^j_i({J_\a}^\b+\tfrac{1}{4}\d_\a^\b D)
                      -\tfrac{1}{2}\d_\a^\b\d_i^j(1-\tfrac{4}{\CN})A+\d_\a^\b
                             R^j_i,\\                               
   [R^i_j,K^{k\a}]\ &=\ -(\d^k_j K^{i\a}-\tfrac{1}{\CN}\d_j^iK^{k\a}),\\
   [A,K^{i\a}]\ &=\ -\tfrac{1}{2}K^{i\a},\qquad
   [D,K^{i\a}]\ =\ -\tfrac{1}{2}K^{i\a},\\
   [J_{\a\b},K^{i\g}]\ &=\ \tfrac{1}{2}\epsilon_{\d(\a}\d^\g_{\b)}K^{i\d},
   \qquad[P_{\a\ad},K^{i\b}]\ =\ -\tfrac{1}{2}\d^\b_\a Q^i_\ad\\
   [R^j_i,Q_{k\a}]\ &=\ \d^j_k Q_{i\a}-\tfrac{1}{\CN}\d^j_iQ_{k\a},\\
   [A,Q_{i\a}]\ &=\ \tfrac{1}{2}Q_{i\a},\qquad
   [D,Q_{i\a}]\ =\ \tfrac{1}{2}Q_{i\a},\\
   [J_{\a\b},Q_{i\g}]\ &=\ -\tfrac{1}{2}\epsilon_{\g(\a}Q_{i\b)},\qquad
   [Q_{i\b},K^{\a\ad}]\ =\ -\tfrac{1}{2}\d_\b^\a K^\ad_i,\\
   [R^j_i,R^l_k]\ &=\ \d^j_k R^l_i-\d^l_i R^j_k,\qquad
   [J_{\a\b},J^{\g\d}]\ =\ -\d^{(\g}_{(\a}{J_{\b)}}^{\d)},\\
   [D,P_{\a\ad}]\ &=\ P_{\a\ad},\qquad[D,K^{\a\ad}]\ =\ -K^{\a\ad},\\
   [J_{\a\b},K^{\g\gd}]\ &=\ \tfrac{1}{2}\epsilon_{\d(\a}\d^\g_{\b)}K^{\d\gd},
   \qquad
   [J_{\a\b},P_{\g\gd}]\ =\ -\tfrac{1}{2}\epsilon_{\g(\a}P_{\b)\gd},\\
}
where, as before, parentheses mean normalized symmetrization of the
enclosed indices.
As is well known, there is a representation of this algebra 
in terms of vector fields on 
the anti-chiral superspace $\IR^{4|2\CN}$ according to
\eqna{scgen}{
           P_{\a\ad} \ &=\ \partial_{\a\ad}^R,\qquad
           Q_{i\a}   \  =\ -2\eta^\ad_i\partial_{\a\ad}^R,\qquad
           Q^i_\ad   \  =\ \partial_\ad^i,\\
           D         \ &=\ -x^{\a\ad}\partial_{\a\ad}^R-\tfrac{1}{2}
                           \eta^\ad_i\partial^i_\ad,\\
           K^{\a\ad} \ &=\ \tfrac{1}{4}x^{\a\bd}_R(x^{\b\ad}_R\partial_{\b\bd}^R+
                           \eta^\ad_i\partial^i_\bd),\\
           K^{i\a}   \ &=\ -\tfrac{1}{2}x^{\a\ad}_R\partial^i_\ad,\qquad
           K^\ad_i   \  =\ \eta^\bd_i(x^{\a\ad}_R\partial_{\a\bd}^R+
                           \eta^\ad_j\partial^j_\bd),\\
           J_{\a\b}  \ &=\ \tfrac{1}{2}x^{\g\ad}_R\epsilon_{\g(\a}
                           \partial_{\b)\ad}^R,\qquad
           J_{\ad\bd}\  =\ \tfrac{1}{2}\left(x^{\a\gd}_R\epsilon_{\gd(\ad}
                           \partial_{\a\bd)}^R+\eta^\gd_i\epsilon_{\gd(\ad}
                           \partial^i_{\bd)}\right),\\
           R_i^j\ &=\ \eta^\ad_i\partial^j_\ad-\tfrac{1}{\CN}\,\d_i^j
                           \eta^\ad_k\partial_\ad^k,\qquad
           A         \  =\ \tfrac{1}{2}\eta^\ad_i\partial^i_\ad.
}

Infinitesimal transformations of the components $\CA_{\a\ad}$ and $\CA^i_\ad$
of the gauge potential under the action of the superconformal group are 
given by
\eqn{infliea}{\d_{N_I}\CA_{\a\ad}\ =\ \CL_{N_I}\CA_{\a\ad}\qquad{\rm and}\qquad
             \d_{N_I}\CA_\ad^i  \ =\ \CL_{N_I}\CA_\ad^i,}
where $N_I=N_I^{\a\ad}\partial_{\a\ad}^R+N^\ad_{I\,i}\partial^i_\ad$ is any 
generator of the superconformal group, and $\CL_{N_I}$ is the Lie 
derivative\footnote{Note that the Lie
derivative is defined as in the purely even setting. In particular,
$\CL_X f:=Xf$ and $\CL_X Y:=[X,Y\}$, where $f$ is some local function
and $X,Y$ are local vector fields. Furthermore, one requires that
$\CL_X$ commutes with the contraction operator. 
For any differential one-form $\o=\dt Z^I\o_I$,
one may readily verify the following result:
$$ Y\lrcorner\CL_X\o\ =\ (-)^{p_Xp_Y}X(Y\lrcorner\o)+[Y,X\}\lrcorner\o.$$
Letting $Y$ be $\partial_I=\partial/\partial Z^I$, one sets 
$(\CL_X\o)_I:=\CL_X\o_I:=(-)^{p_Xp_I}\partial_I\lrcorner\CL_X\o$.
The extra sign has been chosen for convenience.}
 along the vector field $N_I$.
Explicitly, Eqs. \eqref{infliea} read as
\eqna{inflieai}{
    \CL_{N_I}\CA_{\a\ad}\ &=\ N_I\CA_{\a\ad}+\CA_{\b\bd}\partial_{\a\ad}^R
                              N_I^{\b\bd}+(-)^{p_I+1}\CA^i_\bd
                              \partial_{\a\ad}^RN^\bd_{I\,i},\\
    \CL_{N_I}\CA_\ad^i  \ &=\ N_I\CA^i_\ad+(-)^{p_I}\CA_{\b\bd}
                              \partial^i_\ad N_I^{\b\bd}+\CA^j_\bd
                              \partial^i_\ad N^\bd_{I\,j}.
}
It is not too difficult to show that for any generator $N_I$ as given in
\eqref{scgen}, the transformations \eqref{infliea} together with 
\eqref{inflieai} give a symmetry
of the $\CN$-extended self-dual SYM equations \eqref{CE-2}.

So far, we have given the action of the superconformal group on the 
components of the gauge potential on Euclidean superspace. 
The linear system \eqref{linsys-22} is,
however, defined on the supertwistor space $\CP^{3|\CN}$. Therefore, 
the question is how to lift the action of the superconformal group to
supertwistor space but at the same time preserving the linear system 
\eqref{linsys-22}. The answer is at hand.
By our discussion given in Sec. \ref{sec:HandS-ch5}, 
we have to preserve the complex structure on $\CP^{3|\CN}$.
 Recall the diffeomorphism
$\CP^{3|\CN}\cong\IR^{4|2\CN}\times S^2$. 
Complex structures on the body $\IR^{4|0}$ of 
$\IR^{4|2\CN}$ are parametrized by a two-sphere $S^2\cong SO(4)/U(2)$. 
The latter can be viewed as the complex projective line $\IC P^1$ 
parametrized by the coordinates $\l_\pm$. Then a complex structure 
$\CJ=(\CJ_{\a\ad}^{~~~\b\bd})$ on $\IR^{4|0}$, compatible with \eqref{E-INV}
and \eqref{xdefeuc-1}, is given by
\eqn{complstronr}{\CJ_{\a\ad}^{~~~\b\bd}\ =\ -\i\g_\pm\d^\b_\a
                 (\l^\pm_\ad\hl_\pm^\bd+\l_\pm^\bd\hl^\pm_\ad),}
where $\g_\pm=(1+\l_\pm\bl_\pm)^{-1}$.
Notice that in the present case the corresponding K\"ahler two-form 
$\o$ is anti-self-dual, i.e., its components are of the form 
$\o_{\a\ad\b\bd}=\epsilon_{\a\b}\CJ_{\ad\bd}$. If we had chosen the 
$\CN$-extended anti-self-dual SYM equations from the very
beginning, the K\"ahler form would have been self-dual.
On the two-sphere $S^2$ which parametrizes the different complex 
structures of $\IR^{4|0}$, we introduce the standard complex structure 
$\fJ$ which, for instance, on the $U_+$ patch is given by 
$\fJ_{\l_+}^{~~\l_+}=\i=-\fJ_{{\bar\l}_+}^{~~{\bar\l}_+}$. Thus,
the complex structure on the body $\IR^{4|0}\times S^2$ of supertwistor
space $\CP^{3|\CN}$ can be taken as $J=(\CJ,{\frak J})$.

Having introduced a complex structure on the even part 
$\IR^{4|0}\times S^2$ of $\IR^{4|2\CN}\times S^2$, 
we need to extend the above discussion to the full supertwistor space. In order
to define a complex structure on $\IR^{4|2\CN}\times S^2$, recall that only 
an even amount of supersymmetries is possible, i.e., $\CN=0,2$ or $4$. Our 
particular choice of the 
symplectic Majorana condition induced by \eqref{EI-xandeta} allows us to introduce a complex 
structure on the odd part $\IR^{0|2\CN}$ similar to \eqref{complstronr}, that is,
\eqn{complstronri}{{\bf J}_{i~\bd}^{\ad~j}\ =\ -\i\g_\pm\d^j_i
                  (\l^\pm_\bd\hl_\pm^\ad+\l_\pm^\ad\hl^\pm_\bd).}
Therefore, $J=(\CJ,{\bf J},{\frak J})$ will be the proper choice of a complex
structure\footnote{Clearly, this choice of the complex structure does not exhaust  
the space of all admissible complex structures. However, in the present case it
is enough to restrict ourselves to this class of complex structures. For more
details, see \cite{Wolf:2004hp}.}
on supertwistor space.

We can now answer the initial question: the generators $N_I$ 
given by \eqref{scgen} of the superconformal group should be lifted 
to vector fields 
$\tN_I=\tN_I^{\a\ad}\partial_{\a\ad}+\tN^\ad_{I\,i}\partial^i_\ad+
\tN_I^{\l_\pm}\partial_{\l_\pm}+\tN_I^{{\bar\l}_\pm}\partial_{{\bar\l}_\pm}$ 
on supertwistor space such that the Lie derivative of the complex 
structure $J$ along the lifted vector fields $\tN_I$ vanishes, i.e.,
\eqn{cononcompstr}{\CL_{\tN_I} J\ =\ 0.} 
Letting $\CJ_\ad^{~\bd}={\g_\pm\over 2\i}(\l^\pm_\ad\hl_\pm^\bd+
\l_\pm^\bd\hl^\pm_\ad)$, we can write \eqref{cononcompstr} explicitly as
\eqna{cononcompstri}{
     2\,\tN_I\CJ_\ad^{~\bd}+\CJ_\gd^{~\ad}\partial_{\a\bd}
     \tN_I^{\a\gd}-\CJ_\bd^{~\gd}\partial_{\a\gd}\tN_I^{\a\ad}\ &=\ 0,\\
     \CJ_\dd^{~\gd}\partial_{\a\ad}\tN^\dd_{I\,i}-\CJ_\ad^{~\bd}
     \partial_{\a\bd}\tN^\gd_{I\,i}\ &=\ 0,\\
     \CJ_\ad^{~\bd}\partial^i_\gd\tN_I^{\a\ad}-\CJ_\gd^{~\dd}\partial^i_\dd
     \tN^{\a\bd}_I\ &=\ 0,\\     
     \d^i_j\,\tN_I\CJ_\ad^{~\bd}-(-)^{p_I}\CJ_\ad^{~\gd}\partial^j_\gd
     \tN^\bd_{I\,i}
     +(-)^{p_I}\CJ_\gd^{~\bd}\partial^j_\bd\tN^\gd_{I\,i}\ &=\ 0,
}
whereas the equations involving $\fJ$ imply that the components
$\tN^{\l_\pm}_I$ and $\tN^{{\bar\l}_\pm}_I$ are holomorphic in $\l_\pm$ and
${\bar\l_\pm}$, respectively. The final expressions for the generators  \eqref{scgen}
lifted to supertwistor space $\CP^{3|\CN}$ and obeying \eqref{cononcompstri} are
\eqna{scgenlifted}{
           \tP_{\a\ad}\ &=\ P_{\a\ad},\qquad
           \tQ_{i\a}\ =\  Q_{i\a},
           \qquad
           \tQ^i_\ad\ =\ Q^i_\ad,\\
           \tD\ &=\ D,\\
           \tK^{\a\ad}\ &=\ K^{\a\ad}+\tfrac{1}{4}x^{\a\bd}Z_\bd^{~\ad},\qquad
           \tK^{i\a}\ =\ K^{i\a},\qquad
           \tK^\ad_i\ =\ K^\ad_i+\eta^\bd_i Z_\bd^{~\ad},\\
           \tJ_{\a\b}\ &=\ J_{\a\b},\qquad
           \tJ_{\ad\bd}\ =\ J_{\ad\bd}-\tfrac{1}{2}Z_{\ad\bd},\\
           \widetilde{R}_i^j\ &=\ R_i^j,\qquad
           \tA\ =\ A,
}
where 
\eqn{scgenliftedi}{Z_{\ad\bd}\ :=\ \pm\l^\pm_\ad\l^\pm_\bd\partial_{\l_\pm}
                  \pm\hl^\pm_\ad\hl^\pm_\bd\partial_{{\bar\l}_\pm}.}

Now we can give the infinitesimal transformation of 
$\psi=\{\psi_+,\psi_-\}\in C^0(\fU,\fP)$ which participates in 
\eqref{linsys-22}
\eqn{infliepsi}{\d_{\tN_I}\psi_\pm\ =\ \CL_{\tN_I}\psi_\pm\ =\ \tN_I\psi_\pm,}
where $\tN_I$ is any of the generators given in \eqref{scgenlifted}. 
It is a 
straightforward exercise to verify explicitly that the linear system 
\eqref{linsys-22} 
is invariant under the transformations \eqref{infliea} and 
\eqref{infliepsi}.

Having collected all necessary ingredients, we can now
start with constructing
affine symmetry algebras related to superconformal symmetries. In
order to keep formulas simple, we shall in the remainder of
this section work in the complexified gauge algebra, that is,
we drop the last point from the list given at the beginning
of Sec. \ref{sec:HandS-ch5}. It should
be stressed, however, that the subsequent derivation can be
modified for a real gauge algebra without any problems.
Let $\tN_I$ be any vector field of \eqref{scgenlifted}, with
\eqn{}{[\tN_I,\tN_J\}\ =\ {f_{IJ}}^K\tN_K,}
where the ${f_{IJ}}^K$s are the structure constants of the 
superconformal group. Then we define the following perturbation
of $f=\{f_{+-}\}$:
\eqn{affineEV}{\d^m_I f_{+-}\ :=\ \l_+^m\tN_I f_{+-}
               \qquad{\rm for}\qquad m\in\IN_0.}
Note that the antiholomorphic $\l$-derivative appearing in
$\tN_I$ drops out as $f_{+-}$ is holomorphic. Therefore,
Eq. \eqref{affineEV} defines a zero-cochain 
$\chi=\{\chi_+,\chi_-\}\in C^0(\fU,T\CP^{3|\CN})$ given 
by\footnote{\label{foot:rc}Note that for a real gauge algebra, $\chi_-$
 would not be zero but instead be given by
$\chi_-=\overline{\tau_E(\chi_+)}$,
where $\tau_E$ represents the antiholomorphic involution
\eqref{E-INV} corresponding to Euclidean signature.}
$$ \chi_+\ =\ \l_+^m\tN_I|_{T\CP^{3|\CN}}\qquad{\rm and}\qquad 
   \chi_-\ =\ 0. $$
Furthermore, one may readily check that
\eqn{affineEVI}{[\d_I^m,\d_J^n\}\ =\ \left({f_{IJ}}^K+n\,g_I\,\d_J^K-
               (-)^{p_Ip_J}m\,g_J\,\d^K_I\right)\d_K^{m+n}}
upon action on $f_{+-}$. Here, we have introduced the shorthand
notation
\eqn{affineEIII}{g_I\ :=\ \l_+^{-1}\tN^{\l_+}_I.}
Generally speaking, \eqref{affineEVI} can be seen as a centerless
Kac-Moody-Virasoro-type algebra. 

Next we need to find the action of $\d^m_I$ on the 
components $\CA_{\a\ad}$ and $\CA^i_\ad$. The derivation
\eqref{eq:adeform-ch5} gets modified according to
\eqn{eq:adeform2-ch5}{ 
      \phi^m_{+-I}\ =\ \phi^m_{+I}-\phi^m_{-I}
                  \ =\ \l_+^m\phi^0_{+I}-\l_+^m\phi^0_{-I},
}
where this time
\eqn{affineEIX}{\phi^0_{\pm I}\ =\ -(\tN_I\psi_\pm)\psi_\pm^{-1}+
          (\tN_I\psi_+)\psi_+^{-1}|_{\l_+=0}.} 
Clearly, the second term of this expression is $\l$-independent,
i.e., it belongs to $H^0(\CP^{3|\CN},{\rm Lie}\,\fP)$. In fact, we have simply 
shifted all the ``zero-modes" into $\phi_{-I}^0$. Recall 
that such shifts are always possible and eventually result in 
gauge transformations. Upon substituting
\eqref{affineEIX} into 
\eqref{infpwtrans} and using Eq. \eqref{phiexpKMY}, we 
find\footnote{Note that $\d^0_I$ is a composition of a superconformal
and a particular gauge transformation.}
\eqna{perturbationAKMYpi-ch5}{
     \d^0_I\CA_{\a{\dot1}}\ &=\ \nabla_{\a{\dot1}}\phi^{0(0)}_{-I}
                             \ =\ - \nabla_{\a{\dot2}}\phi^{0(1)}_{+I},\\
     \d^0_I\CA_{\a{\dot2}}\ &=\ -\nabla_{\a{\dot1}}\phi^{0(1)}_{-I}
                                +\nabla_{\a{\dot2}}\phi^{0(0)}_{-I}
                            \ =\ 0,
}
since $\phi^{0(0)}_+=0$. Similar expressions hold for $\CA^i_\ad$.
Eq. \eqref{eq:adeform2-ch5} yields, after expanding $\phi_{\pm I}^m$ 
in powers of $\l_\pm$ and comparing with $\phi_{\pm I}^0$, the
following coefficient functions: 
\eqn{phiexpKMY2}{\phi_{+I}^{m(n)}\ =\ \begin{cases}
                       0&{\rm for}\quad n=0\\
                       \phi^{0(n-m)}_{+I}-\phi^{0(m-n)}_{-I} &{\rm for}\quad n>0
                        \end{cases}
              \qquad{\rm and}\qquad
               \phi_{-I}^{m(n)}\ =\ \phi_{-I}^{0(m+n)}.
               }
Note that $\phi^{0(-n)}_{\pm I}=0$ for $n>0$. 
These coefficients together with the integral formulas \eqref{infpwtrans}  
give the action of $\d^m_I$ on $\CA_{\a\ad}$ and 
$\CA_\ad^i$ according to
\eqna{perturbationAKMYpi3}{
     \d^m_I\CA_{\a{\dot1}}\ &=\ \nabla_{\a{\dot1}}\phi^{m(0)}_{-I}
                           \ =\ -   \nabla_{\a{\dot2}}\phi^{m(1)}_{+I},\\      
     \d^m_I\CA_{\a{\dot2}}\ &=\ -\nabla_{\a{\dot1}}\phi^{m(1)}_{-I}
                                +\nabla_{\a{\dot2}}\phi^{m(0)}_{-I}
                           \ =\ 0,
}
and similarly for $\CA^i_\ad$.

Next we are interested in the underlying algebraic structure. Thm. 
\ref{thm:symtrafo-ch5} is not directly applicable, as we now have to deal
with structure functions. Recall that the $g_I$s appearing in
\eqref{affineEVI} are holomorphic functions on $\CU_+\cap\CU_-$. 
However, there are only minor modifications to be made. In proving
Thm. \ref{thm:symtrafo-ch5}, we have introduced the functions
$\Sigma^{mn}_{\pm IJ}$.
Their difference is now given by
\eqn{eq:bla-ch5}{\Sigma^{mn}_{+IJ}-\Sigma^{mn}_{-IJ}\ =\ {C_{IJ}}^K 
        (\phi^{m+n}_{+K}-\phi^{m+n}_{-K}),}
where
\eqn{}{{C_{IJ}}^K\ :=\ {f_{IJ}}^K+n\,g_I\,\d_J^K-
               (-)^{p_Ip_J}m\,g_J\,\d^K_I.}
In order to determine $\Sigma^{mn}_{\pm IJ}$, 
one expands both sides of Eq. \eqref{eq:bla-ch5} in powers
of $\l_+$ and compares coefficients. We are only interested
in $\Sigma^{mn(0)}_{\pm IJ}$ and $\Sigma^{mn(1)}_{\pm IJ}$,
as they determine the action of $[\d^m_I,\d^n_J\}$ onto
the components of the gauge potential. We find
\eqna{}{\Sigma^{mn(0)}_{\pm IJ}\ &=\ {C_{IJ}}^{K(0)}\phi^{m+n(0)}_{\pm K}
        +{C_{IJ}}^{K(\mp1)}\phi^{m+n(1)}_{\pm K}-
         {C_{IJ}}^{K(-1)}\phi^{m+n(1)}_{+K},\\                     
        \Sigma^{mn(1)}_{\pm IJ}\ &=\ {C_{IJ}}^{K(\pm1)}(\phi^{m+n(0)}_{\pm K}
         -\phi^{m+n(0)}_{\mp K})+{C_{IJ}}^{K(0)}\phi^{m+n(1)}_{\pm K}+
         {C_{IJ}}^{K(1)}\phi^{m+n(2)}_{\mp K},       
}
where ${C_{IJ}}^{K}=\sum_{k=-1}^1\l_+^k{C_{IJ}}^{K(k)}$ has been inserted.
Note that last term in the first line represents again a gauge transformation.
It has been adjusted such that $\Sigma^{mn(0)}_{+ IJ}=0$. 
By virtue of \eqref{infpwtrans} and \eqref{phiexpKMY2}, we finally arrive at
\eqna{perturbationAKMYpi4}{
     {[\d^m_I,\d^n_J\}}\CA_{\a{\dot1}}\ &=\ \nabla_{\a{\dot1}}
             \sum_{k=-1}^1{C_{IJ}}^{K(k)}\phi_{-K}^{0(m+n+k)},\\
     {[\d^m_I,\d^n_J\}}\CA_{\a{\dot2}}\ &=\ 0
}
and similarly for $\CA^i_\ad$. In deriving this result, the explicit
dependence of ${C_{IJ}}^K$ on $m$ and $n$ has been used. Obviously, the
algebra closes if and only if the coefficient functions are independent
of $(x_R^{\a\ad},\eta^\ad_i)$ --
a fact which we have already encountered at the end of Sec. 
\ref{sec:HandS-ch5}. Therefore, we are left with only a subalgebra
of the superconformal algebra. Indeed, we need to exclude the generators
$\tK^{\a\ad}$ and $\tK^\ad_i$ of special conformal transformations.
Let us consider the maximal subalgebra of the superconformal
algebra which contains neither $\tK^{\a\ad}$ nor $\tK^\ad_i$. Furthermore,
let ${h_{IJ}}^K$ denote the corresponding structure constants. 
Then we end up with 
\eqn{}{[\d^m_I,\d^n_J\}\ =\ {h_{IJ}}^K\d^{m+n}_K+\sum_{k=-1}^1
         \left(ng^{(k)}_I\d_J^K-(-)^{p_Ip_J}mg_J^{(k)}\d^K_I\right)
           \d^{m+n+k}_K,}
where $g_I=\sum_{k=-1}^1\l_+^kg_I^{(k)}$
is the expansion of $g_I$ defined in \eqref{affineEIII}.

\section{Hierarchies}\label{sec:hierarchies-ch5}

In Refs. \cite{Mason,Takasaki,Mason:1992vd,Ablowitz,MasonRF,IvanovaZT} it was shown
that a solution to the self-dual YM equations can be embedded 
into an infinite-parameter family of new solutions by moving
it along commuting flows of a so-called self-dual YM
hierarchy. The lowest generators of the latter are the generators
of space-time translations. Putting it differently, the self-dual 
YM hierarchy describes infinitely many Abelian symmetries
of the self-dual YM equations associated with translational
symmetries. Our next topic is the generalization of these
ideas to the self-dual SYM equations. For the sake of clarity,
we shall again work in a complex setting, that is, we consider
the double fibration \eqref{DF-21}. If desired, reality
conditions can be imposed.

\paragraph{Generalized supertwistor space.}
Consider the Abelian subalgebra of the superconformal algebra 
which is spanned by the translation
generators $\tP_{\a\ad}$ and $\tQ^i_\ad$. On supertwistor space, we
may use the coordinates $z^\a_\pm=x^{\a\ad}_R\l^\pm_\ad$, 
$\pi^\pm_\ad=\l^\pm_\ad$ and $\eta^\pm_i=\eta^\ad_i\l^\pm_\ad$. 
Expressing the generators $\tP_{\a\ad}$ and $\tQ^i_\ad$ in terms 
of these coordinates, we obtain
\eqn{indactp}{\tP_{\a\ad}\ =\ 
              \pi^\pm_\ad\der{z^\a_\pm}
             \qquad{\rm and}\qquad
             \tQ^i_{\ad}\ =\ 
             \pi^\pm_\ad\der{\eta^\pm_i},}
when acting on holomorphic functions of 
$(z^\a_\pm,\pi^\pm_\ad,\eta_i^\pm)$ on $\CP^{3|\CN}$. Then
we may define local holomorphic
vector fields $\tP_{\a\ad_1\cdots\ad_{m_\a}}^\pm$ and 
$\tQ^i_{\pm\ad_1\cdots\ad_{n_i}}$ by
\eqn{defofps}{\tP^\pm_{\a\ad_1\cdots\ad_{m_\a }}\ :=\ 
               \pi^\pm_{\ad_1}\cdots\pi^\pm_{\ad_{m_\a }}
             \der{z^\a_\pm}\qquad{\rm and}\qquad
             \tQ^i_{\pm\ad_1\cdots\ad_{n_i}}\ :=\ 
             \pi^\pm_{\ad_1}\cdots\pi^\pm_{\ad_{n_i}}
              \der{\eta^\pm_i},}
for $m_\a,n_i\in\IN$. Clearly, the vector fields
$\tP^\pm_{\a\ad_1\cdots\ad_{m_\a}}$ and 
$\tQ^i_{\pm\ad_1\cdots\ad_{n_i}}$ are totally symmetric under an 
exchange of their dotted indices. As before,
we define a perturbation
of the transition function $f=\{f_{+-}\}$ of a smoothly
trivial holomorphic vector bundle $\CE\to\CP^{3|\CN}$ which is
trivial along $\IC P^1_{x_R,\eta}\hookrightarrow\CP^{3|\CN}$
according to\footnote{Clearly, such a deformation does
not preserve the reality condition 
$f_{+-}(\cdots)=[f_{+-}(\tau_E(\cdots))]^\dagger$. See also
footnote \ref{foot:rc}.}
\eqn{defoftrafosi}{f_{+-}\ \mapsto\ \tP^\pm_{\a\ad_1\cdots\ad_{m_\a }}f_{+-}
                   \qquad{\rm and}\qquad
                   f_{+-}\ \mapsto\ \tQ^i_{\pm\ad_1\cdots\ad_{n_i}}f_{+-}.}
Next one could pull back $f'=f+\d f$ to the correspondence
space and solve the corresponding infinitesimal
Riemann-Hilbert problem (which we have already done in the
preceding section) to construct the symmetry 
transformations of the components $\CA_{\a\ad}$ and
$\CA_\ad^i$ of the gauge potential. We shall, however,
proceed differently and instead consider the following
dynamical system:
\eqna{dynsystrans}{
                 \der{t^{\a\ad_1\cdots\ad_{m_\a }}}f_{+-}\ 
                 &=\ \d_{\a\ad_1\cdots\ad_{m_\a }}f_{+-}\ =\  
                  \pi^\pm_{\ad_1}\cdots\pi^\pm_{\ad_{m_\a }}
                 \der{z^\a_\pm}f_{+-},\\
                 \der{\xi^{\ad_1\cdots\ad_{n_i}}_i}f_{+-}\ 
                 &=\ \d^i_{\ad_1\cdots\ad_{n_i }}f_{+-}\ =\  
                   \pi^\pm_{\ad_1}\cdots\pi^\pm_{\ad_{n_i}}
                 \der{\eta^\pm_i}f_{+-},
}
where $t^{\a\ad_1\cdots\ad_{m_\a }}$ and $\xi^{\ad_1\cdots\ad_{n_i}}_i$ are 
parameters. These equations can easily be
solved. In fact, the solution to \eqref{dynsystrans} reads as
\eqn{dynsystransi}{f_{+-}\ =\ f_{+-}(z^\a_\pm+ t^{\a\ad_1\cdots\ad_{m_\a }}
                 \l^\pm_{\ad_1}\cdots\l^\pm_{\ad_{m_\a }},\pi^\pm_\ad=\l^\pm_\ad,
                  \eta^\pm_i+\xi^{\ad_1\cdots\ad_{n_i}}_i
                  \l^\pm_{\ad_1}\cdots\l^\pm_{\ad_{n_i}}).}
Note that any point of $\IC^{4|2\CN}$ can be obtained by a shift of the
origin and hence we may put, without loss of generality, $x^{\a\ad}_R$ and 
$\eta^\ad_i$ to zero. Therefore, \eqref{dynsystransi} simplifies to
\eqn{dynsystransii}{f_{+-}\ =\ f_{+-}(t^{\a\ad_1\cdots\ad_{m_\a }}
                  \l^\pm_{\ad_1}\cdots\l^\pm_{\ad_{m_\a }},\l^\pm_\ad,
                  \xi^{\ad_1\cdots\ad_{n_i}}_i
                  \l^\pm_{\ad_1}\cdots\l^\pm_{\ad_{n_i}}),}
where now 
$$(t^{\a{\dot1}\cdots{\dot1}},t^{\a{\dot2}{\dot 1}\cdots{\dot1}},
\xi^{{\dot1}\cdots{\dot1}}_i,\xi^{{\dot2}{\dot 1}\cdots{\dot1}}_i)$$ 
are interpreted as coordinates on the anti-chiral superspace $\IC^{4|2\CN}$
whereas the others are additional moduli sometimes also referred to
as  ``higher times''. 

For finite sums in \eqref{dynsystransii}, $m_\a,n_i<\infty$, the polynomials
\eqn{defofpol}{z^\a_\pm\ =\ t^{\a\ad_1\cdots\ad_{m_\a }}
              \l^\pm_{\ad_1}\cdots\l^\pm_{\ad_{m_\a }}\qquad{\rm and}\qquad
              \eta_i^\pm\ =\  \xi^{\ad_1\cdots\ad_{n_i}}_i
              \l^\pm_{\ad_1}\cdots\l^\pm_{\ad_{n_i}}}
can be regarded as holomorphic sections of the bundle
\eqn{}{\CO_{\IC P^1}(m_1)\oplus\CO_{\IC P^1}(m_2)\oplus\bigoplus_{i=1}^\CN
  \Pi\CO_{\IC P^1}(n_i)\ \to\ \IC P^1.}
In the following,
we shall call this space generalized supertwistor space and denote it 
by $\CP^{3|\CN}_{m_1,m_2|n_1,\ldots,n_\CN}$. Note that it
can be understood as an open subset in the weighted projective superspace
$W\IC P^{3|\CN}[m_1,m_2,1,1|n_1,\ldots,n_\CN]$,
$$\begin{aligned}\CP^{3|\CN}_{m_1,m_2|n_1,\ldots,n_\CN}\ &=\ \\
       &\kern-4.5cm =\ W\IC P^{3|\CN}[m_1,m_2,1,1|n_1,\ldots,n_\CN]\setminus
       W\IC P^{1|\CN}[m_1,m_2|n_1,\ldots,n_\CN].\end{aligned}$$
See also our discussion presented in Sec. \ref{sec:OSDM-ch2}. 
Thus, $W\IC P^{3|\CN}[m_1,m_2,1,1|n_1,\ldots,n_\CN]$ is a 
compactified version of $\CP^{3|\CN}_{m_1,m_2|n_1,\ldots,n_\CN}$.
Moreover,
for the particular combination
$$m_1+m_2-(n_1+\cdots+n_\CN)+2\ =\ 0,$$
it becomes a formal Calabi-Yau supermanifold. As we shall discuss
in detail below, 
\eqref{dynsystransii} can then be interpreted as transition function of
a holomorphic vector bundle 
$\CE$ over $\CP^{3|\CN}_{m_1,m_2|n_1,\ldots,n_\CN}$.

\paragraph{Penrose-Ward transform.}
Like supertwistor space, also generalized supertwistor space is a
part of a double fibration. For notational
convenience, let us denote the generalized supertwistor space
by $\CP^{3|\CN}_{m,n}$. Furthermore, $H^0(\IC P^1,\CP^{3|\CN}_{m,n})\cong\IC^{M|N}$,
where
$M:=m_1+m_2+2$ and $N:=n_1+\cdots +n_\CN+2\CN$. Then
we find
\eqna{eq:DF1-ch5}{
 \begin{picture}(50,40)
  \put(0.0,-2.0){\makebox(0,0)[c]{$\CP_{m,n}^{3|\CN}$}}
  \put(64.0,-2.0){\makebox(0,0)[c]{$\IC^{M|N}$}}
  \put(34.0,35.0){\makebox(0,0)[c]{$\CF^{M+1|N}_R$}}
  \put(7.0,18.0){\makebox(0,0)[c]{$\pi_1$}}
  \put(55.0,18.0){\makebox(0,0)[c]{$\pi_2$}}
  \put(25.0,25.0){\vector(-1,-1){18}}
  \put(37.0,25.0){\vector(1,-1){18}}
 \end{picture}
}
where the correspondence space is again a direct product
$\CF^{M+1|N}_R\cong
  \IC^{M|N}\times\IC P^1.$
The natural choice of coordinates on $\CF^{M+1|N}_R$ is
\eqn{defofcoordsonets}{(x_R^{\a\ad_1\cdots\ad_{m_\a}},\l^\pm_\ad,
                      \eta_i^{\ad_1\cdots\ad_{n_i}}).}
Generalized supertwistor space $\CP^{3|\CN}_{m,n}$ can be 
covered by two coordinate patches, $\fU=\{\CU_+,\CU_-\}$, and 
equipped with local coordinates $(z^\a_\pm,z^3_\pm,\eta_i^\pm)$.
On the intersection $\CU_+\cap\CU_-$, they are related by
\eqn{}{z^\a_+\ =\ \frac{1}{(z^3_-)^{m_\a}}z^\a_-,\qquad
       z^3_+\ =\ \frac{1}{z^3_-},\qquad{\rm and}\qquad
       \eta^+_i\ =\ \frac{1}{(z^3_-)^{n_i}}\eta^-_i.}
Hence, the projections $\pi_1$ and $\pi_2$ in the fibration
\eqref{eq:DF1-ch5} act as follows: 
\eqna{defofpisdh}{
              &\pi_1\,:\,(x^{\a\ad_1\cdots\ad_{m_\a}}_R,\l^\pm_\ad, 
                \eta_i^{\ad_1\cdots\ad_{n_i}})\ \mapsto\ 
             (z^\a_\pm=x_R^{\a\ad_1\cdots\ad_{m_\a}}\l^\pm_{\ad_1}
                      \cdots\l^\pm_{\ad_{m_\a}},z^3_\pm=\l_\pm,\\ 
               &\kern7cm\eta_i^\pm=\eta_i^{\ad_1\cdots\ad_{n_i}}\l^\pm_{\ad_1}
                      \cdots\l^\pm_{\ad_{n_i}}),\\
            &\pi_2\,:\,(x^{\a\ad_1\cdots\ad_{m_\a}}_R,\l^\pm_\ad, 
                       \eta_i^{\ad_1\cdots\ad_{n_i}})\ \mapsto\
             (x^{\a\ad_1\cdots\ad_{m_\a}}_R, \eta_i^{\ad_1\cdots\ad_{n_i}}).
}
By virtue of these projections, we obtain the following proposition:
{\Pro There exist the following geometric correspondences:
\vspace*{-2mm}
\begin{center} 
 \begin{tabular}{cccc}
   {\rm (i)} & point $p$ in $\CP^{3|\CN}_{m,n}$ & $\quad\longleftrightarrow\quad$ & 
                                    $\IC^{M-2|N-\CN}_p$ in $\IC^{M|N}$\\
   {\rm (ii)} & $\IC P^1_{x_R,\eta}\hookrightarrow\CP^{3|\CN}_{m,n}$
               &$\quad\longleftrightarrow\quad$ & 
                          point $(x_R,\eta)$ in $\IC^{M|N}$
 \end{tabular}
\end{center}}\vskip 4mm

Let $\CE$ be a rank $r$ holomorphic vector bundle over $\CP^{3|\CN}_{m,n}$ 
and $\pi_1^*\CE$ be the pull-back of $\CE$ to the correspondence space 
$\CF^{M+1|N}_R$. The covering of the latter is 
denoted by $\hfU=\{\hCU_+,\hCU_-\}$.
These bundles are defined by transition functions\footnote{Here, we again use 
the same letter $f$ for both bundles.} $f=\{f_{+-}\}$ which are annihilated
by the vector fields
\eqn{defofvfsdh}{
                D^\pm_{\a\ad_1\cdots\ad_{m_\a-1}}\ =\ 
                \l^\ad_\pm\partial_{\a(\ad\ad_1\cdots\ad_{m_\a-1})}^R\quad{\rm and}\quad
                D^i_{\pm\ad_1\cdots\ad_{n_i-1}}
                 \ =\ \l^\ad_\pm\partial^i_{(\ad\ad_1\cdots\ad_{n_i-1})}
}
as they freely generate the relative tangent sheaf
$(\O^1(\CF^{M+1|N}_R)/\pi_1^*\O^1(\CP^{3|\CN}_{m,n}))^*$. The
derivatives $\partial_{\a(\ad_1\cdots\ad_{m_\a})}^R$ and
$\partial^i_{(\ad_1\cdots\ad_{n_i})}$ are defined according to
\eqna{}{\partial_{\a(\ad_1\cdots\ad_{m_\a})}^R\,x_R^{\b\bd_1\cdots\bd_{m_\b}}
       \ &:=\ \d_\a^\b\d_{(\ad_1}^{\bd_1}\cdots\d_{\ad_{m_\a})}^{\bd_{m_\b}},\\
       \partial^i_{(\ad_1\cdots\ad_{n_i})}\,\eta_j^{\bd_1\cdots\bd_{n_j}}\ &:=\   
         \d_j^i\d_{(\ad_1}^{\bd_1}\cdots\d_{\ad_{n_i})}^{\bd_{n_j}}
}
and can be understood as generalizations of \eqref{eq:someder_ch3}.

The requirement of smooth triviality of the bundle 
$\CE\to\CP_{m,n}^{3|\CN}$
allows us to split the transition function $f_{+-}$ according to
\eqn{eqforfbig}{f_{+-}\ =\ \psi^{-1}_+\psi_-,}
whereas $\IC^{M|N}$-triviality
ensures that there exists a $\psi=\{\psi_+,\psi_-\}$ which
belongs to $C^0(\hfU,\fP)$. Therefore, we may introduce a
Lie-algebra valued one-form such that
\eqna{defofasdh}{
          \CA^\pm_{\a\ad_1\cdots\ad_{m_\a-1}}\  &:=\ 
                \l^\ad_\pm\CA_{\a\ad\ad_1\cdots\ad_{m_\a-1}} \ = \ 
                 \psi_\pm D^\pm_{\a\ad_1\cdots\ad_{m_\a-1}}
                    \psi_\pm^{-1},\\
             \CA^i_{\pm\ad_1\cdots\ad_{n_i-1}}\ &:=\ 
             \l_\pm^\ad\CA^i_{\ad\ad_1\cdots\ad_{n_i-1}}\ =\  
               \psi_\pm D^i_{\pm\ad_1\cdots\ad_{n_i-1}}\psi_\pm^{-1},
} 
and therefore
\eqna{linsyssdh}{
   (D^\pm_{\a\ad_1\cdots\ad_{m_\a-1}}+
              \CA^\pm_{\a\ad_1\cdots\ad_{m_\a-1}})\psi_\pm\ &=\ 0,\\
            ( D^i_{\pm\ad_1\cdots\ad_{n_i-1}}+
               \CA^i_{\pm\ad_1\cdots\ad_{n_i-1}})\psi_\pm\ &=\ 0. 
} 
We note that for $m_\a=n_i=1$ this system reduces, of
course, to the old one given by \eqref{linsys-2}. 
Moreover, we have the following symmetry properties
\eqn{soioa}{\CA_{\a\ad\ad_1\cdots\ad_{m_\a-1}}\ =\ 
           \CA_{\a\ad(\ad_1\cdots\ad_{m_\a-1})}\quad{\rm and}\quad
           \CA^i_{\ad\ad_1\cdots\ad_{n_i-1}}\ =\ 
           \CA^i_{\ad(\ad_1\cdots\ad_{n_i-1})}.}
The compatibility conditions for \eqref{linsyssdh} read as
\eqna{compconsdh}{
             {[ D_{\a\ad\ad_1\cdots\ad_{m_\a-1}}^R,
              D_{\b\bd\bd_1\cdots\bd_{m_\b-1}}^R]}+
             [ D_{\a\bd\ad_1\cdots\ad_{m_\a-1}}^R,
              D_{\b\ad\bd_1\cdots\bd_{m_\b-1}}^R]\ &=\ 0,\\
             [ D^i_{\ad\ad_1\cdots\ad_{n_i-1}},
              D_{\b\bd\bd_1\cdots\bd_{m_\b-1}}^R]+
             [ D^i_{\bd\ad_1\cdots\ad_{n_i-1}},
              D_{\b\ad\bd_1\cdots\bd_{m_\b-1}}^R]\ &=\ 0,\\
             \{ D^i_{\ad\ad_1\cdots\ad_{n_i-1}},
              D^j_{\bd\bd_1\cdots\bd_{n_j-1}}\}+
             \{ D^i_{\bd\ad_1\cdots\ad_{n_i-1}},
              D^j_{\ad\bd_1\cdots\bd_{n_j-1}}\}\ &=\ 0.
}
Here, we have defined the first order differential operators
\eqna{defofcovdersdh}{
                  D_{\a\ad\ad_1\cdots\ad_{m_\a-1}}^R\ &:=\ 
                 \partial_{\a(\ad\ad_1\cdots\ad_{m_\a-1})}^R+
                 \CA_{\a\ad\ad_1\cdots\ad_{m_\a-1}},\\
                  D^i_{\ad\ad_1\cdots\ad_{n_i-1}}\ &:=\ 
                 \partial^i_{(\ad\ad_1\cdots\ad_{n_i-1})}+
                 \CA^i_{\ad\ad_1\cdots\ad_{n_i-1}}.
}
We remark that the components 
$$\CA_{\a\ad{\dot1}\cdots{\dot1}}\qquad{\rm and}\qquad
  \CA^i_{\ad{\dot1}\cdots{\dot1}}$$ 
coincide with the components $\CA_{\a\ad}$ and $\CA^i_\ad$ of the gauge 
potential on $\IC^{4|2\CN}$. In the sequel, we
shall refer to \eqref{compconsdh} as the truncated $\CN$-extended self-dual SYM
hierarchy of level $(m_1,m_2|n_1,\ldots,n_\CN)$.
The full hierarchy is then obtained by taking the 
limit $m_\a,n_i\to\infty$. 

From Eqs. \eqref{defofasdh} it follows that
\eqna{pwtranssdh}{
             \CA_{\a\ad\ad_1\cdots\ad_{m_\a-1}}\ &=\ 
             {1\over 2\pi\i}\oint_\cC{\rm d}\l_+
             {\CA^+_{\a\ad_1\cdots\ad_{m_\a-1}}\over\l_+\l_+^\ad},\cr
             \CA^i_{\ad\ad_1\cdots\ad_{n_i-1}}\ &=\ 
             {1\over 2\pi\i} \oint_\cC{\rm d}\l_+
             {\CA^i_{+\ad_1\cdots\ad_{n_i-1}}\over\l_+\l_+^\ad} ,
             \cr
}
where $\cC=\{\l_+\in\IC P^1\,|\,|\l_+|=1\}$. As in the previous discussion, 
Eqs. \eqref{pwtranssdh} make the Penrose-Ward transform explicit.

\paragraph{Field expansions and field equations.}
So far, we have written down the truncated self-dual SYM 
hierarchies \eqref{compconsdh} quite abstractly as compatibility conditions of 
a linear system. The next step in our discussion is to construct the equations 
of motion on superfield level equivalent to \eqref{compconsdh}. To do this, 
we need to identify the field content. At first sight, we expect to find 
as fundamental field content (in a covariant formulation)
 the field content of $\CN$-extended self-dual SYM theory 
plus a tower of additional fields depending on the parameters $m_\a$ and 
$n_i$. However, as we shortly realize, this will not entirely be true. 
For $n_i>1$, we instead find that certain combinations of   
$\CA^i_{\ad\ad_1\cdots\ad_{n_i-1}}$ play the role of potentials for a lot 
of the naively expected fields, such that those combinations should be regarded
as fundamental fields. 

In the remainder of this paragraph, we shall for simplicity consider the case 
when $m_1=m_2=:m$ and $n_1=\cdots=n_\CN=:n$. To simplify the subsequent formulas,
let us also introduce a shorthand index notation
\eqn{shorthandI}{
                \CA_{\a\ad\Ad}\ :=\ \CA_{\a\ad\ad_1\cdots\ad_{m-1}}\qquad{\rm and}\qquad
                \CA^I_\ad\ :=\ \CA_{\ad\ad_1\cdots\ad_{n-1}}^i.}

First, we point out that Eqs. \eqref{compconsdh} can concisely be rewritten
as
\eqna{compconsdhI}{
             {[ D_{\a\ad\Ad}^R, D_{\b\bd\Bd}^R]}+
             [ D_{\a\bd\Ad}^R, D_{\b\ad\Bd}^R]
             \ &=\ 0,\qquad
             [ D^I_\ad, D_{\b\bd\Bd}^R]+[ D^I_\bd, D_{\b\ad\Bd}^R]
             \ =\ 0,\\
        &\kern-.8cm
           \{ D^I_\ad, D^J_\bd\}+\{ D^I_\bd, D^J_\ad\}\ =\ 0,
}
which translates to the following superfield definitions 
\eqna{defofsfI}{
 {[ D_{\a\ad\Ad}^R, D_{\b\bd\Bd}^R]}\ &:=\ \epsilon_{\ad\bd}
             f_{\a\Ad\b\Bd},\\
             [ D^I_\ad, D_{\b\bd\Bd}^R]\ &:=\ \epsilon_{\ad\bd}
             \chi^I_{\b\Bd},\\
             \{ D^I_\ad, D^J_\bd\}\ &:=\ \epsilon_{\ad\bd}
             W^{IJ}.
}
Note that quite generally we have
\eqna{defofcI}{
   \CF_{\a\ad\Ad\b\bd\Bd}\ &=\ [ D_{\a\ad\Ad}^R, D_{\b\bd\Bd}^R]
   \ =\ \tfrac{1}{2}(\CF_{\a\ad\Ad\b\bd\Bd}-\CF_{\b\bd\Bd\a\ad\Ad})\\
    &=\ \epsilon_{\ad\bd}f_{\a\Ad\b\Bd}+\epsilon_{\a\b}f_{\ad\Ad\bd\Bd}+
             \CF_{\a\ad[\Ad\b\bd\Bd]},}
where
\eqna{defofcII}{f_{\a\Ad\b\Bd}\ &:=\ 
           -\tfrac{1}{2}\epsilon^{\ad\bd}\CF_{\a\ad\Ad\b\bd\Bd},\\
           f_{\ad\Ad\bd\Bd}\ &:=\ 
           \tfrac{1}{2}\epsilon^{\a\b}\CF_{\a\ad\Ad\b\bd\Bd},\\
            \CF_{\a\ad[\Ad\b\bd\Bd]}\ &:=\ 
           \tfrac{1}{2}(\CF_{\a\ad\Ad\b\bd\Bd}-\CF_{\a\ad\Bd\b\bd\Ad}).}
Eq. \eqref{defofcI} can be simplified further to
\eqn{defofcIII}{\CF_{\a\ad\Ad\b\bd\Bd}\ =\ 
               \epsilon_{\ad\bd}f_{\a\Ad\b\Bd}+
               \epsilon_{\a\b}f_{\ad(\Ad\bd\Bd)}+\CF_{(\a\ad[\Ad\b)\bd\Bd]}.}
Therefore, the first equation of \eqref{defofsfI} implies that
\eqn{superfieldeqhI}{f_{\ad(\Ad\bd\Bd)}\ =\ 0\qquad{\rm and}\qquad
                    \CF_{(\a\ad[\Ad\b)\bd\Bd]}\ =\ 0,}
which are the first two of the superfield equations of motion. We point out 
that for the choice $\Ad=\Bd=({\dot1}\cdots{\dot1})$ the set
\eqref{superfieldeqhI} represents nothing but the self-dual YM equations.

Next we consider the Bianchi identity for the triple 
$( D_{\a\ad\Ad}^R, D^I_\bd, D_{\g\gd\Cd}^R)$. We find
\eqn{superfieldeqhII}{ D^I_\ad f_{\a\Ad\b\Bd}\ =\ 
                      D_{\a\ad\Ad}^R\chi^I_{\b\Bd}.}
From this equation we deduce another two field equations, 
\eqn{superfieldeqhIII}{\epsilon^{\a\b} D_{\a\ad(\Ad}^R\chi^I_{\b\Bd)}\ =\ 0
                     \qquad{\rm and}\qquad
                      D_{(\a\ad[\Ad}^R\chi^I_{\b)\Bd]}\ =\ 0.} 
The Bianchi identity for $( D_{\a\ad\Ad}^R, D^I_\bd, D^J_\gd)$ 
implies
\eqn{superfieldeqhIV}{ D_{\a\ad\Ad}^RW^{IJ}\ =\  D^I_\ad
                     \chi^J_{\a\Ad}.}
Applying $ D_{\b\bd\Bd}^R$ to \eqref{superfieldeqhIV}, we obtain upon 
(anti)symmetrization the following two equations of motion: 
\eqna{superfieldeqhV}{
     \tfrac{1}{2}\epsilon^{\a\b}\epsilon^{\ad\bd} D_{\a\ad(\Ad}^R D_{\b\bd\Bd)}^RW^{IJ}
     +\epsilon^{\a\b}\{\chi^I_{\a(\Ad},\chi^J_{\b\Bd)}\}\ &=\ 0,\\
     \tfrac{1}{2}\epsilon^{\ad\bd} D_{(\a\ad[\Ad}^R D_{\b)\bd\Bd]}^RW^{IJ}+
     \{\chi^I_{(\a[\Ad},\chi^J_{\b)\Bd]}\}\ &=\ 0.
}
Furthermore, the Bianchi identity for the combination 
$( D^I_\ad, D^J_\bd, D^K_\gd)$ shows that $ D^I_\ad W^{JK}$ 
determines a superfield which is totally antisymmetric in the indices
$IJK$, i.e.,
\eqn{superfieldeqhVI}{ D^I_\ad W^{JK}\ =:\ \tfrac{1}{2}\chi^{IJK}_\ad.}
Upon acting on both sides by $ D_{\a\ad\Ad}^R$ and contracting the dotted
indices, we obtain the field equation for $\chi^{IJK}_\ad$
\eqn{superfieldeqhVII}{\epsilon^{\ad\bd} D_{\a\ad\Ad}^R\chi^{IJK}_\bd-
                      6[W^{[IJ},\chi^{K]}_{\a\Ad}]\ =\ 0.}
The application of $ D^I_\ad$ to $\chi^{JKL}_\bd$ and symmetrization in 
$\ad$ and $\bd$ leads by virtue of \eqref{superfieldeqhVI} to a new superfield which
is totally antisymmetric in $IJKL$,
\eqn{superfieldeqhVIII}{ D^I_{(\ad}\chi^{JKL}_{\bd)}\ =:\ 
                       -G_{\ad\bd}^{IJKL}.}
Some algebraic manipulations show that
\eqn{superfieldeqhIX}{ D^I_\ad\chi^{JKL}_\bd\ =\ 
        D^I_{(\ad}\chi^{JKL}_{\bd)}+ D^I_{[\ad}\chi^{JKL}_{\bd]}
       \ =\ -G_{\ad\bd}^{IJKL}+3\epsilon_{\ad\bd}[W^{I[J},W^{KL]}],}
where equation \eqref{superfieldeqhVI} and the definition \eqref{superfieldeqhVIII} 
have been used. From this equation, the equation of motion for the superfield
$G_{\ad\bd}^{IJKL}$ can readily be derived. We obtain
\eqn{superfieldeqhX}{\epsilon^{\ad\gd} D_{\a\ad\Ad}^RG_{\bd\gd}^{IJKL}-
     4\{\chi^{[I}_{\a\Ad},\chi^{JKL]}_\bd\}-6[W^{[JK}, D_{\a\bd\Ad}^RW^{LI]}]
     \ =\ 0.}
As \eqref{superfieldeqhVI} implies the existence of the superfield
$G_{\ad\bd}^{IJKL}$, definition \eqref{superfieldeqhVIII} determines a new
superfield $\psi^{IJKLM}_{\ad\bd\gd}$ being totally antisymmetric in
$IJKLM$ and totally symmetric in $\ad\bd\gd$, i.e.,
\eqn{superfieldeqhXI}{ D^I_{(\ad}G^{JKLM}_{\bd\gd)}\ =:\
                     \psi^{IJKLM}_{\ad\bd\gd}.}
It is easily shown that
\eqn{superfieldeqhXII}{ D^I_{\ad}G^{JKLM}_{\bd\gd}\ =\
                       D^I_{(\ad}G^{JKLM}_{\bd\gd)}-\tfrac{2}{3}
         \epsilon_{\ad(\bd}\epsilon^{\dd\ed} D^I_\dd G^{JKLM}_{\ed\gd)}.}
After some tedious algebra, we obtain from \eqref{superfieldeqhXII} the formula
\eqn{superfieldeqhXIII}{ D^I_{\ad}G^{JKLM}_{\bd\gd}\ =\ 
    \psi^{IJKLM}_{\ad\bd\gd}-\tfrac{2}{3}\epsilon_{\ad(\bd}
    \left(4[W^{I[J},\chi^{KLM]}_{\gd)}]+3[\chi^{I[JK}_{\gd)},W^{LM]}]\right),}
where definition \eqref{superfieldeqhXI} has been substituted. This
equation in turn implies the equation of motion for $\psi^{IJKLM}_{\ad\bd\gd}$,
\eqna{superfieldeqhXIV}{\epsilon^{\ad\dd} D_{\a\ad\Ad}^R
     \psi^{IJKLM}_{\bd\gd\dd}&+5[\chi^{[I}_{\a\Ad},G^{JKLM]}_{\bd\gd}]\ -\\ 
     &-\tfrac{20}{3}[ D_{\a(\bd\Ad}^RW^{[IJ},\chi^{KLM]}_{\gd)}]+
     \tfrac{10}{3}[W^{[IJ}, D_{\a(\bd\Ad}^R\chi^{KLM]}_{\gd)}]\ =\ 0,
}
which follows after a somewhat lengthy calculation. 

Now one can continue this procedure of defining superfields via the action of
$ D^I_\ad$ and of finding the corresponding equations of motion. 
Generically, the number of fields one obtains in this way is determined by the
parameter $n$, i.e., the most one can get is
$$\psi^{I_1\cdots I_{\CN n}}_{\ad_1\cdots\ad_{\CN n-2}},$$
which is, as before, totally antisymmetric in $I_1\cdots I_{\CN n}$ and 
totally symmetric in $\ad_1\cdots\ad_{\CN n-2}$. 

Let us collect the superfield equations of motion for the $\CN$-extended 
self-dual truncated SYM hierarchy:
\eqna{superfieldeqhXV}{
  f_{\ad(\Ad\bd\Bd)}\ =\ 0 \qquad{\rm and}\qquad
  \CF_{(\a\ad[\Ad\b)\bd\Bd]}\ &=\ 0,\\
  \epsilon^{\a\b} D_{\a\ad(\Ad}^R\chi^I_{\b\Bd)}\ =\ 0 \qquad{\rm and}\qquad
   D_{(\a\ad[\Ad}^R\chi^I_{\b)\Bd]}\ &=\ 0,\\
  \tfrac{1}{2}
  \epsilon^{\a\b}\epsilon^{\ad\bd} D_{\a\ad(\Ad}^R D_{\b\bd\Bd)}^RW^{IJ}
  +\epsilon^{\a\b}\{\chi^I_{\a(\Ad},\chi^J_{\b\Bd)}\}\ &=\ 0,\\
  \tfrac{1}{2}
  \epsilon^{\ad\bd} D_{(\a\ad[\Ad}^R D_{\b)\bd\Bd]}^RW^{IJ}+
  \{\chi^I_{(\a[\Ad},\chi^J_{\b)\Bd]}\}\ &=\ 0,\\
  \epsilon^{\ad\bd} D_{\a\ad\Ad}^R\chi^{IJK}_\bd-
  6[W^{[IJ},\chi^{K]}_{\a\Ad}]\ &=\ 0,\\
  \epsilon^{\ad\gd} D_{\a\ad\Ad}^RG_{\bd\gd}^{IJKL}-
  4\{\chi^{[I}_{\a\Ad},\chi^{JKL]}_\bd\}-6[W^{[JK}, D_{\a\bd\Ad}^RW^{LI]}]
  \ &=\ 0,\\
  &\kern-2cm\vdots\\
  \epsilon^{\ad\bd} D_{\a\ad\Ad}^R
  \psi^{I_1\cdots I_{\CN n}}_{\bd\ad_1\cdots\ad_{\CN n-3}}+ 
  J^{I_1\cdots I_{\CN n}}_{\a\Ad\ad_1\cdots\ad_{\CN n-3}}\ &=\ 0,
}
where the currents
$J^{I_1\cdots I_{\CN n}}_{\a\Ad\ad_1\cdots\ad_{\CN n-3}}$ are determined in 
an obvious manner. Clearly, the system \eqref{superfieldeqhXV} contains as a 
subset the $\CN$-extended self-dual SYM equations. In particular, for the choice 
$m=n=1$ it reduces to the latter.
Altogether, we have obtained the 
field content of $\CN$-extended self-dual SYM theory plus a number of 
additional fields together with their superfield equations of motion. 

However, as we have already indicated, this is not the end of the story. The 
system \eqref{superfieldeqhXV}, though describing the truncated hierarchy, 
contains a lot of 
redundant information. For that reason, 
it should not be regarded as the fundamental 
system displaying the truncated hierarchy. In fact, the use of
the shorthand index notation \eqref{shorthandI} does not entirely 
reflect all of the
possible index symmetry properties of the appearing superfields. In 
order to incorporate all possibilities, we instead need to write out the 
explicit form of $I,J,K,\ldots$.

As before, let us impose the transversal gauge condition
\eqn{tgH}{
    \eta^{\ad\ad_1\cdots\ad_{n-1}}_i\CA^i_{\ad\ad_1\cdots\ad_{n-1}}\ =\ 0,}
which again reduces super gauge transformations to ordinary ones. Note that
in \eqref{tgH} only $\CA^i_{(\ad\ad_1\cdots\ad_{n-1})}$ contributes, since 
the fermionic coordinates are totally symmetric under an exchange of their 
dotted indices. The condition \eqref{tgH} then allows to define the recursion 
operator $\cD$ according to
\eqn{defofrec}{\cD\ :=\ \eta^{\ad\ad_1\cdots\ad_{n-1}}_i
               D^i_{\ad\ad_1\cdots\ad_{n-1}}\ =\ 
              \eta^{\ad\ad_1\cdots\ad_{n-1}}_i
              \partial^i_{(\ad\ad_1\cdots\ad_{n-1})}.}

The third equation of \eqref{defofsfI} yields
\eqna{recursionI}{
                 (1+\cD)\CA^i_{(\ad\ad_1\cdots\ad_{n-1})}\ &=\ 
                 -\epsilon_{\bd(\ad}\eta^{\bd\bd_1\cdots\bd_{n-1}}_j
                 W^{ij}_{\ad_1\cdots\ad_{n-1})\,\bd_1\cdots\bd_{n-1}},\\
                 \cD\CA^i_{[\ad\bd]\ad_1\cdots\ad_{n-2}}\ &=\ \tfrac{1}{2} 
                 \epsilon_{\ad\bd}\eta^{\gd\bd_1\cdots\bd_{n-1}}_j
                 W^{ij}_{\gd\ad_1\cdots\ad_{n-2}\,\bd_1\cdots\bd_{n-1}},
}
which states that $\CA^i_{(\ad\ad_1\cdots\ad_{n-1})}$ does not
have a zeroth order component in the $\eta$-expansion while 
$\CA^i_{[\ad\bd]\ad_1\cdots\ad_{n-2}}$ does. Therefore, we obtain as a
fundamental superfield
\eqn{fundamentfieldsI}{\phi^i_{\ad_1\cdots\ad_{n-2}}\ :=\ 
                      2\CA^i_{[\dot1\dot2]\ad_1\cdots\ad_{n-2}}
                      \qquad{\rm for}\qquad n>1.}
Note that as $\phi^i_{\ad_1\cdots\ad_{n-2}}=\phi^i_{(\ad_1\cdots\ad_{n-2})}$
it defines for each $i$ a spin $\frac{n}{2}-1$ superfield of odd parity. 

The second equation of \eqref{defofsfI} reads explicitly as
$$[ D^i_{\ad\ad_1\cdots\ad_{n-1}}, D_{\b\bd\bd_1\cdots\bd_{m-1}}^R]
  \ =\ 
 \epsilon_{\ad\bd}\chi^i_{\ad_1\cdots\ad_{n-1}\,\b\,\bd_1\cdots\bd_{m-1}}.$$
The contraction with $\epsilon^{\ad\ad_1}$ shows that
\eqn{prepotentialI}{\chi^i_{\ad_1\cdots\ad_{n-1}\,\b\,\bd_1\cdots\bd_{m-1}}
  \ =\  D_{\b(\ad_1\bd_1\cdots\bd_{m-1}}^R\phi^i_{\ad_2\cdots\ad_{n-1})},}
where the symmetrization is only meant between the $\ad_1\cdots\ad_{n-1}$.
Therefore, the superfield 
$\chi^i_{\ad_1\cdots\ad_{n-1}\,\b\,\bd_1\cdots\bd_{m-1}}$ cannot be regarded
as a fundamental field -- the superfield \eqref{fundamentfieldsI} plays the role
of a potential for the former. 

Next we discuss the superfield 
$W^{ij}_{\ad_1\cdots\ad_{n-1}\,\bd_1\cdots\bd_{n-1}}$. To decide which
combinations of it are really fundamental, we need some preliminaries. Consider
the index set
$$ \ad_1\cdots\ad_n\bd_1\cdots\bd_n, $$
which separately is totally symmetric in $\ad_1\cdots\ad_n$ and 
$\bd_1\cdots\bd_n$, respectively. Then we have the useful formula
\eqna{usefulformula}{
                   \ad_1\cdots\ad_n\bd_1\cdots\bd_n\ &=\ 
                   (\ad_1\cdots\ad_n\bd_1\cdots\bd_n)\ +\ 
                   \sum\ {\rm all\ possible\ contractions}\\
                    &=\ (\ad_1\cdots\ad_n\bd_1\cdots\bd_n)\ +\\
                    &\kern2cm +\ A_1\sum_{i,j}
                 \ad_1\cdots\!\!\contra{70}{\ \ad_i\cdots\ad_n\bd_1\cdots\bd_j}
                  \cdots\bd_n\ +\ \cdots,
}
where the parentheses denote, as before, symmetrization of the enclosed 
indices and ``contraction'' means antisymmetrization in the respective index 
pair. The $A_i$s for $i=1,\ldots,n$ are combinatorial coefficients, whose
explicit form is not needed in the sequel. The proof of \eqref{usefulformula} is 
quite similar to the one of the Wick theorem and we thus leave it to the 
interested reader. 

The third equation of \eqref{defofsfI} is explicitly given by
$$ \{ D^i_{\ad\ad_1\cdots\ad_{n-1}}, D^j_{\bd\bd_1\cdots\bd_{n-1}}\}
\ =\ \epsilon_{\ad\bd}W^{ij}_{\ad_1\cdots\ad_{n-1}\,\bd_1\cdots\bd_{n-1}}.$$
After contraction with $\epsilon^{\ad\ad_1}$ we obtain
\eqn{fundamentfieldsII}{W^{ij}_{\ad_1\cdots\ad_{n-1}\,\bd_1\cdots\bd_{n-1}}
  \ =\ - D^j_{\ad_1\bd_1\cdots\bd_{n-1}}\phi^i_{\ad_2\cdots\ad_{n-1}},}
where the definition \eqref{fundamentfieldsI} has been inserted. Contracting this
equation with $\epsilon^{\ad_1\bd_1}$, we get
\eqn{fundamentfieldsIII}{\epsilon^{\ad_1\bd_1}
         W^{ij}_{\ad_1\cdots\ad_{n-1}\,\bd_1\cdots\bd_{n-1}}\ =\ 
         -\{\phi^i_{\ad_2\cdots\ad_{n-1}},\phi^j_{\bd_2\cdots\bd_{n-1}}\}.}
Thus, we conclude that  
$W^{ij}_{\ad_1\cdots\ad_{n-2}[\ad_{n-1}\bd_1]\bd_2\cdots\bd_{n-1}}$ is
a composite field and hence not a fundamental one. Using formula
\eqref{usefulformula}, we may schematically write
\eqn{fundamentfieldsIV}{W^{ij}_{\ad_1\cdots\ad_{n-1}\,\bd_1\cdots\bd_{n-1}}
       \ =\ W^{ij}_{(\ad_1\cdots\ad_{n-1}\,\bd_1\cdots\bd_{n-1})}
            +\sum\ {\rm all\ possible\ contractions}.}
The contraction terms in \eqref{fundamentfieldsIV}, however, solely consist of
composite 
expressions due to \eqref{fundamentfieldsIII}. Therefore, only the superfield
\eqn{fundamentfieldsV}{W^{ij}_{(\ad_1\cdots\ad_{n-1}\,\bd_1\cdots\bd_{n-1})}
              \ =\ W^{[ij]}_{(\ad_1\cdots\ad_{n-1}\,\bd_1\cdots\bd_{n-1})}}
is fundamental. For each combination $[ij]$ it represents an even
superfield with spin $n-1$.

Next we need to consider the superfield defined in \eqref{superfieldeqhVI},
$$ \chi^{ijk}_{\ad\,\ad_1\cdots\ad_{n-1}\,\bd_1\cdots\bd_{n-1}\,
   \gd_1\cdots\gd_{n-1}}\ =\  D^i_{\ad\ad_1\cdots\ad_{n-1}}
   W^{ij}_{\bd_1\cdots\bd_{n-1}\,\gd_1\cdots\gd_{n-1}}. $$
By extending the formula \eqref{usefulformula} to the index triple 
$$\ad_1\cdots\ad_{n-1}\,\bd_1\cdots\bd_{n-1}\,\gd_1\cdots\gd_{n-1}$$
and by utilizing the symmetry properties of 
$\chi^{ijk}_{\ad\,\ad_1\cdots\ad_{n-1}\,\bd_1\cdots\bd_{n-1}\,\gd_1\cdots
\gd_{n-1}}$, one can show, by virtue of the above arguments, that only the
combination
\eqn{fundamentfieldsVI}{\chi^{ijk}_{(\ad\,\ad_1\cdots\ad_{n-1}\,
   \bd_1\cdots\bd_{n-1}\,\gd_1\cdots\gd_{n-1})}\ =\ 
  \chi^{[ijk]}_{(\ad\,\ad_1\cdots\ad_{n-1}\,
   \bd_1\cdots\bd_{n-1}\,\gd_1\cdots\gd_{n-1})}}
of $\chi^{ijk}_{\ad\,\ad_1\cdots\ad_{n-1}\,\bd_1\cdots\bd_{n-1}\,
\gd_1\cdots\gd_{n-1}}$ remains as a fundamental superfield. It defines
for each $[ijk]$ an odd superfield with spin $\frac{3}{2}n-1$.

Repeating this procedure, we deduce from the definition \eqref{superfieldeqhVIII} 
that 
\eqn{fundamentfieldsVII}{
  G^{ijkl}_{(\ad\bd\,\ad_1\cdots\ad_{n-1}\,
   \bd_1\cdots\bd_{n-1}\,\gd_1\cdots\gd_{n-1}\,\dd_1\cdots\dd_{n-1})}\ =\  
  G^{[ijkl]}_{(\ad\bd\,\ad_1\cdots\ad_{n-1}\,
   \bd_1\cdots\bd_{n-1}\,\gd_1\cdots\gd_{n-1}\,\dd_1\cdots\dd_{n-1})}
}
is fundamental and it represents one\footnote{Note that $\CN\leq4$.}
spin $2n-1$ superfield which is even. All higher order fields,
such as \eqref{superfieldeqhXI}, yield no further fundamental fields due to the
antisymmetrization of $ijklm$, etc. In summary, the fundamental field 
content of the truncated self-dual SYM hierarchies is given by
\eqna{fundamentfieldsVIII}{
           \CA_{\a\ad\ad_1\cdots\ad_{n-1}},\quad
            \phi^i_{\ad_1\cdots\ad_{n-2}},\quad
            W^{[ij]}_{(\ad_1\cdots\ad_{n-1}\,\bd_1\cdots\bd_{n-1})},\\
           \chi^{[ijk]}_{(\ad\,\ad_1\cdots\ad_{n-1}\,
   \bd_1\cdots\bd_{n-1}\,\gd_1\cdots\gd_{n-1})},\quad
         G^{[ijkl]}_{(\ad\bd\,\ad_1\cdots\ad_{n-1}\,
   \bd_1\cdots\bd_{n-1}\,\gd_1\cdots\gd_{n-1}\,\dd_1\cdots\dd_{n-1})},
}
where we assume that $n>1$.\footnote{For $n=1$, the field 
$\phi^i_{\ad_1\cdots\ad_{n-2}}$ must be replaced by $\chi^i_\a$.} All other 
naively expected fields, which for instance appear in \eqref{superfieldeqhXV}, are 
composite expressions of the above fields. Note that the field
$\CA_{\a\ad\ad_1\cdots\ad_{n-1}}$ can be decomposed according to
\eqn{}{\CA_{\a\ad\ad_1\cdots\ad_{n-1}}\ =\ \CA_{\a(\ad\ad_1\cdots\ad_{n-1})}+
        \epsilon_{\ad(\ad_1}\Phi_{\a\ad_2\cdots\ad_{n-1})}.}
Therefore, $\CA_{\a(\ad_1\cdots\ad_{n})}$ can be interpreted as a gauge potential 
while $\Phi_{\a\ad_1\cdots\ad_{n-2}}$ represents a collection of Higgs fields.

It remains to find the superfield equations of motion for the fields 
\eqref{fundamentfieldsVIII}. This, however, is easily done since we have already 
derived \eqref{superfieldeqhXV}. By following the lines which led to \eqref{superfieldeqhXV}
and by taking into account the definition \eqref{fundamentfieldsI}, the system 
\eqref{superfieldeqhXV} reduces for $n>1$ to
\eqna{superfieldeqhXVII}{
 f_{\ad(\ad_1\cdots\ad_{m-1}\bd\bd_1\cdots\bd_{m-1})}\ =\ 0 
 \qquad{\rm and}\qquad
 \CF_{(\a\ad[\ad_1\cdots\ad_{m-1}\b)\bd\bd_1\cdots\bd_{m-1}]}\ &=\ 0,\\
 \epsilon^{\a\b}\epsilon^{\ad\bd} D_{\a\ad(\ad_1\cdots\ad_{m-1}}^R
  D^R_{\b\bd\bd_1\cdots\bd_{m-1})}\phi^i_{\gd_1\cdots\gd_{n-2}}\ &=\ 0,\\
  D^R_{(\a\ad[\ad_1\cdots\ad_{m-1}} D^R_{\b)(\gd_1\bd_1\cdots\bd_{m-1}]}
 \phi^i_{\gd_2\cdots\gd_{n-1})}\ &=\ 0,\\
 &\kern-10cm\epsilon^{\ad\bd_1} D^R_{\a\ad\ad_1\cdots\ad_{m-1}}
 W^{[ij]}_{(\bd_1\cdots\bd_{n-2})}\ -\\ 
 -\ \{\phi^{[i}_{(\bd_2\cdots\bd_{n-1}}, D^R_{\a\bd_{n}\ad_1\cdots
 \ad_{m-1}}\phi^{j]}_{\bd_{n+1}\cdots\gd_{2n-2})}\}\ &=\ 0,\\
 &\kern-10cm\epsilon^{\ad\bd} D^R_{\a\ad\ad_1\cdots\bd_{m-1}}
 \chi^{[ijk]}_{(\bd\bd_1\cdots\bd_{3n-3})}\ -\\
 -\ 6[W^{[ij}_{(\bd_1\cdots\bd_{2n-2}}, D^R_{\a\bd_{2n-1}\ad_1\cdots
  \ad_{m-1}}\phi^{k]}_{\bd_{2n}\cdots\bd_{3n-3})}]\ &=\ 0,\\
 &\kern-10cm\epsilon^{\ad\gd} D^R_{\a\ad\ad_1\cdots\ad_{m-1}}
 G_{(\bd\gd\bd_1\cdots\bd_{4n-4})}^{[ijkl]}\ +\\
 &\kern-9.5cm\ +4\{ D^R_{\a(\bd_1\ad_1\cdots\ad_{m-1}}\phi^{[i}_{\bd_2\cdots
 \bd_{n-1}},\chi^{ijk]}_{\bd\bd_{n}\cdots\bd_{4n-4})}\}\ +\\
 +\ 6[W^{[ij}_{(\bd_1\cdots\bd_{2n-2}},
  D^R_{\a\bd\ad_1\cdots\ad_{m-1}}W^{kl]}_{\bd_{2n-1}\cdots\bd_{4n-4})}
 ]\ &=\ 0.
}
These are the superfield equations of motion for the truncated
$\CN$-extended self-dual SYM hierarchy.

Above we have derived the superfield equations of motion. What remains is to 
show how the superfields \eqref{fundamentfieldsVIII} are expressed in terms of their 
zeroth order components
\eqn{zerothorderfundamentfieldsI}{
    \Ac_{\a\ad\ad_1\cdots\ad_{m-1}},\quad
    \pc\ \!\!^i_{\ad_1\cdots\ad_{n-2}},\quad
    \Wc\ \!\!^{[ij]}_{(\ad_1\cdots\ad_{2n-2})},\quad
    \cc\ \!\!^{[ijk]}_{(\ad_1\cdots\ad_{3n-2})},\quad
    \Gc\ \!\!^{[ijkl]}_{(\ad_1\cdots\ad_{4n-2})},
}
and furthermore that the field equations on $\IC^4$ (or on $\IR^4$
after reality conditions have been imposed), i.e., those 
equations which are obtained from the set \eqref{superfieldeqhXVII} by projecting
onto the zeroth order components \eqref{zerothorderfundamentfieldsI} of the 
superfields \eqref{fundamentfieldsVIII}, imply the compatibility conditions 
\eqref{compconsdhI}. We will, however, be not too explicit in showing this
equivalence, since the argumentation goes along similar lines as those 
given for the $\CN$-extended self-dual SYM equations. Here, 
we just want to give the outline.

In order to write down 
the superfield expansions, remember that we have imposed the gauge
\eqref{tgH} which led to the recursion operator $\cD$ according to \eqref{defofrec}.
Using the formulas \eqref{defofsfI}, \eqref{superfieldeqhIV}, \eqref{superfieldeqhVI},
\eqref{superfieldeqhVIII}, \eqref{superfieldeqhXI} and \eqref{prepotentialI}, we obtain the 
following recursion relations:
\eqna{recursionrelations}{
     (1+\cD)\CA^i_{(\ad_1\cdots\ad_{n})}\ &=\ 
     -\epsilon_{\bd(\ad_1}\eta^{\bd\bd_1\cdots\bd_{n-1}}_j
     W^{ij}_{\ad_2\cdots\ad_{n})\,\bd_1\cdots\bd_{n-1}},\\
     \cD\CA_{\a\ad\ad_1\cdots\ad_{m-1}}\ &=\ -\epsilon_{\ad\bd}
     \eta_i^{\bd\bd_1\cdots\bd_{n-1}} D^R_{\a\bd_1\ad_1\cdots\ad_{m-1}}
     \phi^i_{\bd_2\cdots\bd_{n-1}},\\
     \cD\phi^i_{\ad_1\cdots\ad_{n-2}}\ &=\ 
     \eta^{\gd\bd_1\cdots\bd_{n-1}}_j
     W^{ij}_{\gd\ad_1\cdots\ad_{n-2}\,\bd_1\cdots\bd_{n-1}},\\
     \cD W^{[ij]}_{(\ad_1\cdots\ad_{2n-2})}\ &=\ \tfrac{1}{2}
    \eta^{\bd\bd_1\cdots\bd_{n-1}}_k\chi^{[ij]k}_{\bd(\ad_1\cdots\ad_{2n-2})\,
    \bd_1\cdots\bd_{n-1}},\\
     \cD\chi^{[ijk]}_{(\ad_1\cdots\ad_{3n-2})}\ &=\ -
    \eta^{\bd\bd_1\cdots\bd_{n-1}}_lG^{[ijk]l}_{\bd(\ad_1\cdots\ad_{3n-2})\,
    \bd_1\cdots\bd_{n-1}}\ +\\
    &\kern.5cm+\ 3\epsilon_{\bd(\ad_1}\eta^{\bd\bd_1\cdots\bd_{n-1}}_l
    [W^{l[i}_{\bd_1\cdots\bd_{n-1}\,\ad_2\cdots\ad_{n}},
    W^{jk]}_{\ad_{n+1}\cdots\ad_{3n-2})}],\\
    \cD G^{[ijkl]}_{(\ad_1\cdots\ad_{4n-2})}\ &=\ 
    \eta^{\bd\bd_1\cdots\bd_{n-1}}_m\psi^{[ijkl]m}_{\bd
     (\ad_1\cdots\ad_{4n-2})\,\bd_1\cdots\bd_{n-1}}\ - \\
    &\kern.5cm - \ \tfrac{2}{3}\epsilon_{\bd(\ad_1}
    \eta^{\bd\bd_1\cdots\bd_{n-1}}_m
    \left(4[W^{m[i}_{\bd_1\cdots\bd_{n-1}\,\ad_2\cdots\ad_{n}},
    \chi^{jkl]}_{\ad_{n+1}\cdots\ad_{4n-2})}]\ +\right.\\
    &\kern1cm\left.+\ 3[\chi^{m[ij}_{\ad_2\,\bd_1\cdots\bd_{n-1}\,\ad_3\cdots
    \ad_{2n-4}},W^{kl]}_{\ad_{2n-3}\cdots\ad_{4n-2})}]\right).
}
An explanation of these formulas is in order. The right hand sides of 
Eqs. \eqref{recursionrelations} depend not only on the fundamental fields
\eqref{fundamentfieldsVIII} but also on composite expressions of those fields: 
For instance, consider the recursion relation of the 
field $\phi^i_{\ad_1\cdots\ad_{n-2}}$. The right hand side of
this equation depends on the superfield 
$W^{ij}_{\ad_1\cdots\ad_{n-1}\,\bd_1\cdots\bd_{n-1}}$. However, as we
learned in \eqref{fundamentfieldsIV}, it can be rewritten as the fundamental field
$W^{[ij]}_{(\ad_1\cdots\ad_{n-1}\,\bd_1\cdots\bd_{n-1})}$ plus contraction
terms which are of the form \eqref{fundamentfieldsIII}. Similar arguments hold
for the other recursion relations. Therefore, the right hand 
sides of \eqref{recursionrelations} can solely be written in terms of the
fundamental fields. However, as these formulas in terms of the fundamental 
fields look rather messy, we refrain from writing them down. 
Note that the field
$\psi^{[ijkl]m}_{\bd(\ad_1\cdots\ad_{4n-2})\,\bd_1\cdots\bd_{n-1}}$ 
appearing in the last recursion relation consists
only of composite expressions of the fields \eqref{fundamentfieldsIII}. Using
Eqs. \eqref{recursionrelations}, one can now straightforwardly determine
the superfield expansions by a successive application of the recursion
operator $\cD$, since 
if one knows the expansions to $n$-th order in the 
fermionic coordinates, the recursions \eqref{recursionrelations} yield them to 
next order. 
But again, this procedure will lead to both unenlightening and
complicated looking expressions, so we do not present them here.
Finally, the recursion operator can be used to show the equivalence between
the field equations and the constraint equations \eqref{compconsdh}.
This can be done
inductively, i.e., one first assumes that Eqs. \eqref{superfieldeqhXVII}
hold to $n$-th order in the fermionic coordinates, then one applies $k+\cD$ to
\eqref{superfieldeqhXVII}, where $k\in\IN_0$ is some properly chosen integer, and 
shows that they also hold to $(n+1)$-th order. For more details, see Ref.
\cite{Wolf:2004hp}.

\paragraph{Light-cone gauge.}\label{par-LCG-ch5}
Let us now give an alternative interpretation of the hierarchy
equations. First, we rewrite the constraint equations \eqref{CE-2}
of $\CN$-extended self-dual SYM theory in light-cone gauge.
One of the interesting issues of this (non-covariant) gauge is that
all the equations for all the fields reduce to equations on a
single Lie-algebra valued superfield $\Psi$. 

To be explicit, assume the following expansion
\eqn{sac}{\psi_+\ =\ 1+\l_+\Psi+\cdots}
on $\hCU_+$. Here, all 
$\l$-dependence has been made explicit, i.e., $\Psi$
is defined on $\IC^{4|2\CN}$.
Note that the
expansion \eqref{sac} can be obtained from some general $\psi_+$ by performing
the gauge transformation  $\psi_+\mapsto (\psi^{(0)}_+)^{-1}\psi_+$.
Upon substituting \eqref{sac} 
into \eqref{linsys-2}, we obtain
\eqn{leznovgauge}{\CA_{\a{\dot1}}\ =\ \partial_{\a{\dot2}}\Psi,\qquad
                 \CA_{\a{\dot2}}\ =\ 0\qquad{\rm and}\qquad
                 \CA^i_{\dot1}\ =\ \partial^i_{{\dot2}}\Psi,\qquad
                 \CA^i_{\dot2}\ =\ 0.}
Plugging \eqref{leznovgauge}
into the constraint equations \eqref{CE-2}, we find the following set of 
equations:
\eqna{compconii}{
               \partial_{1{\dot1}}\partial_{2{\dot2}}\Psi-
               \partial_{2{\dot1}}\partial_{1{\dot2}}\Psi+
               [\partial_{1{\dot2}}\Psi,\partial_{2{\dot2}}\Psi]\ &=\ 0,\\
               \partial^i_{{\dot1}}\partial_{\a{\dot2}}\Psi-
               \partial_{\a{\dot1}}\partial^i_{{\dot2}}\Psi+
               [\partial^i_{{\dot2}}\Psi,\partial_{\a{\dot2}}\Psi]\ &=\ 0,\\
               \partial^i_{{\dot1}}\partial^j_{{\dot2}}\Psi+
               \partial^j_{{\dot1}}\partial^i_{{\dot2}}\Psi+
               \{\partial^i_{{\dot2}}\Psi,\partial^j_{{\dot2}}\Psi\}\ &=\ 0.\\
}

Let us now come back to the linear system \eqref{linsyssdh} and the
constraint equations \eqref{compconsdh} of the truncated
self-dual SYM hierarchy. Upon imposing light-cone gauge, that is,
upon assuming an expansion of the form 
\eqn{}{\psi_+\ =\ 1+\l_+\hat\Psi+\cdots,}
where now $\hat\Psi$ is defined on $\IC^{M|N}$, we find
\eqna{leznovgaugesdh}{
                    \CA_{\a{\dot1}\ad_1\cdots\ad_{m-1}}\ =\ 
                    \partial_{\a{\dot2}\ad_1\cdots\ad_{m-1}}\hat\Psi
                    \qquad&{\rm and}\qquad
                    \CA_{\a{\dot2}\ad_1\cdots\ad_{m-1}}\ =\ 0,\\
                    \CA^i_{{\dot1}\ad_1\cdots\ad_{n-1}}\ =\ 
                    \partial^i_{{\dot2}\ad_1\cdots\ad_{n-1}}\hat\Psi
                    \qquad&{\rm and}\qquad
                    \CA^i_{{\dot2}\ad_1\cdots\ad_{n-1}}\ =\ 0.
}
Therefore, \eqref{compconsdh} turns into the following 
system:
\eqna{compconsdhii}{
               \partial_{\a{\dot1}\ad_1\cdots\ad_{m-1}}
               \partial_{\b{\dot2}\bd_1\cdots\bd_{m-1}}\hat\Psi&-
               \partial_{\b{\dot1}\bd_1\cdots\bd_{m-1}}
               \partial_{\a{\dot2}\ad_1\cdots\ad_{m-1}}\hat\Psi\ +\\
               &\kern1cm+\ [\partial_{\a{\dot2}\ad_1\cdots\ad_{m-1}}\hat\Psi,
               \partial_{\b{\dot2}\bd_1\cdots\bd_{m-1}}\hat\Psi]\ =\ 0,\\
               \partial^i_{{\dot1}\ad_1\cdots\ad_{n-1}}
               \partial_{\b{\dot2}\bd_1\cdots\bd_{m-1}}\hat\Psi&-
               \partial_{\b{\dot1}\bd_1\cdots\bd_{m-1}}
               \partial^i_{{\dot2}\ad_1\cdots\ad_{n-1}}\hat\Psi\ +\\
               &\kern1cm+\ [\partial^i_{{\dot2}\ad_1\cdots\ad_{n-1}}\hat\Psi,
               \partial_{\b{\dot2}\bd_1\cdots\bd_{m-1}}\hat\Psi]\ =\ 0,\\
               \partial^i_{{\dot1}\ad_1\cdots\ad_{n-1}}
               \partial^j_{{\dot2}\bd_1\cdots\bd_{n-1}}\hat\Psi&+
               \partial^j_{{\dot1}\bd_1\cdots\bd_{n-1}}
               \partial^i_{{\dot2}\ad_1\cdots\ad_{n-1}}\hat\Psi\ +\\
               &\kern1cm+\ \{\partial^i_{{\dot2}\ad_1\cdots\ad_{n-1}}\hat\Psi,
               \partial^j_{{\dot2}\bd_1\cdots\bd_{n-1}}\hat\Psi\}\ =\ 0.
}
Clearly, if $\hat\Psi$ solely depends on space-time coordinates, i.e.,
$\hat\Psi$ is independent of ``higher time moduli", all the extra fields
disappear and the above system reduces to \eqref{compconii}. 

If we define
\eqn{symmetries}{\d_{\a\ad_1\cdots\ad_{m}}\hat\Psi\ :=\ 
               \partial_{\a\ad_1\cdots\ad_{m}}\hat\Psi
               \qquad{\rm and}\qquad
               \d_{\ad_1\cdots\ad_{n}}^i\hat\Psi\ :=\ 
               \partial^i_{\ad_1\cdots\ad_{n}}\hat\Psi,}
the system \eqref{compconsdhii} implies
\eqna{compconsdhiii}{
               \partial_{\a{\dot1}\cdots{\dot1}}
               \d_{\b{\dot2}\bd_1\cdots\bd_{m-1}}\hat\Psi-
               \partial_{\a{\dot2}{\dot1}\cdots{\dot1}}
               \d_{\b{\dot1}\bd_1\cdots\bd_{m-1}}\hat\Psi\ +
               [\partial_{\a{\dot2}{\dot1}\cdots{\dot1}}\hat\Psi,
               \d_{\b{\dot2}\bd_1\cdots\bd_{m-1}}\hat\Psi]\ &=\ 0,\\
               \partial^i_{{\dot1}\cdots{\dot1}}
               \d_{\b{\dot2}\bd_1\cdots\bd_{m-1}}\hat\Psi-
               \partial^i_{{\dot2}{\dot1}\cdots{\dot1}}
               \d_{\b{\dot1}\bd_1\cdots\bd_{m-1}}\hat\Psi\ +
               [\partial^i_{{\dot2}{\dot1}\cdots{\dot1}}\hat\Psi,
               \d_{\b{\dot2}\bd_1\cdots\bd_{m-1}}\hat\Psi]\ &=\ 0,\\
               \partial^i_{{\dot1}\cdots{\dot1}}
               \d^j_{{\dot2}\bd_1\cdots\bd_{n-1}}\hat\Psi-
               \partial^i_{{\dot2}{\dot1}\cdots{\dot1}} 
               \d^j_{{\dot1}\bd_1\cdots\bd_{n-1}}\hat\Psi\ +
               \{\partial^i_{{\dot2}{\dot1}\cdots{\dot1}}\hat\Psi,
               \d^j_{{\dot2}\bd_1\cdots\bd_{n-1}}\hat\Psi\}\ &=\ 0.
}
If one differentiates these equations with respect to 
the space-time coordinates, one realizes that the resulting equations
coincide with the linearized versions of \eqref{compconii}.
Putting it differently, some equations of the self-dual
SYM hierarchy can be interpreted as equations on symmetries
for the self-dual SYM equations. 

\paragraph{Summary.}
As for $\CN$-extended self-dual SYM theory, 
we may now summarize the above discussion as follows:
{\Thm\label{SDYMhier-HCS} There is a one-to-one correspondence between gauge 
equivalence classes of local
solutions to the truncated $\CN$-extended self-dual SYM hierarchy of
level $(m_1,m_2|n_1,\ldots,n_\CN)$ 
on four-dimensional space-time and equivalence classes of
holomorphic vector bundles $\CE$ over  generalized supertwistor space 
$\CP^{3|\CN}_{m_1,m_2|n_1,\ldots,n_\CN}$ 
which are smoothly trivial and holomorphically trivial on any projective line 
$\IC P^1_{x_R,\eta}\hookrightarrow
\CP^{3|\CN}_{m_1,m_2|n_1,\ldots,n_\CN}$.}\vskip 4mm

\paragraph{Example.}
Let us now give an explicit example of a truncated hierarchy which also
makes contact with our discussion presented in Sec. \ref{sec:OSDM-ch2}.
Consider the truncated $\CN=2$ 
self-dual SYM hierarchy of level
$(m_1,m_2|n_1,n_2)=(1,1|2,2)$. Its field equations 
are given by 
\eqna{examplehierarchyI}{\fc_{\ad\bd}\ &=\ 0,\\
            \epsilon^{\a\b}\epsilon^{\ad\bd}\cnab\ \!\!^R_{\a\ad}\cnab\ \!\!^R_{\b\bd}
            \pc\ \!\!^i\ &=\ 0,\\
            \epsilon^{\ad\bd}\cnab\ \!\!^R_{\a\ad}\Wc\ \!\!^{[ij]}_{\bd\gd}-
            \{\pc\ \!\!^{[i},\cnab\ \!\!^R_{\a\gd}\pc\ \!\!^{j]}\}\ &=\ 0,
}
and follow from \eqref{superfieldeqhXVII}. The $\CN=2$ self-dual SYM equations, 
which are the first three equations of \eqref{fieldeqn-2} (with $i,j=1,2$), are by 
construction a ``subset'' of \eqref{examplehierarchyI}. That is, we apply 
to the last equation of \eqref{examplehierarchyI} the operator $\cnab\ \!\!^R_{\b\dd}$ and 
contract with $\epsilon^{\a\b}$ to obtain
$$    \epsilon^{\ad\bd}\epsilon^{\a\b}\cnab\ \!\!^R_{\b\dd}
      \cnab\ \!\!^R_{\a\ad}\Wc\ \!\!^{[ij]}_{\bd\gd}-\epsilon^{\a\b}
      \{\cnab_{\b\dd}\ \!\!^R\pc\ \!\!^{[i},\cnab\ \!\!^R_{\a\gd}\pc\ \!\!^{j]}\}\ =\ 0.
$$   
Eq. \eqref{prepotentialI} together with
$$ \epsilon^{\a\b}\cnab\ \!\!_{\a\ad}^R\cnab\ \!\!_{\b\bd}^R\ =\
  \tfrac{1}{2}\epsilon_{\ad\bd}\epsilon^{\a\b}\epsilon^{\gd\dd}
   \cnab\ \!\!_{\a\gd}^R\cnab\ \!\!_{\b\dd}^R$$
imply
\eqna{examplehierarchyIII}{\fc_{\ad\bd}\ &=\ 0,\cr
      \epsilon^{\a\b}\cnab\ \!\!^R_{\a\ad}\cc\ \!\!^i_{\bd\b}\ &=\ 0,\\
      \tfrac{1}{2}\epsilon^{\ad\bd}\epsilon^{\a\b}\cnab\ \!\!^R_{\a\ad}\cnab\ \!\!^R_{\b\bd}
      \Wc\ \!\!^{[ij]}_{\gd\dd}+\epsilon^{\a\b}
      \{\cc\ \!^{[i}_{\gd\a},\cc\ \!^{j]}_{\dd\b}\}\ &=\ 0.
}
These equations
reduce to the $\CN=2$ self-dual SYM equations when the dotted indices of
$\cc\ \!\!^i_{\ad\a}$ and $\Wc\ \!\!^{[ij]}_{\ad\bd}$ are chosen to be one.
Note that as $i,j$ run only from one to two, the last equation of 
\eqref{examplehierarchyI} can be rewritten in the form
\eqna{examplehierarchyII}{\fc_{\ad\bd}\ &=\ 0,\\
            \epsilon^{\a\b}\epsilon^{\ad\bd}\cnab\ \!\!^R_{\a\ad}\cnab\ \!\!^R_{\b\bd}
            \pc\ \!\!^i\ &=\ 0,\\
            \epsilon^{\ad\bd}\cnab\ \!\!^R_{\a\ad}\Gc_{\bd\gd}-
            \epsilon_{ij}\{\pc\ \!\!^{i},
            \cnab\ \!\!^R_{\a\gd}\pc\ \!\!^{j}\}\ &=\ 0,\\
}
where we have defined an anti-self-dual differential two-form according to
$\Gc_{\ad\bd}:=\epsilon_{ij}\Wc\ \!\!^{[ij]}_{\ad\bd}$.
Note that these are exactly the field equations given in
\eqref{eq:hier-ch2}. Hence, \eqref{eq:hier-ch2} can be interpreted
to describe the $(1,1|2,2)$ hierarchy
of $\CN=2$ self-dual SYM theory. Recall also that in this case
we have an appropriate action principle leading to 
\eqref{examplehierarchyII}. It is an interesting fact 
that even though it is not possible to write down an action
functional of $\CN=2$ self-dual SYM theory ($\CP^{3|2}$ is
not formally Calabi-Yau), it is possible to find one for
a certain hierarchy thereof.  

\section{Nonlocal conservation laws}

What remains is to give the nonlocal conservation laws
associated with the symmetry transformations discussed above.
In what follows, we use the ideas of \cite{Ablowitz}.

\paragraph{Conserved nonlocal currents.}\label{par:CC-ch5}
The compatibility conditions of the linear system \eqref{linsys-2}
can equivalently be written as
\eqn{eq:bla2-ch5}{{[\bnab_\a^\pm,\bnab_\b^\pm]}\ =\ 0,\qquad
        [\bnab_\a^\pm,\bnab^i_\pm]\ =\ 0\qquad{\rm and}\qquad
        \{\bnab^i_\pm,\bnab^j_\pm\}\ =\ 0.}
Recall further that $\phi_{+-}$ as given in \eqref{defofphi}
satisfies
\eqn{}{\bnab_\a^\pm\phi_{+-}\ =\ 0\qquad{\rm and}\qquad
       \bnab^i_\pm\phi_{+-}\ =\ 0,}
where the covariant derivatives act in the adjoint representation.
Then we may define
\eqn{eq:defofj-ch5}{J_\a^{+-}\ :=\ {\rm tr}\left\{\phi_{+-}\d\CA_\a^+\right\}\ =\ {\rm tr}\left\{\phi_{+-}\nabla_\a^+\phi_+\right\},}
where $\d\CA_\a^+=\nabla_\a^+\phi_+$ is a symmetry of \eqref{eq:bla2-ch5} given
according to \eqref{deforeqai}. Here,``tr" denotes the matrix trace.
Notice that this expression is
gauge invariant as both, $\phi_{+-}$ and $\d\CA_\a^+$ transform
in the adjoint representation. Now it is a straightforward
exercise to show that
\eqn{eq:cl-ch5}{\epsilon^{\a\b}\bV_\a^+ J^{+-}_\b\ =\ 0.}
Indeed, we have
$$\begin{aligned}
     \bV^+_\a J^{+-}_\b\ &=\ {\rm tr}\left\{-[\CA^+_\a,\phi_{+-}]\d\CA_\b^+\right\}
                                   +{\rm tr}\left\{\phi_{+-}\bV^+_\a\d\CA_\b^+\right\}\\
       &=\ {\rm tr}\left\{-[\CA^+_\a,\phi_{+-}]\d\CA_\b^+\right\}+{\rm tr}\left\{\phi_{+-}\left(
             \bV^+_\b\d\CA^+_\a-[\d\CA^+_\a\CA^+_\b]-[\CA^+_\a\d\CA^+_\b]\right)\right\}\\
       &=\ {\rm tr}\left\{\phi_{+-}\bV^+_\b\d\CA^+_\a\right\}+{\rm tr}\left\{-[\CA^+_\b,\phi_{+-}]
              \d\CA_\a^+\right\}\\
       &=\ \bV^+_\b J^{+-}_\a.
  \end{aligned}
$$
This allows us to associate with any symmetry transformation a conserved current
according to
\eqn{eq:current-ch5}{j_{\a\ad}\ :=\ \frac{1}{2\pi\i}\oint_\cC{\rm d}\l_+\,
               J^{+-}_\a\l^+_\ad,}
since
\eqn{}{\partial^{\a\ad}j_{\a\ad}\ =\ 0}
by virtue of Eq. \eqref{eq:cl-ch5}. In \eqref{eq:current-ch5},
the contour $\cC=\{\l_+\in\IC P^1\,|\,|\l_+|=1\}$ encircles $\l_+=0$.
It is important to stress that \eqref{eq:current-ch5} is a superfield.
Moreover, Eq. \eqref{eq:current-ch5} can slightly be rewritten to get
\eqna{eq:current2-ch5}{j_{\a\ad}\ &=\ \frac{1}{2\pi\i}\oint_\cC{\rm d}\l_+\,
                {\rm tr}\left\{\phi_{+-}\bnab^+_\a\phi_+\right\}\l^+_\ad,\\
                         &=\ \frac{1}{2\pi\i}\oint_\cC{\rm d}\l_+\,
                \l_\ad^+\bV_\a^+\left({\rm tr}\left\{\phi_{+-}\phi_+\right\}\right)
}
being more in spirit of Penrose's integral formulas. Recall also that due to
Eqs. \eqref{eq:c-1-ch5} and \eqref{phiexpKMY2}, certain combinations of
the coefficients functions $\phi_{\pm}^{0(m)}$ determine the explicit
transformations $\d\CA_{\a\ad}$ and $\d\CA^i_\ad$ and hence contribute to the
integrals \eqref{eq:current-ch5} and \eqref{eq:current2-ch5}, respectively.

Finally, we stress that one may also associate currents with symmetries
in the following way.
Let $\Upsilon_{+-}$ be any $\mathfrak{gl}(r,\IC)$-valued function
which 
is annihilated by $\bV_\a^\pm$, $\bV^i_\pm$ and $\partial_{\bl_\pm}$.\footnote{See
also our discussion given in \ref{par:NASCI-ch3}\kern6pt.}
Then it is not too difficult to see that $\psi_+\Upsilon_{+-}\psi_-^{-1}$
satisfies
\eqn{}{\bnab^\pm_\a(\psi_+\Upsilon_{+-}\psi_-^{-1})\ =\ 0
       \qquad{\rm and}\qquad\bnab^i_\pm(\psi_+\Upsilon_{+-}\psi_-^{-1})\ =\ 0,}
the covariant derivatives act in the adjoint representation.
Therefore, we may take 
\eqn{}{\tilde{j}_{\a\ad}\ :=\ \frac{1}{2\pi\i}\oint_\cC{\rm d}
           \l_+\,{\rm tr}\left\{\psi_+\Upsilon_{+-}\psi_-^{-1}\d\CA_\a^+\right\}\l_\ad^+,}
which is conserved by the above reasoning.

\clearemptydoublepage


\pagenumbering{roman}
\chapter*{Summary and discussion}
\addcontentsline{toc}{chapter}{Summary and discussion}

\lhead[\fancyplain{}{\bfseries \thepage}]{\fancyplain{}{\bfseries Summary and discussion}}
\rhead[\fancyplain{}{\bfseries Summary and discussion}]{\fancyplain{}{\bfseries\thepage}}
\cfoot{}

In this thesis, we have reported on various aspects of supersymmetric
gauge theories within the supertwistor approach. In particular, we first
gave a detailed twistor description of $\CN$-extended self-dual SYM theory.
We in addition discussed some related self-dual models which follow from self-dual 
SYM theory by suitable reductions. Their twistor formulation requires
certain weighted projective superspaces as twistor manifolds. As we have
shown, all these manifolds are formally Calabi-Yau thus being naturally
equipped with globally well-defined holomorphic volume forms. This property
enabled us to also present appropriate action functionals for these
models. Besides four-dimensional self-dual models, we also discussed 
a dimensional reduction to three dimensions. As a result, we obtained
certain supersymmetric Bogomolny models. In fact, we generalized
Hitchin's twistor construction of non-Abelian monopoles 
to a (maximally) supersymmetric setting. In addition,
appropriate action principles on mini-supertwistor space and Cauchy-Riemann
twistor space were given and shown to be equivalent to the action functional
reproducing the field equations of the Bogomolny model. In connection to
this, we were naturally led to the notion of Cauchy-Riemann supermanifolds
and to partially hCS theory. 
Moreover, we discussed certain complex structure deformations on 
mini-supertwistor space. As a result, we obtained a supersymmetric Bogomolny
model in three space-time dimensions with massive fermionic and scalar fields.
Similar to the massless case, we also derived appropriate action principles
on the twistor manifolds in question. We furthermore developed novel
solution generating techniques by studying infinitesimal deformations
of vector bundles on mini-supertwistor and Cauchy-Riemann twistor spaces,
respectively. The algorithms were then exemplified in the case where only fields with
helicity $\pm 1$ and the Higgs field were nontrivial. As we argued,
the corresponding Abelian configurations give rise to the Dirac
monopole-antimonopole systems. The fifth chapter of this thesis was
devoted to the studies of hidden symmetries of $\CN$-extended self-dual 
SYM theory. We first gave a detailed
cohomological interpretation of hidden symmetries of self-dual SYM
theory. We saw how general deformation algebras on the twistor side
are mapped to corresponding symmetry algebras in self-dual SYM theory.
Kac-Moody algebras as affine extensions of internal
symmetries were constructed. In addition, we discussed affine extensions
of space-time symmetries, that is, we obtained an affinization of
the superconformal algebra. The algebra in question turned out to be
of Kac-Moody-Virasoro-type. As was argued, the existence of such 
infinite-dimensional algebras of hidden nonlocal symmetries
originates from the fact that the full group of continuous transformations
acting on the space of holomorphic vector bundles over supertwistor space
is a semi-direct product of the group of local holomorphic automorphisms
of the supertwistor space and of the group of one-cochains with respect
to a certain covering with values in the sheaf of holomorphic
maps of supertwistor space into the gauge group. Besides symmetry algebras, we
constructed certain self-dual SYM
hierarchies. The basis of this construction was a generalization of
twistor space. As we saw, such a hierarchy consists
of an infinite system of partial differential equations, where the
self-dual SYM equations are embedded in. As was shown, the lowest level
flows of the hierarchies in question represent supertranslations. Indeed, 
the existence of such hierarchies allows us
to embed a given solution into an
infinite-parameter family of new solutions. Moreover, a detailed
derivation of the field equations together with the corresponding 
superfield expansions for the truncated hierarchies was presented. 

However, there are certainly a lot of open issues and questions which deserve 
further investigations:

\begin{itemize}
\item 
An obvious and challenging task is the generalization of the constructions
given in Chap. \ref{HS-CHAPTER} to the full SYM theory.
In Chap. \ref{ch:SYMT}, we saw how solutions to the field equations of  
$\CN=3$ SYM theory are related to certain holomorphic vector bundles over 
superambitwistor space $\CL^{5|6}$.
In principle, 
the algorithms relating infinitesimal 
deformations of vector bundles over supertwistor space to symmetries 
being developed for self-dual SYM theory in Chap. \ref{HS-CHAPTER} can
also be applied to full $\CN=3$ SYM theory. This is basically because
of the vanishing of appropriate cohomology groups. Furthermore, as a matter
of fact, Thm. \ref{thm:symtrafo-ch5} also applies to the full case which is
due to the structural similarity of the corresponding linear system. 
Hence, once given a suitable deformation algebra (of Lie-algebra type) on 
the twistor side, one automatically has a hidden symmetry algebra in full
$\CN=3$ SYM theory.\footnote{After having finished this thesis, an alternative
approach of discussing hidden symmetries of $\CN=4$ SYM theory in terms of
integrable hierarchies has been proposed in \cite{Popov:2006qu}.}
 
\item 
Besides questions associated with symmetry transformations, etc., one
in addition needs to write down the related conserved
nonlocal currents and charges -- not only as superfields but also in 
components. In Chap. \ref{HS-CHAPTER}, we considered particular 
nonlocal conservation laws associated with any symmetry of
the field equations. It is desirable, to generalize such constructions
to the full $\CN=3$ theory. Presumably, the construction of fermionic
conserved currents\footnote{See, e.g., \cite{Abdalla:1986xq} for
such constructions in ten-dimensional SYM theory.}, that is, currents obeying 
equations of the form
$D^{i\ad}j_{i\ad}=0=D_i^\a j^i_{\a}$ will simplify the discussion, since
with the help of those one may derive conservation laws of the
form $\partial^{\a\ad}j_{\a\ad}=0$ by way of
$$ j_{\a\ad}\ :=\ \int\dt^6\te\dt^6\eta\, \te^i_\a j_{i\ad} +\ {\rm h.c.} $$
However, to proceed further 
(also in view of passing to the quantum theory)
and to clarify the physical significance of such currents, one 
clearly needs to find their superfield expansions. In addition, one
should then compute the classical
Poisson brackets among the corresponding charges. After passing to the
quantum regime, a challenging task will be to make
contact with the quantum symmetry algebras considered
in \cite{Dolan:2003uh,Dolan:2004ps}.
Moreover, it would also be interesting to see 
how such quantum symmetry algebras fit
into the context of twistor string theory \cite{Witten:2003nn} and
how they can be understood within the recently proposed twistor 
approach to $\CN=4$ SYM theory \cite{Boels:2006ir}.

\item
Another issue also worthwhile to explore is the construction
of hidden symmetry algebras and hierarchies of gravity theories, in 
particular of self-dual supergravity (see Ref. \cite{Siegel:1992wd} 
for the case of maximal $\CN=8$ supersymmetry).\footnote{In 
\cite{Merkulov:1990hs}--\cite{Merkulov:1992gw}, an extension of 
Penrose's nonlinear graviton construction \cite{Penrose:1976js} 
to a supersymmetric setting has been discussed.} By applying similar 
techniques as those presented in Chap. \ref{HS-CHAPTER}, one will 
eventually obtain hidden infinite-dimensional symmetry algebras of
the self-dual supergravity equations, generalizing the 
results known for self-dual gravity. See Refs. \cite{BoyerAJ,
TakasakiCG,ParkVI,PopovUU,JunemannHI,LechtenfeldIK,DunajskiIQ,Dunajski:2003gp},
for instance. 

\item 
Recall that it was conjectured by Ward \cite{WardGZ,Ward86,Ward90} 
that all integrable models in less than four space-time dimensions
can be obtained from self-dual YM theory in four dimensions.
Examples are the nonlinear Schr\"odinger equation, the
Korteweg-de Vries equation, the sine-Gordon model, etc.
In particular, they all follow 
from the self-dual YM equations upon implementing suitable 
algebraic ans\"atze for the gauge potential followed by a dimensional 
reduction. In a similar spirit, the Ward conjecture can be ``supersymmetrically'' 
extended in order to derive the supersymmetric versions of the above-mentioned 
models. Therefore, it would be of interest to take the $\CN$-extended 
self-dual SYM hierarchy presented in Chap. \ref{HS-CHAPTER} and to derive the 
corresponding super hierarchies of these integrable systems in less than
four dimensions.

\item
Finally, we address additional points for further investigations being
related to noncommutative field theories. In \cite{Saemann:2004cf}, we
(in collaboration with Christian S\"amann)
considered $\CN=4$ SYM theory on a nonanticommutative superspace,
that is, instead of $\IC^{m|n}=(\IC^m,\CO_{\rm red}(\Lambda^\bullet\IC^n))$
we took $\IC^{m|n}_\hbar=(\IC^m,\CO_{\rm red}({\rm Cliff}(\IC^n)))$, where
${\rm Cliff}(\IC^n)$ denotes the Clifford algebra of $\IC^n$
$$ \{\eta_i,\eta_j\}\ =\ \hbar C_{ij}\qquad{\rm for}\qquad i,j=1,\ldots,n. $$
Upon introducing an involution corresponding to Euclidean signature,
we derived the superfield expansions and the field equations of deformed
$\CN=4$ SYM theory to first order in the deformation parameter $\hbar$. 
In showing this, we proposed an extension of the Seiberg-Witten map \cite{Seiberg:1999vs}
to superspace. Our derivation was based on the $\CN=4$ formulation of the
constraint system.
Clearly, one may straightforwardly translate our results into the
$\CN=3$ formulation of $\CN=4$ SYM theory. It would then be of interest to
establish a nonanticommutative version of the superambitwistor 
correspondence. In particular, Eqs. \eqref{eq:IR-ch4}
induce a nonanticommutative structure on superambitwistor space.
In addition,
generalizing the results of \cite{Mason:2005kn}, it should in
principle be possible to derive a corresponding action functional
of deformed $\CN=4$ SYM theory which is still lacking. Moreover, splitting and dressing 
methods obtained from twistor theory have successfully
been applied to the construction of solitons and instantons in
noncommutative field theories. See, e.g., Refs.
\cite{Lechtenfeld:2001aw,Lechtenfeld:2001gf,
Lechtenfeld:2001ie,Wolf:2002jw,Horvath:2002bj,Ihl:2002kz,
Lechtenfeld:2003vv,Lechtenfeld:2004qh,Domrin:2004pg,Wimmer:2005bz}. 
It would also be interesting to see how the solution
generating techniques as considered in, e.g., Chap. \ref{ch:SBME-ch3}
need to be generalized to noncommutative field theories. Partial
results on that matter have already been given in \cite{Lechtenfeld:2005xi}.
In addition,
the methods used in this thesis might also shed light 
on the question of
 hidden symmetries in noncommutative/nonanticommutative field
theories.

\end{itemize}

\clearemptydoublepage


\chapter*{List of symbols}
\addcontentsline{toc}{chapter}{List of symbols}

\lhead[\fancyplain{}{\bfseries \thepage}]{\fancyplain{}{\bfseries List of symbols}}
\rhead[\fancyplain{}{\bfseries List of symbols}]{\fancyplain{}{\bfseries\thepage}}
\cfoot{}

\begin{flushleft}
\renewcommand{\arraystretch}{.9}
 \begin{tabular}{ll}
  $\IR$ ($\IC$) & real (complex) numbers\\
  $\IZ_n$ &$\IZ/n\IZ$, where $\IZ$ is the integers\\
  $R$ & supercommutative ring\\
  $M,\ N$ & $R$-modules\\
  $X,\ Y,\ Z,\ \ldots$ & (super)manifolds\\
  $TX$ & tangent bundle (respectively, sheaf) of a (super)manifold $X$\\
  $N_{Y|X}$, $N_Y$  & normal bundle (respectively, sheaf) of a (super)manifold $Y$\\
                    &  in a (super)manifold $X$\\
  $\CA,\ \CB,\ \CC,\ \ldots$ & sheaves\\
  $\CE,\ \CF,\ \CG,\ \ldots$ & vector bundles\\
  $\CI$ & ideal sheaf\\
  $\CN$ & sheaf of nilpotents\\
  $\CU,\ \CV,\ \CW,\ \ldots$ & open sets\\
  $\fU,\ \fV,\ \fW,\ \ldots$ & (open) coverings\\
  $\cDb$ & Cauchy-Riemann structure\\
  $\cT$ & $\cT$-structure, distribution\\
  $\CO_X$ & sheaf of holomorphic functions\\
  $\CS_X$ & sheaf of smooth functions\\
  $\CC_X$ & sheaf of $\cT$-functions\\
  $\CO_{\rm red}=\CO_X/\CN$ & reduced sheaves \\
  $\CS_{\rm red}=\CS_X/\CN$ & \\
  $\CC_{\rm red}=\CC_X/\CN$ & \\
  $(Y,\CO^{(k)}_Y=\CO_X/\CI^k)$ & $k$-th formal neighborhood of $Y$ in $X$\\
   $\CH_X$ & group of local biholomorphisms on a (super)manifold $X$\\
   $\Pi$ & parity map\\
\end{tabular}
\end{flushleft}

\newpage
\begin{flushleft}
\renewcommand{\arraystretch}{.9}
\begin{tabular}{ll}
   $\CO_{\IC P^{m|n}}(-1)$ & tautological sheaf on $\IC P^{m|n}$, 
                            $\CO_{\IC P^{m|n}}(1)=\CO_{\IC P^{m|n}}(-1)^*$,\\
                          &   $\CO_{\IC P^{m|n}}(d)=\CO_{\IC P^{m|n}}(1)^{\otimes d}$\\
  $C^q(\fU,\CA)$ & set of $q$-cochains with values in $\CA$\\
  $Z^q(\fU,\CA)$ & set of $q$-cocycle with values in $\CA$\\
  $H^q(\fU,\CA)$, $H^q(X,\CA)$ & $q$-th \v Cech cohomology set with values in $\CA$\\
  End$\,\CE$ & endomorphism bundle\\
  sdet$\,\CE$ & superdeterminant line bundle\\
  $c_k(\CE)$ ($ch_k(\CE)$) & $k$-th Chern class (character)\\
  $\O^k(X)$ & sheaf of differential $k$-forms\\
  $\O^k_0(X)$ & sheaf of differential $k$-forms with compact support\\
  $\O^k_\cT(X)$ & sheaf of $\cT$-differential $k$-forms\\
  $\O^{p,q}(X)$ & sheaf of differential $(p,q)$-forms\\
  $\O^{p,q}_{\rm CR}(X)$ & sheaf of CR differential $(p,q)$-forms\\
  Ber$(X)$ & sheaf of volume forms\\
  Ber$_0(X)$ & sheaf of volume forms with compact support\\
  $\Sigma^k(X)$ & sheaf of integral $k$-forms\\
   $\Sigma^{p,q}(X)$ & sheaf of integral $(p,q)$-forms\\ 
  $\Lambda^\bullet$ & exterior algebra\\
  Hom$(M,N)$ & group of morphisms between $R$-modules $M$ and $N$\\
  $\cH om(\CA,\CB)$ & sheaf of local morphisms between sheaves $\CA$ and $\CB$\\
  Mat$(m|n,p|q,R)$ & morphisms between free $R$-modules of rank $m|n$ and $p|q$\\
  $GL(m|n,R)$ & Lie group of invertible even automorphisms of $R^{m|n}$\\
  $\mathfrak{gl}(m|n,R)$ & Lie algebra of $GL(m|n,R)$\\
  $\fH=GL(m|n,\CO_X)$ & sheaf of $GL$-valued holomorphic functions\\
  Lie$\,\fH=\mathfrak{gl}(m|n,\CO_X)$ & sheaf of $\mathfrak{gl}$-valued holomorphic functions\\
  $\fS=GL(m|n,\CS_X)$ & sheaf of $GL$-valued smooth functions\\
  Lie$\,\fS=\mathfrak{gl}(m|n,\CS_X)$ & sheaf of $\mathfrak{gl}$-valued smooth functions\\
  $\fC=GL(m|n,\CC_X)$ & sheaf of $GL$-valued $\cT$-functions\\
  Lie$\,\fC=\mathfrak{gl}(m|n,\CC_X)$ & sheaf of $\mathfrak{gl}$-valued $\cT$-functions\\
\end{tabular}
\end{flushleft}

\clearemptydoublepage


\lhead[\fancyplain{}{\bfseries\thepage}]{\fancyplain{}{\bfseries\rightmark}}
\rhead[\fancyplain{}{\bfseries\leftmark}]{\fancyplain{}{\bfseries\thepage}}
\cfoot{}


\clearemptydoublepage

\end{document}